\documentclass[12pt]{article}

\usepackage[dvips]{color}
\usepackage{amsmath}
\usepackage{graphicx} 
\usepackage[dvips]{color}
\usepackage{graphicx}
\usepackage{tabularx}
\usepackage{subfig}
\usepackage[outercaption]{sidecap}

\newcommand{\toolboxNameAcronym}{BGEA}
\newcommand{\toolboxLink}{{\ttfamily{www.brainarchitecture.org/allen-atlas-brain-toolbox}}}

\newcommand{\widthParamFig}{0.98}
\newcommand{\widthParamTable}{0.82}
\newcommand{\numTypesPerTable}{6 }
\newcommand{\widthForBig}{1.6in}
\newcommand{\bigTwelve}{{\ttfamily{'big12'}}}
\newcommand{\bigTwelveSpace}{{\ttfamily{'big12'}}$\;$}
\newcommand{\fine}{{\ttfamily{'fine'}}}
\newcommand{\fineSpace}{{\ttfamily{'fine'$\;$}}}
\newcommand{\widthCoeff}{0.77}
\newcommand{\rhoNew}{\rho^{\mathrm{new}}}
\newcommand{\Ctrue}{C_{\mathrm{true}}}
\newcommand{\widthParamForTable}{0.88}
\newcommand{\numTypesPerReTable}{3$\;$}
\newcommand{\CPyr}{C^{\mathrm{pyr}}}
\newcommand{\CNew}{C^{\mathrm{new}}}
\newcommand{\GammaNew}{\Gamma^{\mathrm{new}}}
\newcommand{\GammaOld}{\Gamma^{\mathrm{old}}}
\newcommand{\rhoPyr}{\rho^{\mathrm{pyr}}}
\newcommand{\COffset}{C_{\mathrm{offset}}} 
\newcommand{\rhoOffset}{\rho_{\mathrm{offset}}}

\begin{document}
\title{Cell-type-specific microarray data and the Allen atlas:
quantitative analysis of brain-wide patterns of correlation and density}
\author{Pascal Grange$^{1,\ast}$, Michael Hawrylycz$ ^{2}$, Partha P. Mitra$ ^{1}$\\
{\normalsize{$ ^{1}$ Cold Spring Harbor Laboratory,
 Cold Spring Harbor, New York 11724, United States}}\\
{\normalsize{$ ^{2}$ Allen Institute for Brain Science,
Seattle, Washington 98103, United States}}\\
\normalsize{$^\ast$E-mail: {\ttfamily{pascal.grange@polytechnique.org}}}}

\date{}
\maketitle

\begin{abstract}
The Allen Atlas of the adult mouse brain is used to estimate the
region-specificity of 64 cell types whose transcriptional profile in
the mouse brain has been measured in microarray experiments. We
systematically analyze the preliminary results presented in
[arXiv:1111.6217], using the techniques implemented in the Brain Gene
Expression Analysis toolbox. In particular, for each
cell-type-specific sample in the study, we compute a brain-wide
correlation profile to the Allen Atlas, and estimate a brain-wide
density profile by solving a quadratic optimization problem at each
voxel in the mouse brain. We characterize the neuroanatomical
properties of the correlation and density profiles by ranking the
regions of the left hemisphere delineated in the Allen Reference
Atlas. We compare these rankings to prior biological knowledge of the
brain region from which the cell-type-specific sample was extracted.
\end{abstract}

\clearpage

\tableofcontents

\section{Introduction and background}

Neuroanatomy is experiencing a renaissance thanks to molecular biology
and computational imaging \cite{GeneToBrain}. The Allen Brain Atlas
(ABA), the first Web-based, genome-wide atlas of gene expression in
the adult mouse brain (eight-week old C57BL/6J male mouse brain), was
obtained using an unified automated experimental pipeline
\cite{AllenGenome,images,BrainAtlasInsights,
  neufoAllen,digitalAtlasing,AllenFiveYears,corrStructureAllen}. The resulting data
set consists of {\emph{in situ}} hybridization (ISH) digitized image
series for thousands of genes.  These image series are co-registered
to the Allen Reference Atlas (ARA) \cite{AllenAtlas}, which allows to
compare ISH data to classical neuroanatomy.

 The gene
 expression data were aggregated into a volumetric grid:
The reference mouse brain is partitioned into $V = 49,742$ cubic voxels of
side 200 microns. For a voxel $v$, the {\it{expression energy}} of
the gene labeled $g$ is defined \cite{AllenAtlasMol} as a weighted sum of the greyscale-value
intensities of pixels $p$ intersecting the voxel:
\begin{equation}
E(v,g) := \frac{\sum_{p\in v} M( p ) I(p)}{\sum_{p\in v} 1},
\label{voxelByGeneData}
\end{equation}
 where $I( p )$ is the intensity image and $M( p )$ is a Boolean mask,
 worked out in the image-processing pipeline,
 with value 1 if the pixel is expressing the gene of interest.
 The ABA has led to the development of software for data exploration
 and analysis such as the Web-based Neuroblast \cite{NeuroBlast} and
 Anatomic Gene Expression Atlas (AGEA) \cite{AllenAtlasMol}. NeuroBlast
 allows users to explore the correlation structure between genes, while the AGEA
 is based on spatial correlation between voxels.\\

  More recently, we developed the Brain Gene Expression Analysis (BGEA)
 MATLAB toolbox \cite{toolboxManual,BGEA}, which
 allows to manipulate the gene-expression energies of the 
 brain-wide ABA on the desktop as matrices \cite{qbCoExpression,markerGenes}. In particular,
 data corresponding to different ISH experiments (i.e. different genes)
 can be combined for computational analysis, and used to study 
 other data sets.\\
 
  A complementary (cell-based) approach to the study of gene-expression 
 energy in the brain uses microarray
 experiments to study co-expression patterns in a small set of brain
 cells of the same type. We studied cell-type-specific microarray gathered from
  different studies
  \cite{OkatyCells,RossnerCells,CahoyCells,DoyleCells,ChungCells,ArlottaCells,HeimanCells,foreBrainTaxonomy}, 
   already analyzed in \cite{OkatyComparison}, for $T=64$ cell
  types\footnote{The studies differ in the way cells are
    visually-identified are separated from their environments. The
    methods are laser-capture microdissection (LCM)
    \cite{ChungCells,RossnerCells}, translating ribosome affinity
    purification (TRAP) \cite{DoyleCells,HeimanCells},
    fluorescence-activated cell-sorting (FACS) \cite{ArlottaCells},
    immuno panning (PAN) \cite{CahoyCells}, and manual sorting
    \cite{OkatyCells,foreBrainTaxonomy}. See Tables
    \ref{metadataTable1} and \ref{metadataTable2} for references to
    the study from which each of the cell-type-specific samples was drawn.}. 
   Each of these cell types is characterized by
  the expression of $G_T = 14,580$ genes.\\

 The region specificity of the transcriptional profiles of
cell types is an open problem (preliminary results
 were presented in \cite{preprintFirst}, without a 
 systematic anatomical analysis). Given a cell type extracted from a
given brain region, it is hard to know where else in the brain cells
with a similar transcriptional profile can be found.  We computed
brain-wide correlation profiles between each of the cell-type-specific
samples and the Allen Atlas. Moreover, we propose a linear model
decomposing the gene-expression energy of the collection of all genes
in the Allen Atlas over the set of cell-type-specific data. The model
provides an estimate of the brain-wide density of the cell types.
 For each cell type, brain regions are then ranked according to 
the average value of correlation, and to their contribution to the total density of the 
cell type.

\section{Methods}

We determined the set of genes that are represented in
\textit{both} data sets (there are $G=2,131$ such genes).  We extracted
the columns of the matrices $E$ and $C$ corresponding to these genes,
and rearranged them to reflect the same order in each data set. $E$
is assumed to be a voxel-by-gene matrix denoted by $E$ (with $V$ rows and $G$ columns),
 and type-by-gene matrix denoted by $C$ (with $T$ rows and $G$ columns).
 The columns of both matrices corresponding to the same set of
genes, ordered in the same way:
\begin{equation}
E(v,g) = {\mathrm{expression\;of\;gene\;labeled\;}}g\;{\mathrm{in\;voxel\;labeled\;}}v,
\label{voxelByGene}
\end{equation}
\begin{equation}
C(t,g) = {\mathrm{expression\;of\;gene\;labeled\;}}g\;{\mathrm{in\;cell\;type\;labeled\;}}t.
\label{typeByGene}
\end{equation}
The computational analysis of these two data sets 
 was undertaken using the Brain Gene Expression Atlas MATLAB toolbox (\toolboxNameAcronym)
  we developed in MATLAB \cite{BGEA}.\\

The present analysis is focused on 4,104 genes for which experiments
conducted using both sagittal and coronal sections were available at
the time of analysis. To minimize reproducibility issues, we computed
the correlation coefficients between sagittal and coronal data volumes
(see \cite{qbCoExpression} for a plot of values) and selected the
genes in the top-three quartiles of correlation (3,041 genes) for
further analysis.\\
 
\subsection{Correlations between cell-type-specific microarray data and the Allen Atlas}
%cellTypesCorrels = cell_types_correls( Ref, D, M, colsToUseInAllen, colsToUseInTypes )
For a fixed cell type labeled $t$, the $t$-th
row of the type-by-gene matrix $C$ defined in equation \ref{typeByGene}
corresponds to a vector in a $G$-dimensional gene space.
  On the other hand, for a fixed voxel labeled $v$ in the mouse brain,
 the $v$-th row of the the voxel-by-gene matrix  $E$ defined in equation \ref{voxelByGene}
  gives rise to another vector in the same gene space.
  We computed the correlation $\mathrm{Corr}(v,t)$ between voxel labeled $v$
 and cell type labeled $t$:
\begin{equation}
\mathrm{Corr}(v,t) = \frac{\sum_{g=1}^G( C(t,g) - \bar{C}(g))( E( v,g ) - \bar{E}(g) )}{\sqrt{\sum_{g=1}^G( C(t,g) - \bar{C}(g))^2}\sqrt{\sum_{g=1}^G ( E( v,g ) - \bar{E}(g) )^2}},
\label{corrDefinition}
\end{equation}
\begin{equation}
\bar{C}(g) = \frac{1}{T}\sum_{t=1}^T C(t,g),
\label{CBarDefinition}
\end{equation}
\begin{equation}
\bar{E}(g) = \frac{1}{V}\sum_{v=1}^V E(v,g).
\label{EBarDefinition}
\end{equation}
For a cell type labeled $t$, the correlation values defined
across all voxels in the ABA, as defined in Equations
\ref{corrDefinition}, \ref{CBarDefinition} and \ref{EBarDefinition}
give rise to a brain-wide \emph{correlation profile} between this cell
type and the Allen Atlas, with all genes taken into account,
 since the genes are summed over in Equation \ref{corrDefinition}. 
The correlation at each voxel reflects a
measure of similarity between the expression profile of the cell type
$t$ and the gene expression profile defined
within a local cube of side 200 microns in the ABA.  See Tables
\ref{tableCorrels1}--\ref{tableCorrels11} for maximum-intensity
projection images and images of individual sections of these
volumetric profiles for each of the $T=64$ cell types in this
study. The next section contains an analysis of the 
 brain-wide correlation profiles, based on the average 
 correlation across brain regions defined by the ARA.

%The next section contains an analysis of the results based on the
%average correlation between cell types and the voxels in the brain
%regions defined within the left hemisphere of the mouse brain.

\subsection{A linear model relating cell-type-specific microarray data to the Allen Atlas}
%fitVoxelsToTypesBis = fit_voxels_to_types_bis( Ref, D, M, colsToUseInAllen, colsToUseInTypes )
%load( 'fitVoxelsToTypesBis0701.mat' );
%cellTypesTable = cell_types_table( Ref, fitVoxelsToTypes );
The above-defined analysis of correlations between the ABA and cell-type-specific
 data is not a decomposition of the signal into  
 cell-type-specific components. In this section, we propose a linear model 
 to attempt such a decomposition, using the cell-type-specific samples
 as a base of gene space.\\
 Let us denote by $C_t$ the vector in gene space
 obtained by taking the $t$-th
row of the type-by-gene matrix $C$ defined in equation \ref{typeByGene}:
\begin{equation}
C_t(g) = C(t,g), \;\;\;1\leq g \leq G.
\label{typeVector}
\end{equation} 
To decompose the gene expression at a voxel of the mouse brain
 into its cell-type-specific components, let us introduce the positive quantity $\rho_t( v )$
 denoting the contribution of cell-type $t$ at voxel $v$, 
and propose the following linear model:
\begin{equation}
E(v,g) = \sum_{ t = 1 }^T \rho_t( v )C_t(g) + {\mathrm{Residual}}(v,g).
\label{linearModel}
\end{equation}
Both sides are estimators of the amount of mRNA for gene $g$ at voxel
$v$.  The residual term in Equation \ref{linearModel} reflects the
fact that $T=64$ cell types are not sufficient to sample the whole
diversity of cell types in the mouse brain, as well as noise in the
measurements, reproducibility issues, and non-linearities in the
relations between numbers of mRNAs, expression energies and microarray
data \cite{ISHVsMicroarray}.\\

To find the parameters $\rho$ that provide the best fit of the model
\ref{linearModel}, we have to minimize the residual term by solving
the following problem:
\begin{equation}
\left(\rho_t( v )\right)_{1\leq t \leq T, 1\leq v \leq V} = {\mathrm{argmin}}_{\phi\in {\mathbf{R}}_+(T,V)}\mathcal{ E}_{E,C}(\phi ),
\label{fittingProblem}
\end{equation}
where 
\begin{equation}
\mathcal{E}_{E,C}(\phi) = \sum_{v=1}^V\left( \sum_{g=1}^G \left( E(v,g) - \sum_{t=1}^T\phi(t,v)C_t(g)\right)^2 \right).
\label{errorTerm}
\end{equation}
The right-hand side of Equation \ref{errorTerm} is a sum
 of positive quadratic functions of $\phi(.,v)$, one per voxel $v$.
 The minimization
problem \ref{fittingProblem} can therefore be solved voxel by voxel.
 For each voxel $v$ we have to
minimize a quadratic function of a vector with $T$ positive
components:
\begin{equation}
\forall v \in [1..V],\;\;(\rho_t( v ))_{1\leq t \leq T}= {\mathrm{argmin}}_{\nu\in {\mathbf{R}}_+^T}\sum_{g=1}^G \left( E(v,g) - \sum_{t=1}^T\nu(t) C_t(g)\right)^2.
\label{voxelByVoxel}
\end{equation}
We solved these quadratic optimization problems (one per voxel),
using the CVX toolbox for convex optimization \cite{cvxLink,cvxLecture}.\\

For each cell type $t$, the coefficients $(\rho_t( v ))_{1\leq v \leq
  V}$ yield an estimated brain-wide density profile for this cell
type. See Tables \ref{tableFittings1}--\ref{tableFittings11} for
maximum-intensity projection images and individual image sections of
these density profiles for all the cell types in this study. The next
section contains an analysis of results related to the anatomical
brain regions defined by the ARA.\\

\section{Results}
\subsection{Rankings of brain regions induced by correlation and density profiles}

 To analyze the neuroanatomical properties of the results, we made
  use of two non-hierarchical systems of annotation, available for
 the left hemisphere at a resolution of 200 microns.  We will refer to them as:
\begin{itemize}

\item the \bigTwelveSpace annotation, consisting of 12 regions of the left
  hemisphere (together with a more patchy group of voxels termed
  'basic cell groups of regions') whose names, sizes and shapes are
  shown in Table \ref{referenceTableBig12}.\\

\item the \fine annotation, a refinement of \bigTwelveSpace into 94 regions,
  reflecting structures further down the ARA hierarchy.\\
\end{itemize}

%code: region_profiles_plot.m
%regionProfiles = region_profiles_plot( Ref );
\begin{table}
\centering

\begin{tabular}{|m{0.3\textwidth}|m{0.09\textwidth}|m{0.15\textwidth}|m{0.4\textwidth}|}
\hline
\textbf{Brain region}&\textbf{Symbol}&\textbf{{\small{Percentage of hemisphere}}}&\textbf{{\small{Profile of region (maximal-intensity projection)}}}\\\hline
{\small{Basic cell groups and regions}}& Brain &4.6&\includegraphics[width=\widthForBig,keepaspectratio]{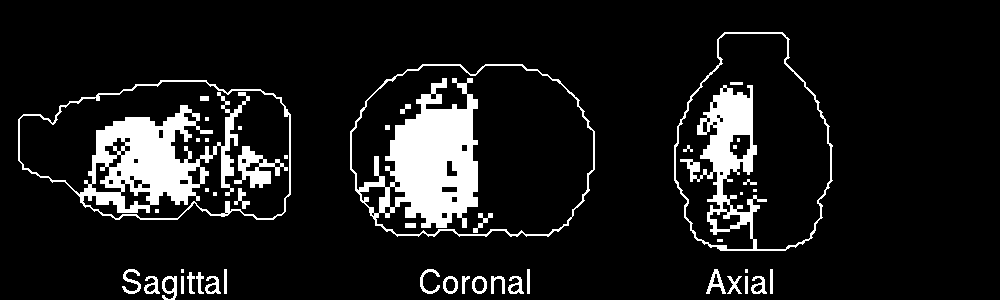}\\
Cerebral cortex& CTX &29.5&\includegraphics[width=\widthForBig,keepaspectratio]{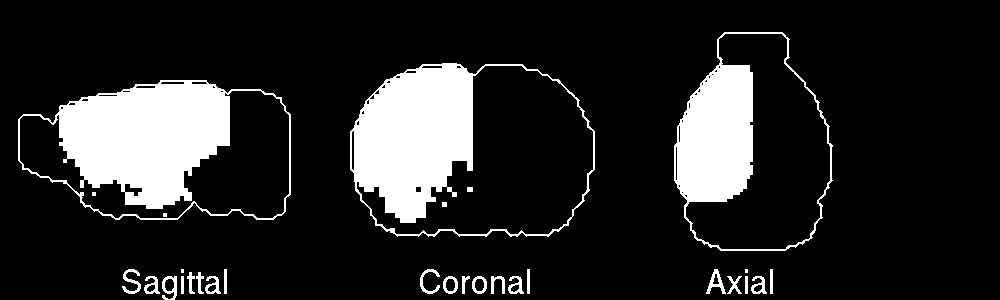}\\
Olfactory areas& OLF &9.2&\includegraphics[width=\widthForBig,keepaspectratio]{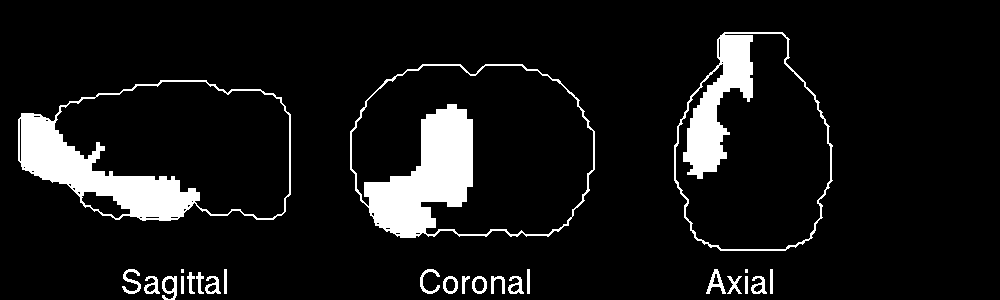}\\
Hippocampal region& HIP &4.3&\includegraphics[width=\widthForBig,keepaspectratio]{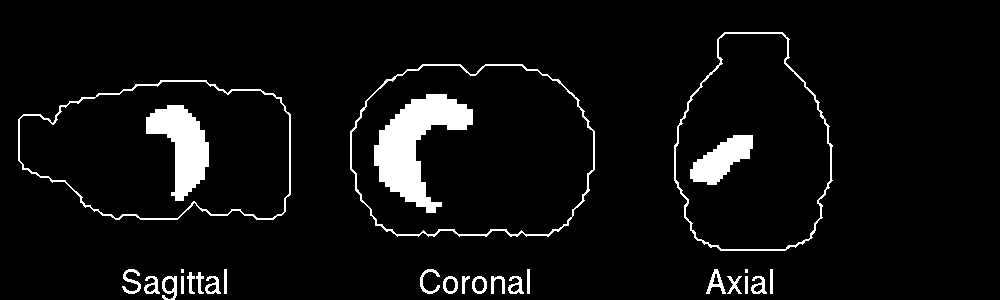}\\
{\small{Retrohippocampal region}}& RHP &4&\includegraphics[width=\widthForBig,keepaspectratio]{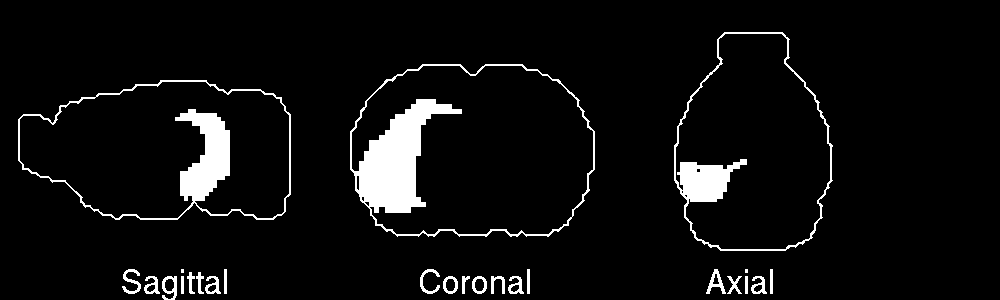}\\
Striatum& STR &8.6&\includegraphics[width=\widthForBig,keepaspectratio]{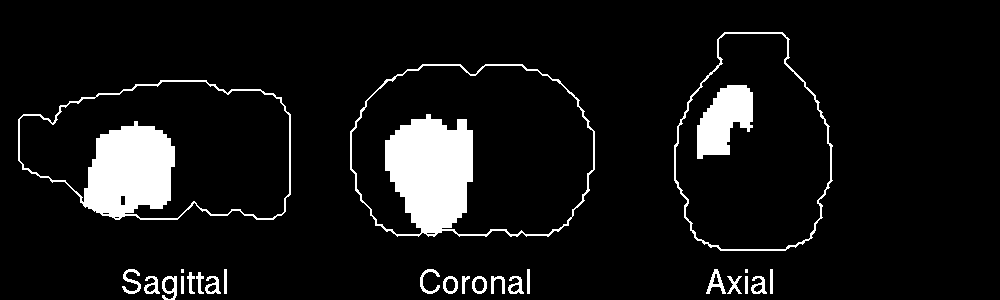}\\
Pallidum& PAL &1.9&\includegraphics[width=\widthForBig,keepaspectratio]{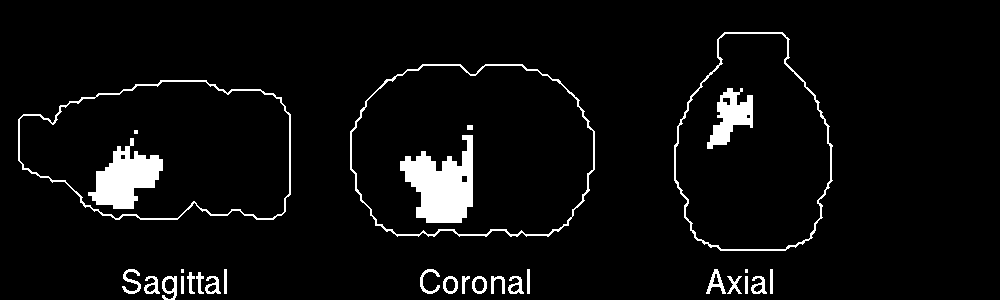}\\
Thalamus& TH &4.3&\includegraphics[width=\widthForBig,keepaspectratio]{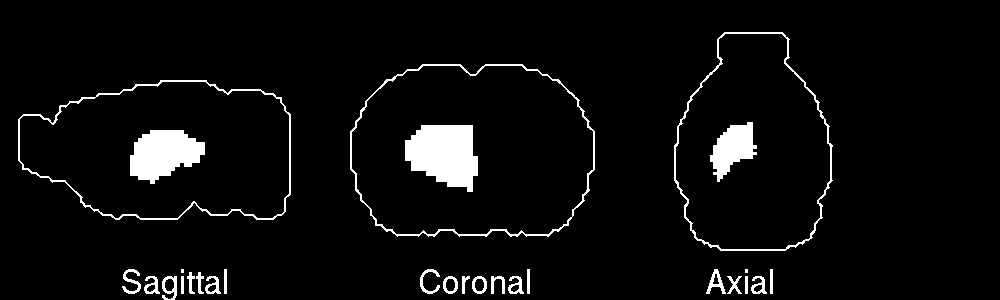}\\
Hypothalamus& HY & 3.5&\includegraphics[width=\widthForBig,keepaspectratio]{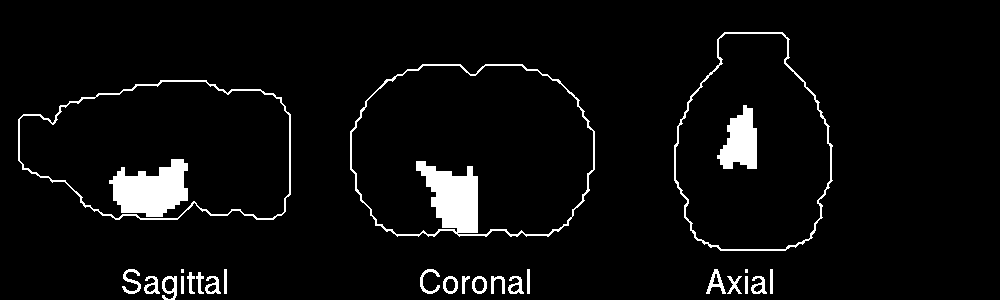}\\
Midbrain& MB &7.8&\includegraphics[width=\widthForBig,keepaspectratio]{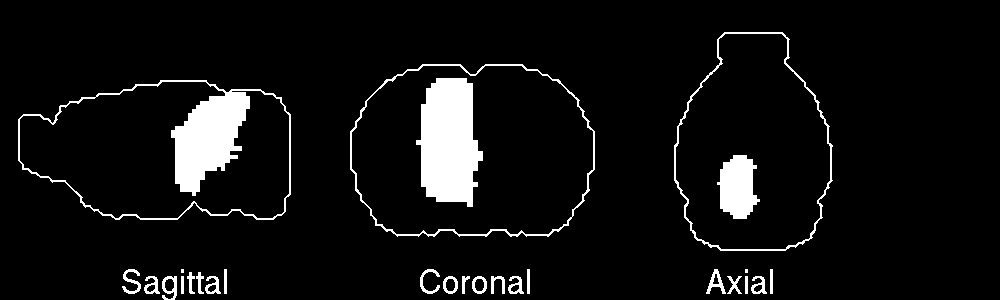}\\
Pons& P &4.6&\includegraphics[width=\widthForBig,keepaspectratio]{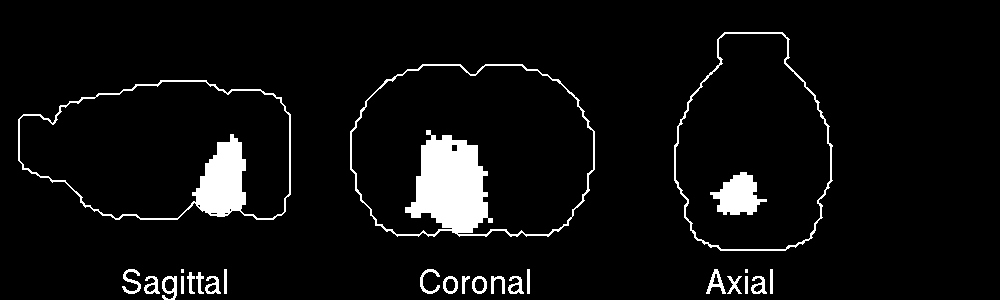}\\
Medulla& MY &6.2&\includegraphics[width=\widthForBig,keepaspectratio]{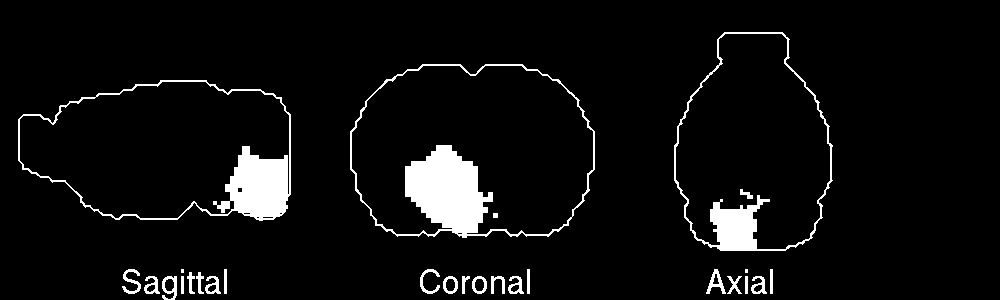}\\
Cerebellum& CB &11.5&\includegraphics[width=\widthForBig,keepaspectratio]{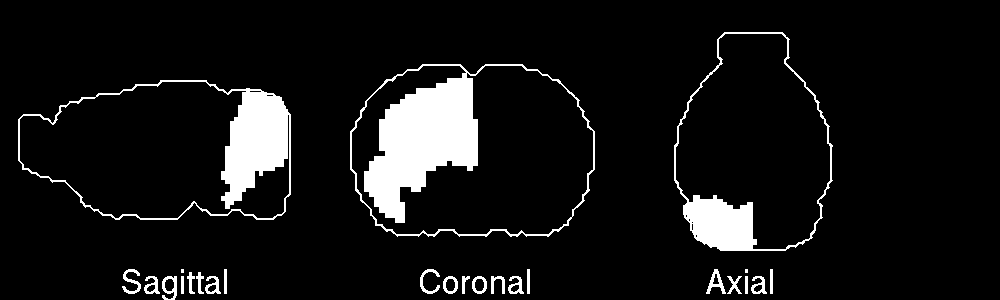}\\\hline
\end{tabular}

\caption{Brain regions in the coarsest annotation of the left hemisphere
in the ARA (referred to as the \bigTwelveSpace annotation).}
\label{referenceTableBig12}
\end{table}

For each cell-type-specific sample in this study, anatomical
metadata specify the brain region from which the sample was extracted
(see Tables \ref{metadataAnatomyTable1} and
\ref{metadataAnatomyTable2}). For
each brain-wide quantity computed (i.e., correlation and density
profiles, defined in Equations \ref{corrDefinition} and \ref{linearModel}), we
can determine how these quantities vary across brain
regions. Specifically, these profiles can be used to produce
\textit{rankings} of brain regions in the Allen Reference Atlas.\\

We have to choose a non-hierarchical partition of the brain according to
the ARA, with $R$ regions, available in a digitized form,
co-registered with the voxel-based gene-expression energies (for
definiteness and ease of presentation we consider the \bigTwelveSpace
annotation of the ARA, Table \ref{referenceTableBig12}).  Let $V_r$
denote the set of voxels belonging to region labeled $r$ in this
partition.\\

\subsubsection{Ranking brain regions by correlation profile}
\begin{figure}
\includegraphics[width=\widthParamFig\textwidth]{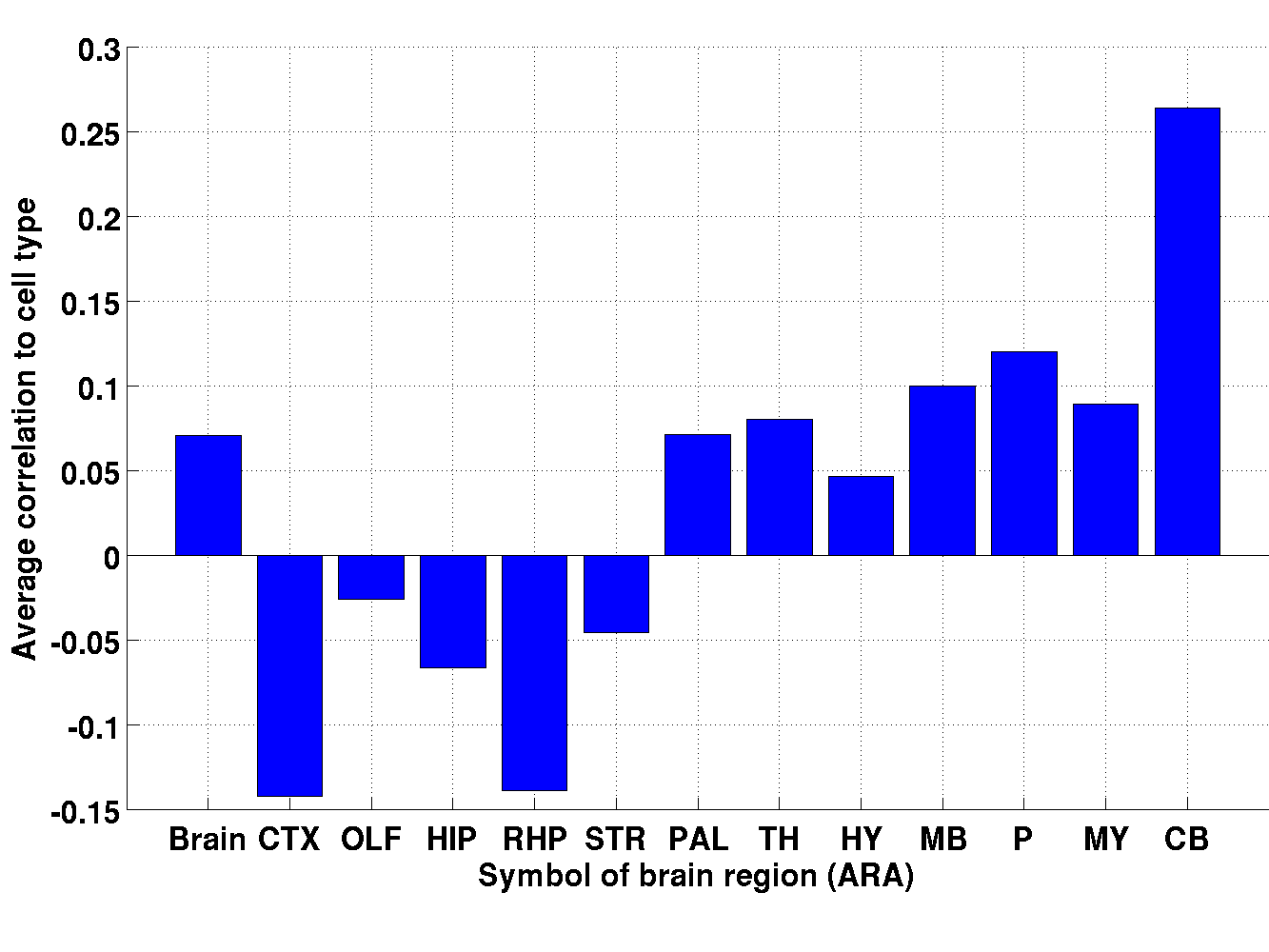}
\caption{Bar diagram of the average correlations between granule cells
  and the voxels in the regions of the \bigTwelveSpace annotation
  of the ARA, as defined in Equation \ref{correlRegion} (granule cells
  correspond to the value $t=20$ for the cell-type index). The symbols
  of the regions in the \bigTwelveSpace annotation read as in Table
  \ref{referenceTableBig12}: Basic cell groups and regions = Brain,
Cerebral cortex = CTX, Olfactory areas = OLF, Hippocampal region =
HIP, Retrohippocampal region = RHP, Striatum = STR, Pallidum = PAL,
Thalamus = TH, Hypothalamus = HY, Midbrain = MB, Pons = P, Medulla =
MY, Cerebellum = CB.}
\label{correlBarIndex20}
\end{figure}

 For a cell type labeled $t$, we can compute the
average correlation with the Allen Atlas in each region (labeled $r$)
 of the ARA:
\begin{equation}
{\overline{\mathrm{Corr}}}( r, t ) = \frac{1}{|V_r|}\sum_{v \in V_r}{\mathrm{Corr}}(v,t).
\label{correlRegion}
\end{equation}
For each cell type, the brain regions in the ARA can be ranked by
decreasing values of the average correlation
${\overline{\mathrm{Corr}}}( ., t )$ defined in Equation
\ref{correlRegion}.  The region with highest average correlation is
called the {\emph{top region by correlation}} for the cell type
labeled $t$.  A bar diagram of the average correlations between
granule cells (index $t$=20, illustrated in Figure 2(a) of the main
text), and the regions of the \bigTwelveSpace annotation, is
shown on Figure \ref{correlBarIndex20} (the symbols of the brain
regions can be found in Table \ref{referenceTableBig12}).  The top
region by correlation for granule cells is the cerebellum.\\

In the figures derived from the brain-wide correlation 
 profiles (Tables \ref{tableCorrels1}--\ref{tableCorrels11}),
 the maximal-intensity projections of each
 correlation profile are supplemented by a section 
 through the top region by correlation.\\

%plotClassificationBig12 = plot_classification_big12( Ref, cellTypesCorrelations, 
% fitVoxelsToTypes, 1, '../writeUp' );

\subsubsection{Ranking brain regions by density profile}
\begin{figure}
\includegraphics[width=\widthParamFig\textwidth]{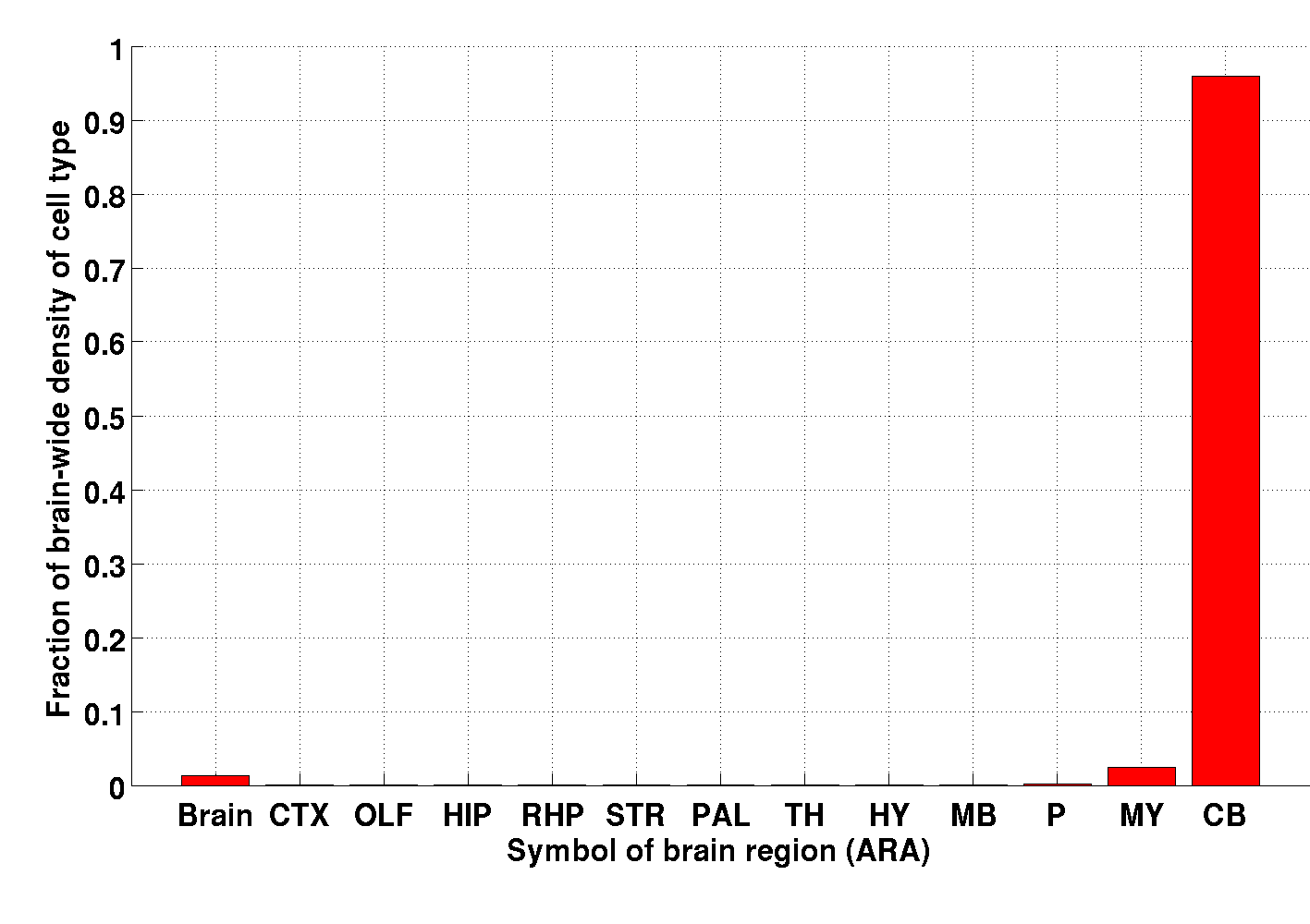}
\caption{Fractions of density of granule cells in the regions 
of the \bigTwelveSpace annotation of the ARA, as defined in Equation
   \label{densityRegion}, for $t=20$.}
\label{densityBarIndex20}
\end{figure}

 For a cell type labeled $t$,
  we can compute the
  fraction  of the total brain-wide density density contributed by voxels of each region
 labeled $r$ in the ARA:
\begin{equation}
{\overline{\rho}}( r, t ) = \frac{1}{|\sum_{v \in {\mathtt{{Brain\;Annotation}}}}\rho_t(v) |}
                                  \sum_{v \in V_r}\rho_t(v),
\label{fittingRegion}
\end{equation}
where {\ttfamily{Brain\;Annotation}} is the set of voxels included in
the annotation (for the \bigTwelveSpace annotation this set
consists of the left hemisphere, as can be seen from the
 projections in Table \ref{referenceTableBig12}). The brain regions in the ARA are
ranked according to the fractions of density defined in Equation
\ref{fittingRegion}.  For each cell type $t$, the fractional densities
supported by the brain regions sum to 1:
\begin{equation}
\forall t \in [1..T],\;\;\; \sum_{r=1}^R {\overline{\rho}}( r, t )=1.
\end{equation}
The region with highest fraction of density for a given cell type 
 is called the {\emph{top region by density}} for this type. A
 bar diagram of the fractions of the density profile for granule cells
 (cell type index $t$=20),
 is shown on Figure \ref{densityBarIndex20}. The top region by density 
 for granule cells is the cerebellum.\\ 

 In the figures derived from the brain-wide correlation density
 profiles (Tables \ref{tableFittings1}--\ref{tableFittings11}),
 the maximal-intensity projections of each
  density profile are supplemented by a section 
 through the top region by density.\\

\subsection{Anatomical data for cell-type-specific samples}
For each of the $T=64$ cell-type-specific samples in this study,
anatomical metadata indicate the brain region from which the sample
was extracted (see Tables \ref{metadataAnatomyTable1} and
\ref{metadataAnatomyTable2}).  Eight of the twelve regions of the
coarsest version of the Allen Reference Atlas (the \bigTwelveSpace
annotation) are represented in this data set (see Table
\ref{metadataAnatomy}), with an over-representation of the cerebral
cortex (41 of the 64 samples come from the cerebral cortex, whereas
the region occupies $\sim$29.5 percent of the volume of the brain, see
Table \ref{referenceTableBig12}). The cerebellum is the only other
brain region in the \bigTwelveSpace annotation to be
over-represented by cell-type-specific samples compared to the volume
of the brain it occupies. The regions that are not represented are the
olfactory areas, the retrohippocampal region and the hypothalamus. The
group of voxels labeled 'Basic cell groups of regions', which includes
the white matter, is also unrepresented.\\

\begin{table}
\centering
\begin{tabular}{|m{0.25\textwidth}|m{0.15\textwidth}|m{0.5\textwidth}|}
\hline
\textbf{Brain region}&\textbf{Number of cell-type-specific 
                               samples}&\textbf{List of sample indices in the list of 64 samples
                  in this study} \\\hline
Basic cell groups and regions & 0 & $\emptyset$\\\hline
Cerebral cortex & 41 & $\{ 2,3,[6,\dots,10], 14, 22, 24, 26,$\\ 
                &    &     $[29,\dots,48], 50, [53,\dots,56], 58, [60,\dots,64] \}$\\\hline
Olfactory areas & 0 & $\emptyset$\\\hline
Hippocampal region & 2 & $\{ 49, 57 \}$ \\\hline
Retrohippocampal region & 0 & $\emptyset$\\\hline
Striatum & 3 & $\{ 13, 15, 16\} $\\\hline
Pallidum & 1 & $\{ 11 \} $\\\hline
Thalamus& 1 & $\{ 59 \}$\\\hline
Hypothalamus& 0 & $\emptyset$\\\hline
Midbrain& 3 & $ \{ 4, 5, 10 \} $ \\\hline
Pons& 1 &$ \{ 51 \} $ \\\hline
Medulla & 1 &  $ \{ 12  \} $ \\\hline
Cerebellum & 11 & $ \{1, [17,\dots,21], 23, 25, 27, 28, 52  \} $ \\
\hline
\end{tabular}
\caption{{\bf{Anatomical data for the cell-type-specific samples.}} Eight of the
 regions defined by the coarsest version of the Allen Reference Atlas are represented in our data set. See Tables \ref{metadataAnatomyTable1} and \ref{metadataAnatomyTable2} for a more detailed
 account of the anatomy of the of the cell-type-specific samples.}
\label{metadataAnatomy}
\end{table}

 For each of the cell-type-specific samples, we computed the ranks of
 the brain regions in the ARA according to correlation and density
 profiles, as defined in Equations \ref{correlRegion} and
 \ref{fittingRegion}. It is interesting to compare the computed
 \emph{top region by density} and the \emph{top region by correlation}
 to the brain region from which the cell-type-specific sample was
 extracted (listed in Tables in S9). In the rest of this section, we
 group the cell-type-specific samples that were extracted from a given
 brain region, and compare this region to the top region by
 correlation and to the top region by density (except for the set of
 voxels called 'Basic cell groups', which is discussed first as it
 appears as the top region in the results for a number of cell types).

\subsection{Neuroanatomical patterns of results,
 grouped by the regions in the \bigTwelveSpace annotation of the left
 hemisphere}

\subsubsection{Basic cell groups and regions}
\begin{figure}
\centering
\includegraphics[width=0.9\textwidth]{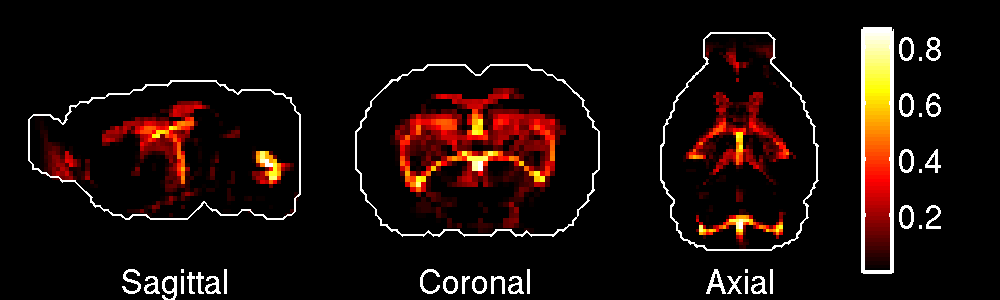}
\caption{Maximum-intensity projection of the sum of density profiles
  of cell-type-specific samples that have 'Basic cell groups and
  regions' as their top region by density, listed in Table
  \ref{whiteMatterTable}.}
\label{sumFitWhiteMatter}
\end{figure}

The patchy group of voxels assigned the label 'Basic cell groups and
regions' in the digitized version of the Allen Atlas at a resolution
of 200 microns (see Table \ref{referenceTableBig12}) is not found in
the list of brain regions from which the cell-type-specific samples
analyzed here were extracted (see Table
\ref{metadataAnatomy}). However, this set of voxels is the top region
by correlation and/or density for several cell types, and an
inspection of the maximum-intensity projection of the correlation and
density profiles for these cell types reveals an anatomical pattern
that resembles the brain's white matter structures, including the
{\emph{arbor vitae}}. As can be seen in Table \ref{whiteMatterTable},
most of these cell types were extracted from the cerebral cortex,
except the astroglia (sample index $t$=28, \cite{DoyleCells}), which
were extracted from the cerebellum). The sum of the density profiles
of these cell types is illustrated on Figure \ref{sumFitWhiteMatter}.
All the cell types whose top region by density is 'Basic cell groups
and regions' are oligodendrocytes, astroglia or astrocytes.\\

\begin{figure}
\centering
 \subfloat[]{\includegraphics[width=\textwidth]{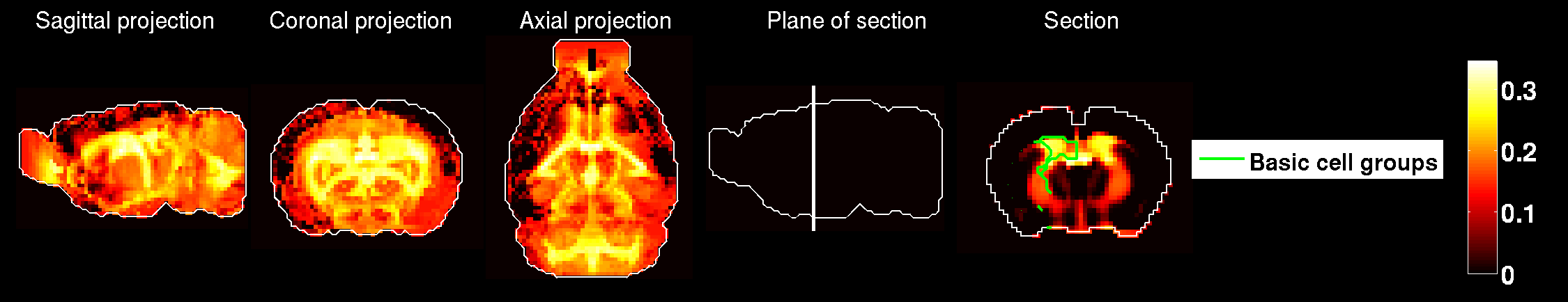}}\\
 \subfloat[]{\includegraphics[width=\textwidth]{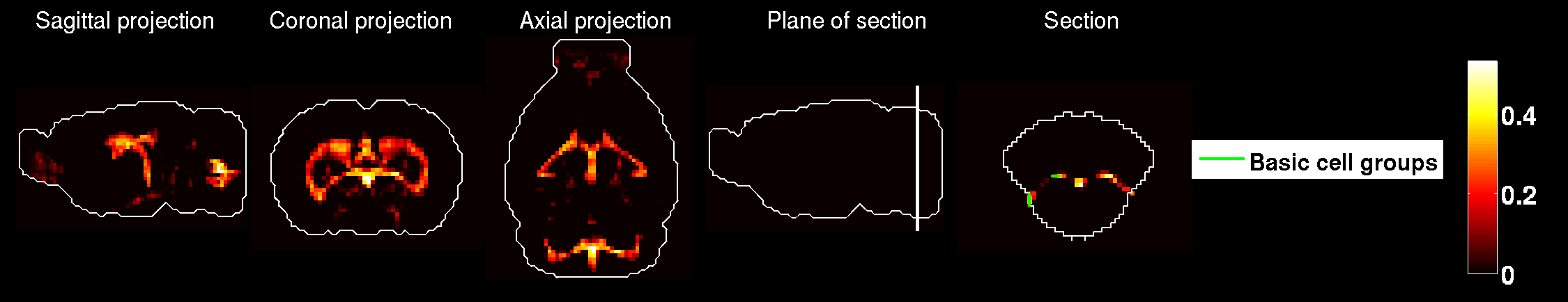}}\\
 \caption{ {\bf{Astrocytes (cell-type index 31).}} (a) Heat map of the brain-wide correlation profile. (b) Heat map of the estimated brain-wide density profile. 'Basic cell groups and regions' is the top region by correlation and by density, hence the choice of section and its legend.}
  \label{type31CorrDensity}
\end{figure}

\begin{figure}
\centering
 \subfloat[]{\includegraphics[width=\widthCoeff\textwidth]{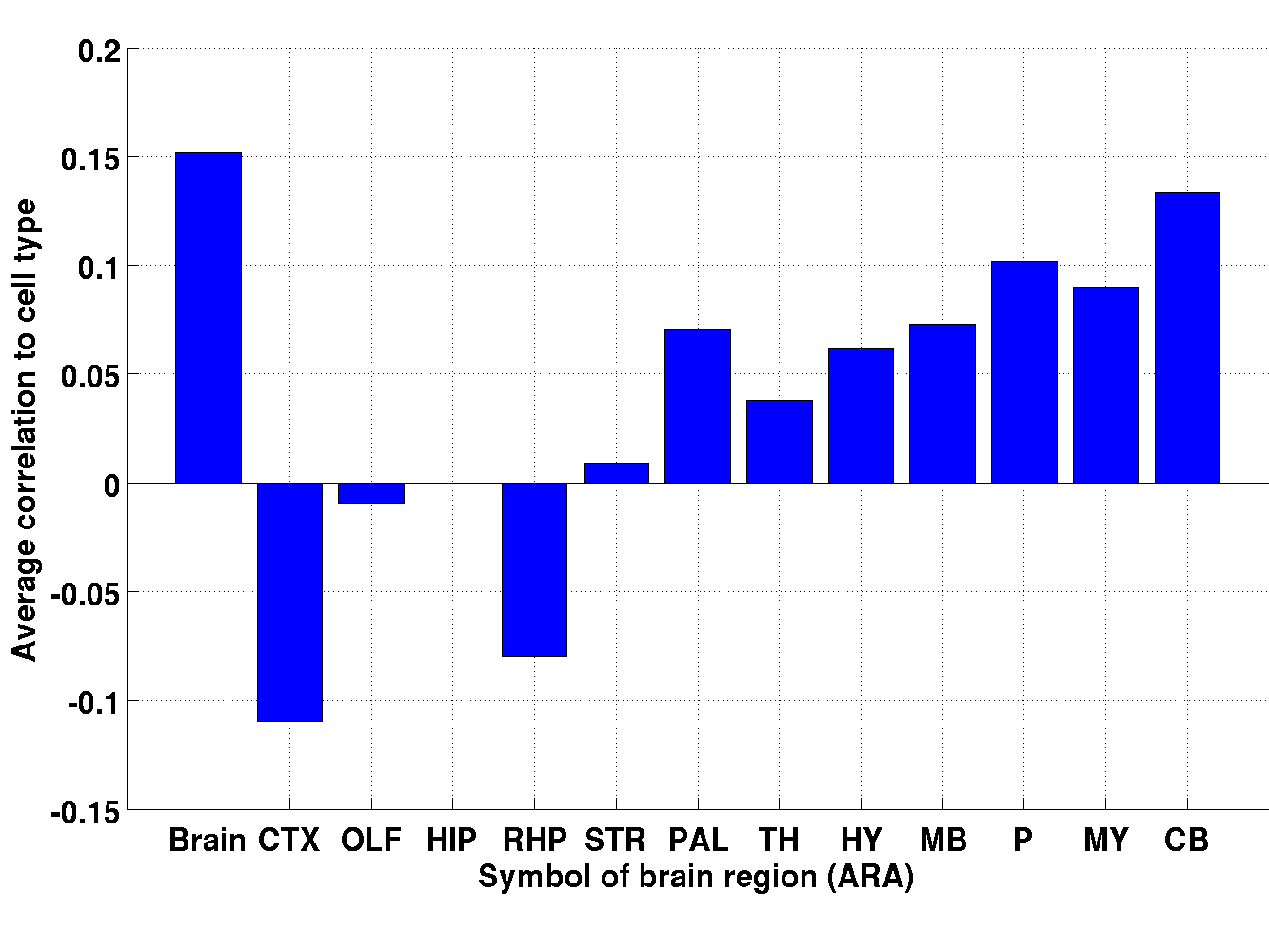}}\\
 \subfloat[]{\includegraphics[width=\widthCoeff\textwidth]{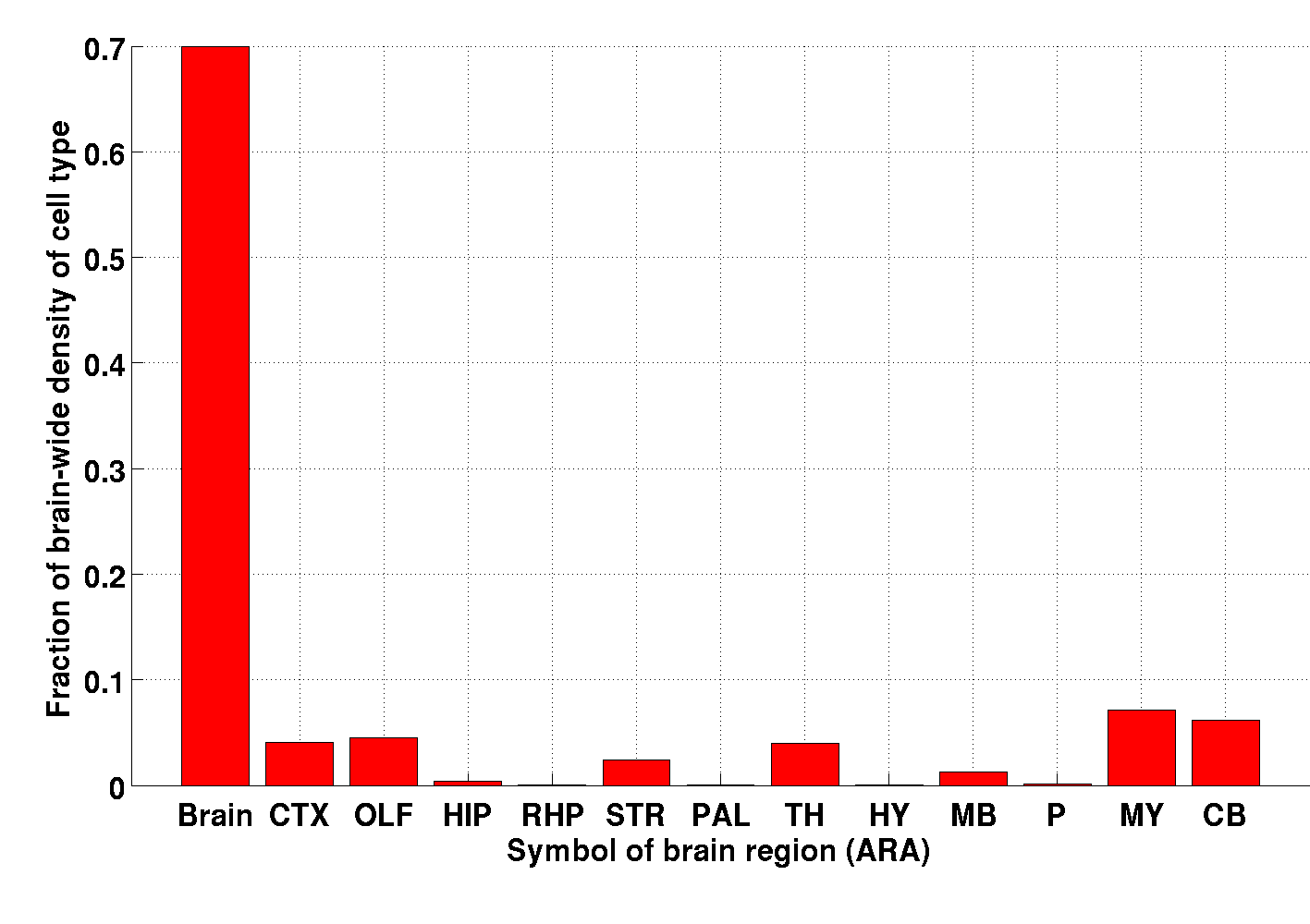}}\\
 \caption{{\bf{Astrocytes (cell-type index 31).}} Average correlation between the cell type and the Allen Atlas, in the regions 
of the \bigTwelveSpace annotation of the ARA. Even though the astrocytes were extracted 
 from the cerebral cortex, their expression profile is negativaly correlated 
 on average with the voxels of the cerebral cortex.}
  \label{type31Anatomy}
\end{figure}

%classifyPatternBasic = classify_pattern_basic( Ref, cellTypesCorrelations, fitVoxelsToTypes )
\begin{table}
\centering
\begin{tabular}{|m{0.28\textwidth}|m{0.25\textwidth}|m{0.23\textwidth}|}
\hline
\textbf{Description (index)}&\textbf{Origin of sample}&\textbf{Fraction of density in 'Basic cell groups and regions' (\%)}\\ \hline 
 Mature Oligodendrocytes (22) &  Cerebral cortex &     53.6   \\\hline 
Astroglia (29) & Cerebral cortex & 56.0\\\hline 
Astrocytes (30) & Cerebral cortex & 19.7\\\hline 
Astrocytes (31)  & Cerebral cortex & 69.9\\\hline 
Astrocytes (32) & Cerebral cortex & 51.4 \\\hline 
Oligodendrocytes (36) & Cerebral cortex & 56.8\\\hline 
Astroglia (28) & Cerebellum & 74.3 \\
\hline
\end{tabular}
\caption{Cell-type-specific samples that have 'Basic cell groups and
  regions' as their top region by density.}
\label{whiteMatterTable}
\end{table}
Figures \ref{type31CorrDensity} and \ref{type31Anatomy} show results
for a class of astrocytes \cite{CahoyCells} (cell-type index 31)
extracted from the cerebral cortex.  The brain-wide correlation and
density profiles exhibit a  pattern resembling white-matter
 structures, with the most caudal component corresponding to the
{\emph{arbor vitae}}.

\subsubsection{Cerebral cortex}
\begin{figure}
\centering
 \subfloat[]{\includegraphics[width=\textwidth]{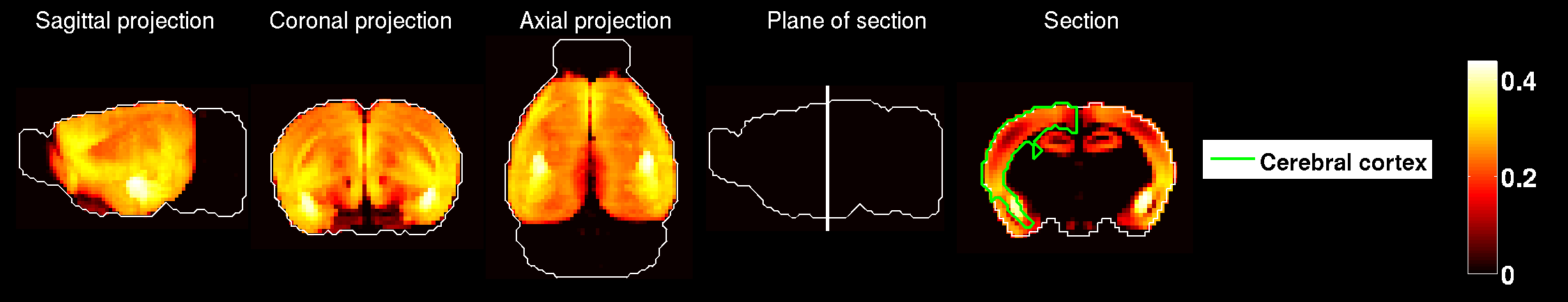}}\\
 \subfloat[]{\includegraphics[width=\textwidth]{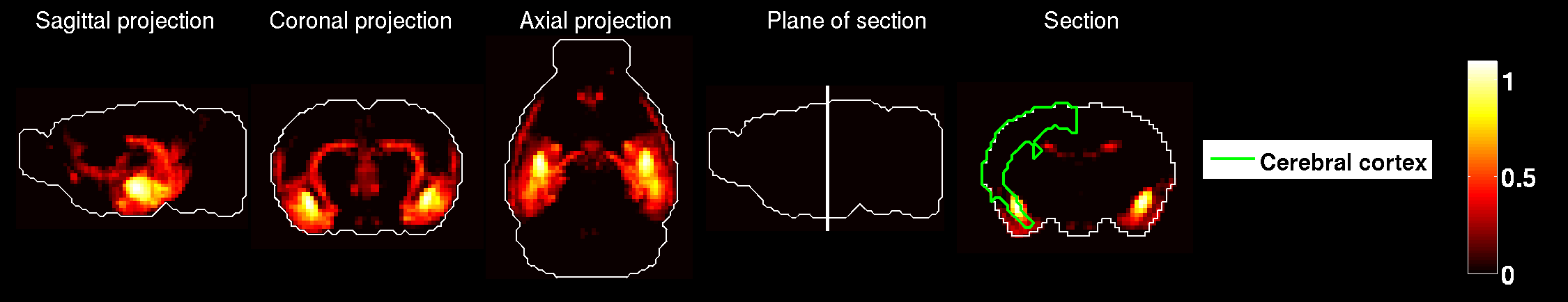}}\\
 \caption{ {\bf{Pyramidal neurons (cell-type index 48, studied in \cite{foreBrainTaxonomy}).}} (a) Heat map of the brain-wide correlation profile. (b) Heat map of the estimated brain-wide density profile.}
  \label{type48CorrDensity}
\end{figure}

\begin{figure}
\centering
 \subfloat[]{\includegraphics[width=\textwidth]{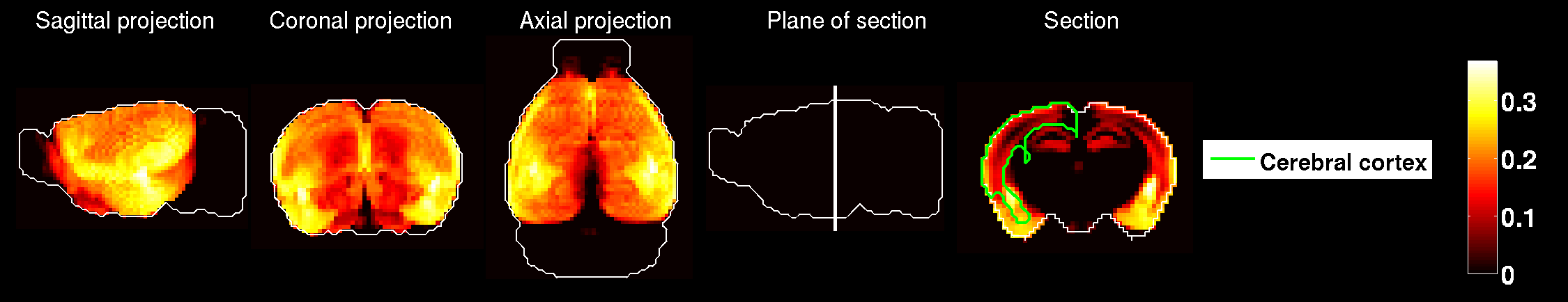}}\\
 \subfloat[]{\includegraphics[width=\textwidth]{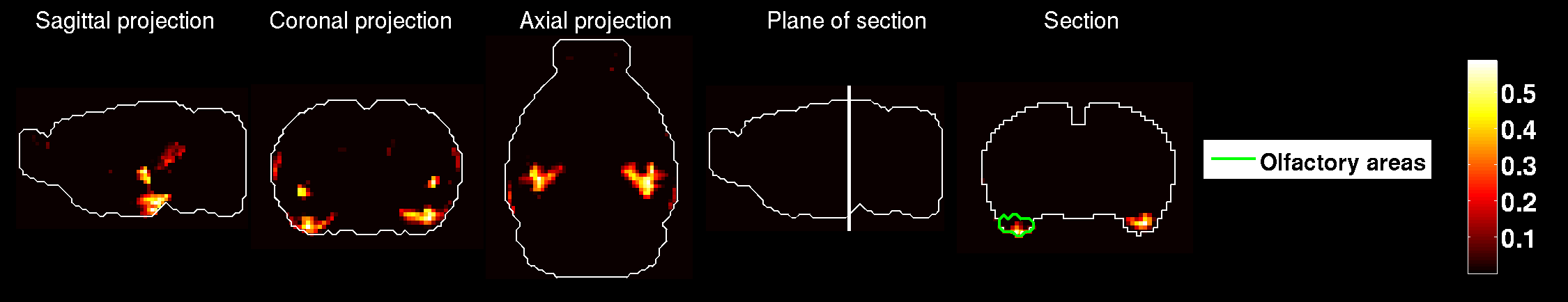}}\\
 \caption{ {\bf{Glutamatergic neurons (cell-type index 53, studied in \cite{foreBrainTaxonomy}).}} (a) Heat map of the brain-wide correlation profile. (b) Heat map of the estimated brain-wide density profile.}
  \label{type53CorrDensity}
\end{figure}

%\subsubsubsection{The white-matter pattern}
  Some of the cell-type-specific
   samples extracted from the cerebral cortex have a correlation
   and/or density profile resembling white matter (which
   is included within the 'Basic cell groups and regions' in the ARA).
   See the previous subsection, Table \ref{whiteMatterTable} and Figure
   \ref{sumFitWhiteMatter} for a separate analysis of these cell types.\\

%\subsubsubsection{Amygdalar cell types} Two cell types (pyramidal
   neurons, index 48, and glutamatergic neurons, index 53, both
   studied in \cite{foreBrainTaxonomy}) were extracted from the
   amygdala (see Tables
   \ref{metadataAnatomyTable1}--\ref{metadataAnatomyTable2}). The
   amygdala in not one of the brain regions in the \bigTwelveSpace
   annotation of the ARA. The amygdala is split between the
   subcortical plate (which is included in the cerebral cortex in the
   numerical version of the ARA at a resolution of 200 microns) and
   the olfactory areas \cite{AllenAtlas}. A visual inspection of the
   correlation and density profiles for both these cell types (Figures
   \ref{type48CorrDensity} and \ref{type53CorrDensity}) allows to
   detect a pattern resembling the amygdala, and indeed the cerebral
   cortex and the olfactory areas rank first and second by the
   fraction of the density profile they support (see Table
   \ref{amygdalaPatternTable}).  It is interesting to examine the
   fraction of the density profile supported by the various
   subdivisions of the olfactory areas (according to the \fine
   annotation), especially for glutamatergic neurons, for which
   olfactory areas support more than 64 \% of the total density
   profile (see Table \ref{amygdalaPatternTableOlfactory}). The
   Cortical amygdalar area and Piriform-amygdalar area are among the
   main subregions of the olfactory areas contributing to the density
   profile of both cell types, which confirms the visual impression 
   of an amygdalar pattern.\\


\begin{table}
\centering
\begin{tabular}{|m{0.25\textwidth}|m{0.15\textwidth}|m{0.15\textwidth}|m{0.15\textwidth}|m{0.15\textwidth}|}
\hline
\textbf{Description (index)}&\textbf{Fraction of density in cerebral cortex (\%)}& \textbf{Fraction of density in olfactory areas(\%)} & \textbf{Rank of Cerebral cortex in \bigTwelve}&
   \textbf{Rank of Olfactory areas in \bigTwelve} \\ \hline 
Pyramidal neurons (48) & 38.5 & 29.3 & 1 & 2 \\\hline
Glutamatergic neurons (53) & 30.2 & 64.9 & 2 & 1\\
\hline
\end{tabular}
\caption{Cell-type-specific samples extracted from the amygdala, with fractions of their density profiles
 supported by the cerebral cortex and the olfactory areas, which are the two brain regions in the ARA 
that intersect the amygdala.}
\label{amygdalaPatternTable}
\end{table}

\begin{table}
\centering
\begin{tabular}{|m{0.20\textwidth}|m{0.35\textwidth}|m{0.15\textwidth}|m{0.15\textwidth}|}
\hline
\textbf{Description (index)}&\textbf{Subregion of Olfactory areas in the ARA ('fine' annotation)}&\textbf{Fraction of density in the region(\%) } &\textbf{Fraction of Olfactory areas occupied by the region}\\ \hline 
Pyramidal neurons (48) &  Piriform area &  40.3 &    28.4 \\
        &    Cortical amygdalar area&    29.3 &     5.5 \\
        & Postpiriform transition area &  15.4 &   2.2 \\
   &  Piriform-amygdalar area  & 9.1   &  1.9 \\
   &  Nucleus of the lateral olfactory tract & 4.0   &  1.0 \\
    & Anterior olfactory nucleus & 1.4   &  9.5\\
   &  Taenia tecta  &0.5   &  3.2 \\
   &  Main olfactory bulb    &   0  &  41.4 \\
   &   Accessory olfactory bulb   &   0  &   1.6 \\ \hline 
Glutamatergic neurons (53) & Cortical amygdalar area & 59.9 &   5.5\\
  & Piriform-amygdalar area & 19.1  &   1.9\\
  & Postpiriform transition area  & 13.2   &  2.2\\
  &  Piriform area  & 7.4  &  28.4\\
  &  Main olfactory bulb  & 0.4  &  41.4\\
   &  Accessory olfactory bulb   &   0  &   1.6\\
   &    Anterior olfactory nucleus &   0   &  9.5\\
   &    Taenia tecta  &  0  &   3.2\\
   &   Nucleus of the lateral olfactory tract  &   0  &   1.0\\
\hline
\end{tabular}
\caption{Subregions of the olfactory areas ranked by the fraction of the density profile they support,
 for the two cell-type-specific samples extracted from the amygdala. The amygdalar regions are over-represented 
 for both cell types, compared to the fraction of the olfactory areas they occupy.}
\label{amygdalaPatternTableOlfactory}
\end{table}
\clearpage

\begin{figure}
\centering
 \subfloat[]{\includegraphics[width=\textwidth]{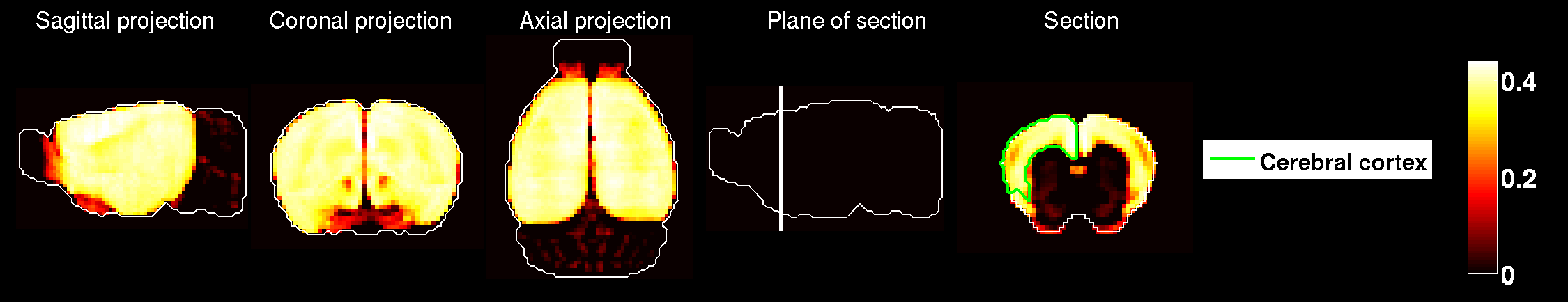}}\\
 \subfloat[]{\includegraphics[width=\textwidth]{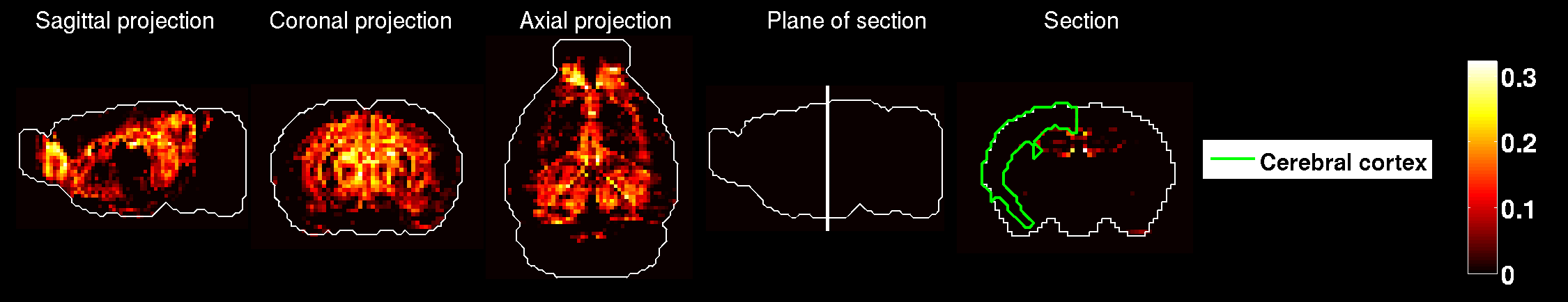}}\\
 \subfloat[]{\includegraphics[width=\textwidth]{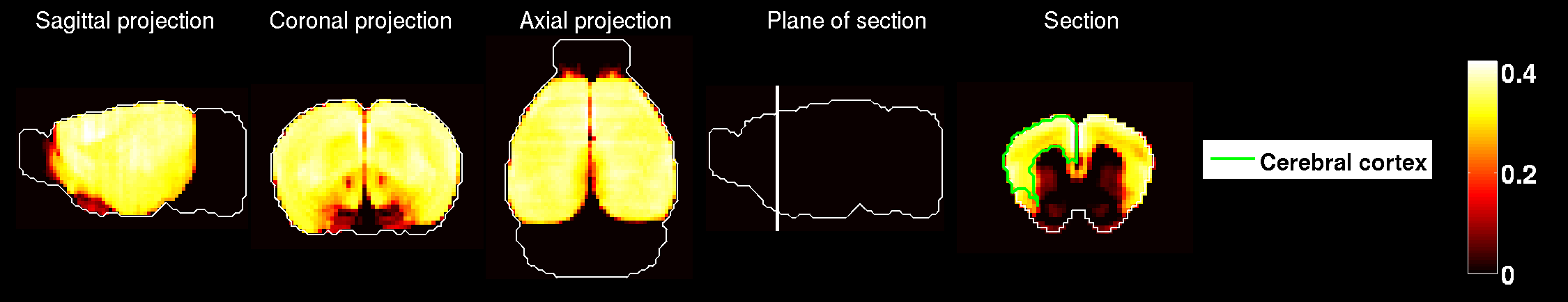}}\\
 \subfloat[]{\includegraphics[width=\textwidth]{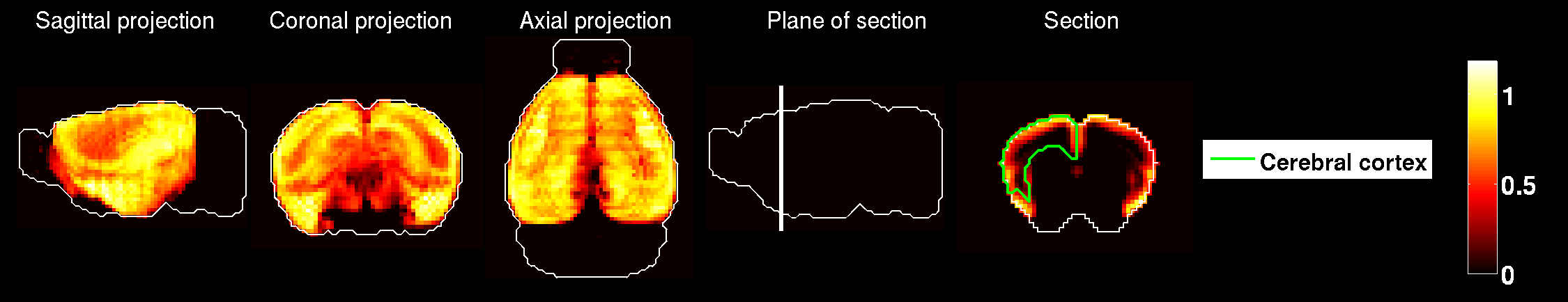}}\\
 \caption{ {\bf{Pyramidal neurons for which the cerebral cortex is the top region both by correlation and by density (I).}} (a) Heat map of the brain-wide correlation profile (cell-type index 7). (b) Heat map of the estimated brain-wide density profile (cell-type index 7). (c) Heat map of the brain-wide correlation profile (cell-type index 40). (d) Heat map of the estimated brain-wide density profile (cell-type index 40).}
  \label{pyr1}
\end{figure}

\begin{figure}
\centering
 \subfloat[]{\includegraphics[width=\textwidth]{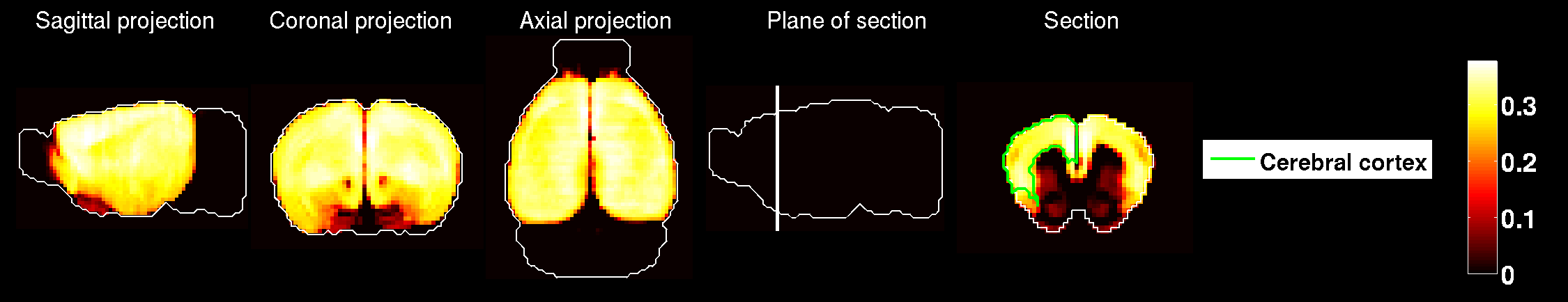}}\\
 \subfloat[]{\includegraphics[width=\textwidth]{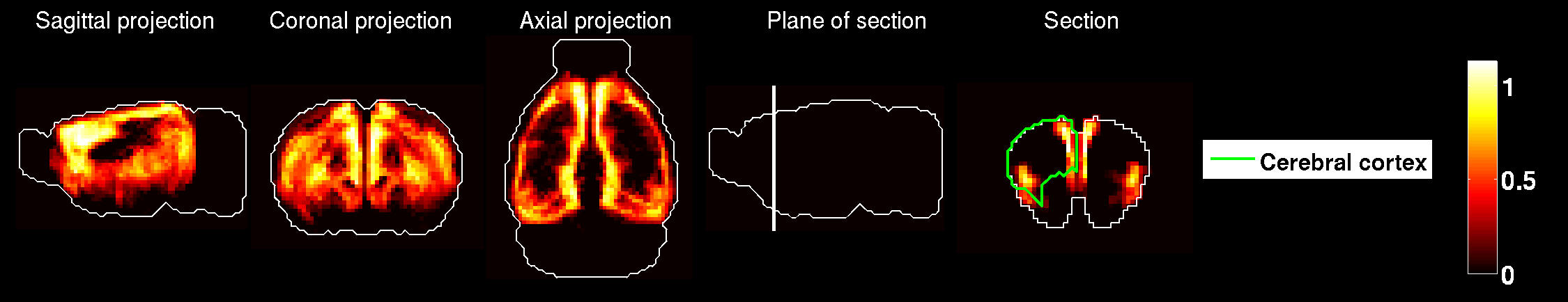}}\\
 \subfloat[]{\includegraphics[width=\textwidth]{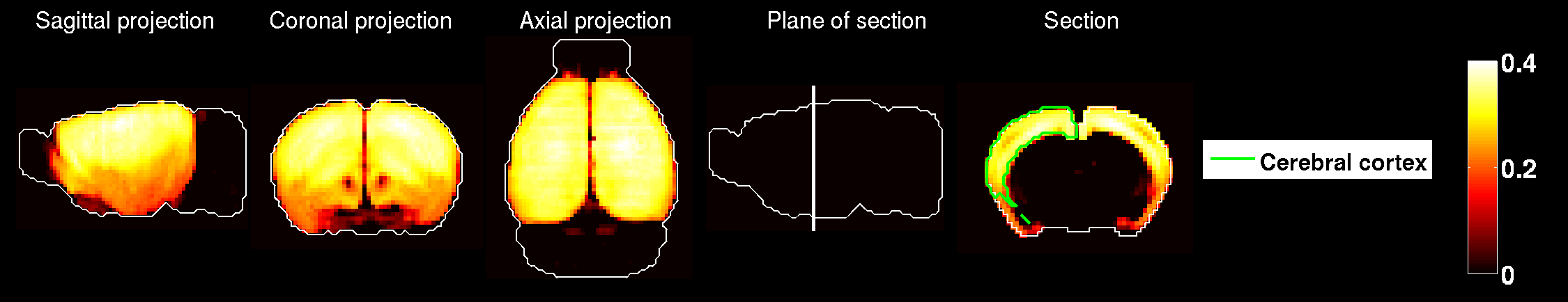}}\\
 \subfloat[]{\includegraphics[width=\textwidth]{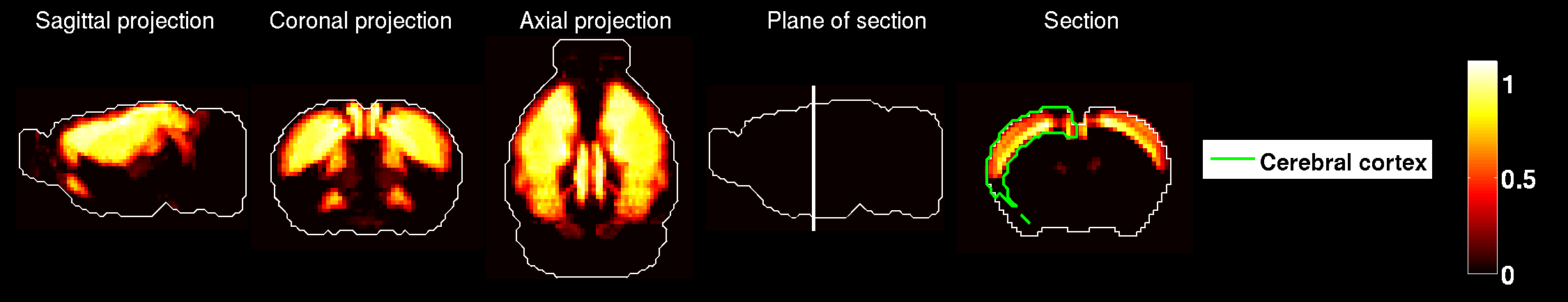}}\\
 \caption{ {\bf{Pyramidal neurons for which the cerebral cortex is the top region both by correlation and by density (II).}} (a) Heat map of the brain-wide correlation profile (cell-type index 45). (b) Heat map of the estimated brain-wide density profile (cell-type index 45). (c) Heat map of the brain-wide correlation profile (cell-type index 46). (d) Heat map of the estimated brain-wide density profile (cell-type index 46).}
  \label{pyr2}
\end{figure}

\begin{figure}
\centering
 \subfloat[]{\includegraphics[width=\textwidth]{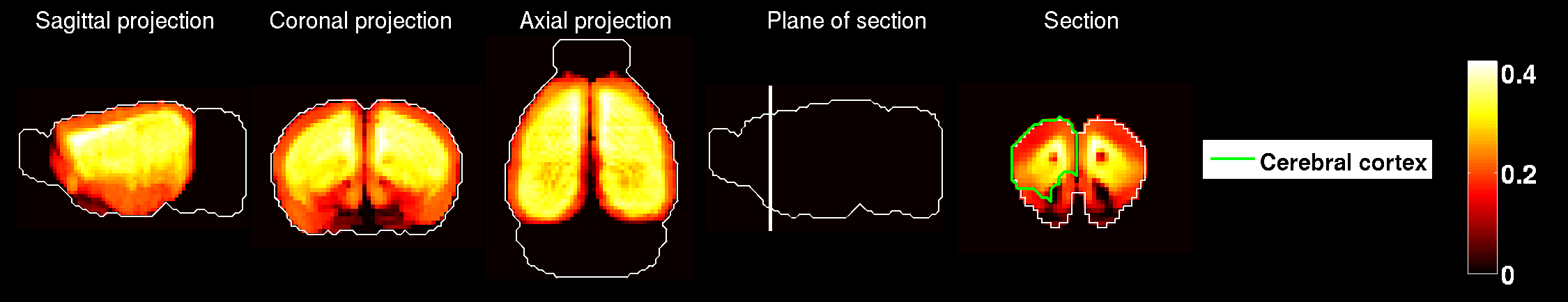}}\\
 \subfloat[]{\includegraphics[width=\textwidth]{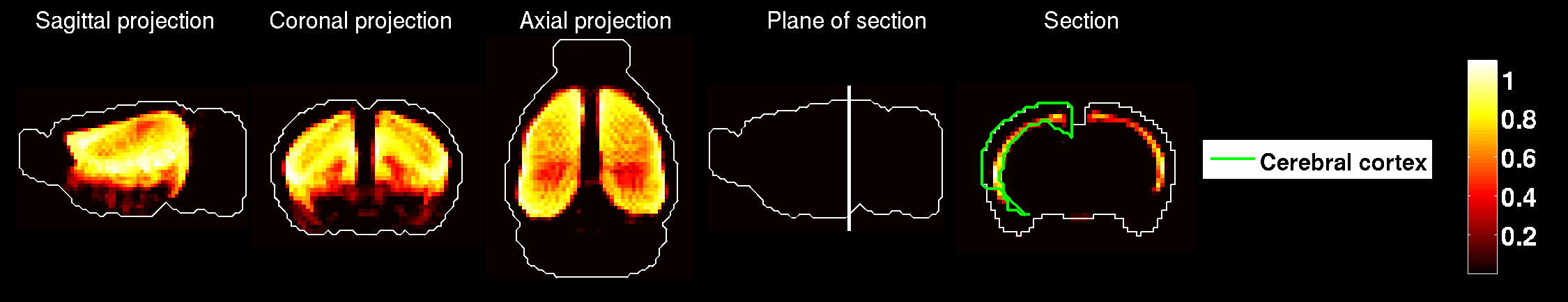}}\\
 \caption{ {\bf{Pyramidal neurons for which the cerebral cortex is the top region both by correlation and by density (III).}} (a) Heat map of the brain-wide correlation profile (cell-type index 47). (b) Heat map of the estimated brain-wide density profile (cell-type index 47).}
  \label{pyr3}
\end{figure}

  Having treated separately the cell types that correlate best or have
  highest density fits to the white matter (see Table
  \ref{whiteMatterTable}), as well as the two amygdalar cell types, we
  are left with 29 cortical cell types. For 5 of these -- all
  pyramidal neurons -- the cerebral cortex is the top region in the
  \bigTwelveSpace annotation of the ARA, both by average correlation and by
  fraction of the density profiles. These pyramidal neurons were all
  extracted from adult animals, except for the cell type labeled 40
  (P14); see Tables \ref{metadataTable1},\ref{metadataTable2} for age
  data. See Figure \ref{pyr1}--\ref{pyr3} for heat maps of the
  brain-wide correlation and density profiles for these cell types.\\


\begin{table}
\centering
\begin{tabular}{|m{0.3\textwidth}|m{0.2\textwidth}|m{0.2\textwidth}|m{0.2\textwidth}|}
\hline
\textbf{Description (index)}&\textbf{Fraction of density supported in cerebral cortex (\%)} & \textbf{Next region in \bigTwelve} &  \textbf{Fraction of density supported in next region (\%)}\\ \hline 
Pyramidal neurons (7) &  25.4 & Olfactory areas & 24.9\\\hline 
Pyramidal neurons, callosally projecting, P14 (40) &  76.6  & Olfactory areas & 15.9\\\hline 
Pyramidal Neurons (45) & 71.8 & Retrohippocampal region & 14.8\\\hline 
Pyramidal Neurons (46) & 93.6 & Retrohippocampal region & 1.8 \\\hline 
Pyramidal Neurons (47) & 84.5 & Retrohippocampal region & 7.2\\
\hline
\end{tabular}
\caption{Cell-type-specific samples extracted from the cerebral cortex, for which the cerebral cortex is
ranked first both by average correlation and fraction of density profile supported.}
\label{corticalPatternBothGood}
\end{table}

% classifyPatternCortexNew = classify_pattern_cortex_new(...
 
 Some cell-type specific samples extracted from the cortex as per the
 anatomical data of Table \ref{metadataAnatomy} have the cerebral
 cortex as top region by correlation, but not as top region by
 density.  There are 8 such cell types (see Table
 \ref{corticalPatternGoodCorrelOnly}), of which 4 have zero density
 profiles in the left hemisphere. The other four cell types
 have Olfactory areas or Retrohippocampal region as their 
 top region by density.\\

\begin{table}
\centering
\begin{tabular}{|m{0.3\textwidth}|m{0.2\textwidth}|m{0.2\textwidth}|m{0.2\textwidth}| }
\hline
\textbf{Description (index)}&\textbf{Top region by density (percentage of density supported)} & \textbf{Rank of the cerebral cortex out of 13 regions} & \textbf{Fraction of density in cerebral cortex (\%)}\\ \hline 
Pyramidal Neurons (2) & Olfactory areas (100) & 2 (and last) & 0 \\\hline 
Pyramidal Neurons (8) & Olfactory areas (100) & 2 (and last) & 0 \\\hline 
Mixed Neurons (9) & Cerebellum (100) & 2 (and last) & 0 \\\hline 
Interneurons (14) & Olfactory areas (100) & 2 (and last) & 0 \\\hline 
Neurons (26) & Olfactory areas (100) & 2 (and last) & 0 \\\hline 
Pyramidal Neurons, Corticospinal, P14 (43) & Olfactory areas (72.4) & 2 & 11.5 \\\hline 
Pyramidal Neurons, Corticotectal, P14 (44) & Olfactory areas (33.0) & 4 & 9.4 \\\hline 
Pyramidal Neurons (50)& Retrohippocampal region (57.6) & 3& 10.7 \\\hline 
GABAergic Interneurons, VIP+ (55) & Olfactory areas (96.7) & 2 & 1.78 \\ 
\hline
\end{tabular}
\caption{Cell-type-specific samples extracted from the cerebral cortex, for which the cerebral cortex is
ranked first  by average correlation but not by the density.}
\label{corticalPatternGoodCorrelOnly}
\end{table}

  Some cortical cell-type-specific samples do not have 
  the cerebral cortex as their top region by correlation or by density. There are 15 such
  cell types, 7 of which have zero density in the left hemisphere (they
   are not detected by the linear model, see
  Table \ref{corticalPatternBothBad}).  Six cell types have the cortex
  as their second ranked region by average correlation, while the top
  region is either the hippocampal region or the retrohippocampal
  region.\\

\begin{table}
\centering
\begin{tabular}{|m{0.24\textwidth}|m{0.18\textwidth}|m{0.08\textwidth}|m{0.17\textwidth}|m{0.12\textwidth}|m{0.1\textwidth}|}\hline
\textbf{Description (index)}& \textbf{Top region by correlation} &\textbf{Rank of cerebral cortex by correlation} & \textbf{Top region by density (percentage of density supported)}& \textbf{Rank of the cerebral cortex by density out of 13 regions} & \textbf{Fraction of density in cerebral cortex (\%)}\\\hline 
Pyramidal Neurons (3) & Hippocampal region & 2 & Olfactory areas (100) & 2 (and last) & 0 \\\hline 
Mixed Neurons (33) & Retrohippocampal region & 2 & Olfactory areas (58.6) & 3 & 4.8 \\\hline 
Oligodendrocyte Precursors (37) &Hypothalamus & 13 & Hypothalamus (33.8) & 7 & 2.6\\\hline 
Pyramidal Neurons, Callosally projecting, P3 (38) & Olfactory areas & 6 & Olfactory areas (100)& 2 (and last)& 0 \\\hline 
Pyramidal Neurons, Callosally projecting, P6 (39)& Retrohippocampal region & 2 & Olfactory areas (99) & 4 (and last) & 0\\\hline 
Pyramidal Neurons, Corticospinal, P3 (41)& Olfactory areas & 5& Olfactory areas (100) & 2 (and last) & 0\\\hline 
Pyramidal Neurons, Corticospinal, P6 (42)& Retrohippocampal region &2 & Olfactory areas (86.6)& 2 & 7.8\\\hline 
GABAergic Interneurons, VIP+ (54)& Retrohippocampal region &2 & Striatum (36.6) & 5 & 6.5\\\hline 
GABAergic Interneurons, SST+ (55) &  Hypothalamus &  10 & Striatum (30.1) & 8 & 2.4 \\\hline 
GABAergic Interneurons, PV+ (58) & Medulla & 7 & Olfactory areas (93.4)   & 4 (and last) & 0 \\\hline 
GABAergic Interneurons, PV+, P7 (60) & Pallidum & 11 & Olfactory areas (99.2) & 2 & 0.7 \\\hline 
GABAergic Interneurons, PV+, P10 (61) & Pallidum & 8 &  Olfactory areas (100) & 2 (and last)& 0 \\\hline 
GABAergic Interneurons, PV+, P13-P15 (62)&  Retrohippocampal region & 2 &  Olfactory areas (100) & 2 (and last)& 0 \\\hline 
GABAergic Interneurons, PV+, P25 (63) & Medulla & 5 & Olfactory areas (56.5) & 6 & 2.1 \\\hline 
GABAergic Interneurons, PV+ (64) & Medulla & 5 & Midbrain (28.1) & 10 & 1.5\\
\hline
\end{tabular}
\caption{Cell-type-specific samples extracted from the cerebral cortex, for which the cerebral cortex is
ranked first neither by average correlation, nor by the  fraction of density profile it supports.}
\label{corticalPatternBothBad}
\end{table}

\clearpage

\subsubsection{Hippocampal region}
  Two cell-type-specific samples that were extracted from the
  hippocampus (see Table \ref{metadataAnatomy}). For one of them
  pyramidal neurons (index 49, studied in \cite{foreBrainTaxonomy}),
  the hippocampal region is the top region both by correlation and by
  density.  We ranked the regions of the \fine annotation in the ARA
  by their contribution to the density profile of this sample (this
  ranking corresponds to Equation \ref{fittingRegion}, with the region
  label $r$ running over the $R=94$ regions in the \fineSpace annotation).
  The first two regions are Ammon's horn (which contributes 48.8
  percent of the total density of this cell type), followed by the
  dentate gyrus (which contributes 25.4 percent of the total density
  of this cell type). These two regions are the two subregions of the
  hippocampal region in the \fine annotation. Moreover, the data of
  \cite{foreBrainTaxonomy} specify that the sample labeled 49 was
  taken from Ammon's horn, which indicates that the brain-wide
  correlation and density profiles are both consistent with prior
  biological knowledge for this cell type.\\

 This sample (index 49) is the only one for which the hippocampus is
 ranked first by density. Another sample (index 3), is ranked first by
 correlation.  This sample was not extracted from the hippocampus, but
 rather from the cerebral cortex (primary somatosensory area, layer
 5).  On the other hand, the cerebral cortex is ranked second by its
 fraction of density defined in Equation \ref{fittingRegion}. This
 sample has an estimated density profile with values close to zero,
 except in a few voxels belonging to the olfactory bulb.\\

 The second sample extracted from the hippocampus consists of
 somatostatin-positive GABAergic interneurons (index 57).  The
 hippocampal region is ranked last by correlations and next-to-last by
 densities, whereas the first region by average correlation is
 hypothalamus, and the first region by density is midbrain (see Figure
 \ref{type57CorrDensity} and \ref{type57Anatomy}). Visual
 inspection of the Tables
 \ref{tableCorrels10},\ref{tableCorrels11},\ref{tableFittings10},\ref{tableFittings11}
 shows that there is a lot of solidarity between the correlation
 profiles of the cell types labeled with indices between 54 and 64,
 which are all GABAergic interneurons.  For these cell types, 
 with the exception of the cell type labeled 55, whose top region by
 correlation is the cerebral cortex, the correlations are higher in
 regions of the brain that are more ventral than the region from which
 the samples where extracted.\\

\begin{table}
\centering
\begin{tabular}{|m{0.14\textwidth}|m{0.15\textwidth}|m{0.15\textwidth}|m{0.14\textwidth}|m{0.14\textwidth}|m{0.14\textwidth}|}
\hline
\textbf{Description (index)}&\textbf{Origin of sample}&\textbf{Rank of region (out of 94) in the 'fine' annotation (by correlation)}&\textbf{Rank of region (out of 94) in the 'fine' annotation (by density)} &\textbf{Fraction of density in the region}& \textbf{Fraction of density supported in the hippocampal region  } \\\hline
Pyramidal neurons (49) & Ammon's horn & 3 & 1 & 48.7\% & 71.5\% \\\hline
GABAergic interneurons, SST+ (57) & Ammon's horn & 91 & 55 & 0.1\% & 0.1\% \\
\hline
\end{tabular}
\caption{Anatomical analysis for the cell-type-specific samples extracted from the hippocampal region.}
\label{metadataAnatomyHippocampal}
\end{table}

%\input{hippocampalPattern.tex}
%\begin{figure}
%\includegraphics[width=\widthParamFig\textwidth]{correlsFig49.png}
%\caption{Brain-wide correlations between pyramidal neurons from the 
%hippocampus and the Allen Atlas.}
%\label{correlsFig49}
%\end{figure}
\begin{figure}
\centering
 \subfloat[]{\includegraphics[width=\textwidth]{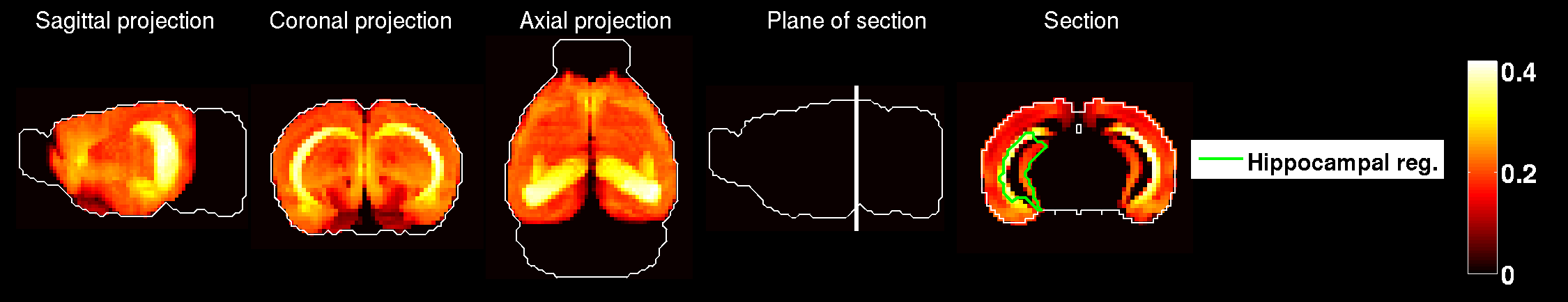}}\\
 \subfloat[]{\includegraphics[width=\textwidth]{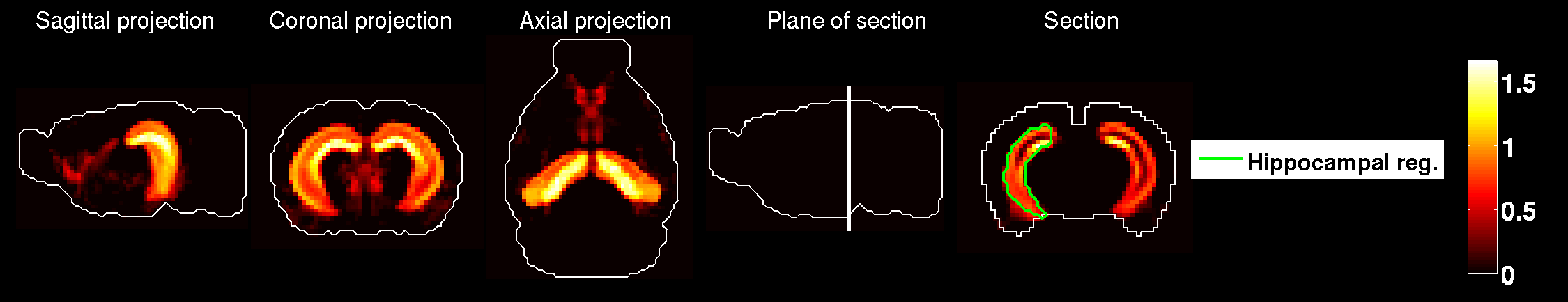}}\\
 \caption{ {\bf{Pyramidal neurons (cell-type index 49).}} (a) Heat map of the brain-wide correlation profile. (b) Heat map of the estimated brain-wide density profile.}
  \label{type49CorrDensity}
\end{figure}
\begin{figure}
\centering
 \subfloat[]{\includegraphics[width=\textwidth]{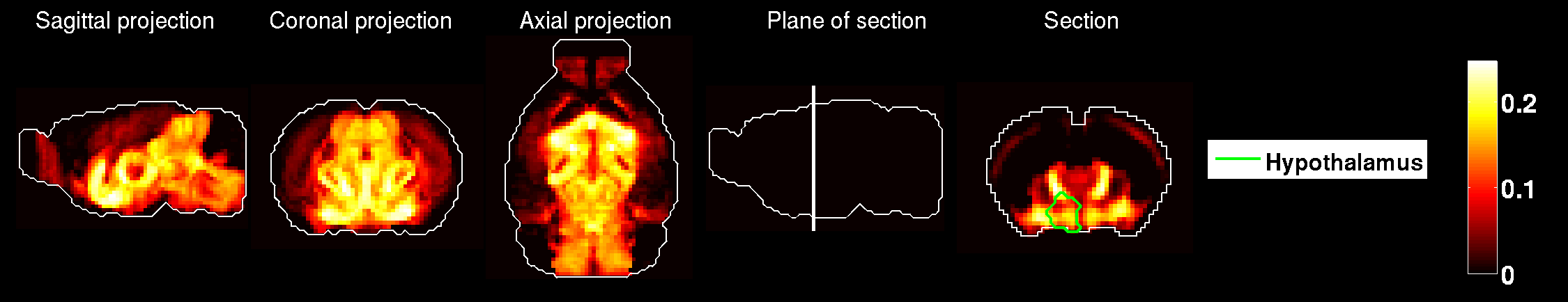}}\\
 \subfloat[]{\includegraphics[width=\textwidth]{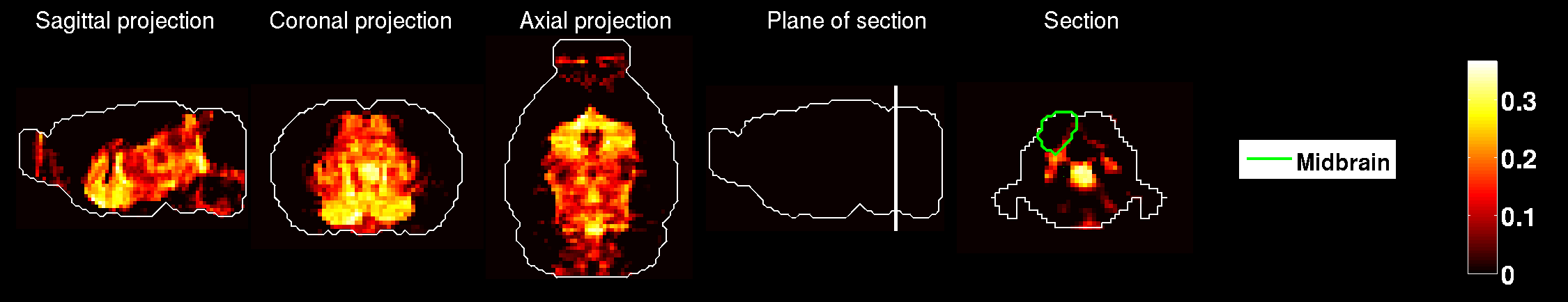}}\\
 \caption{ {\bf{GABAergic interneurons, SST+ (cell-type index 57).}} (a) Heat map of the brain-wide correlation profile. (b) Heat map of the estimated brain-wide density profile.}
  \label{type57CorrDensity}
\end{figure}
\begin{figure}
\centering
 \subfloat[]{\includegraphics[width=\widthCoeff\textwidth]{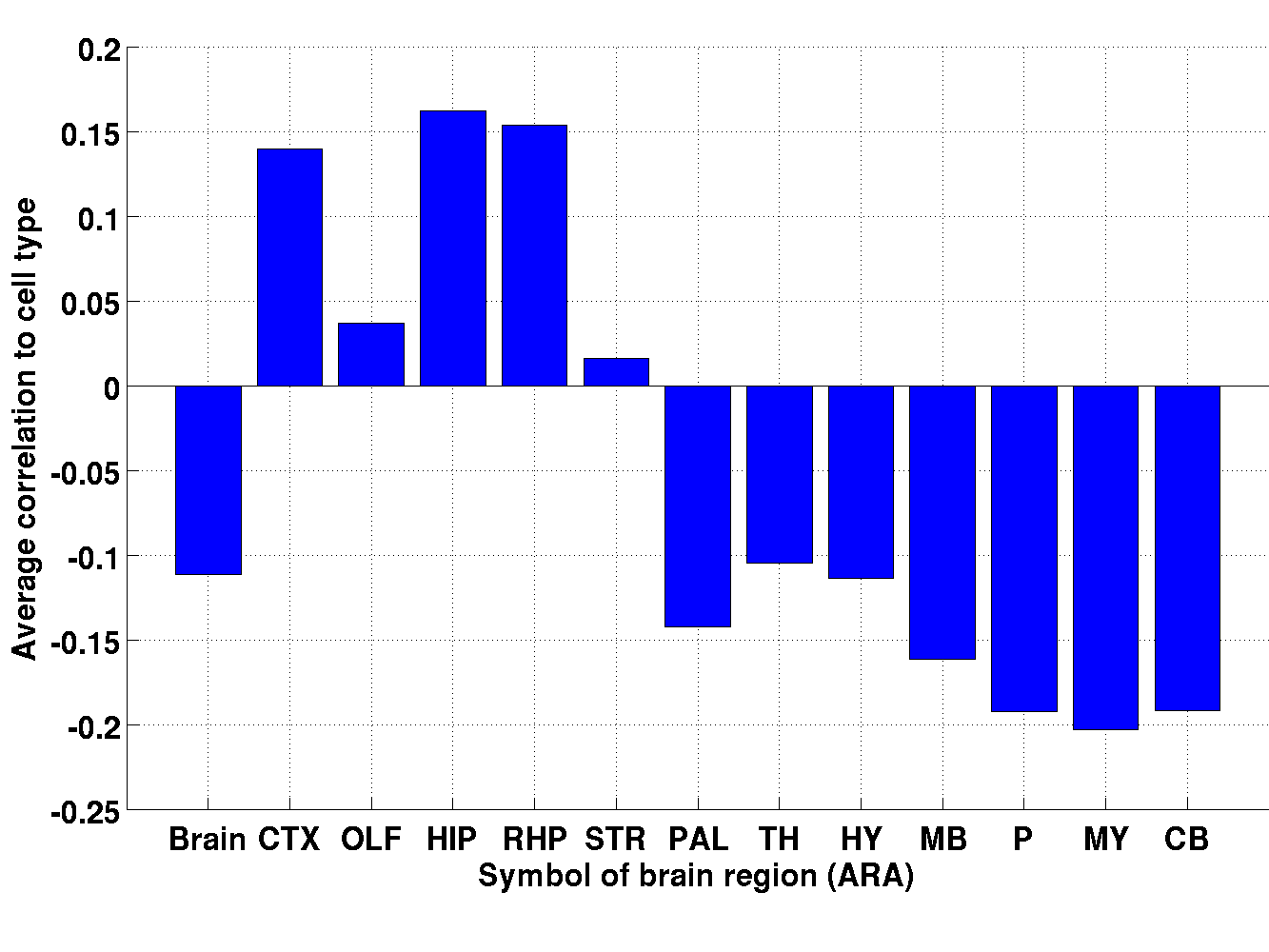}}\\
 \subfloat[]{\includegraphics[width=\widthCoeff\textwidth]{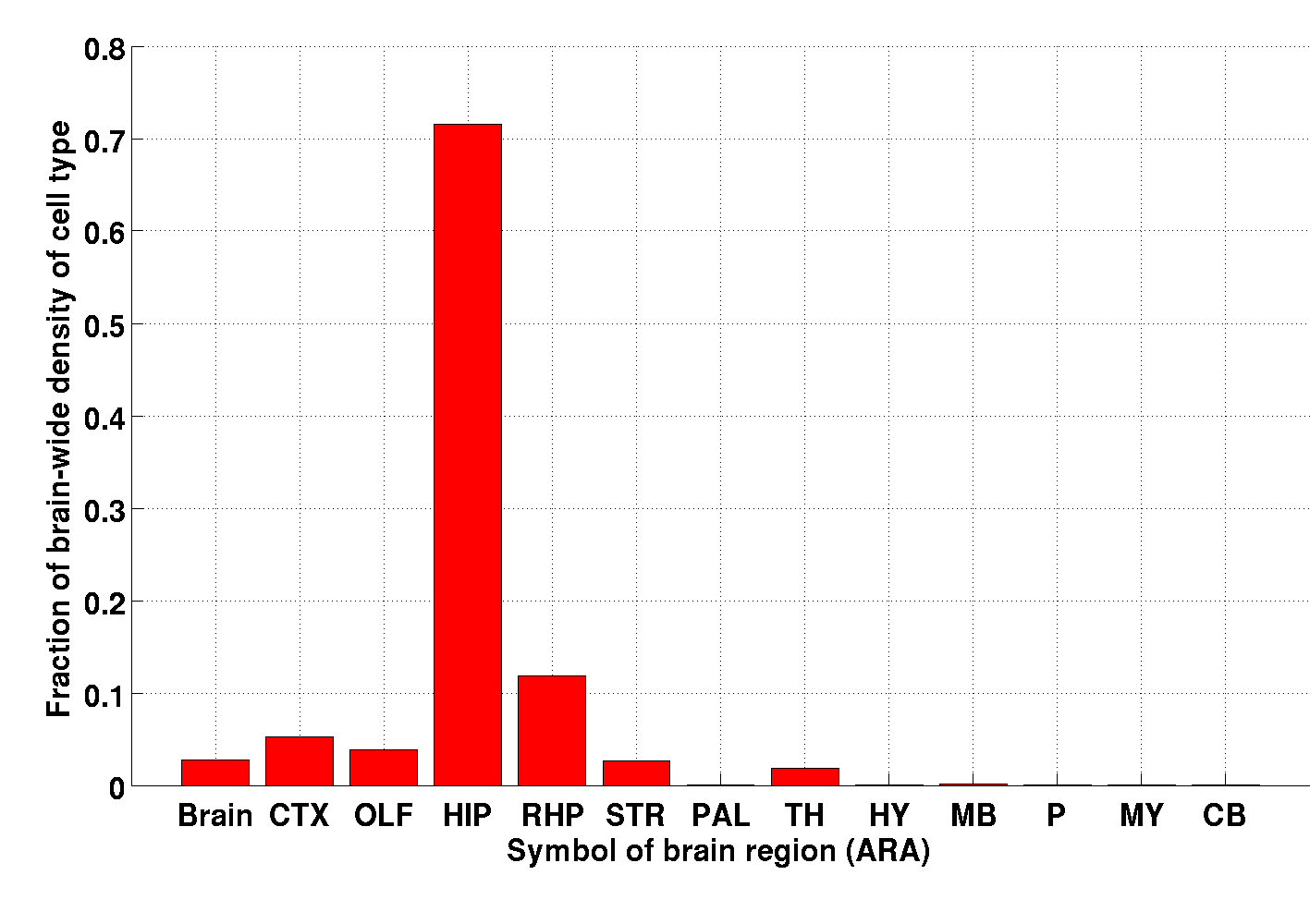}}\\
 \caption{{\bf{Pyramidal neurons (cell-type index 49).}} Average correlation between the cell type and the Allen Atlas, in the regions 
of the \bigTwelveSpace annotation of the ARA. The estimated density of this cell type is zero in the left hemisphere.}
  \label{type49Anatomy}
\end{figure}
\begin{figure}
\centering
 \subfloat[]{\includegraphics[width=\widthCoeff\textwidth]{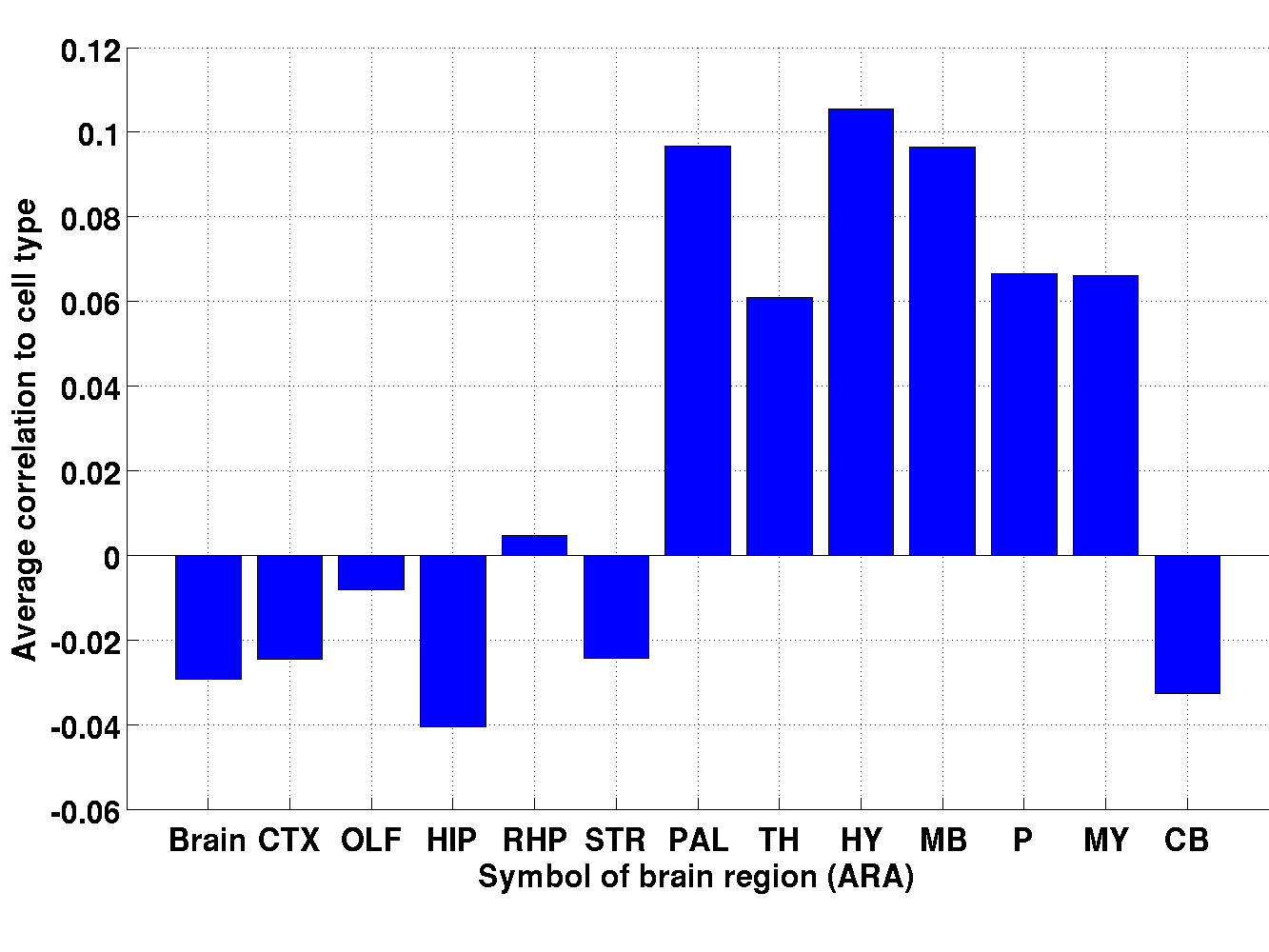}}\\
 \subfloat[]{\includegraphics[width=\widthCoeff\textwidth]{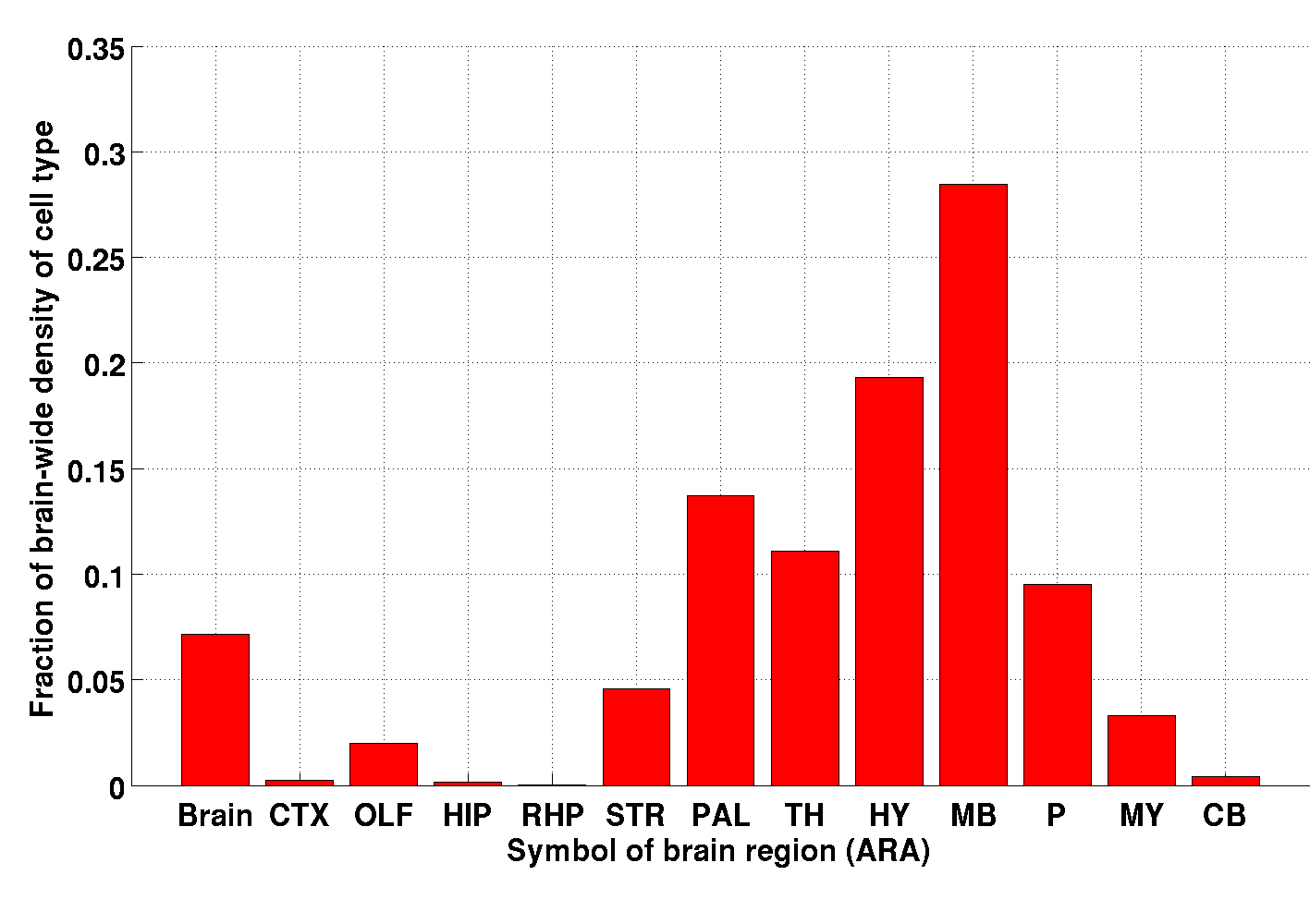}}\\
 \caption{{\bf{GABAergic interneurons, SST+ (cell-type index 57).}} Average correlation between the cell type and the Allen Atlas, in the regions 
of the \bigTwelveSpace annotation of the ARA.}
  \label{type57Anatomy}
\end{figure}

\clearpage
\subsubsection{Striatum}
%DUsed = D( :, colsToUseInAllen );
%MUsed = M( :, colsToUseInTypes );
%classifyPatternStriatum = classify_pattern_striatum( Ref, cellTypesCorrelations, fitVoxelsToTypes )

Three cell-type-specific samples (cholinergic neurons -- index 13,
{\emph{Drd1}}+ medium spiny neurons -- index 15, and {\emph{Drd2}}+
medium spiny neurons -- index 16) in the data set were extracted from
the striatum.

\begin{table}
\centering
\begin{tabular}{|m{0.14\textwidth}|m{0.11\textwidth}|m{0.22\textwidth}|m{0.22\textwidth}|m{0.14\textwidth}|m{0.14\textwidth}|}
\hline
\textbf{Description (index)}&\textbf{Origin of sample}&\textbf{Rank of region (out of 12) in the {\bigTwelveSpace}annotation (by correlation)}& \textbf{Rank of region (out of 12) in the {\bigTwelveSpace} annotation (by density)}&\textbf{Fraction of density supported in the striatum  } \\\hline
Cholinergic neurons (13) & Striatum &  7 & 6  & 3.3\% \\\hline
{\emph{Drd1}}+ medium spiny neurons (15) & Striatum  &  1 &  2& 30\% \\\hline
{\emph{Drd2}}+ medium spiny neurons (16) & Striatum &  1&  1& 91.4\% \\
\hline
\end{tabular}
\caption{Anatomical analysis for the cell-type-specific samples extracted from the striatum.}
\label{metadataAnatomyStriatum}
\end{table}

\begin{figure}
\centering
 \subfloat[]{\includegraphics[width=\textwidth]{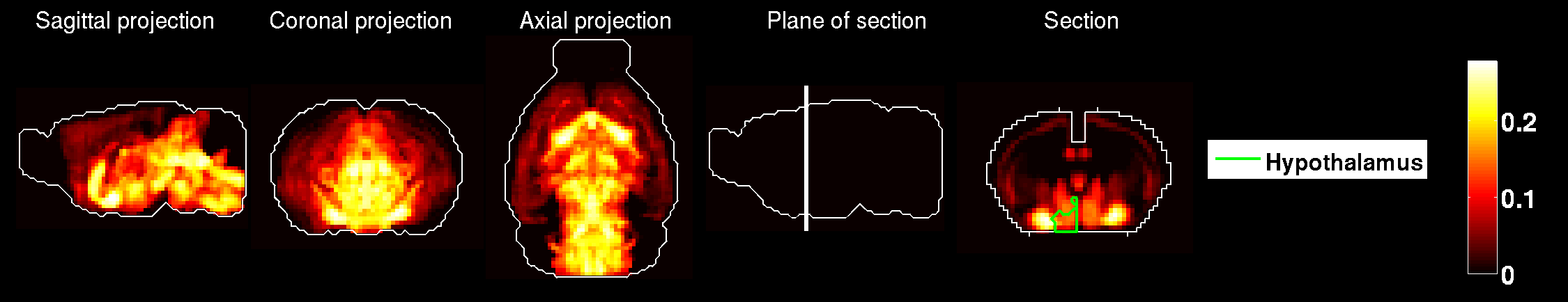}}\\
 \subfloat[]{\includegraphics[width=\textwidth]{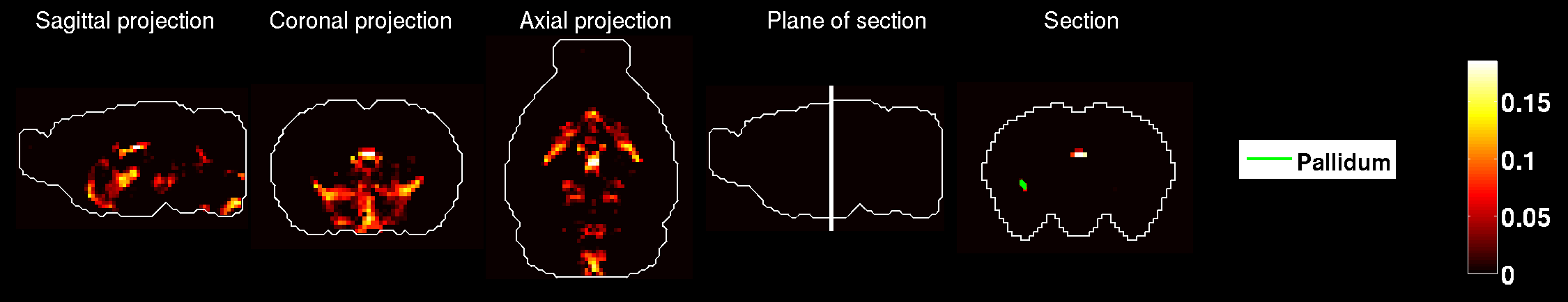}}\\
 \caption{ {\bf{Cholinergic neurons (cell-type index 13).}} (a) Heat map of the brain-wide correlation profile. (b) Heat map of the estimated brain-wide density profile.}
  \label{type13CorrDensity}
\end{figure}

The pallidum is the region in the \bigTwelveSpace annotation that supports the
highest fraction (43\%) of the density of cholinergic neurons (index
13), and the striatum ranks only 7th by correlation and 6th by density
(see Table \ref{metadataAnatomyStriatum} Figures
\ref{type13CorrDensity} and \ref{type13Anatomy}).

\begin{figure}
\centering
 \subfloat[]{\includegraphics[width=\textwidth]{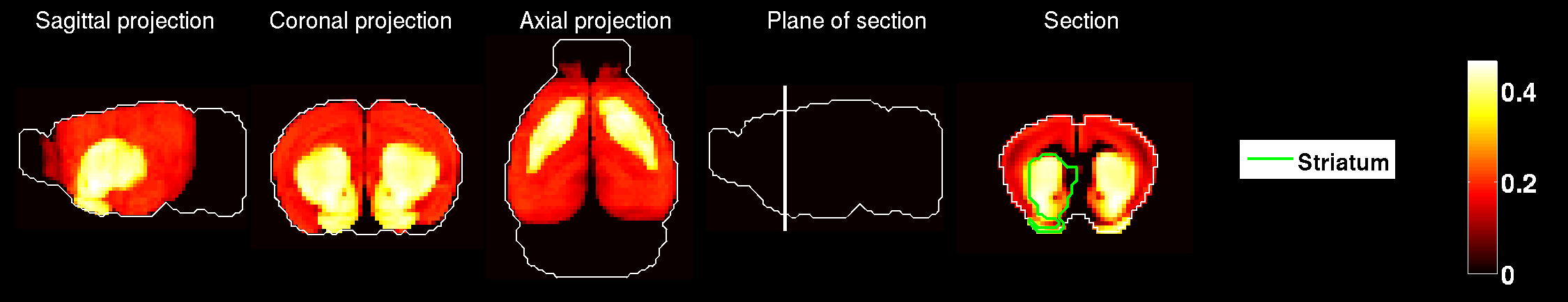}}\\
 \subfloat[]{\includegraphics[width=\textwidth]{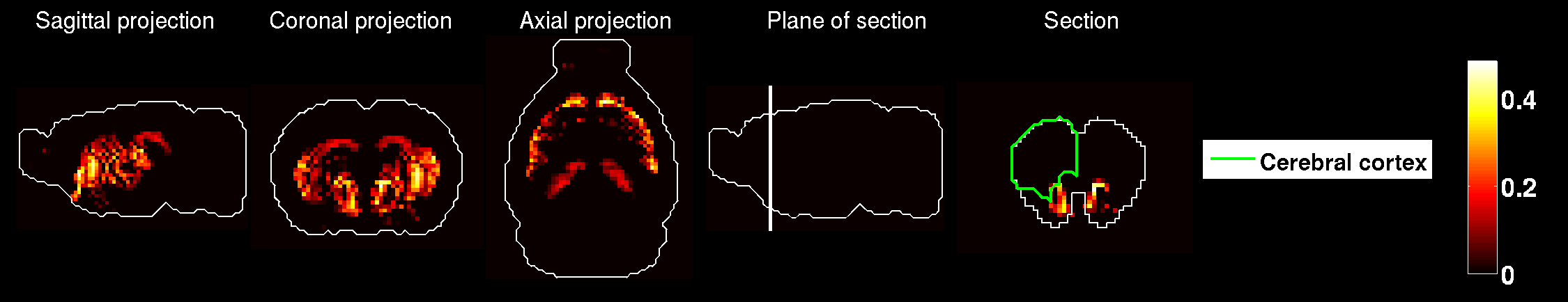}}\\
 \caption{ {\bf{{\emph{Drd1}}+ medium spiny neurons (cell-type index 15).}} (a) Heat map of the brain-wide correlation profile. (b) Heat map of the estimated brain-wide density profile.}
  \label{type15CorrDensity}
\end{figure}

\begin{figure}
\centering
 \subfloat[]{\includegraphics[width=\textwidth]{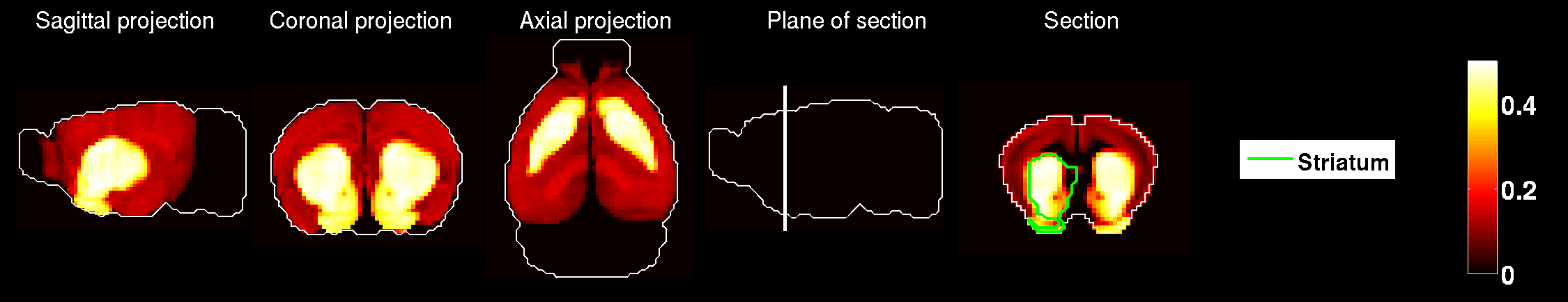}}\\
 \subfloat[]{\includegraphics[width=\textwidth]{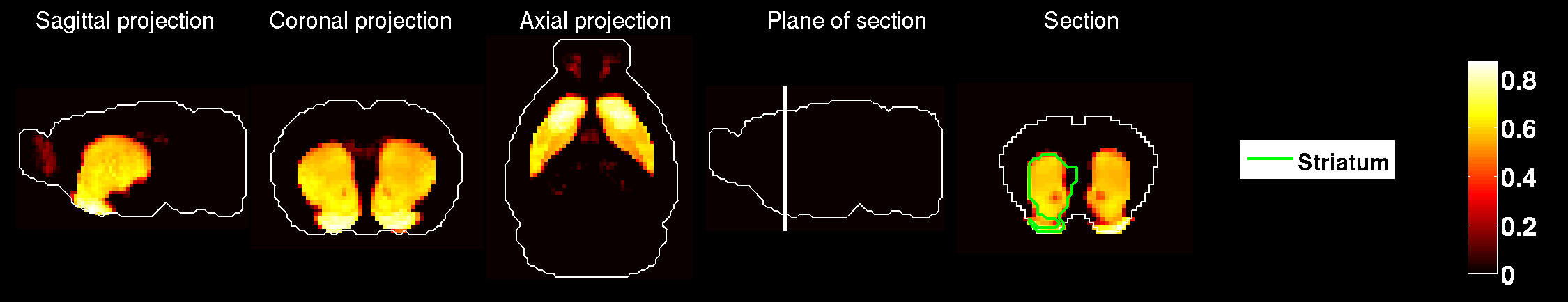}}\\
 \caption{ {\bf{ {\emph{Drd2}}+ medium spiny neurons (cell-type index 16).}} (a) Heat map of the brain-wide correlation profile. (b) Heat map of the estimated brain-wide density profile.}
  \label{type16CorrDensity}
\end{figure}

\begin{figure}
\centering
 \subfloat[]{\includegraphics[width=\widthCoeff\textwidth]{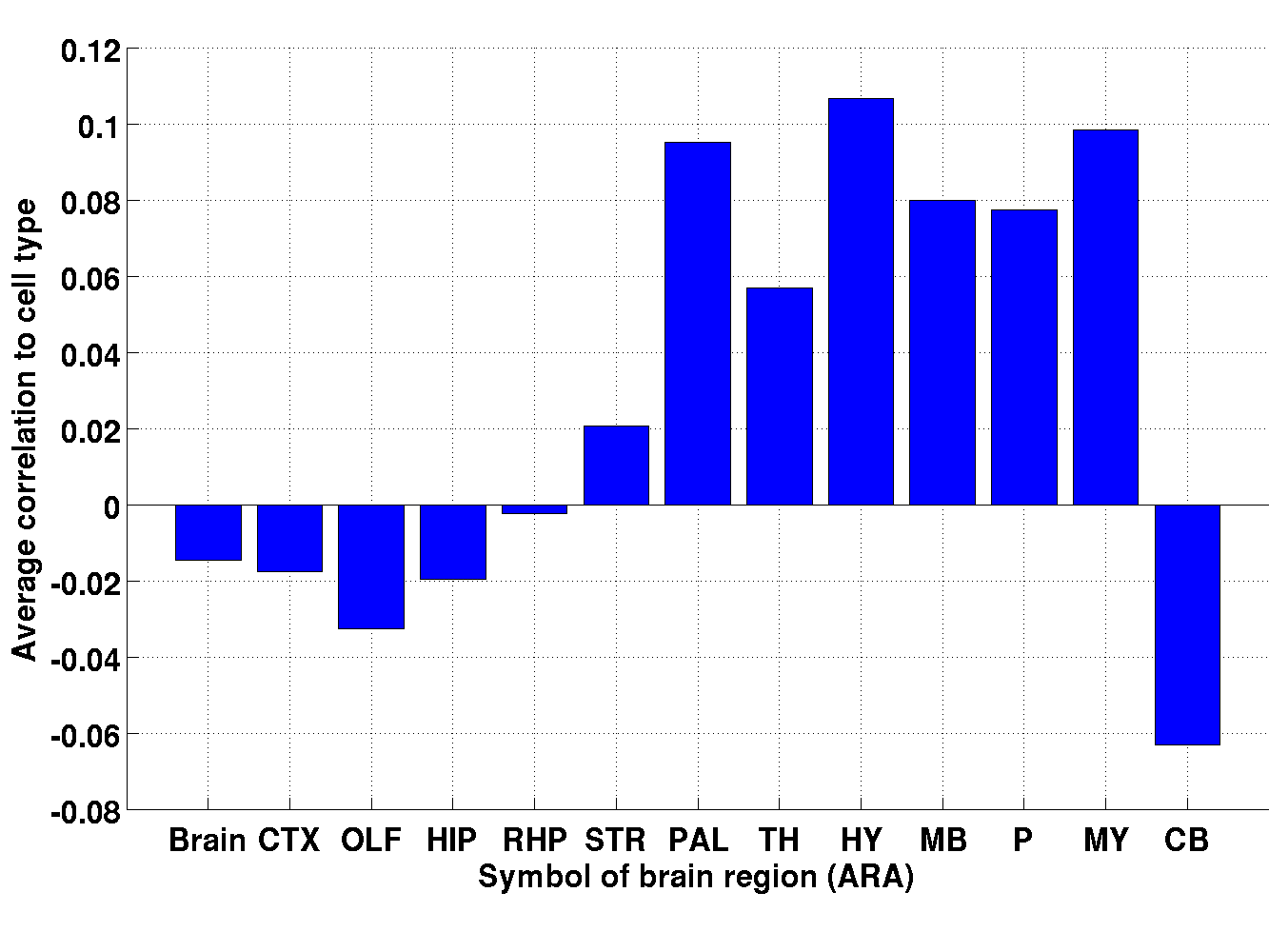}}\\
 \subfloat[]{\includegraphics[width=\widthCoeff\textwidth]{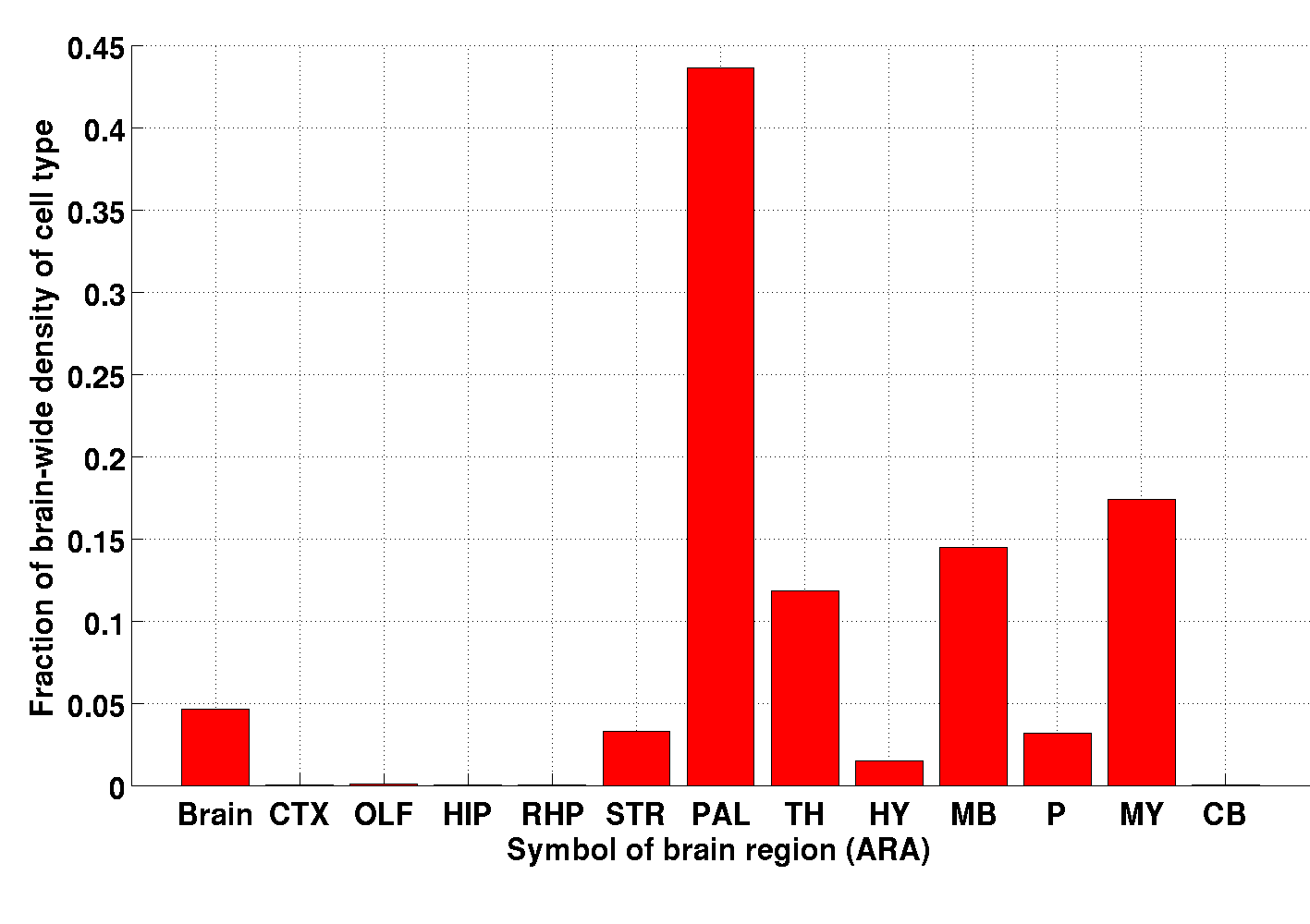}}\\
 \caption{{\bf{ Cholinergic neurons (cell-type index 13).}} (a) Average correlation between the cell type and the Allen Atlas, in the regions 
of the \bigTwelveSpace annotation of the ARA.
 (b) Fractions of density of cell type in the regions 
of the \bigTwelveSpace annotation of the ARA. }
  \label{type13Anatomy}
\end{figure}

\begin{figure}
\centering
 \subfloat[]{\includegraphics[width=\widthCoeff\textwidth]{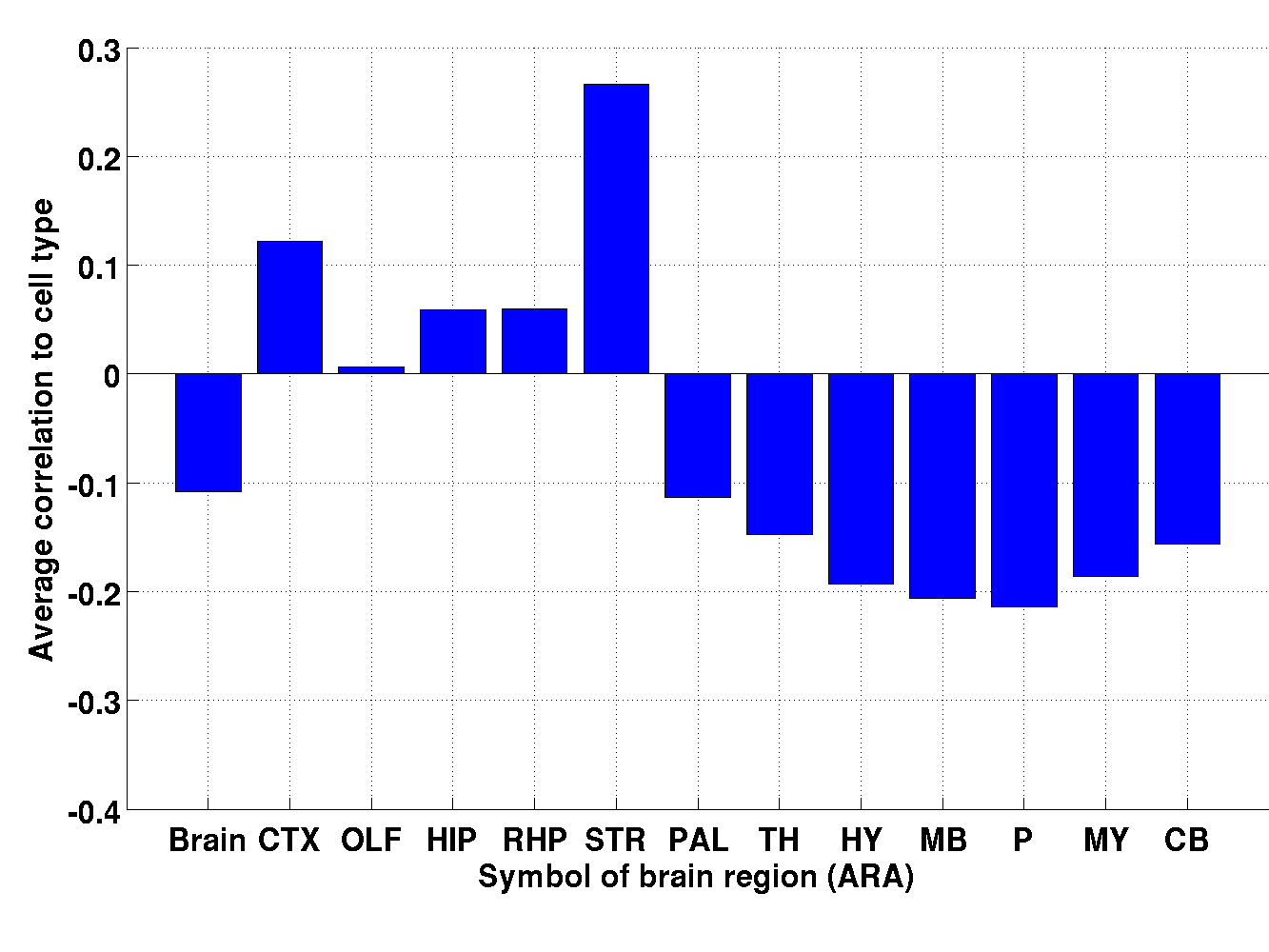}}\\
 \subfloat[]{\includegraphics[width=\widthCoeff\textwidth]{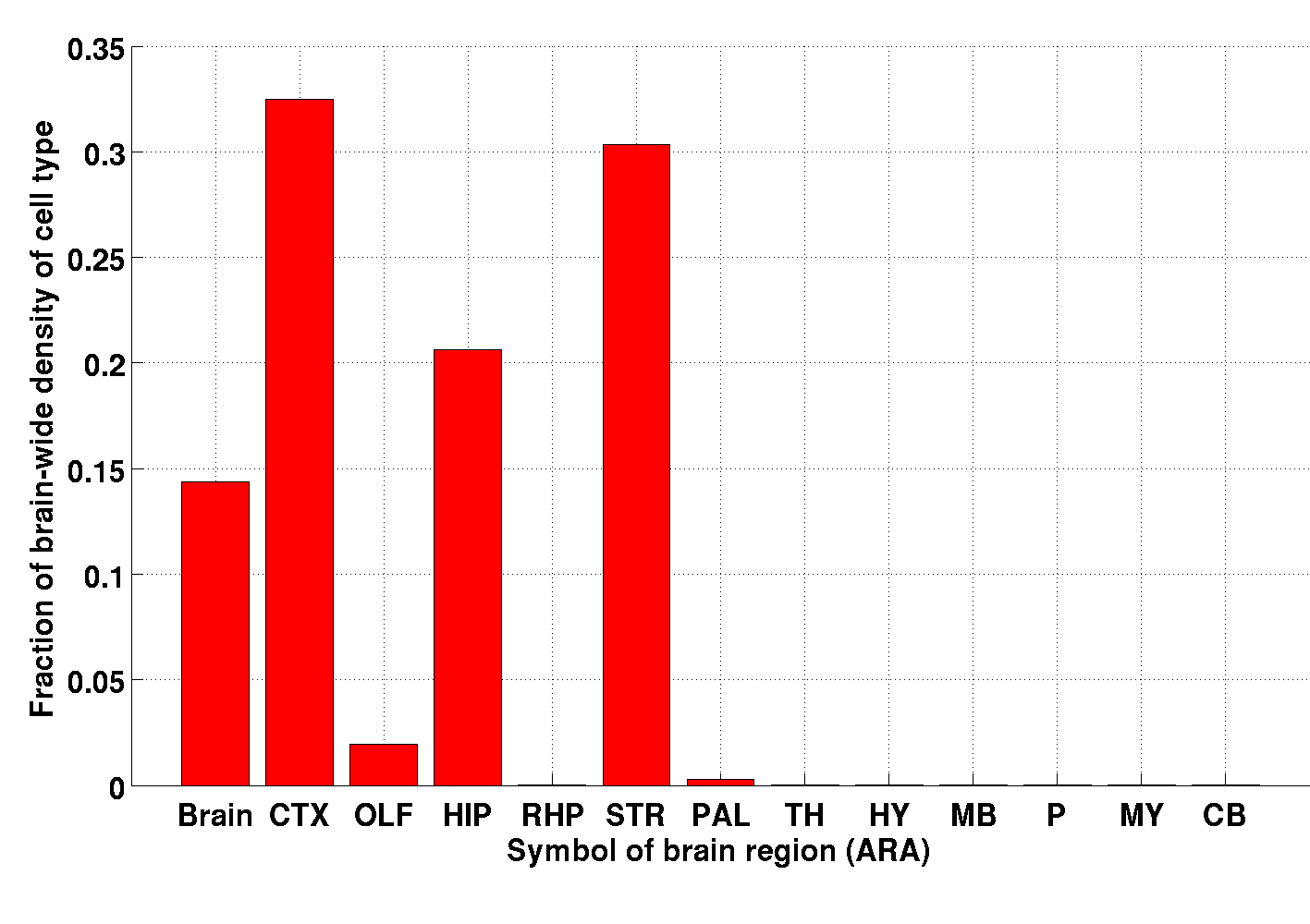}}\\
 \caption{{\bf{ {\emph{Drd1}}+ medium spiny neurons (cell-type index 15).}} (a) Average correlation between the cell type and the Allen Atlas, in the regions 
of the \bigTwelveSpace annotation of the ARA.
 (b) Fractions of density of cell type in the regions 
of the \bigTwelveSpace annotation of the ARA. }
  \label{type15Anatomy}
\end{figure}

\begin{figure}
\centering
 \subfloat[]{\includegraphics[width=\widthCoeff\textwidth]{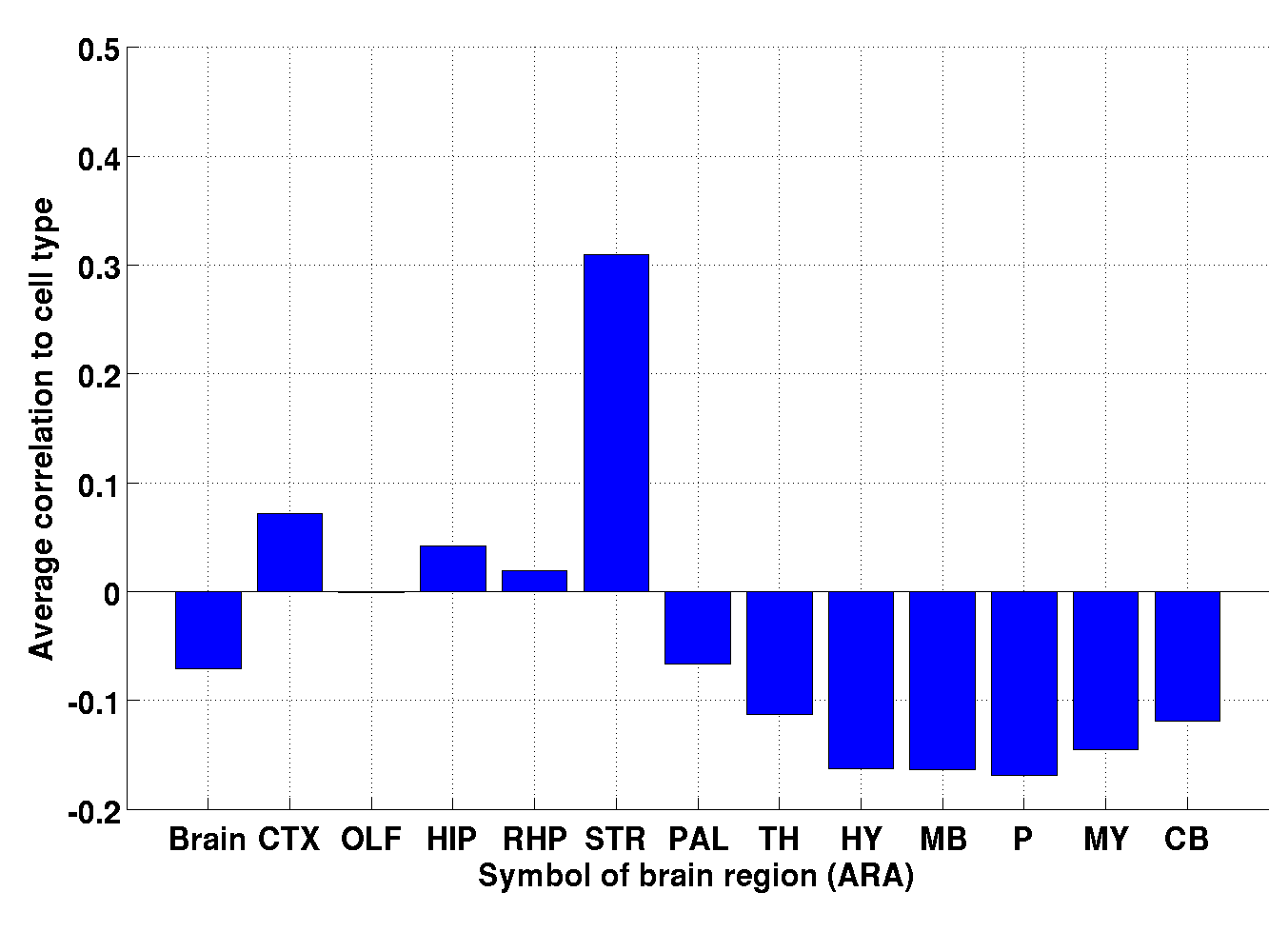}}\\
 \subfloat[]{\includegraphics[width=\widthCoeff\textwidth]{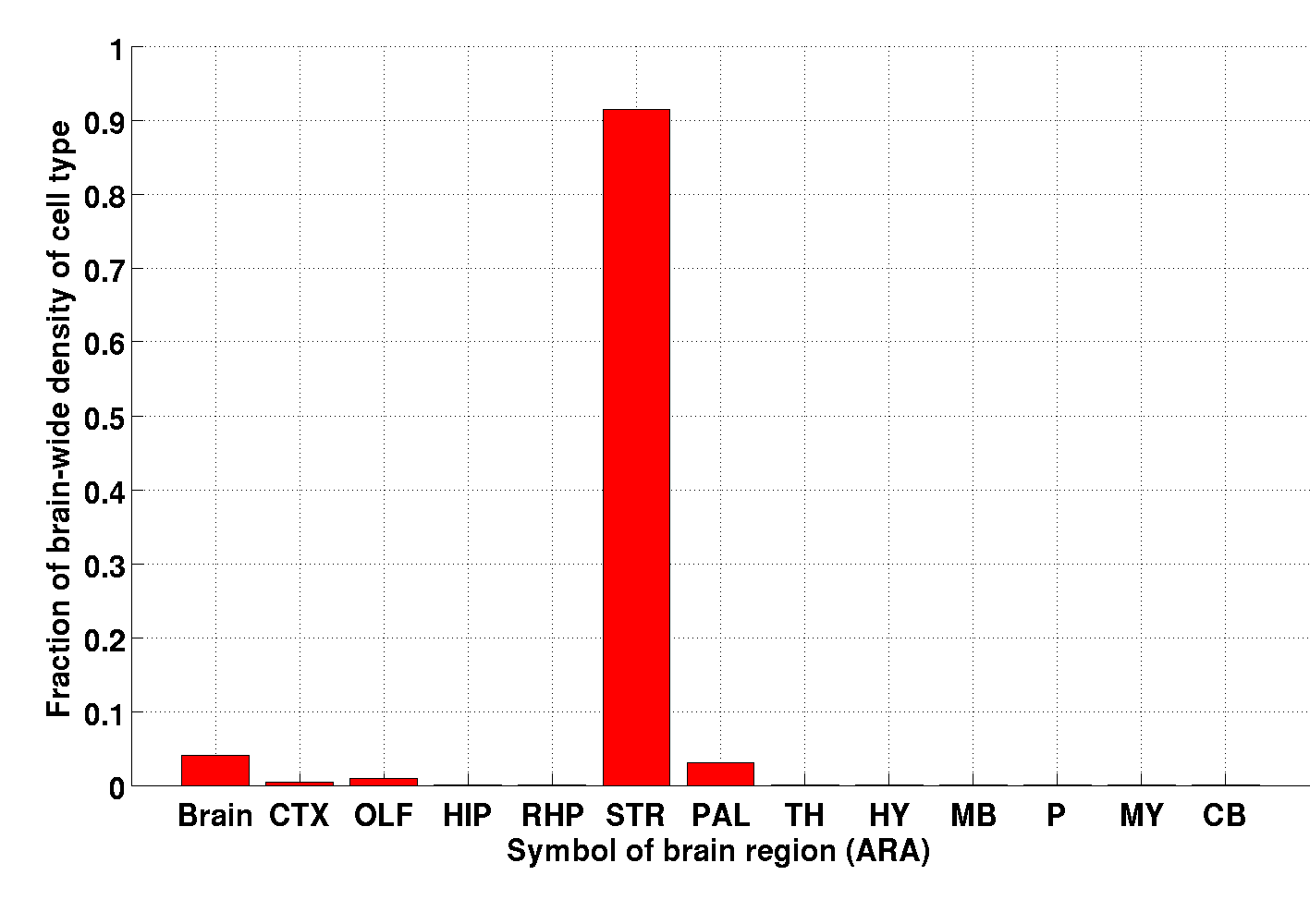}}\\
 \caption{{\bf{ {\emph{Drd2}}+ medium spiny neurons (cell-type index 16).}} (a) Average correlation between the cell type and the Allen Atlas, in the regions 
of the \bigTwelveSpace annotation of the ARA.
 (b) Fractions of density of cell type in the regions 
of the \bigTwelveSpace annotation of the ARA. }
  \label{type16Anatomy}
\end{figure}

%figureStriatum = figure_striatum( Ref, fitVoxelsToTypes, DUsed, MUsed
On the other hand, medium spiny neurons \cite{DoyleCells} expressing
both dopamine receptor types have striatum as their top region by
correlations (see Figures \ref{type15CorrDensity},
\ref{type16CorrDensity}, \ref{type15Anatomy}, and
\ref{type16Anatomy}).  The striatum is the top region by density for
    {\emph{Drd2}}+ medium spiny neurons, and the second region by
    density for {\emph{Drd1}}+ medium spiny neurons, after the
    cerebral cortex.



\clearpage
\subsubsection{Pallidum}
\begin{figure}
\centering
 \subfloat[]{\includegraphics[width=\textwidth]{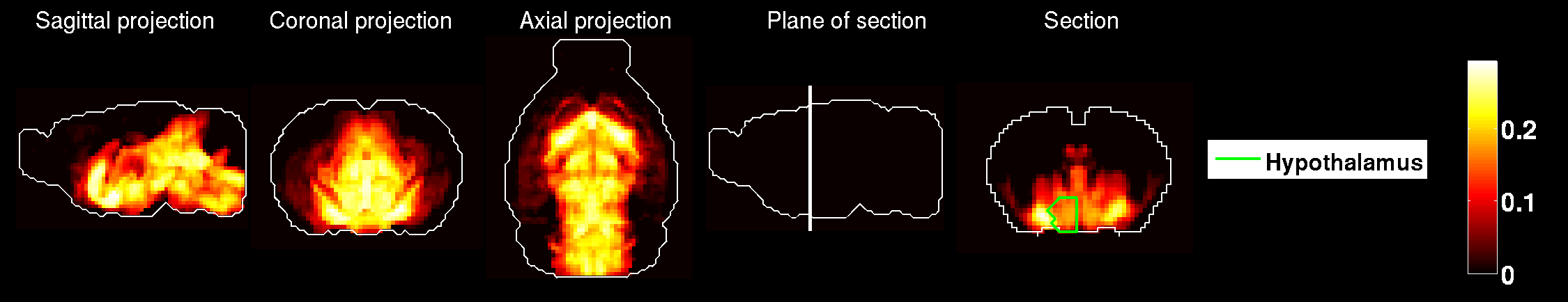}}\\
 \subfloat[]{\includegraphics[width=\textwidth]{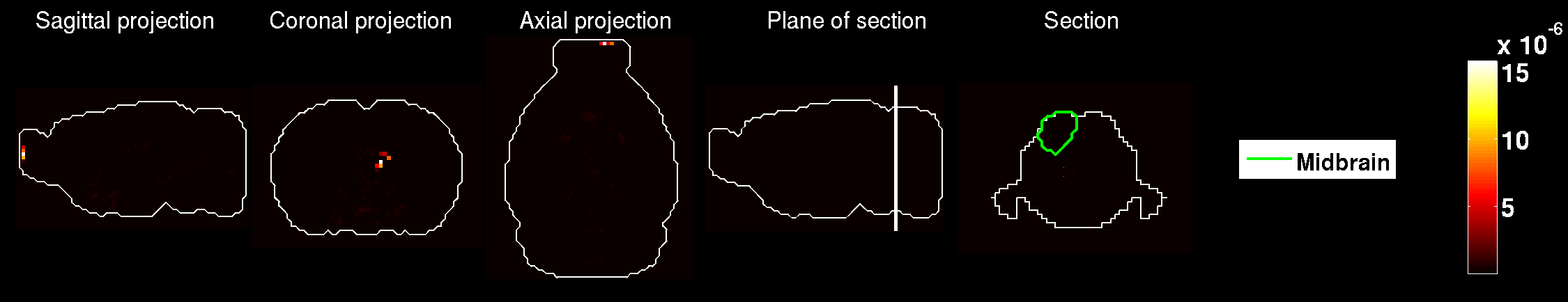}}\\
 \caption{ {\bf{Cholinergic projection neurons (cell-type index 11).}} (a) Heat map of the brain-wide correlation profile. (b) Heat map of the estimated brain-wide density profile.}
  \label{type11CorrDensity}
\end{figure}

\begin{figure}
\centering
 \includegraphics[width=\widthCoeff\textwidth]{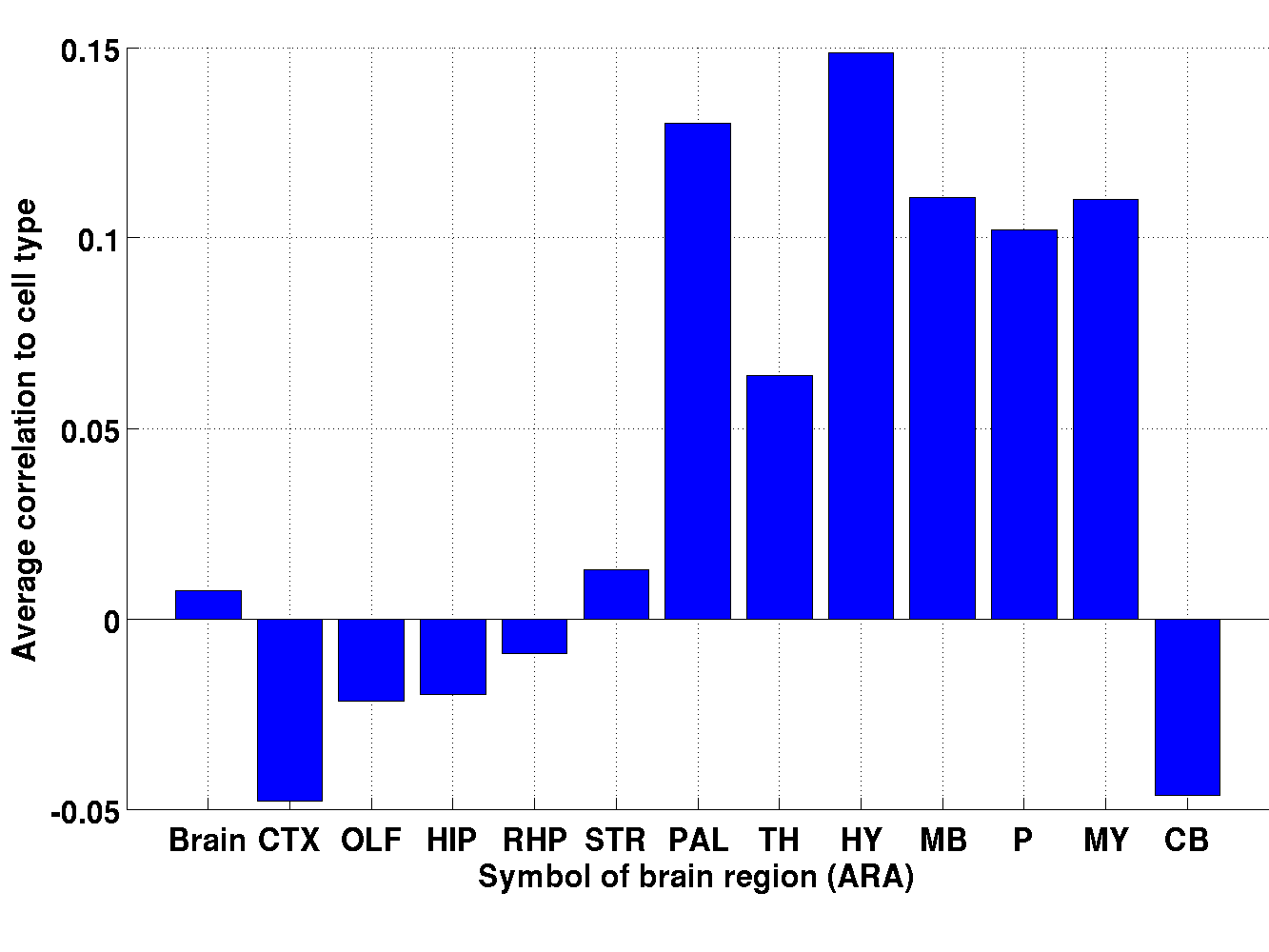}
% \subfloat[]{\includegraphics[width=0.8\textwidth]{densityBig12Type11.png}}\\
 \caption{{\bf{Cholinergic projection neurons (cell-type index 11).}} Average correlation between the cell type and the Allen Atlas, in the regions 
of the \bigTwelveSpace annotation of the ARA. The estimated density of this cell type is zero in the left hemisphere.}
  \label{type11Anatomy}
\end{figure}

 The cell-type-specific sample with index 11 (Cholinergic projection
 neurons), which is the only sample obtained from the pallidum, has
 the hypothalamus as its top region by correlation, followed by
 pallidum (see Figures \ref{type11CorrDensity} and
 \ref{type11Anatomy}).  Pallidum ventral region, which is the region
 in the {\ttfamily{'fine'}} annotation from which the cell-type-specific sample
 was extracted, ranks 30th out of 94 regions by average correlation
 (see Table \ref{metadataAnatomyPallidum}).  This sugests that
 cholinergic projection neurons are not very region specific. Moreover,
 the estimated density of this cell type is zero in the left
 hemisphere.
\begin{table}
\centering
\begin{tabular}{|m{0.14\textwidth}|m{0.14\textwidth}|m{0.14\textwidth}|m{0.14\textwidth}|m{0.14\textwidth}|m{0.14\textwidth}|}
\hline
\textbf{Description (index)}&\textbf{Origin of sample}&\textbf{Rank of region (out of 94) in the 'fine' annotation (by correlation)}&\textbf{Rank of region (out of 94) in the 'fine' annotation (by density)}&\textbf{Fraction of density in the region (\%)}& \textbf{Fraction of density supported in the pallidum}\\ \hline 
 Cholinergic projection neurons (11) &  Pallidum ventral region &  30  &   N/A  &  N/A  &   N/A\% \\
\hline
\end{tabular}
\caption{Anatomical analysis for the cell-type-specific sample extracted from the pallidum.}
\label{metadataAnatomyPallidum}
\end{table}

\clearpage
\subsubsection{Thalamus}
\begin{figure}
\centering
 \subfloat[]{\includegraphics[width=\textwidth]{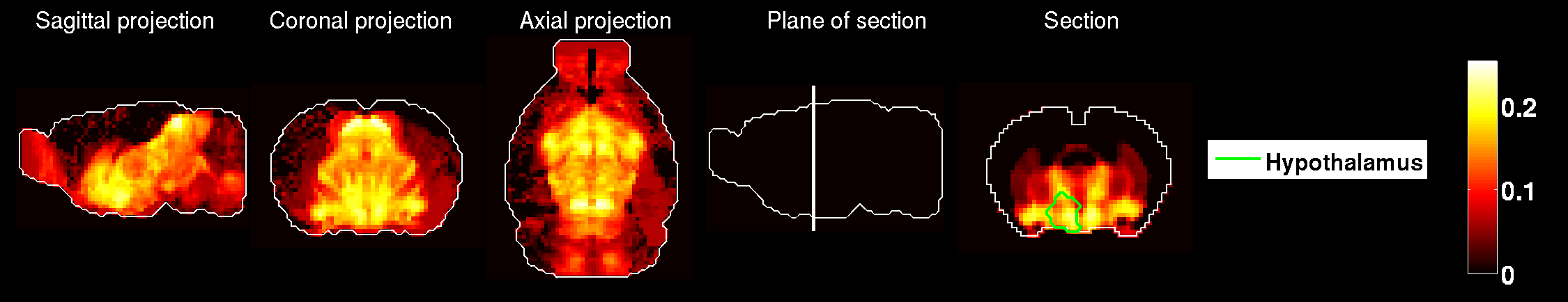}}\\
 \subfloat[]{\includegraphics[width=\textwidth]{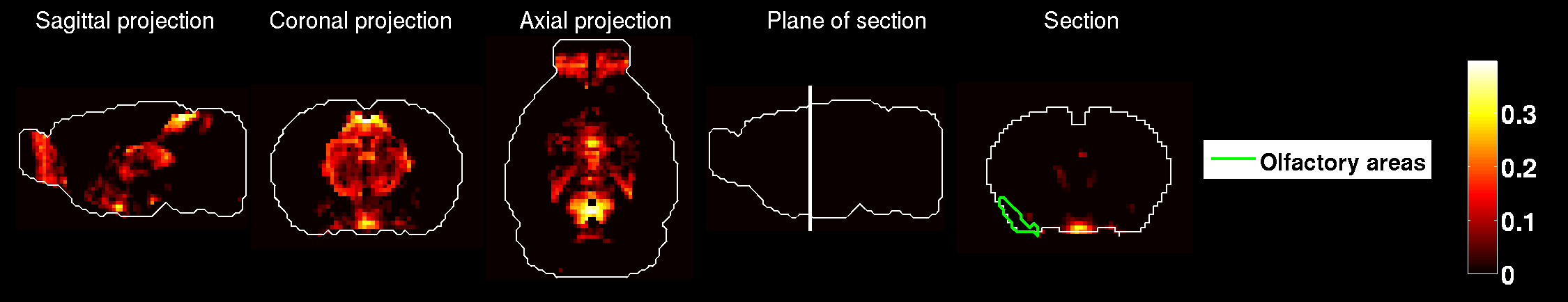}}\\
 \caption{ {\bf{GABAergic Interneurons, PV+ (cell-type index 59).}} (a) Heat map of the brain-wide correlation profile. (b) Heat map of the estimated brain-wide density profile.}
  \label{type59CorrDensity}
\end{figure}

\begin{figure}
\centering
 \subfloat[]{\includegraphics[width=\widthCoeff\textwidth]{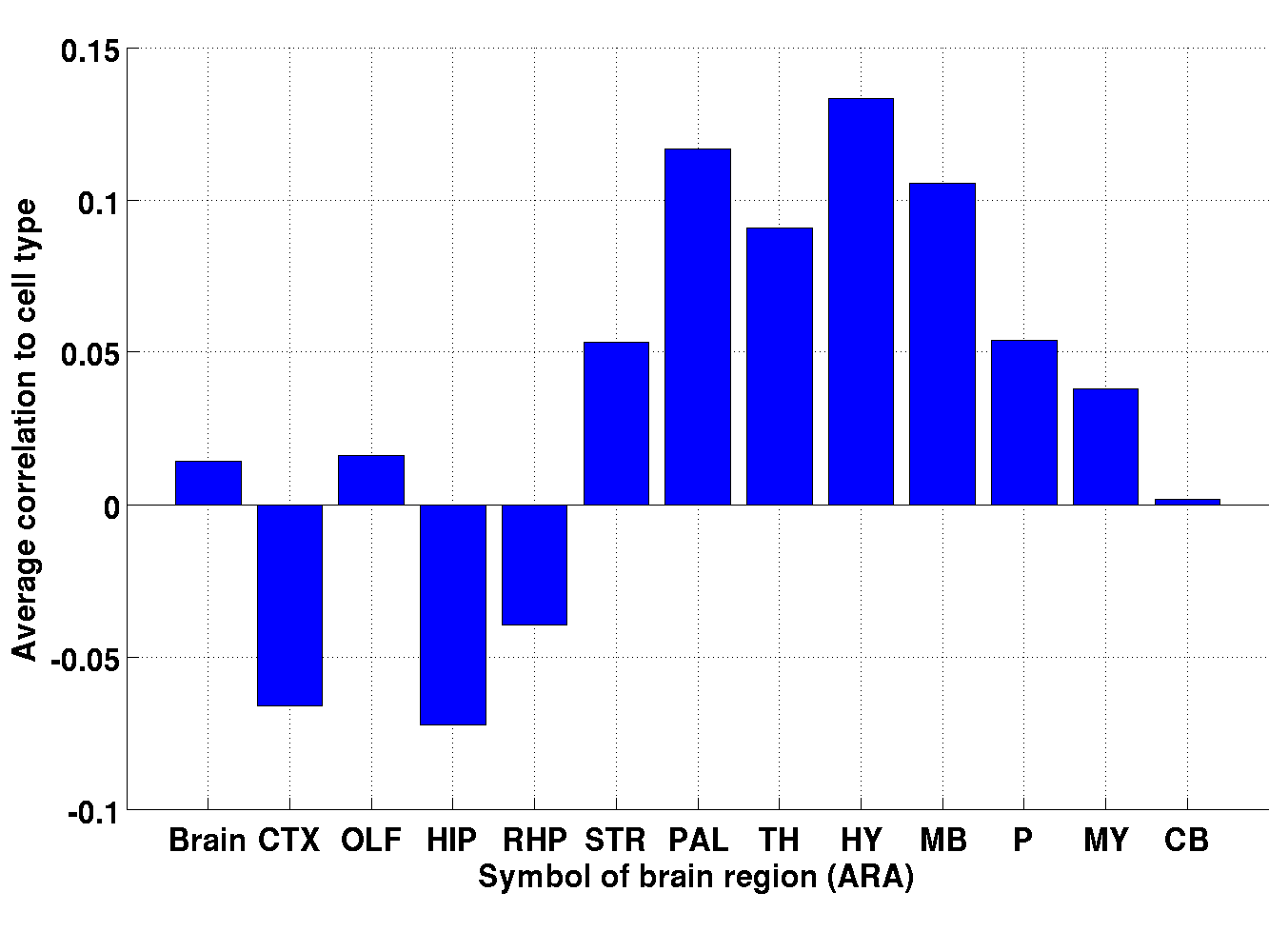}}\\
 \subfloat[]{\includegraphics[width=\widthCoeff\textwidth]{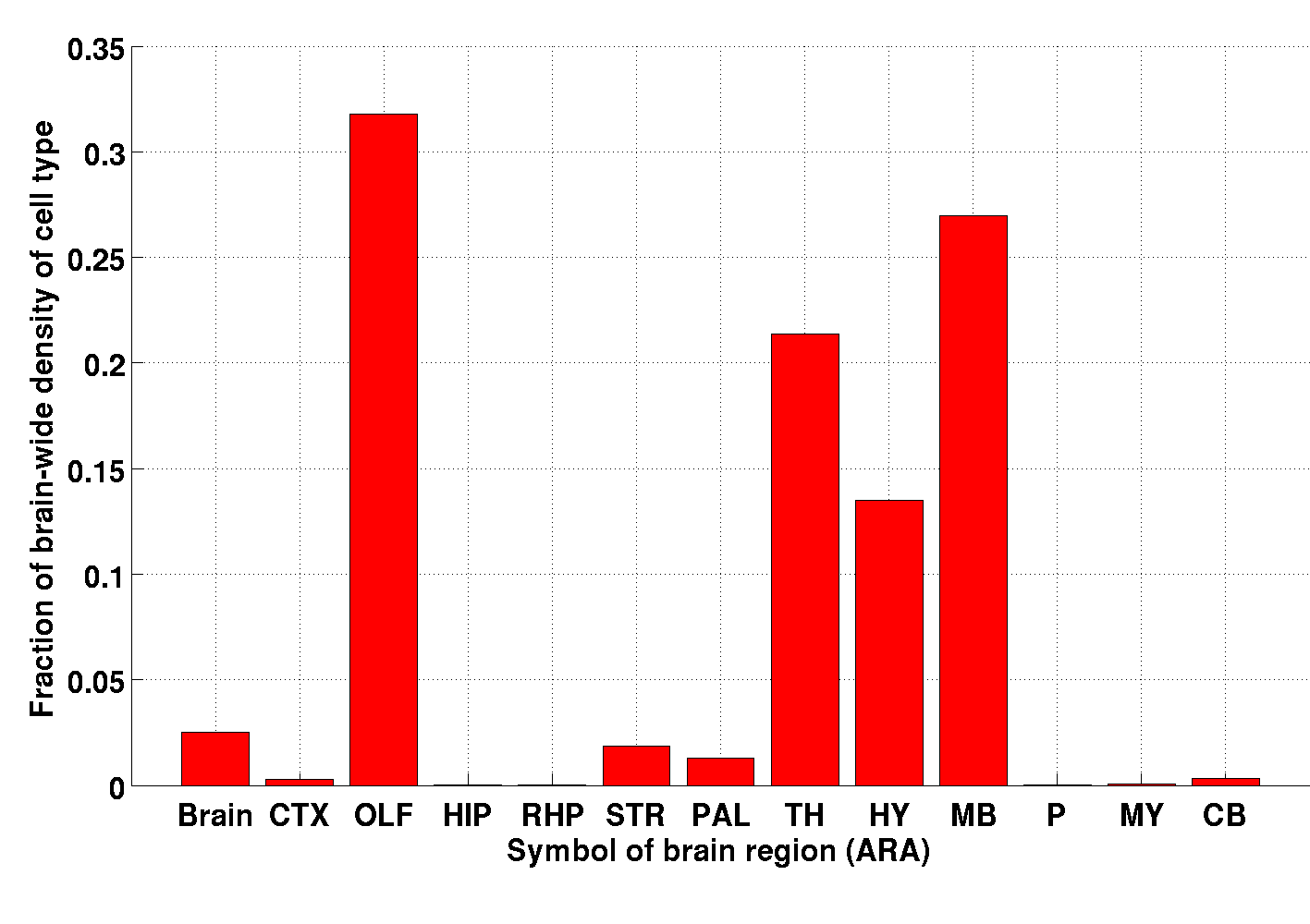}}\\
 \caption{{\bf{GABAergic Interneurons, PV+ (cell-type index 59).}} (a) Average correlation between the cell type and the Allen Atlas, in the regions 
of the \bigTwelveSpace annotation of the ARA.
 (b) Fractions of density of cell type in the regions 
of the \bigTwelveSpace annotation of the ARA.}
  \label{type59Anatomy}
\end{figure}

Table \ref{metadataAnatomyThalamus} summarizes results for the
cell-type-specific sample extracted from the thalamus (GABAergic Interneurons,
PV+, index 59). Thalamus supports less of the density profile than
Olfactory areas and Midbrain (it is ranked third by density and
 fourth by correlation). Moreover, there seems to be little
solidarity between the ranking of regions by correlation and by
density (see Figures \ref{type59CorrDensity} and \ref{type59Anatomy}).
\begin{table}
\centering
\begin{tabular}{|m{0.14\textwidth}|m{0.14\textwidth}|m{0.14\textwidth}|m{0.14\textwidth}|m{0.14\textwidth}|m{0.14\textwidth}|}
\hline
\textbf{Description (index)}&\textbf{Origin of sample}&\textbf{Rank of region (out of 94) in the 'fine' annotation (by correlation)}&\textbf{Rank of region (out of 94) in the 'fine' annotation (by density)}&\textbf{Fraction of density in the region}& \textbf{Fraction of density supported in the thalamus}\\ \hline 
 GABAergic Interneurons, PV+ (59) &   Dorsal part of the lateral geniculate complex &  63  &   31  &  0.2\% &   21.3\% \\
\hline
\end{tabular}
\caption{Anatomical analysis for the cell-type-specific sample extracted from the thalamus.}
\label{metadataAnatomyThalamus}
\end{table}

\clearpage
\subsubsection{Midbrain}

\begin{figure}
\centering
 \subfloat[]{\includegraphics[width=\textwidth]{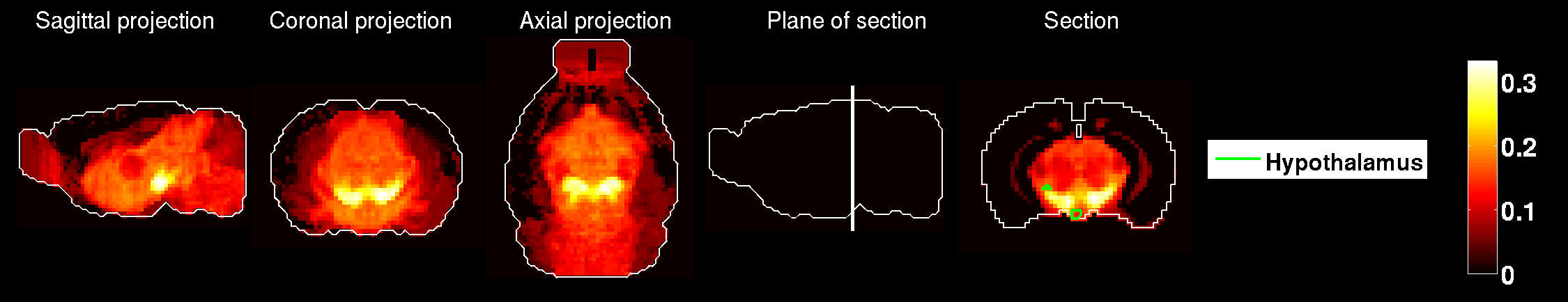}}\\
 \subfloat[]{\includegraphics[width=\textwidth]{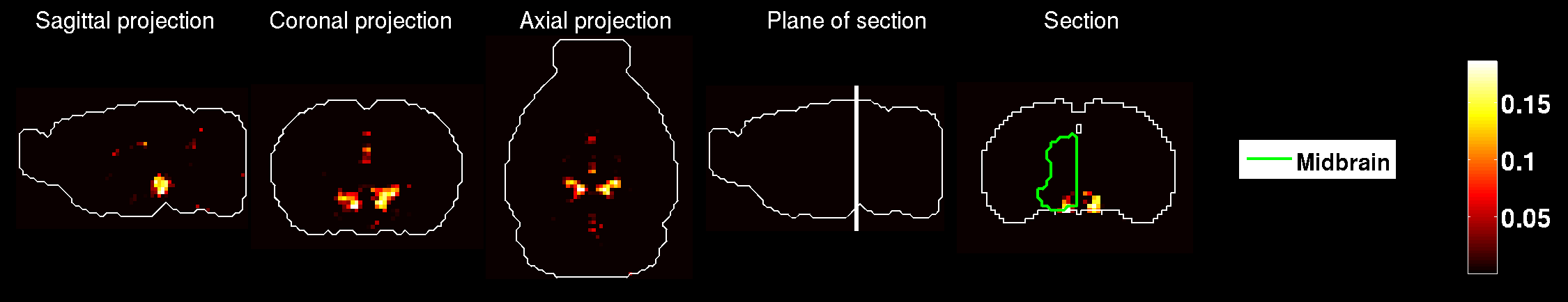}}\\
 \caption{ {\bf{A9 dopaminergic neurons (cell-type index 4).}} (a) Heat map of the brain-wide correlation profile. (b) Heat map of the estimated brain-wide density profile.}
  \label{type4CorrDensity}
\end{figure}

\begin{figure}
\centering
 \subfloat[]{\includegraphics[width=\textwidth]{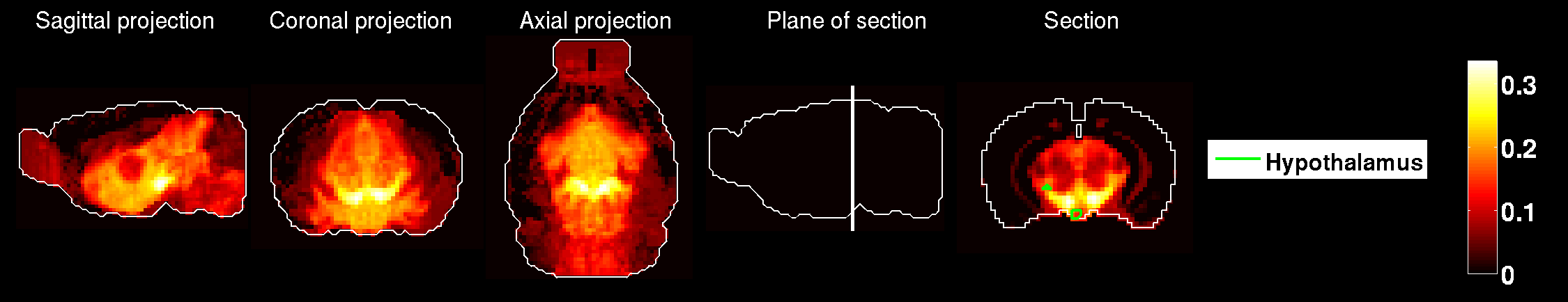}}\\
 \subfloat[]{\includegraphics[width=\textwidth]{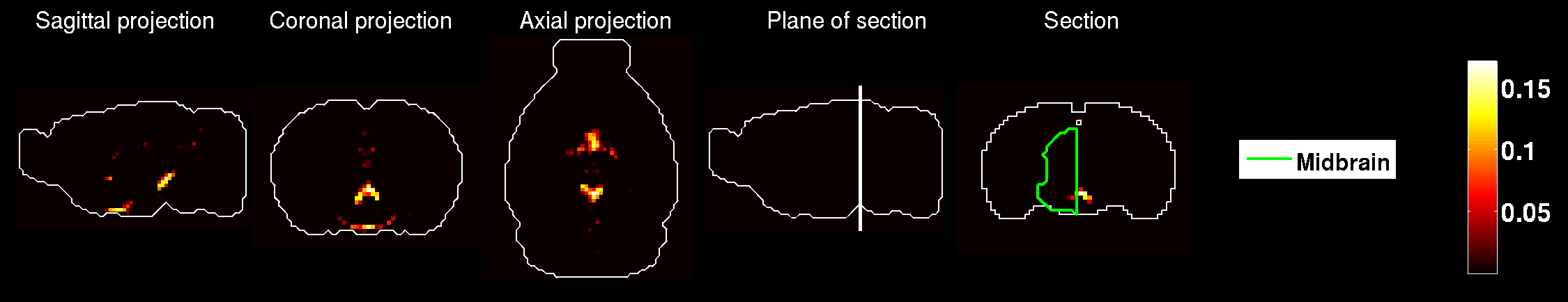}}\\
 \caption{ {\bf{A10 dopaminergic neurons (cell-type index 5).}} (a) Heat map of the brain-wide correlation profile. (b) Heat map of the estimated brain-wide density profile.}
  \label{type5CorrDensity}
\end{figure}

Three cell-type-specific samples were extracted from the midbrain.
Out of them, two (A9 dopaminergic neurons --index 4-- and A10
dopaminergic neurons --index 5) have the midbrain as their top region
by density (see Figures \ref{type4CorrDensity},
\ref{type5CorrDensity}, \ref{type4Anatomy} and
\ref{type5Anatomy}). For both of these samples, the midbrain ranks
second by correlation, after the hypothalamus.\\

 However, a visual inspection of the projections of correlation and density profiles
 of  Figures \ref{type4CorrDensity} and \ref{type5CorrDensity}
  shows some heterogeneity across midbrain, with higher values
 in the ventral region of it. The \bigTwelveSpace annotation 
 is too coarse for this heterogeneity to be detected by our ranking procedures.
  We therefore ranked the regions of the \fineSpace annotation are ranked by correlation and
 density according to Equations \ref{correlRegion} and
 \ref{fittingRegion}. The region that is ranked highest by density for
 A9 dopaminergic neurons is the 'Substantia nigra, compact part',
 which is a subregion of midbrain, and is indeed the finest anatomical
 label available (see Table \ref{metadataAnatomyTable1}). This region is
 ranked second by correlation, the first being 'Ventral tegmental
 area' (which is also a subregion of the midbrain).\\

For A10 dopaminergic neurons the region that is ranked first is the
'Hypothalamus' (which is the 'generic' subregion of hypothalamus in
the \fineSpace annotation, assigned to any voxels that are in hypothalamus
but cannot be reliably assigned to a finer subdivision), but the
second region is the 'ventral tegmental area, which is a subregion of
midbrain, and is indeed the finest anatomical label available from
Tables.  It is ranked second by
correlation, the first being 'Ventral tegmental area'.\\
 
Whereas the midbrain supports a majority of the density for A9
dopaminergic neurons and A10 dopaminergic neurons, it supports only
9.8\% of the density in Motor Neurons, Midbrain Cholinergic Neurons
(whereas pons and medulla support 32.6\% and 51.6\% respectively, as
can be see on Figure \ref{type10Anatomy}).

\begin{table}
\begin{tabular}{|m{0.14\textwidth}|m{0.15\textwidth}|m{0.15\textwidth}|m{0.14\textwidth}|m{0.14\textwidth}|m{0.14\textwidth}|}
\hline
\textbf{Description (index)}&\textbf{Origin of sample}&\textbf{Rank of region (out of 94) in the 'fine' annotation (by correlation)}&\textbf{Rank of region (out of 94) in the 'fine' annotation (by density)} &\textbf{Fraction of density in the region}& \textbf{Fraction of density supported in midbrain } \\\hline
A9 dopaminergic neurons (4) & Substantia nigra\_ compact part & 2 & 1 & 39 \% & 77 \% \\\hline
A10 dopaminergic neurons (5) & Ventral tegmental area & 1 & 2  & 35 \% & 50 \% \\\hline
Motor Neurons, Midbrain Cholinergic Neurons (10) & Pedunculo-- pontine nucleus & 17 & 94 (zero  density) &  0  & 9.8 \% \\
\hline
\end{tabular}
\caption{Anatomical analysis for the cell-type-specific samples extracted from the midbrain.}
\label{metadataAnatomyMidbrain}
\end{table}

\begin{figure}
\centering
 \subfloat[]{\includegraphics[width=\textwidth]{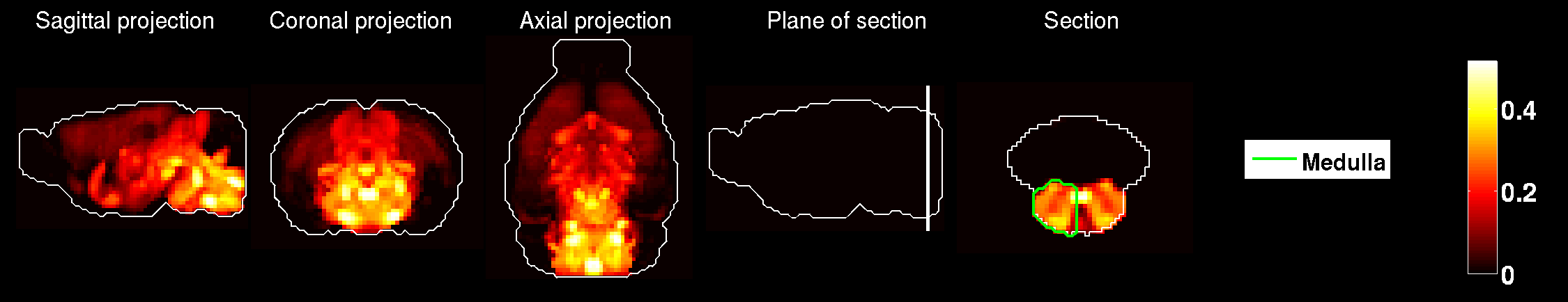}}\\
 \subfloat[]{\includegraphics[width=\textwidth]{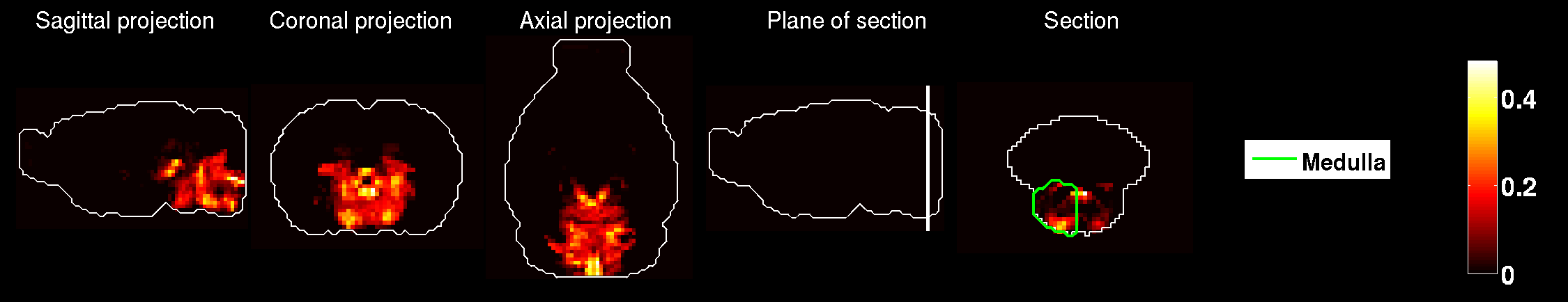}}\\
 \caption{ {\bf{Motor Neurons, Midbrain Cholinergic Neurons (cell-type index 10).}} (a) Heat map of the brain-wide correlation profile. (b) Heat map of the estimated brain-wide density profile.}
  \label{type10CorrDensity}
\end{figure}

\begin{figure}
\centering
 \subfloat[]{\includegraphics[width=\widthCoeff\textwidth]{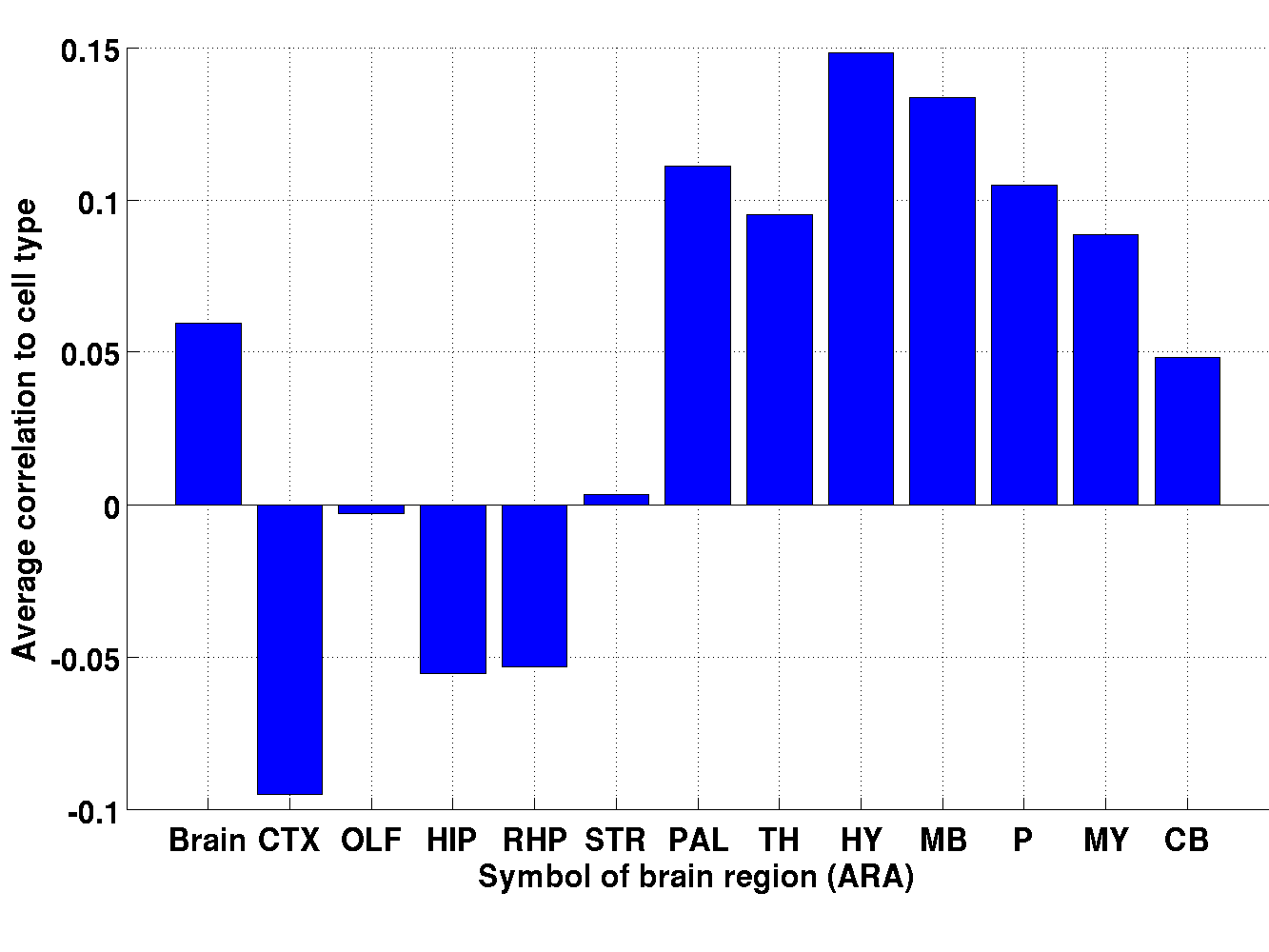}}\\
 \subfloat[]{\includegraphics[width=\widthCoeff\textwidth]{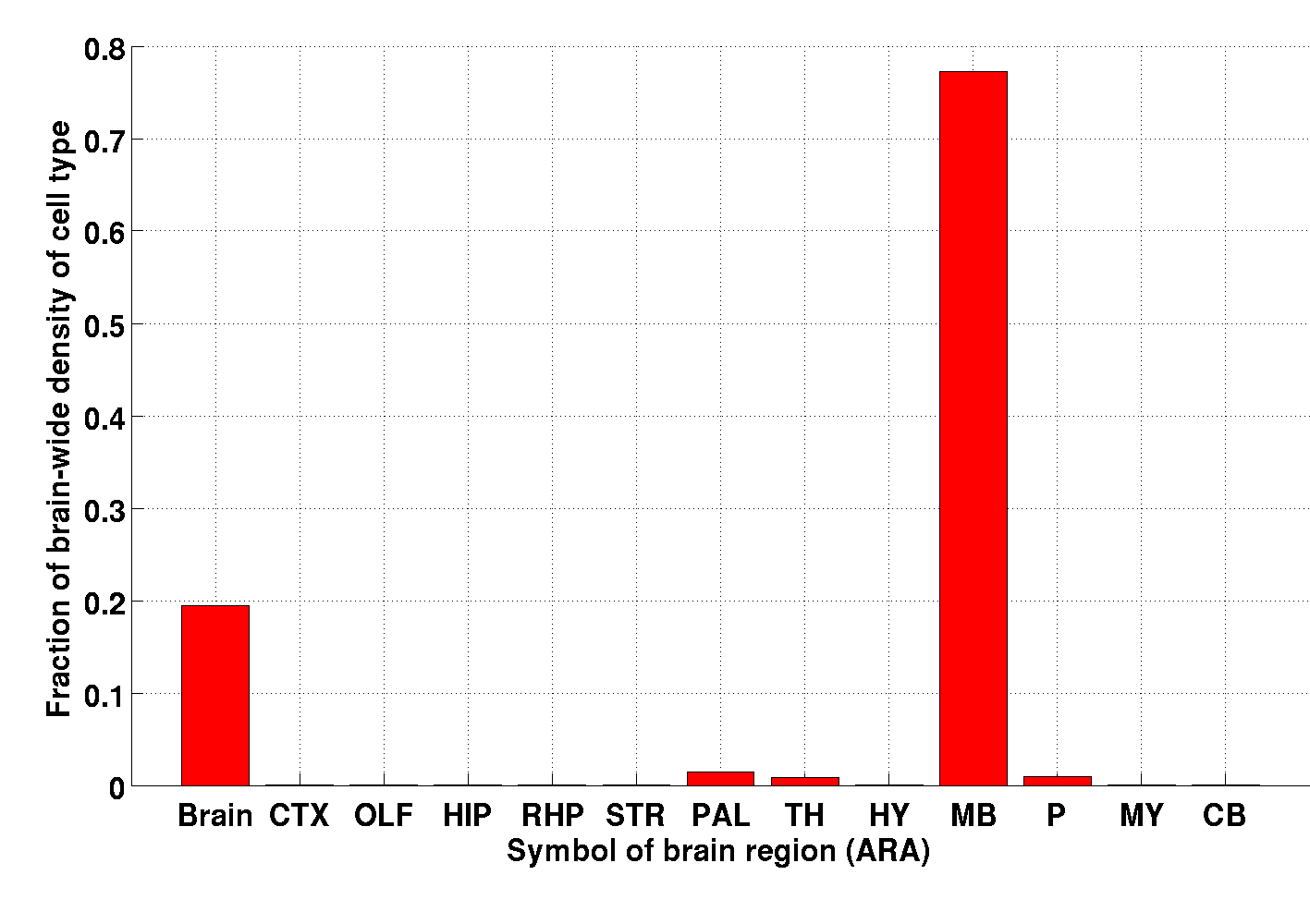}}\\
 \caption{{\bf{A9 dopaminergic neurons (cell-type index 4).}} (a) Average correlation between the cell type and the Allen Atlas, in the regions 
of the \bigTwelveSpace annotation of the ARA.
 (b) Fractions of density of cell type in the regions 
of the \bigTwelveSpace annotation of the ARA. }
  \label{type4Anatomy}
\end{figure}
\begin{figure}
\centering
 \subfloat[]{\includegraphics[width=\widthCoeff\textwidth]{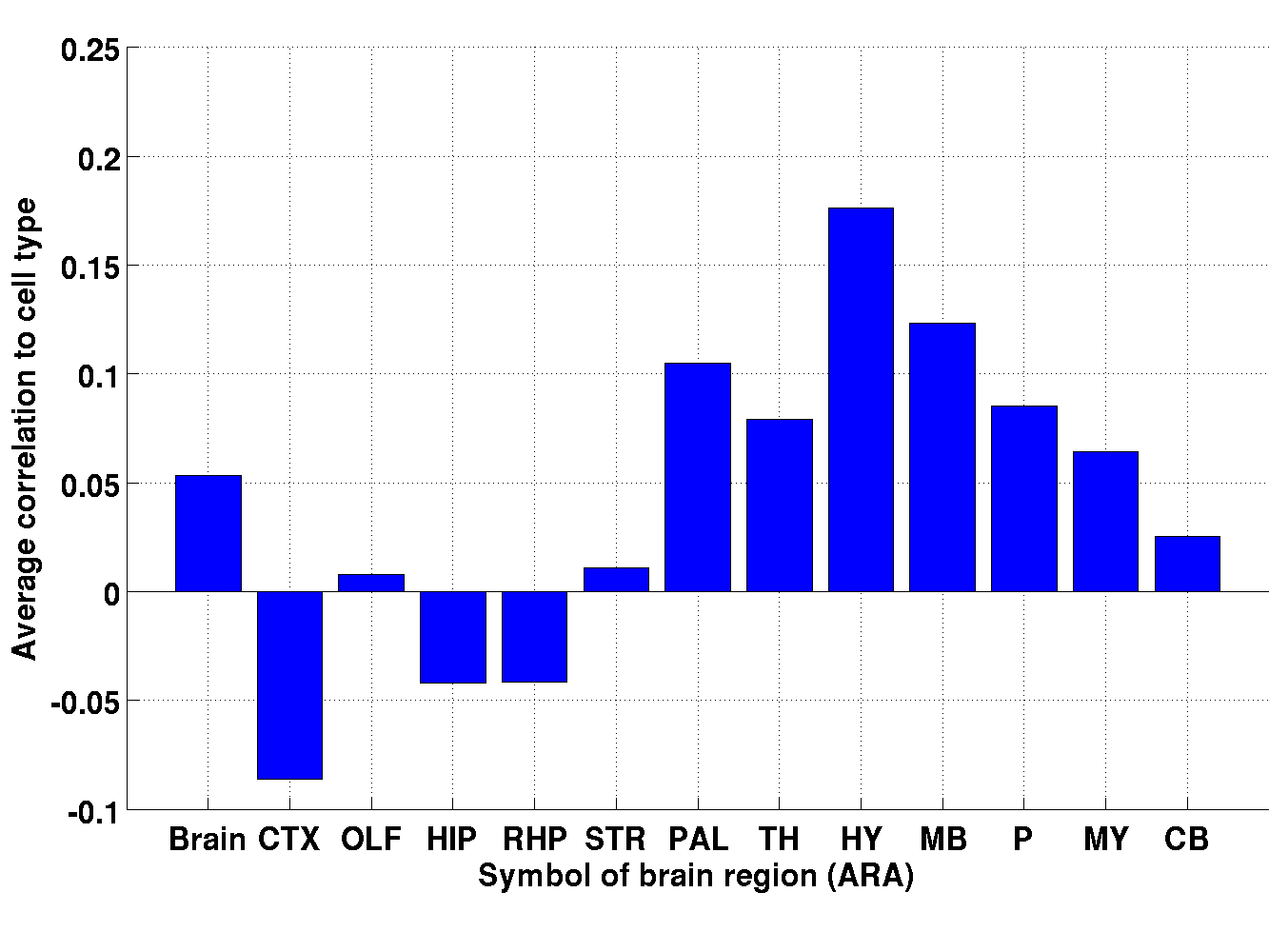}}\\
 \subfloat[]{\includegraphics[width=\widthCoeff\textwidth]{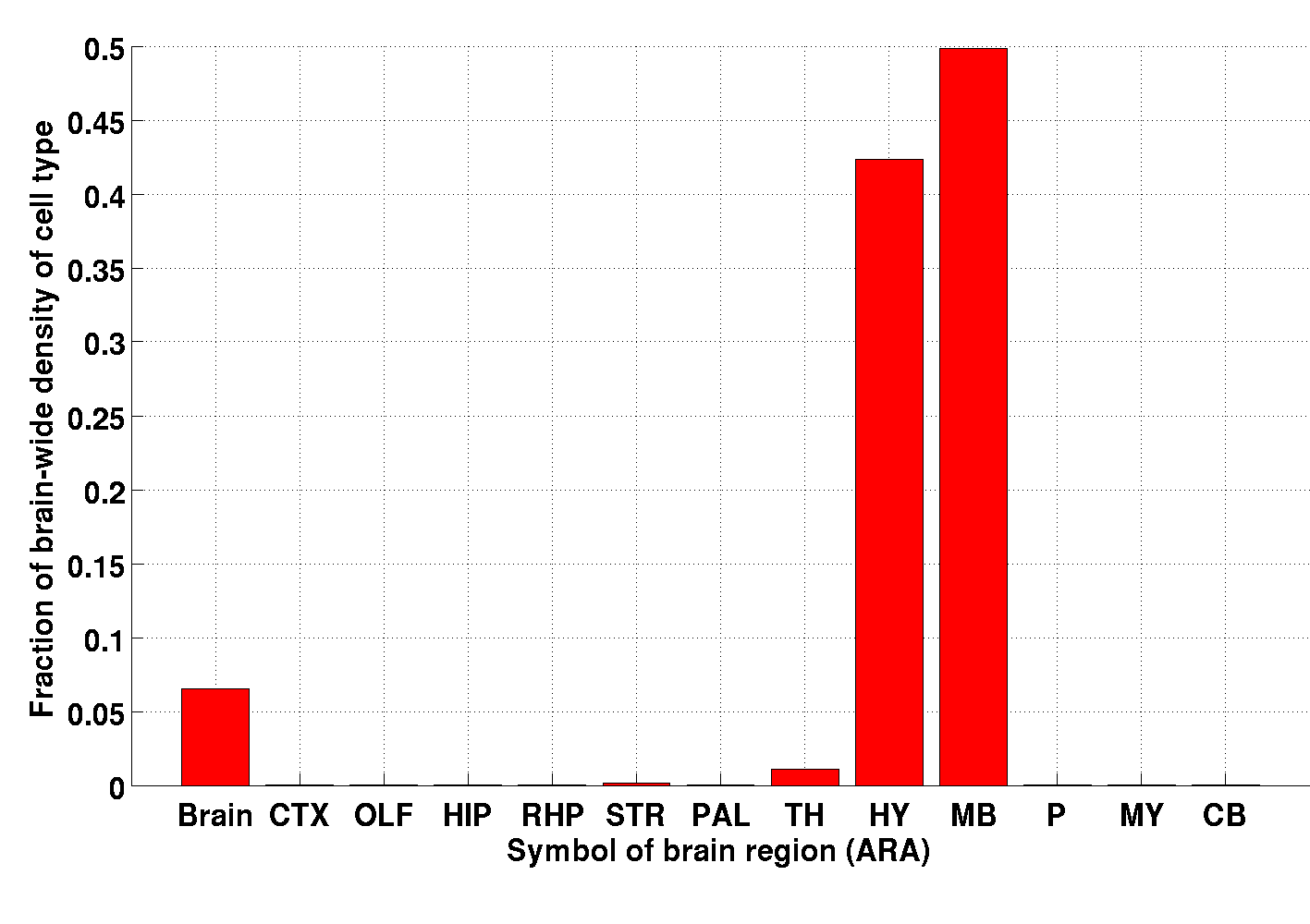}}\\
 \caption{{\bf{A10 dopaminergic neurons (cell-type index 5).}} (a) Average correlation between the cell type and the Allen Atlas, in the regions 
of the \bigTwelveSpace annotation of the ARA.
 (b) Fractions of density of cell type in the regions 
of the \bigTwelveSpace annotation of the ARA. }
  \label{type5Anatomy}
\end{figure}
\begin{figure}
\centering
 \subfloat[]{\includegraphics[width=\widthCoeff\textwidth]{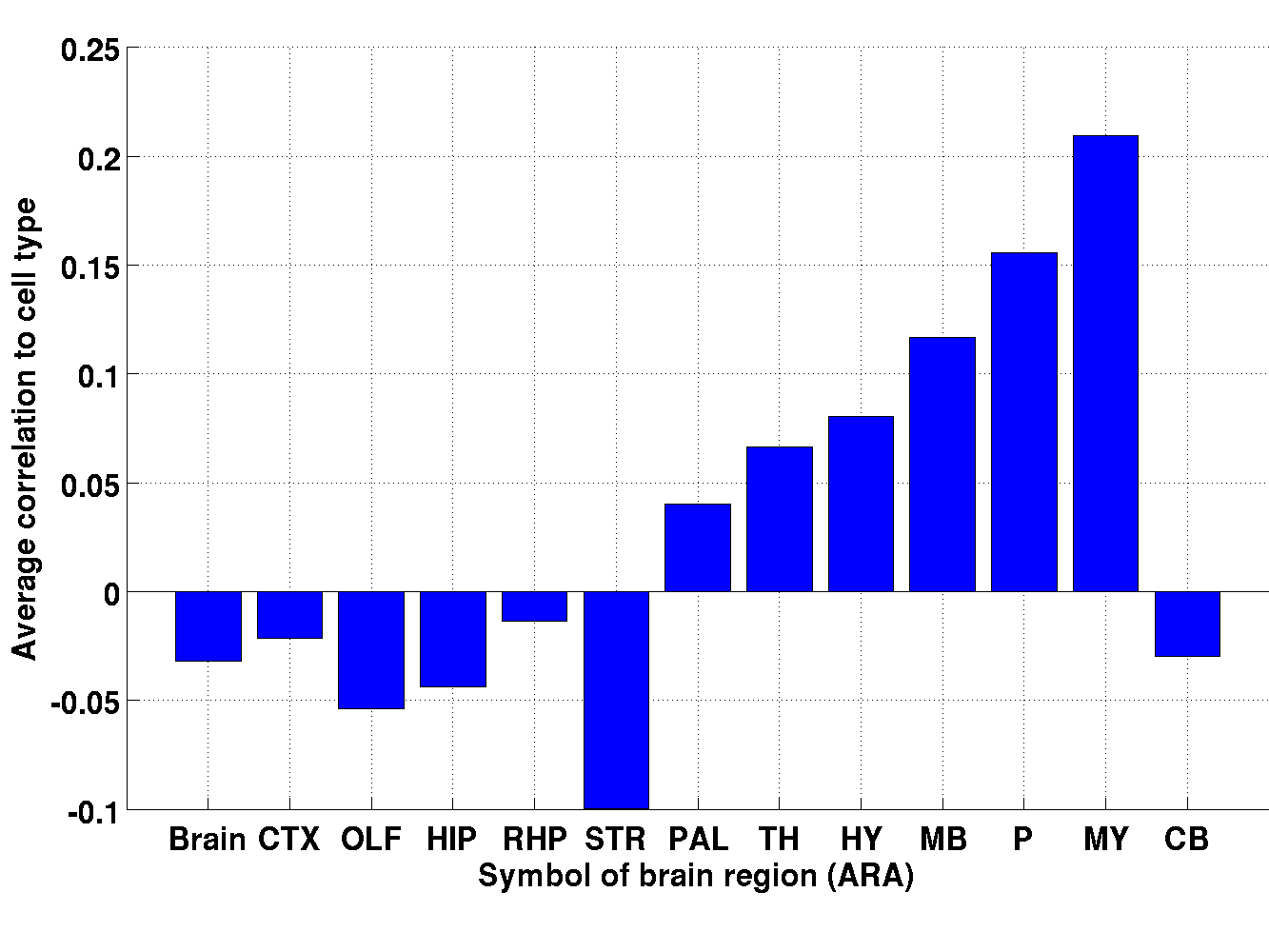}}\\
 \subfloat[]{\includegraphics[width=\widthCoeff\textwidth]{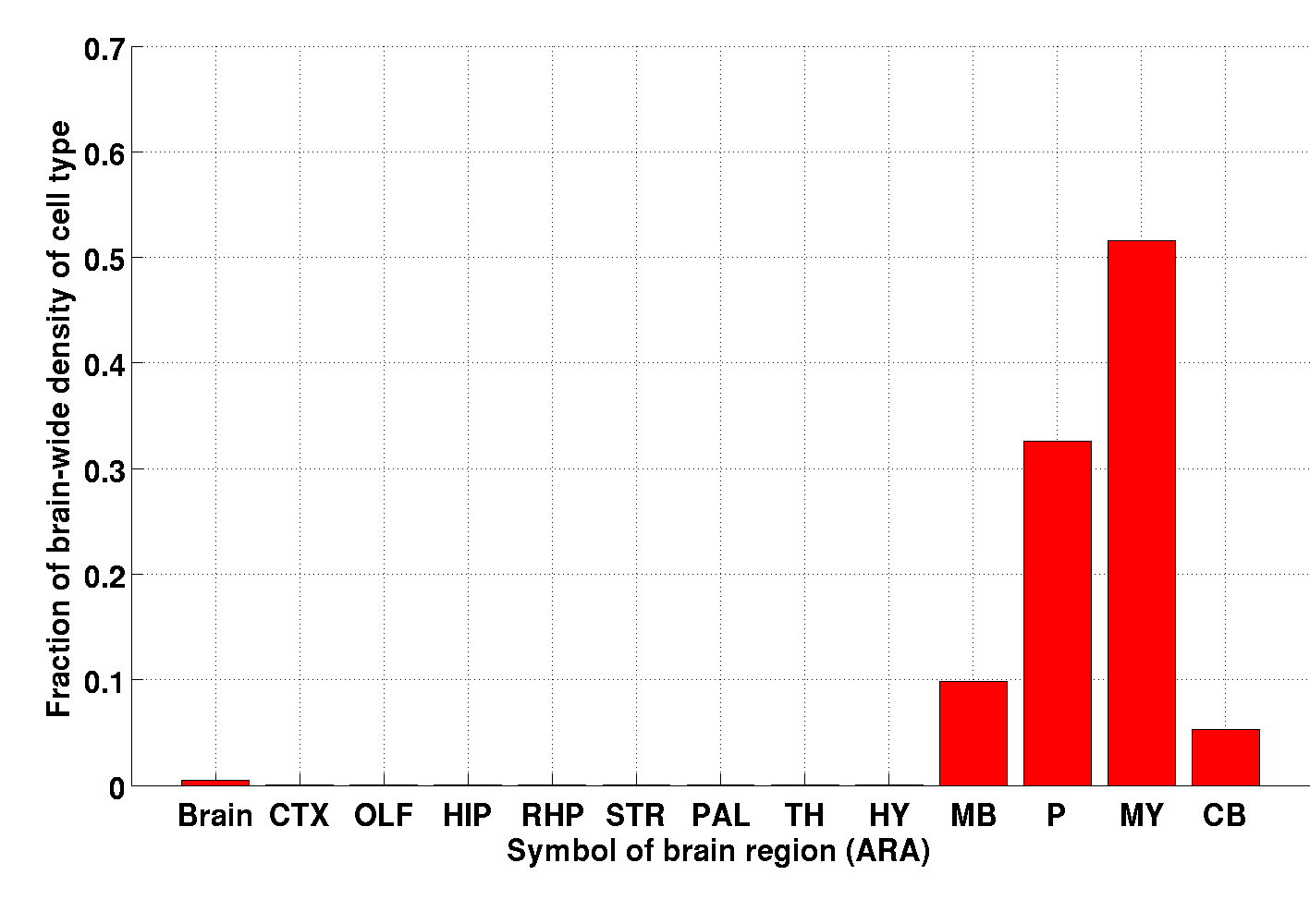}}\\
 \caption{{\bf{Motor Neurons, Midbrain Cholinergic Neurons (cell-type index 10).}} (a) Average correlation between the cell type and the Allen Atlas, in the regions 
of the \bigTwelveSpace annotation of the ARA.
 (b) Fractions of density of cell type in the regions 
of the \bigTwelveSpace annotation of the ARA. }
  \label{type10Anatomy}
\end{figure}

Wheras midbrain supports a majority of the density 
for A9 dopaminergic neurons and A10 dopaminergic
 neurons, it supports only 9.8\% of the density in Motor Neurons, Midbrain Cholinergic Neurons 
(whereas medulla and pons support 32.6\% and 51.6\% respectively). However,
 it should be noted that very few cell types in our data 
 come from medulla. 

\clearpage
\subsubsection{Pons}
%classifyPatternsPons = classify_pattern_pons( Ref, cellTypesCorrelations, fitVoxelsToTypes )
One cell-type-specific sample in this study was extracted from the
pons (index 51, unpublished data).  According to the
{\ttfamily{'fine'}} annotation, it was extracted from the pontine
central gray.\\ Pons is ranked 4th in the \bigTwelveSpace annotation both by
density and correlation (after hypothalamus, midbrain and medulla,
which support 28.9\%, 26.7\% and 12.5\% of the density respectively,
whereas pons supports 10.4\% of the density for this cell type). The
pontine central gray is ranked 26th out of 94 regions in the fine
annotation.  The fraction of the total estimated density of this cell
type cumulated by the first 26 regions in the fine annotation is
81.9\%.\\ Conversely, this cell type is the second most important
detected in the pons (after the GABAErgic interneurons, PV+, index
64).\\
 \begin{table}
\centering
\begin{tabular}{|m{0.15\textwidth}|m{0.14\textwidth}|m{0.12\textwidth}|m{0.12\textwidth}|m{0.12\textwidth}|m{0.14\textwidth}|}
\hline
\textbf{Description (index)}&\textbf{Origin of sample}&\textbf{Rank of region (out of 94) in the 'fine' annotation (by correlation)}&\textbf{Rank of region (out of 94) in the 'fine' annotation (by density)}&\textbf{Fraction of density in the region}& \textbf{Fraction of density supported in pons}\\ \hline 
Tyrosine Hydroxylase Expressing (51) & Pontine central gray & 9 & 26 & 1.3 \% & 10.4 \% \\
\hline
\end{tabular}
\caption{Anatomical analysis for the cell-type-specific sample extracted from the pons.}
\label{metadataAnatomyPons}
\end{table}
\begin{figure}
\centering
 \subfloat[]{\includegraphics[width=\textwidth]{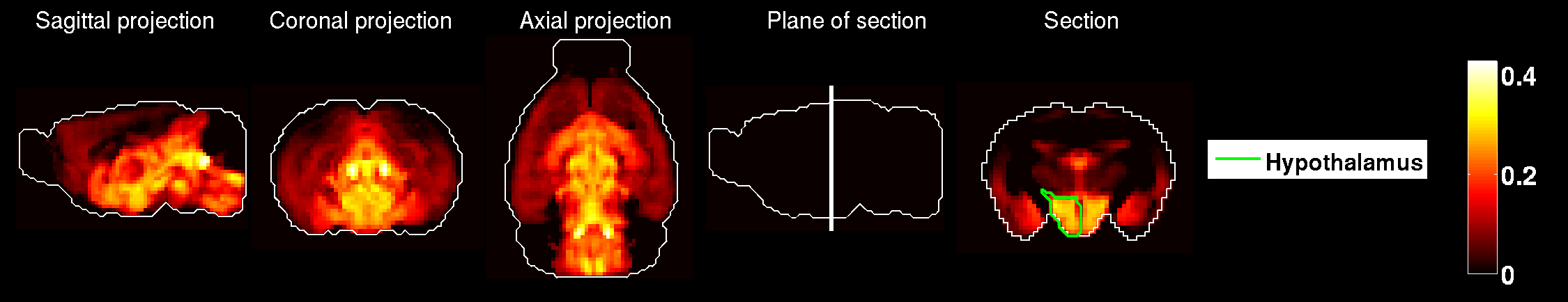}}\\
 \subfloat[]{\includegraphics[width=\textwidth]{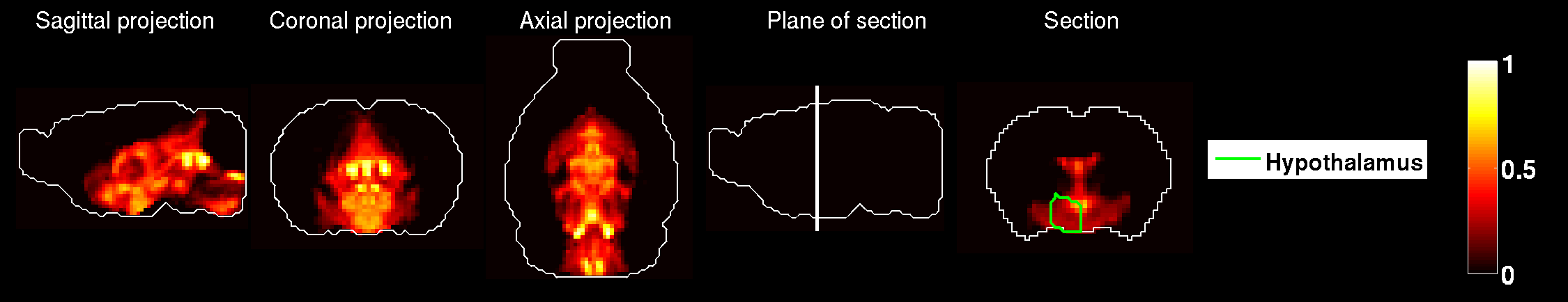}}\\
 \caption{ {\bf{Tyrosine Hydroxylase Expressing (cell-type index 51).}} (a) Heat map of the brain-wide correlation profile. (b) Heat map of the estimated brain-wide density profile.}
  \label{spinalCord}
\end{figure}
\begin{figure}
\centering
 \subfloat[]{\includegraphics[width=\widthCoeff\textwidth]{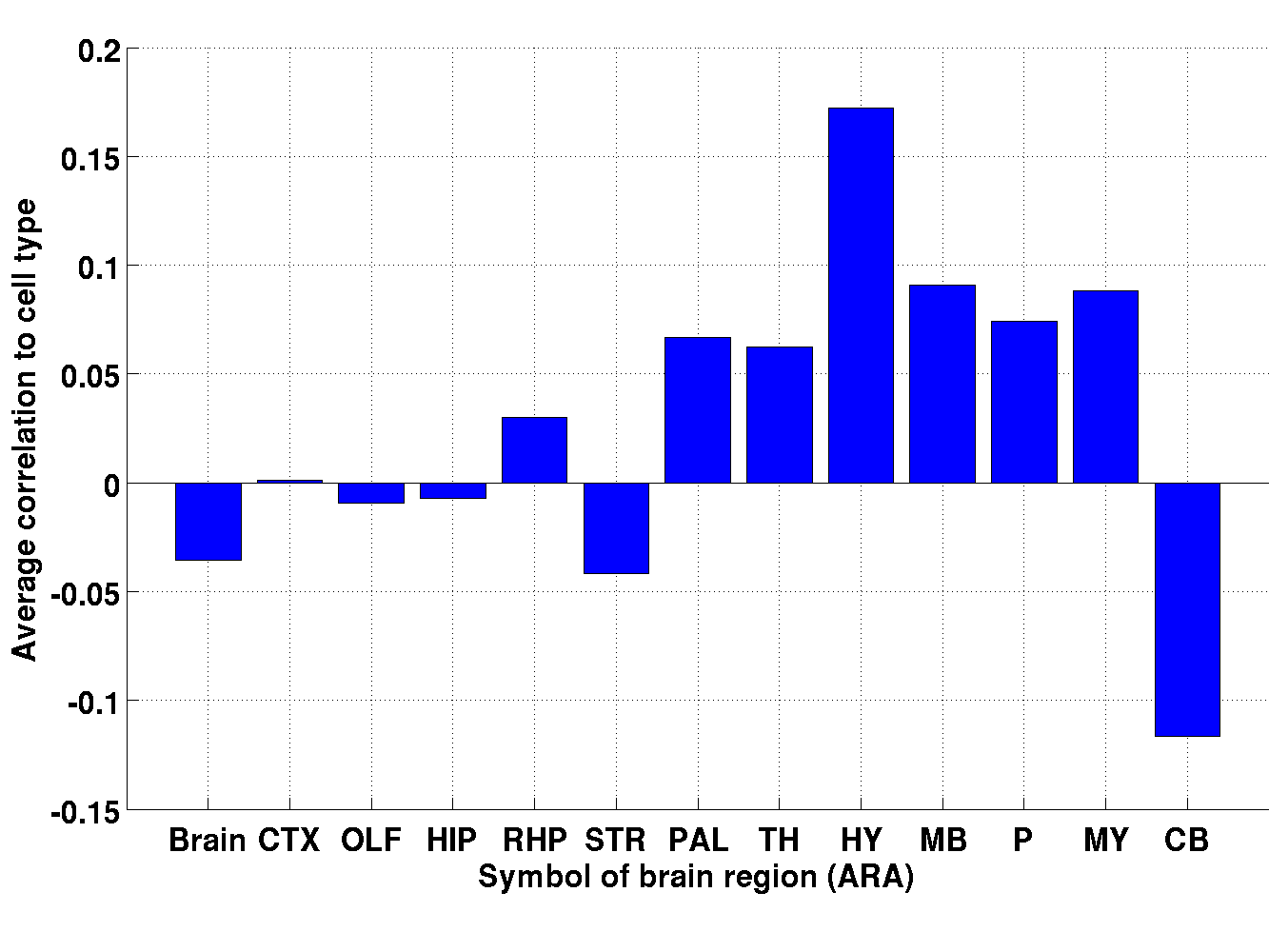}}\\
 \subfloat[]{\includegraphics[width=\widthCoeff\textwidth]{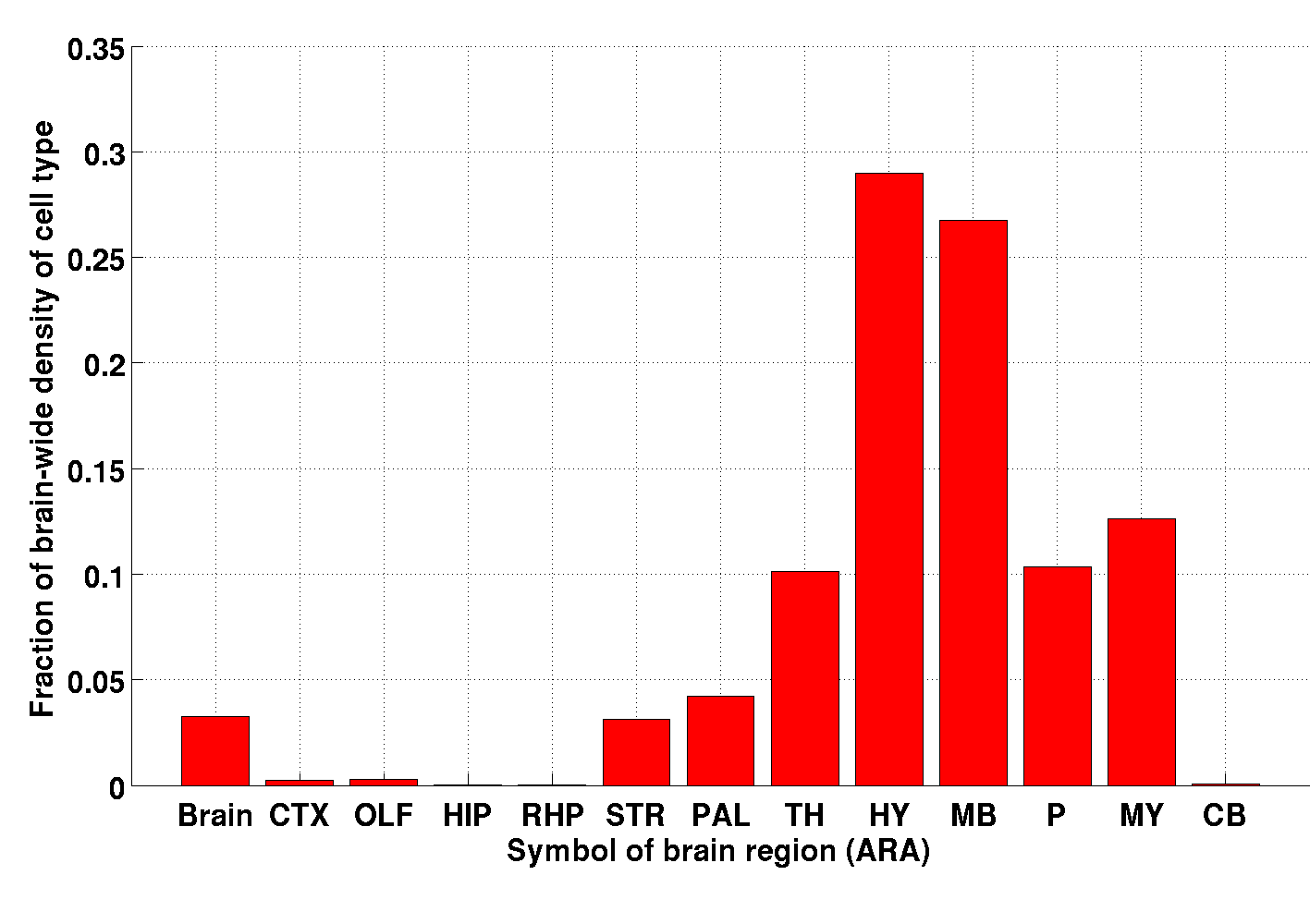}}\\
 \caption{{\bf{Tyrosine Hydroxylase Expressing (cell-type index 51).}} (a) Average correlation between the cell type and the Allen Atlas, in the regions 
of the \bigTwelveSpace annotation of the ARA.
 (b) Fractions of density of cell type in the regions 
of the \bigTwelveSpace annotation of the ARA. }
  \label{spinalCordAnatomy}
\end{figure}
\clearpage
\subsubsection{Medulla}
The cell-type-specific sample with index 12 (motor neurons,
cholinergic interneurons), was extracted from the spinal chord. The
closest region in \bigTwelveSpace is the medulla, which has a
refinement into 15 different regions. Medulla is indeed the top region
in the \bigTwelveSpace annotation for this cell type, both by
correlation and density (see Figures \ref{spinalCord} and \ref{spinalCordAnatomy}).  The second region in \bigTwelveSpace by
  density is the pons (33\%), so the medulla and the
pons support more than 95\% of the estimated density of this cell
type. The breakdown of the density amon the regions of the medulla
 is shown in Table \ref{finerMedulla}.\\
\begin{table}
\centering
\begin{tabular}{|m{0.14\textwidth}|m{0.14\textwidth}|m{0.14\textwidth}|m{0.14\textwidth}|m{0.14\textwidth}|m{0.14\textwidth}|}
\hline
\textbf{Description (index)}&\textbf{Origin of sample}&\textbf{Rank of region (out of 94) in the 'fine' annotation (by correlation)}&\textbf{Rank of region (out of 94) in the 'fine' annotation (by density)}&\textbf{Fraction of density in the region}& \textbf{Fraction of density supported in the medulla}\\ \hline 
 Motor neurons, cholinergic interneurons (12) &  Medulla  &   13  &   1  &  32.5  \% &   62.3\% \\
\hline
\end{tabular}
\caption{Anatomical analysis for the cell-type-specific sample extracted from the medulla.}
\label{metadataAnatomyMedulla}
\end{table}
\begin{figure}
\centering
 \subfloat[]{\includegraphics[width=\textwidth]{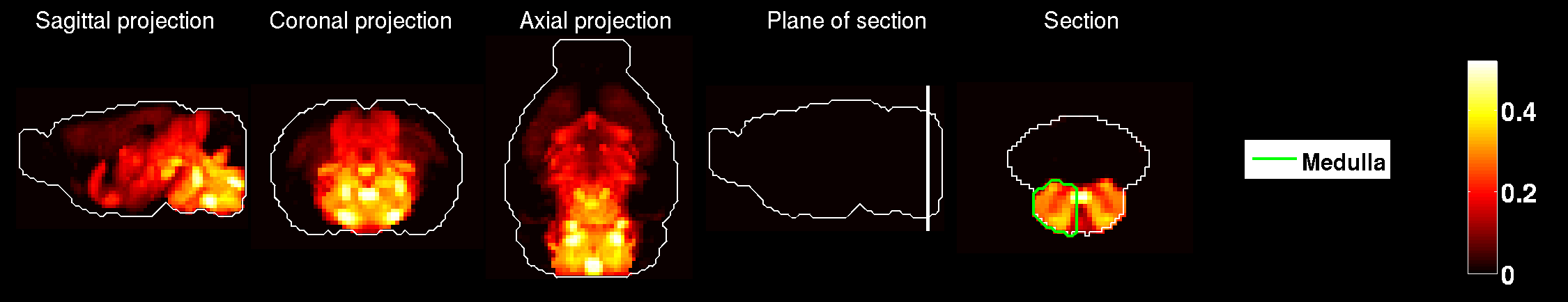}}\\
 \subfloat[]{\includegraphics[width=\textwidth]{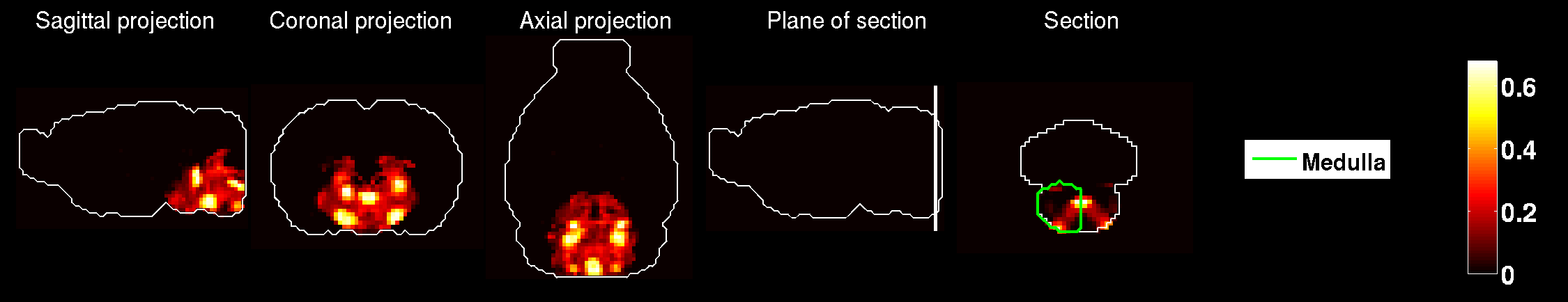}}\\
 \caption{ {\bf{Motor neurons, cholinergic interneurons (cell-type index 12).}} (a) Heat map of the brain-wide correlation profile. (b) Heat map of the estimated brain-wide density profile.}
  \label{spinalCord}
\end{figure}
\begin{figure}
\centering
 \subfloat[]{\includegraphics[width=\widthCoeff\textwidth]{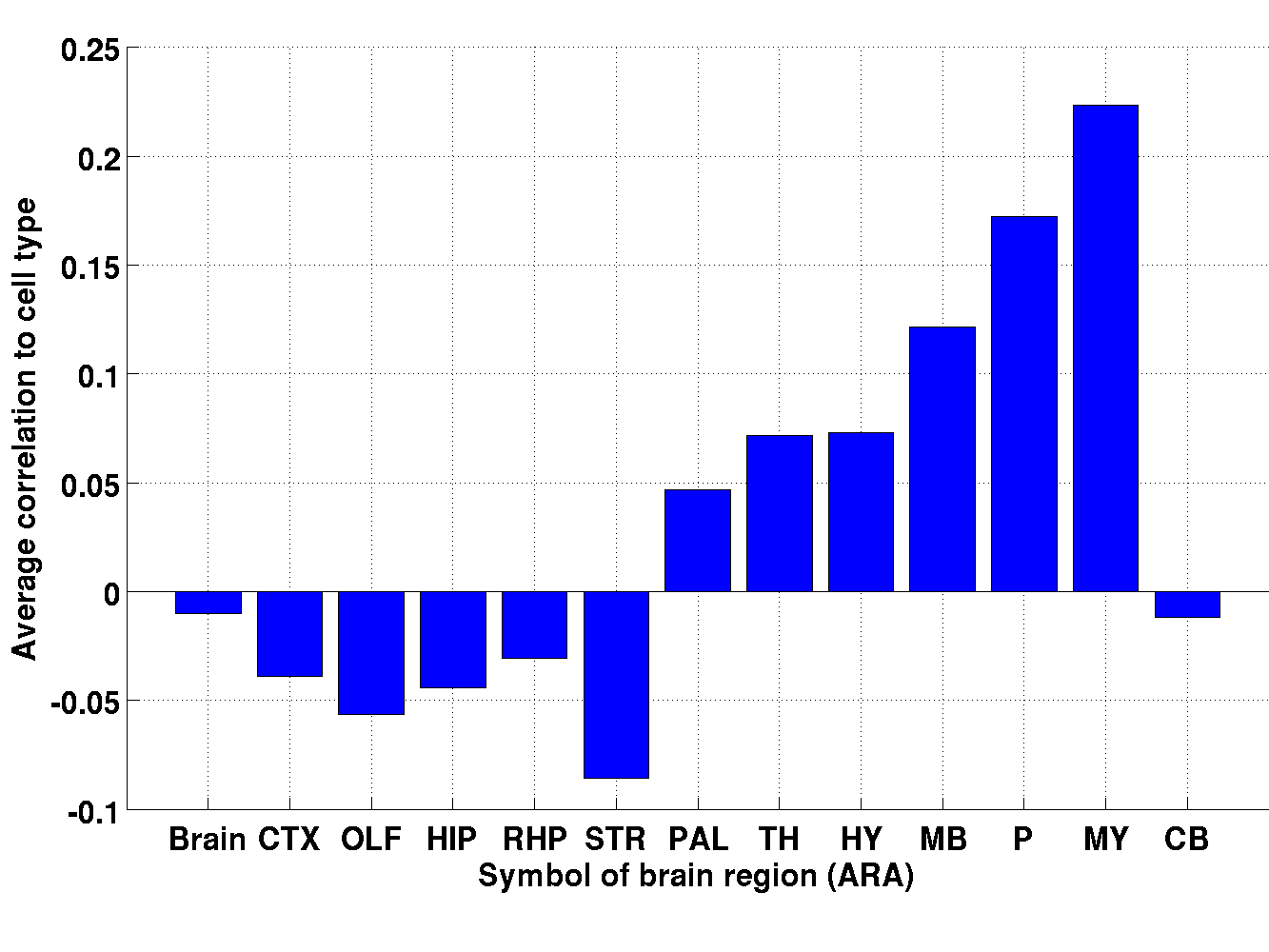}}\\
 \subfloat[]{\includegraphics[width=\widthCoeff\textwidth]{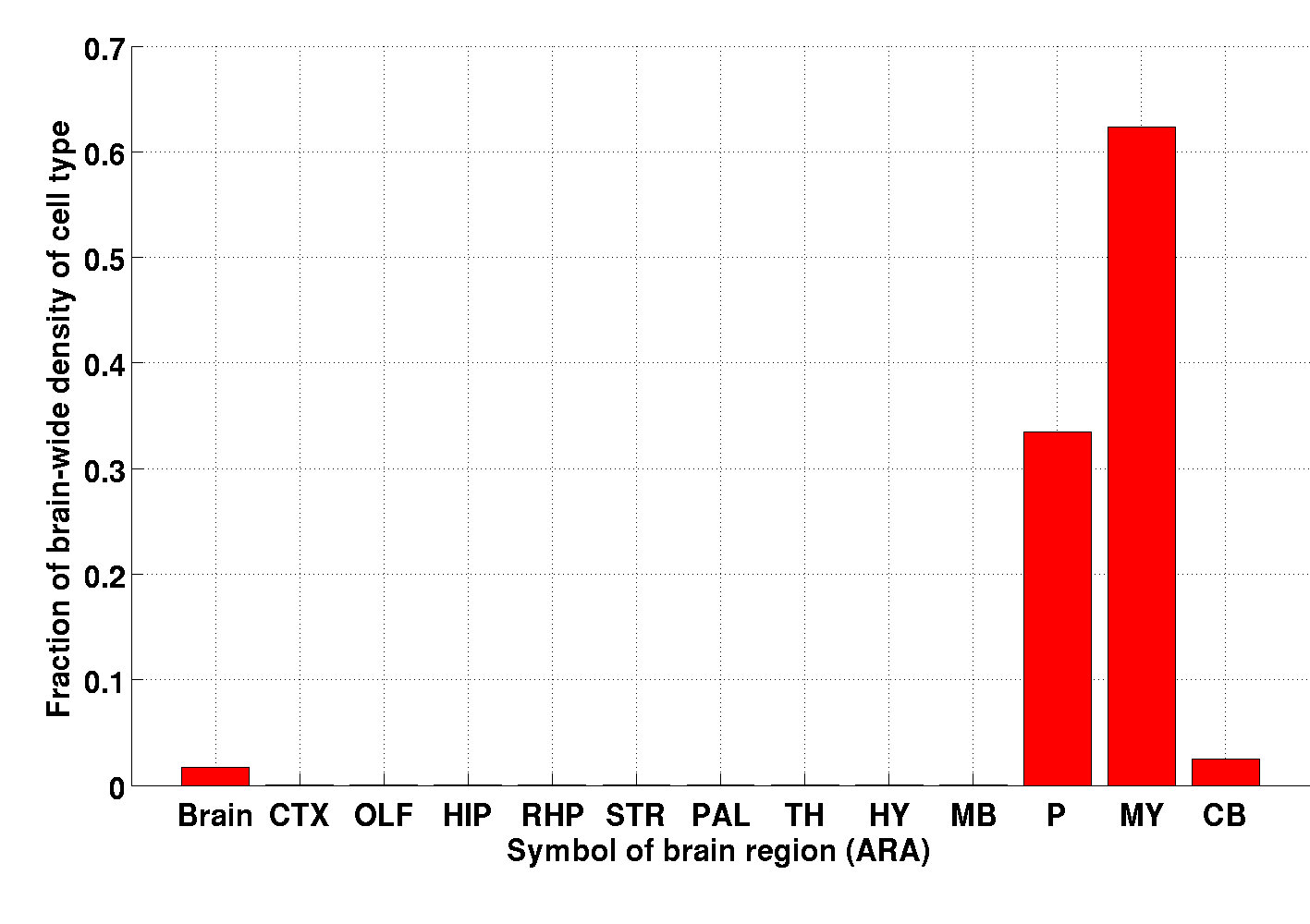}}\\
 \caption{{\bf{Motor neurons, cholinergic interneurons (cell-type index 12).}} (a) Average correlation between the cell type and the Allen Atlas, in the regions 
of the \bigTwelveSpace annotation of the ARA.
 (b) Fractions of density of cell type in the regions 
of the \bigTwelveSpace annotation of the ARA. }
  \label{spinalCordAnatomy}
\end{figure}
\begin{table}
\begin{tabular}{|m{0.55\textwidth}|m{0.3\textwidth}|}
\hline
\textbf{Brain region (in the 'fine' annotation)}& \textbf{Fraction of density supported in the region (\%)}\\ \hline 
     Medulla & 32.5 \\ \hline 
    Facial motor nucleus & 6.9\\ \hline 
    Vestibular nuclei & 5.5\\ \hline 
    Hypoglossal nucleus &3.6\\ \hline 
    Paragigantocellular reticular nucleus &3.3\\ \hline 
    Spinal nucleus of the trigeminal\_ interpolar part& 3.2\\ \hline 
    Lateral reticular nucleus & 2.8\\ \hline 
    Spinal nucleus of the trigeminal\_ oral part & 1.8\\ \hline 
    Magnocellular reticular nucleus & 1.3\\ \hline 
    Cochlear nuclei & 1\\ \hline 
    Medulla\_ behavioral state related & 0.4\\ \hline 
    Spinal nucleus of the trigeminal\_ caudal part & 0.2\\ \hline 
    Dorsal column nuclei & 0.2\\ \hline 
    Inferior olivary complex & 0.2\\ \hline 
    Nucleus of the solitary tract&  0\\
\hline
\end{tabular}
\caption{Fractions of the density profile of cell-type index 12 supported 
 by subsets of medulla in the ARA.}
\label{finerMedulla}
\end{table}

\clearpage
\subsubsection{Cerebellum}
\begin{figure}
\centering
\subfloat[]{\includegraphics[width=\widthParamFig\textwidth]{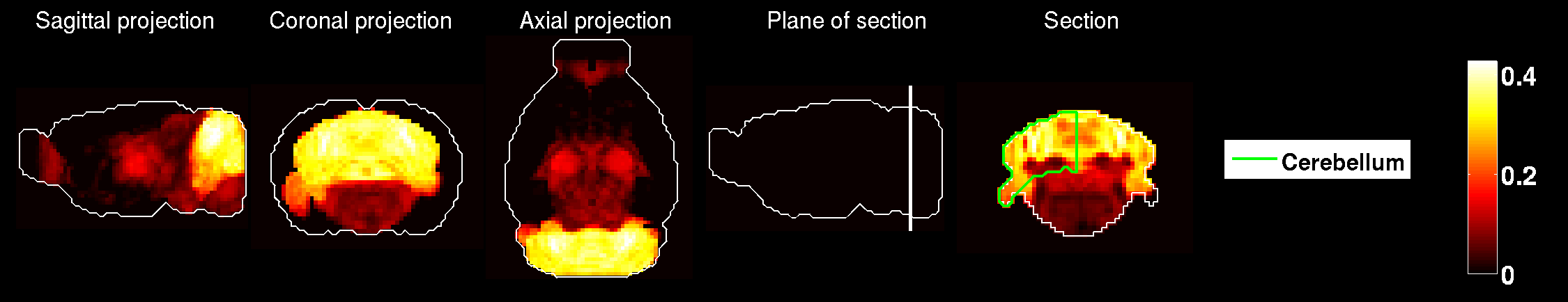}}\\
\subfloat[]{\includegraphics[width=\widthParamFig\textwidth]{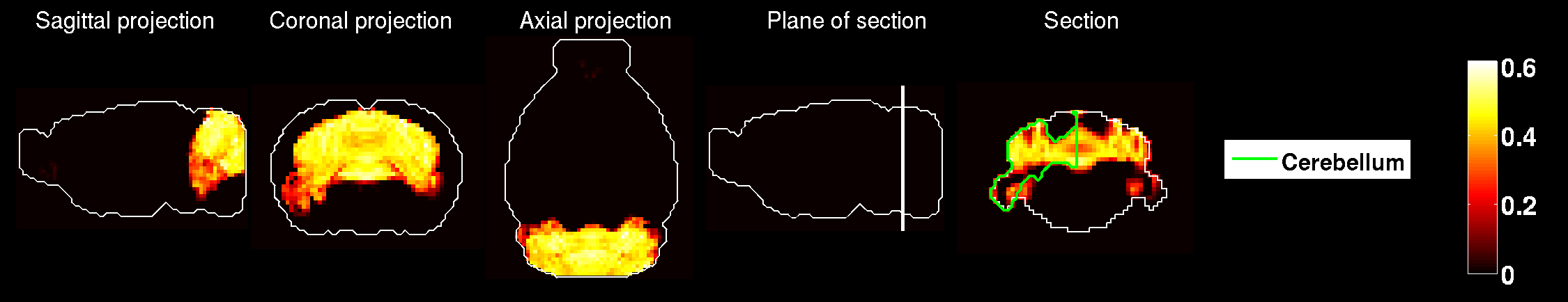}}\\
\caption{{\bf{Purkinje cells (cell-type index $t=1$).}} (a) Average correlation between the cell type and the Allen Atlas, in the regions 
of the \bigTwelveSpace annotation of the ARA.
 (b) Fractions of density of cell type in the regions 
of the \bigTwelveSpace annotation of the ARA.}
\label{fittingsFig1}
\end{figure}

\begin{figure}
\centering
\subfloat[]{\includegraphics[width=\widthParamFig\textwidth]{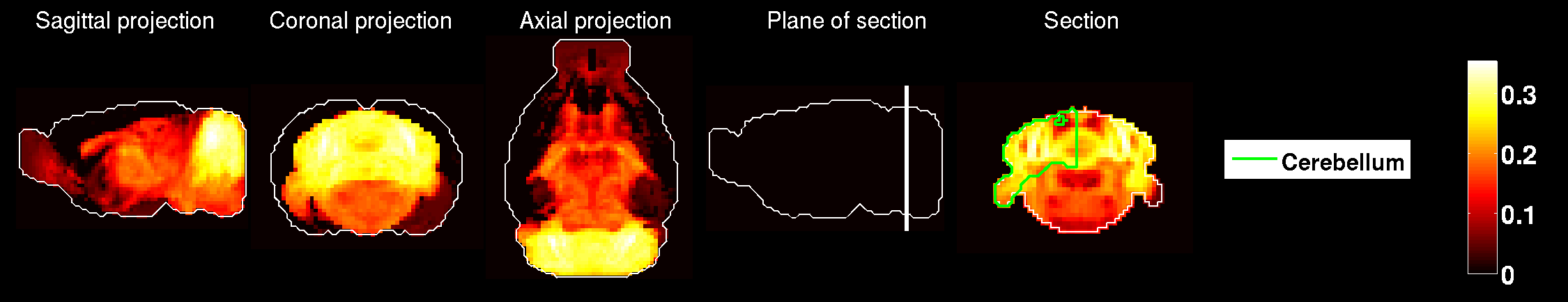}}\\
\subfloat[]{\includegraphics[width=\widthParamFig\textwidth]{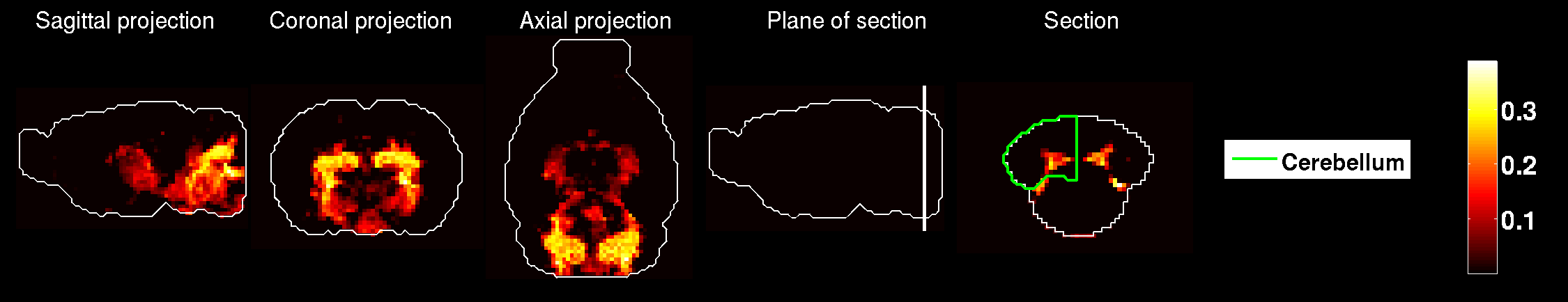}}\\
\caption{{\bf{Mature oligodendrocytes (cell-type index $t=21$).}}(a) Average correlation between the cell type and the Allen Atlas, in the regions 
of the \bigTwelveSpace annotation of the ARA.
 (b) Fractions of density of cell type in the regions 
of the \bigTwelveSpace annotation of the ARA.}
\label{fittingsFig21}
\end{figure}

 Out of the 11 samples that were drawn from the cerebellum, 7 have the
 cerebellum as their top region by correlation, and 4 have the
 cerebellum as their top region by density (all of which also have the
 cerebellum as their top region by correlation, see \ref{cerebellarSamplesAll}). All the samples that
 have the cerebellum as their top region were indeed taken from the
 cerebellum.  See Table \ref{whiteMatterTable} for the Mature
 oligodendrocytes (index 28), whose density profiles follows a
 white-matter pattern that includes the {\emph{arbor vitae}}.
 See Figure \ref{fittingsFig1} for a class of Purkinje cells
\cite{foreBrainTaxonomy} and Figure \ref{fittingsFig21} for a class of
mature oligodendrocytes \cite{DoyleCells}, both extracted from the
cerebellum. Their correlation and estimated density patterns are
indeed mostly localized in cerebellum.\\

\begin{table}
\centering
\begin{tabular}{|m{0.3\textwidth}|m{0.2\textwidth}|m{0.2\textwidth}|m{0.15\textwidth}| }
\hline
\textbf{Description (index)}&\textbf{Top region by correlation}&\textbf{Top region by density (percentage of density supported)} & \textbf{Fraction of density supported in the cerebellum (\%)}\\ \hline 
Purkinje Cells (1) & Cerebellum&  Cerebellum& 95.8 \\ \hline
Golgi Cells (17) & Pons & N/A($\ast$) & 0  \\ \hline
Unipolar Brush cells (some Bergman Glia) (18) & Cerebellum& Thalamus& 0.2\\\hline
Stellate Basket Cells (19) &Cerebellum& Medulla & 18.8\\\hline
Granule Cells (20)& Cerebellum & Cerebellum & 96.0 \\\hline
Mature Oligodendrocytes (21)& Cerebellum & Cerebellum & 39.9 \\\hline
Mixed Oligodendrocytes (23)& Pons &N/A($\ast$) & 0  \\ \hline
Purkinje Cells (25) &Cerebellum & Olfactory areas& 0 \\ \hline
Bergman Glia (27) &Cerebellum & Olfactory area & 5.5 \\ \hline
Purkinje Cells (52)&Cerebellum & Thalamus & 5.9 \\\hline
$\ast${Zero density in the left hemisphere.} 
\end{tabular}
\caption{Cell-type-specific samples extracted from the cerebellum.}
\label{cerebellarSamplesAll}
\end{table}

\begin{figure}
\centering
\subfloat[]{\includegraphics[width=\widthParamFig\textwidth]{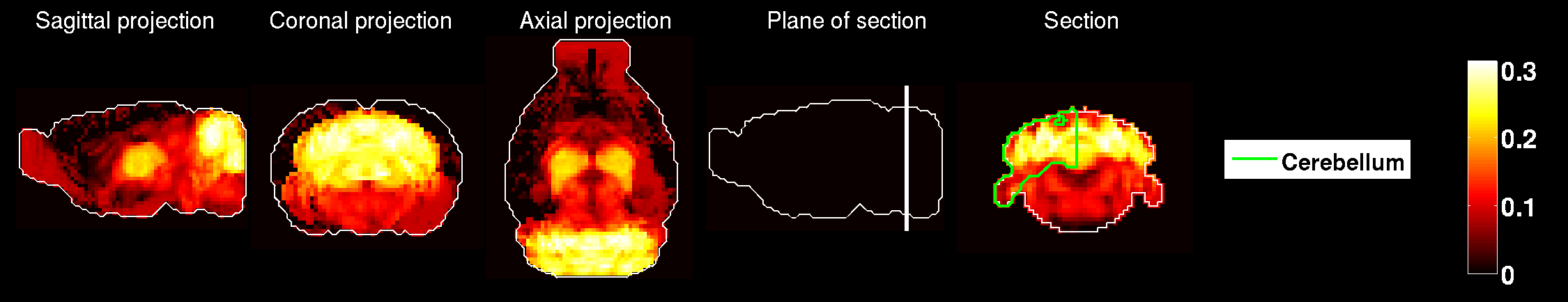}}\\
\subfloat[]{\includegraphics[width=\widthParamFig\textwidth]{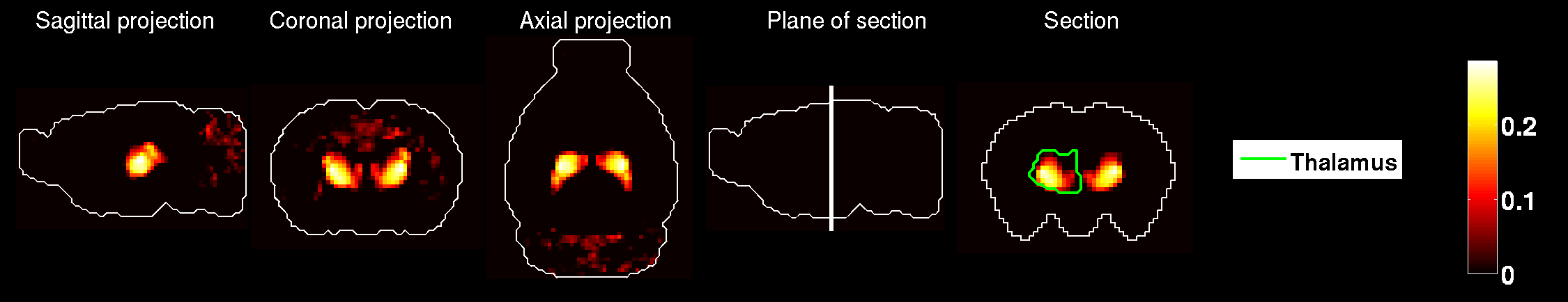}}\\
\subfloat[]{\includegraphics[width=\widthCoeff\textwidth]{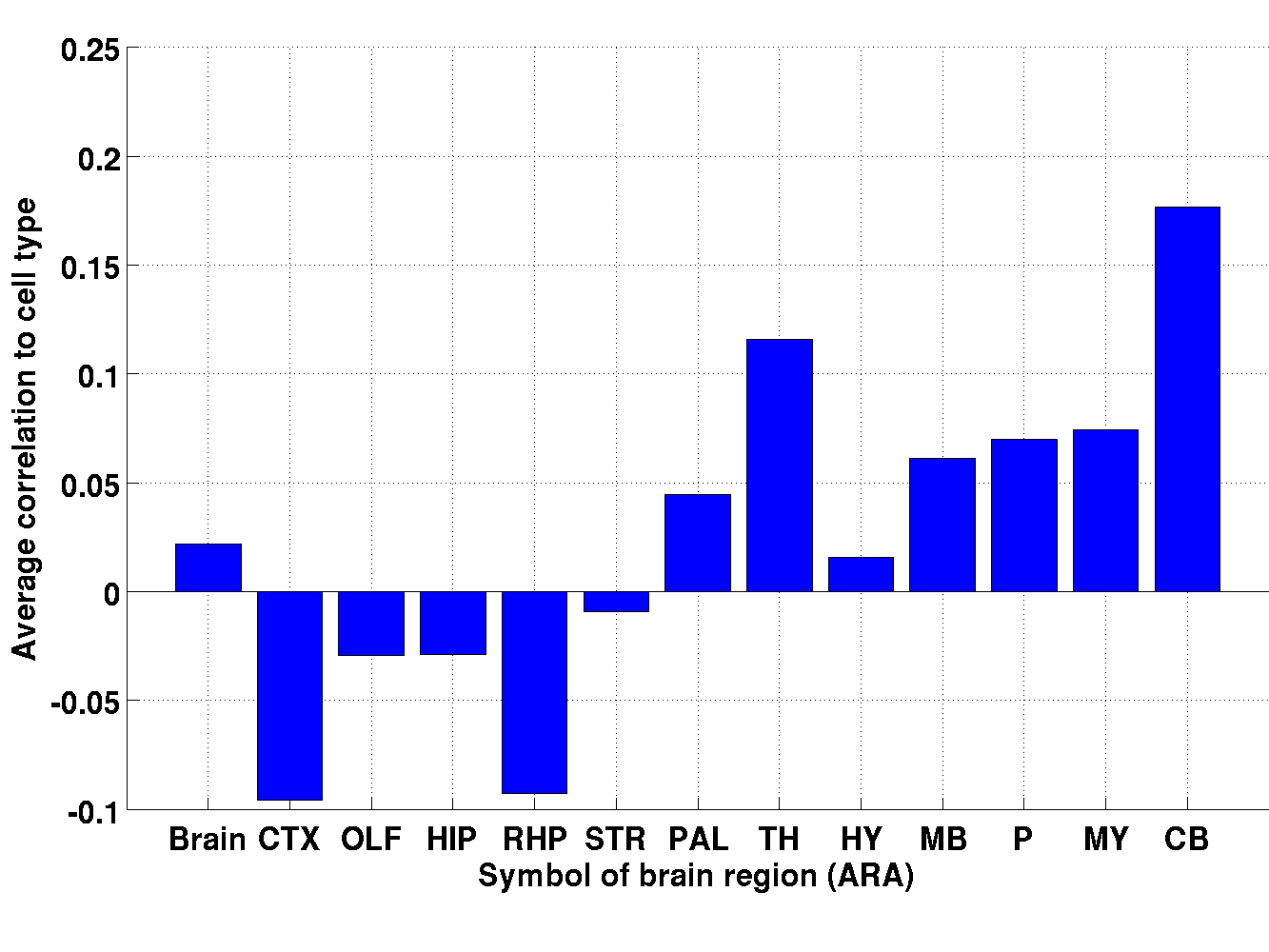}}\\
\caption{{\bf{Purkinje cells (cell-type index $t=52$).}}(a) Average correlation between the cell type and the Allen Atlas, in the regions 
of the \bigTwelveSpace annotation of the ARA.
 (b) Fractions of density of cell type in the regions 
of the \bigTwelveSpace annotation of the ARA. Thalamus supports
 a majority of the estimated density. (c) Average correlations in the regions of the 
 \bigTwelveSpace annotation. Cerebellum is the top region, followed by thalamus.}
\label{fig:PurkinkjeThalamusPattern}
\end{figure}

A remarkable class of Purkinje cells does not have the cerebellum
 as its top region by density. See Figure \ref{fig:PurkinkjeThalamusPattern} for a
class of Purkinje cells (index 52, unpublished) that correlates best
with the Allen Atlas both in the thalamus and in the cerebellum, but
that fits based on density coefficients only in the thalamus. This
indicates that thalamus must contain cell types whose gene expression
profile is closest to Purkinje cells in the present data set, but that
thalamus is not sampled in enough detail by our microarray data set for
these cell types to be distinguished from this class of Purkinje
cells.\\

%classifyPurest = classify_purest( Ref, cellTypesCorrelations, fitVoxelsToTypesPurest );
%stop at cc = 59;   

\section{Discussion of the linear model}
In this section we discuss several limitations of the linear model. We
 address them by refitting the linear model to a modified by panel 
 on cell-type-specific transcription profile:\\
\begin{itemize}
\item In the first subsection we replace pairs of highly 
 similar cell types\footnote{such as the two samples of medium spiny neurons, 
 indices $15$ and $16$ in the original panel} by one of these cell types in the
 fitting panel, and verify that the estimated profile of the remaining cell
 type is close to the sum of the two estimated profiles
 for the pair of cell types in the 
 original model (while the results for other cell types are stable).\\

\item Moreover, we refit the model to a panel where the 18 
 transcription profiles of pyramidal neurons are replaced by one composite cell type
 equal to their average. The estimated density profile of the composite cell type
 is mostly cortical, and correlated to the sum of estimated profiles of pyramidal
 neurons (the rest of the results are stable).\\

\item Finally, we study the response of the estimated densities
 to a correction in the cell-type-specific profiles consisting of a negative 
 term (which offsets the cross-hybridization). The top regions
 of most cell types do not change, which confirms the anatomical 
 conclusions of the linear model, but the residual terms go down,
 suggesting that the accuracy of the model is improved when cross-hybridization 
 is taken into account.\\
\end{itemize}

 Refitting of the model gives rise to a 
new set of estimated density profiles $\rhoNew_t$ of cell types,
 for $t$ in a set of cell types, some of which have corresponding 
 entries in the original model discussed above:
 \begin{equation}
  E(v,g) = \sum_{t=1}^{T'}\rhoNew_t(v) C^{\mathrm{new}}(t,g)+ {\mathrm{Residual^{new}}}(v,g).
 \label{refittingEquation}
 \end{equation}
 If a cell type labeled $t$ is present in both panels,
 we can compare the results of the two models
 quantitatively by computing the correlation coefficient between 
 the two brain-wide density profiles labeled $t$ and $l$ in the respective models:
 \begin{equation}
 \mathcal{C}( \rho_t,\rhoNew_l ) = \frac{\sum_{v=1}^V\left(\rho_t(v)-\overline{\rho_t}\right)
 \left(\rhoNew_l(v)-\overline{\rhoNew_l}\right)}{\sqrt{\left(\sum_{v=1}^V\left(\rho_t(v)-\overline{\rho_t}\right)^2\right)
\left(\sum_{v=1}^V\left(\rhoNew_l(v)-\overline{\rhoNew_l}\right)^2\right)}},
 \label{corrProfiles}
 \end{equation}
where the overlined quantities are cell-type-specific densities averaged
over voxels:
\begin{equation}
\overline{\rho_t} = \frac{1}{V} \sum_{v=1}^{V}\rho_t(v),\;\;\;
\overline{\rhoNew_t} = \frac{1}{V}\sum_{v=1}^{V} \rhoNew_t(v).
\end{equation}

%the ratio of the $L^2$-distance between
% the two density profiles to the norm of the original density profile (assuming this
% original density profile is not zero):
%\begin{equation}
%d_t(\rhoNew_t - \rho_t):= \sqrt{\sum_{v=1}^V \left( \rhoNew_t(v)-\rho_t(v)\right)^2}
%\end{equation}
To study the anatomical properties of the results in a refitted model,
 we can compute the average density in each brain region labeled $r$:
\begin{equation}
{\overline{\rho}}( r, t ) = \frac{1}{|\sum_{v \in {\mathtt{{Brain\;Annotation}}}}\rho_t(v) |}
                                  \sum_{v \in V_r}\rho_t(v),
\label{fittingRegion} 
\end{equation}
and check for a cell type labeled $t$ if the region with the maximum value of 
$\overline{\rhoNew}(., t )$ is identical to the one induced
 by ${\overline{\rho}}(., t)$. The top region is the one that is selected 
 to choose the section when plotting the heat maps of the brain-wide density profile.
It can therefore be read off when plotting the densities $\rho_t$ and $\rhoNew_t$ side by side.

\subsection{Competition between pairs of similar cell types}
\subsubsection{Methods}
 As can be expected from the names of the $T=64$ cell types
 in this study (see Tables \ref{metadataTable1}, \ref{metadataTable2}),
 some pairs of cell-type-specific transcription profiles
 can be very close to each other. This can lead to uniqueness
 problems in the predicted densities.
 Indeed, in aa situation where two rows, labeled say $l_1$ and $l_2$ of the matrix $C$ are identical, there 
 is a one-parameter family of solutions to the optimization problem 
for the brain-wide densities of the two corresponding cell types:
\begin{align}
\forall \phi \in [0,1],\;\forall v \in [1..V]\;\;\;\;\; 
&\rho_{t_1}(v)C(t_1,.) + \rho_{t_2}(v)C(t_2,.)  =\\ \nonumber 
&(1-\phi)\rho_{t_1}(v)C(t_1,.)+ \left(\rho_{t_2}(v)+\phi\rho_{t_1}(v)\right)C(t_2,.).
\label{transferEquation}
\end{align}
 The transfer of any positive amount of signal between the 
 two cell types at any voxel does not change the value 
 of the sum (the density profiles of the two types 
 are anti-correlated across the family of solutions).\\

 %look for pairs of similar profiles (dot-products)
 % and clusters of similar profiles.

 %%%%%%%%%
 % CODE:
 % similarTypes = similar_types( C, colsToUseInTypes );
 %%%%%%%%%

 We looked for families of cell-type-specific samples
 that are very similar to each other. We computed 
 the pairwise type-by-type correlation matrix
  between centered cell-type specific transcription profiles:
\begin{equation}
\mathrm{typeCorr}(t,t') = \frac{\sum_{g=1}^G\left( C(t,g) - \bar{C}(g)\right)\left( C(t',g ) - \bar{C}(g) \right)}{\sqrt{\sum_{g=1}^G\left( C(t,g) - \bar{C}(g)\right)^2 \sum_{h=1}^G \left( C(t',h ) - \bar{C}(h) \right)^2}},\label{corrEquation}
\end{equation}
 The closer the entry $\mathrm{typeCorr}(t,t')$ is to 1, the more similar the 
 two cell types $t$ and $t'$ are.  
 We applied the thresholding procedure that was used for co-expression 
matrices of genes in the brain in \cite{MenasheCoExpr}. For each value of $\tau$ in the interval 
 $[-1,1]$, a threshold can be applied to the matrix $\mathrm{typeCorr}$ 
 by putting to zero all entries lower than $\tau$:

\begin{equation}
\mathrm{typeCorr}_\rho(t,t') = \mathbf{1}\left( \mathrm{typeCorr}(t,t') \geq \rho\right)\left(1 +\mathrm{typeCorr}(t,t') \right)
\label{thresholdedCorr}
\end{equation}
 Applying Tarjan's algorithm
 \cite{Tarjan} to the thresholded matrix $\mathrm{typeCorr}_\rho$ produces a partition
 of the $T$ cell types into strongly connected components at this value
 of the threshold. At $\tau=1$, all the cell types are disconnected (unless
 there are cell types with exactly the same value
 for all genes, which is not the case in our data set). At $\tau=-1$,
 there is only one connected component, and all cell types 
 are connected. When $\tau$ decreases from $1$ to $-1$, the 
 more similar cell types group into connected components.
 We are interested in pairs of cell types that 
 are connected at high values of $\tau$ and do no not connect 
  with other cell types
  for the lowest possible value of $\tau$, 
 when $\tau$ is decreased (such cell types labeled  $t$ and $t'$ are the closest 
 pairs of cell types to the degenerate situation described by Equation \ref{transferEquation}.\\
 % CODE: 
 % pairsOfSimilarTypes = pairs_of_similar_types( C, colsToUseInTypes );
 % persistence length

 Consider a pair of indices $(t_1,t_2)$ singled out by the above analysis.
  Let us  introduce the notation $\hat{C}^{i}$ for the type-by-gene
 matrix obtained from $C$ by leaving the $i$-th row out, 
 we computed the corresponding density profiles 
 denoted by $\hat{\rho}^{t_1}$ and $\hat{\rho}^{t_2}$:\\
   
\begin{equation}
\left(\hat{\rho}^{t_1}_t( v )\right)_{1\leq t \leq T}= {\mathrm{argmin}}_{\nu\in {\mathbf{R}}_+^T}\sum_{g=1}^G \left( E(v,g)-\sum_{t=1}^T \hat{C}^{t_1} (t,g) \nu(t) \right)^2,
\label{voxelByVoxelt1}
\end{equation}

\begin{equation}
\left(\hat{\rho}^{t_2}_t( v )\right)_{1\leq t \leq T}= {\mathrm{argmin}}_{\nu\in {\mathbf{R}}_+^T}\sum_{g=1}^G \left( E(v,g)-\sum_{t=1}^T \hat{C}^{t_2} (t,g) \nu(t) \right)^2,
\label{voxelByVoxelt2}
\end{equation}

Intuitively, refitting the model with just one of the two similar types 
present ($t_1$ or $t_2$ )should result in the remaining cell type inheriting the 
 sum of the two density profiles estimated in the original model:
\begin{equation}
\hat{\rho}^{t_1}_{t_2}\overset{?}{\simeq}\rho_{t_1}+ \rho_{t_2},
\label{conjectureSum12}
\end{equation}
\begin{equation}
\hat{\rho}^{t_2}_{t_1}\overset{?}{\simeq} \rho_{t_1}+ \rho_{t_2}.
\label{conjectureSum21}
\end{equation}
 The profiles of the other cell types are expected to be stable:
\begin{equation}
\forall t\neq t_1, t_2, \hat{\rho}^{t_1}_{t}\overset{?}{\simeq}\rho_{t},
\end{equation}
\begin{equation}
\forall t\neq t_1, t_2, \hat{\rho}^{t_2}_{t}\overset{?}{\simeq}\rho_{t}.
\end{equation}
  To test this idea for each 
 pair of cell types $[t_1, t_2]$ singled out by the analysis of pairwise similarities,
 we computed the type-by-type matrices (of size $T$ by $T$, so that they can be plotted 
 as heat maps and compared more easily to the original $T$ by $T$ matrix of pairwise correlations
 between density profiles in the original model:
\begin{align}
\hat{\Gamma}^{t_1}(t,u)= \mathcal{C}(\hat{\rho}^{t_1}_t,\rho_u)\;\;\hat{\Gamma}^{t_1}(u,t) &=\mathcal{C}(\rho_u,\hat{\rho}^{t_1}_t), \;\;\; t,u\neq t_1,\;\; t,u\neq t_1,\\
\hat{\Gamma}^{t_1}(t_2,t_2)&= \mathcal{C}(\hat{\rho}^{t_1}_{t_2}, \rho_{t_1}+\rho_{t_2}).\\\label{sumDefCorr1}
\hat{\Gamma}^{t_1}(t_1, .)&=\hat{\Gamma}^{t_1}(., t_1) = 0.
\end{align}
\begin{align}
\hat{\Gamma}^{t_2}(t,u)= \mathcal{C}(\hat{\rho}^{t_2}_t,\rho_u)\;\;\hat{\Gamma}^{t_2}(u,t) &=\mathcal{C}(\rho_u,\hat{\rho}^{t_2}_t), \;\;\; t,u\neq t_2,\;\; t,u\neq t_2,\\
\hat{\Gamma}^{t_2}(t_1,t_1)&= \mathcal{C}(\hat{\rho}^{t_2}_{t_1}, \rho_{t_1}+\rho_{t_2}),\\\label{sumDefCorr2}
\hat{\Gamma}^{t_2}(t_2,. ) &= \hat{\Gamma}^{t_2}(.,t_2) =0.
\end{align}
The matrices $\hat{\Gamma}^{t_1}$ and $\hat{\Gamma}^{t_2}$ should have (apart from a ``cross'' of zeroes at the row and column indices $t_1$ and $t_2$ respectively) omitted from the new fitting panel,
 a diagonal of high correlation coefficients, and off-diagonal terms that should 
 be close to the off-diagonal terms of the type-by-type matrix $\Gamma$ of correlations between cell types  
 in the original model defined as follows:
\begin{equation}
\Gamma(t,u)= \mathcal{C}(\rho_t, \rho_u), \;\;\; t,u \in [ 1..T].
\label{GammaOriginal}
\end{equation}
 The matrix $\Gamma$ is symmetric and has its diagonal entries
all equal 1 by construction.

\subsubsection{Results}

The pairs of cell types singled out by the above-described
analysis are the following:\\
$(i)$ indices $(15,16)$, which are both medium spiny neurons ({\emph{Drd1+}} and {\emph{Drd2+}} respectively),\\
$(ii)$ indices $(4,5)$, which are both dopaminergic neurons (A9 and A10 respectively),\\ 
$(iii)$ indices $(2,3)$, which are both pyramidal neurons (but are not detected by the 
 model except at a few voxels in the olfactory bulb).\\

%%%%%%%%%%%%%%%%%%%%%%%%%% OLD %%%%%%%%%%%%%%%%%%%%%%%%%%%%
% CODE:
% studyPairsOfSimilarTypesStudy = study_pairs_of_similar_types( CUsed, fitVoxelsToTypes, pairsToStudy  );
% plot the type_by_type distances
% figure with the largest discrepancies between corresponding types
% study the top regions, note the differences
% CODE FOR FIGURES:
% refittedPairsOfTypes = refitted_pairs_of_types( Ref, E, C, colsToUseInAllen,...
%  colsToUseInTypes, fitVoxelsToTypes, pairOfIndices )
%%%%%%%%%%%%%%%%%%%%%%%%%%%%%%%%%%%%%%%%%%%%%%%%%%%%%%%%

% CODE FOR FIGURES: CORRELATIONS... AND DENSITIES
% figures_mutilated( Ref, fitVoxelsToTypes )
% 
% ... AND DENSITIES (contained in figures_mutilated )
% 
%    figure_densities_mutilated( Ref, fitVoxelsToTypes, fitMutilatedFirst, fitMutilatedSe% cond,firstIndex, secondIndex );

\begin{itemize}

\item{{\bf{The pair of medium spiny neurons (labeled by indices $t_1=15$ and $t_2=16$).}}}
Having computed the refitted profiles $\hat{\rho}^{15}$ and $\hat{\rho}^{16}$, which 
contain one medium spiny neurons instead of two, we can plot the density profiles 
 oh the remaining medium spiny neurons in each of the refitted models ($\hat{\rho}^{15}_{16}$ and $\hat{\rho}^{16}_{15}$, Figures \ref{figureDensitiesMutilated_15_16}(b) and \ref{figureDensitiesMutilated_15_16}(c) respectively), as well as the sum of the densities of the two samples of medium spiny neurons in the original model (i.e. $\rho_{15}+\rho_{16}$, Figure \ref{figureDensitiesMutilated_15_16}(c)). The three plots are hard to distinguish 
 from each other by eye (see Figure \ref{figureDensitiesMutilated_15_16}) and present 
 the same characteristic concentration of signal in the striatum. This supports 
 the conjectures \ref{conjectureSum12} and \ref{conjectureSum12}, as each of the medium spiny neurons seems to inherit the signal of the pair labeled by indices $15$ and $16$ when the model is refitted, without picking up any significant other signal.\\
 
\begin{figure}
\includegraphics[width=\textwidth,keepaspectratio]{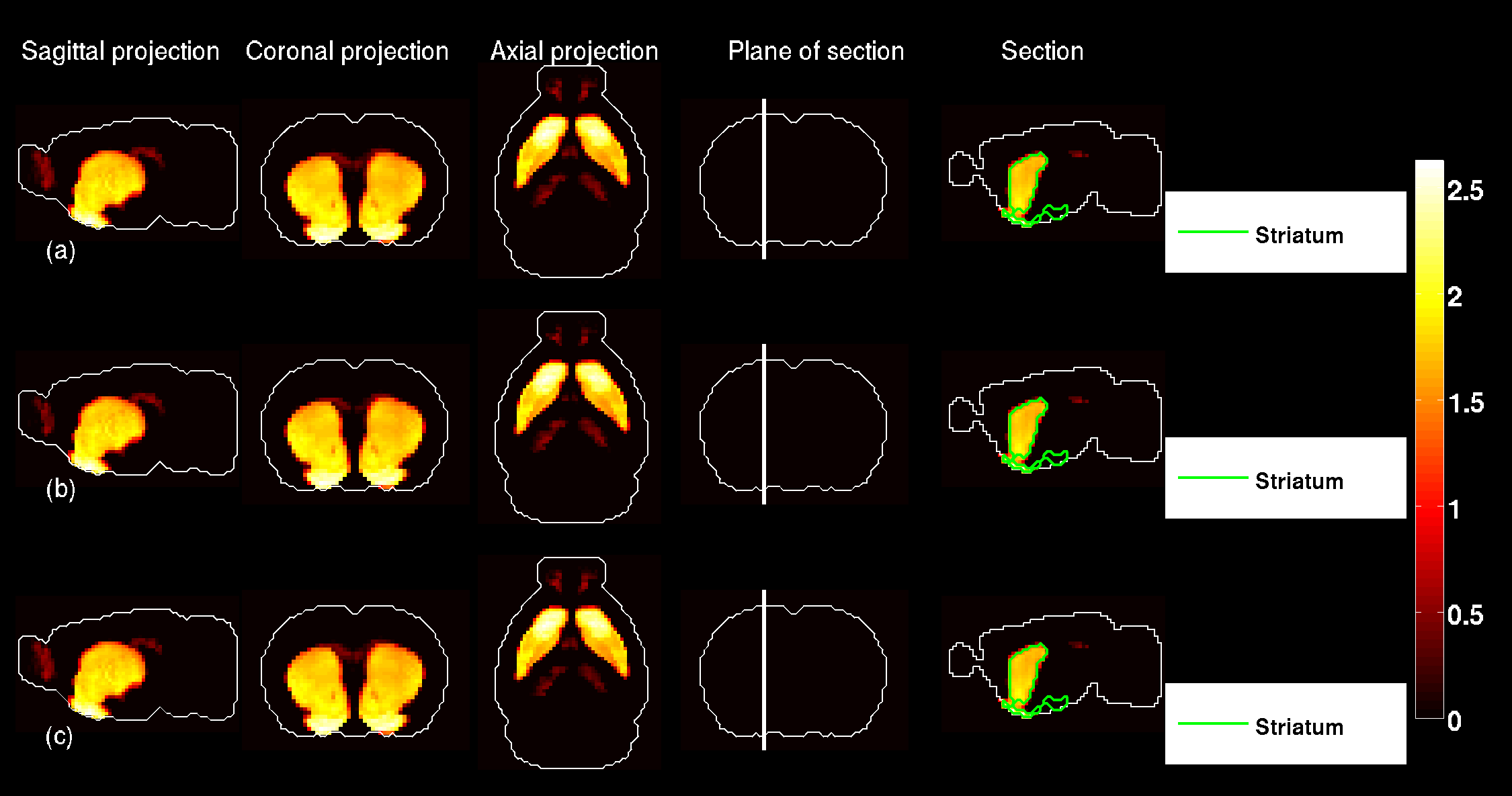}
\caption{{\bf{Medium spiny neurons are labeled by indices $t_1=15$ and $t_2=16$ in the origional fitting panel}}. (a) Heat map of the sum of estimated brain-wide densities in the original model, $\rho_{t_1}+\rho_{t_2}$. (b) Heat map of $\hat{\rho}^{t_2}_{t_1}$, the density of cell type $t_1=15$ when cell type $t_2=16$ is omitted from the fitting panel; it is 
 much closer to the sum $\rho_{t_1}+\rho_{t_2}$ than to the rather sparse profile $\rho_{t_1}$. (c) Heat map of $\hat{\rho}^{t_1}_{t_2}$, the density of cell type $t_2=16$ when cell type $t_1=15$ is omitted from the fitting panel. The three profiles are very close to each other by eye, consistently with the conjectures of Equations \ref{conjectureSum12} and \ref{conjectureSum12}.}
\label{figureDensitiesMutilated_15_16}
\end{figure}

Computing the correlation matrices $\hat{\Gamma}^{15}$ and $\hat{\Gamma}^{16}$
 and reading off the entries 
\begin{equation}
\hat{\Gamma}^{15}_{16}= 0.9997,\;\;\;\;\hat{\Gamma}^{16}_{15}=0.9992
\end{equation}
confirms the visual impression of Figure \ref{figureDensitiesMutilated_15_16},
 due to definitions \ref{sumDefCorr1} and \ref{sumDefCorr2}.\\

% Saved values of correlation matrices:
% Gamma: typeByTypeCorrelationsOriginal.mat 
% \hat{\Gamma}^{15}: correlationsMutilated15.mat
% \hat{\Gamma}^{16}: correlationsMutilated16.mat

 Moreover, plotting the matrices $\hat{\Gamma}^{15}$ and $\hat{\Gamma}^{16}$ as
heat maps (Figures \ref{projectionsPairfit_15_16}b and Figures \ref{projectionsPairfit_15_16}c) shows that the entries are roughly symmetric, and that the two matrices  
 resemble the type-by-type matrix of $\Gamma$ of pairwise correlations
 in the original model (Figure \ref{projectionsPairfit_15_16}a ), apart from the 
 ``cross'' of zeroes at the omitted indices (row 15 and column 15 on
 Figure \ref{projectionsPairfit_15_16}b, row 16 and column 16 on
 Figure \ref{projectionsPairfit_15_16}c). Moreover, the diagonal coefficients
 of $\hat{\Gamma}^{15}$ and $\hat{\Gamma}^{16}$ are visibly close to 1 (their sorted
 values are plotted on Figure \ref{diagCorrels_15_16}),
 with the exception of the cholinergic projection neurons (labeled $t=11$),
 whose density profile $\hat{\rho^{15}}_{11}$ has only a 0.133 correlation coefficient 
 with $\rho_{11}$. However, this cell type has very little signal 
 in any of the models ($\rho_{11}$ has only 7 voxels with positive density,
 $\hat{\rho}^{15}_{11}$ has 8 and  $\hat{\rho}^{16}_{11}$ has 7), and these three profiles
 all represent less than $10^{-8}$ times the sum of all the 
 densities in the respective models:
\begin{equation}
\frac{\sum_v\rho_{11}(v)}{\sum_t\sum_v\rho_{11}(v)}=1.53 \times 10^{-9},\;\;\;
\frac{\sum_v\hat{\rho}^{15}_{11}(v)}{\sum_t\sum_v\hat{\rho}^{15}_{11}(v)}= 6.65 \times 10^{-9},\;\;\;
\frac{\sum_v\hat{\rho}^{16}_{11}(v)}{\sum_t\sum_v\hat{\rho}^{16}_{11}(v)}= 1.53 \times 10^{-9}.
\end{equation}
 The  low value of the correlation for cell-type index 11 
 is therefore compatible with the claim that density profiles
 are generally stable when replacing the pair of medium spiny 
 neurons by just one of them, and that the remaining 
  medium spiny neuron in the fitting panel inherits the sum of
 the density profiles predicted in the original model.

\begin{SCfigure}   
\includegraphics[width=0.6\textwidth,keepaspectratio]{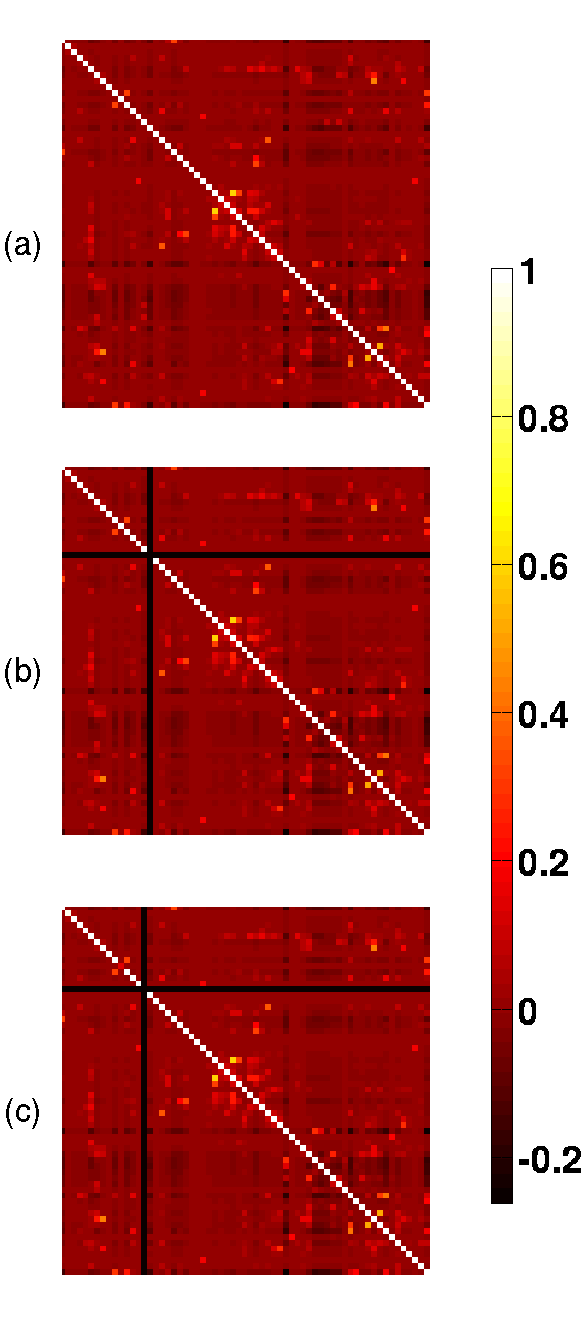}
\caption{{\bf{Heat maps of the type-by-type matrices of pairwise
      correlations between density profiles}}. (a) Using density
  profiles in the original model, or $\Gamma$, Equation
  \ref{GammaOriginal}. (b) Using density profiles $\hat{\rho}^{15}$
  from the model using only medium spiny neuron labeled 16, and
  densities from the original model, except at column 16 where the sum
  $\rho^{15}+\rho^{16}$ is used (see Equation \ref{sumDefCorr1}). (c) Using density profiles $\hat{\rho}^{16}$
  from the model using only medium spiny neuron labeled 15, and
  densities from the original model, except at column 15 where the sum
  $\rho^{15}+\rho^{16}$ is used (see Equation \ref{sumDefCorr2}). The three matrices
 look similar (apart from the zeroes at row and column indices 15 and 16 on (b) and (c) respectively). This suggests that the results of the model are stable upon using 
 one sample of medium spiny neuron instead of two.}
\label{projectionsPairfit_15_16}
\end{SCfigure}

\begin{figure}
\includegraphics[width=\textwidth,keepaspectratio]{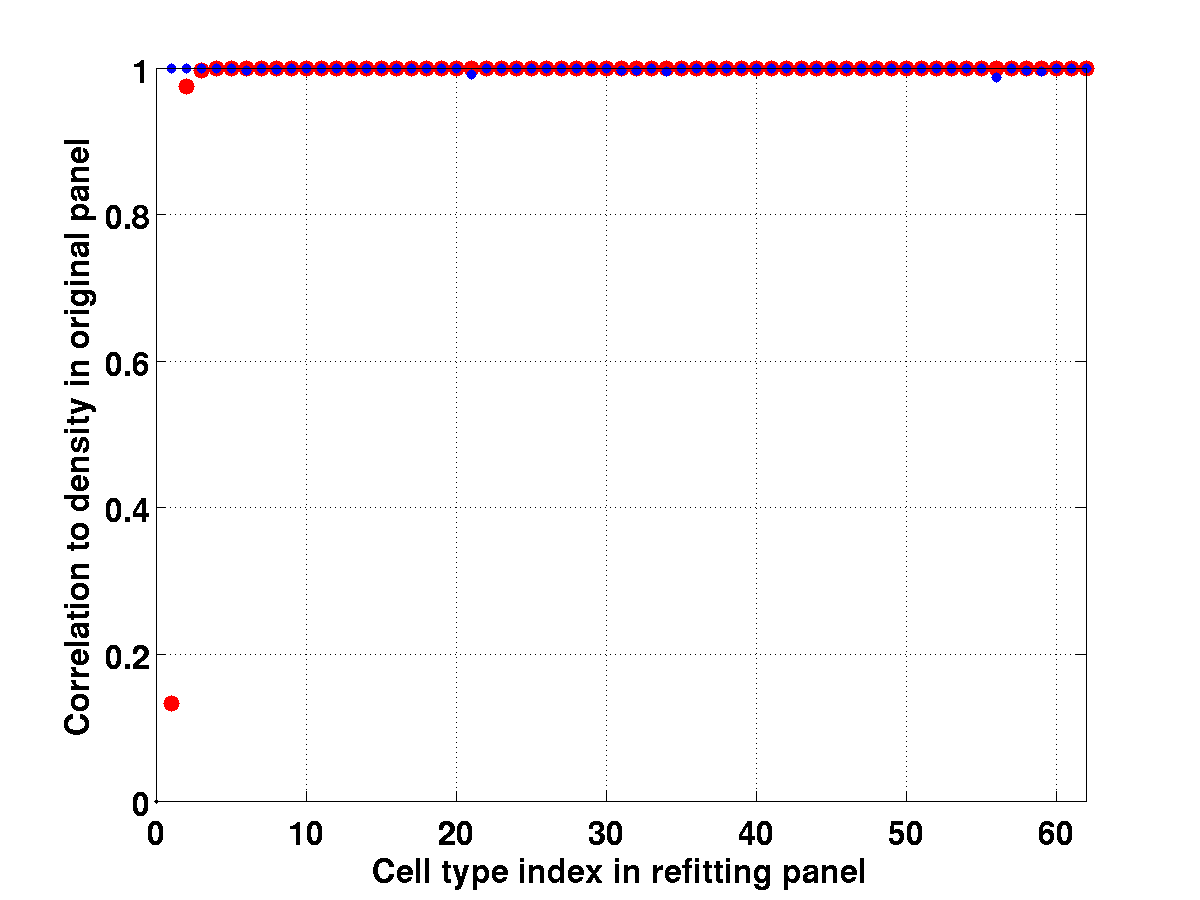}
\caption{Sorted diagonal elements of the type-by-type correlation
  matrix $\hat{\Gamma}^{15}$ (in red), and $\hat{\Gamma}^{16}$ (in blue,
  plotted in the same order as the diagonal coefficients of
  $\hat{\Gamma}^{15}$). All diagonal entries of $\hat{\Gamma}^{15}$ and
  $\hat{\Gamma}^{16}$ are above $0.97$, except the entry of
  $\hat{\Gamma}^{15}$ corresponding to index 11 (cholinergic projection
  neurons), but this cell type has very low density in all three
  models $\rho$, $\hat{rho}^{15}$, $\hat{\rho}^{16}$.}
\label{diagCorrels_15_16}
\end{figure}

\item{{\bf{The pair (A9 dopaminergic neurons, A10 dopaminergic neurons) labeled by indices
 $t_1=4$ and $t_2=5$.}}}
 We computed the refitted profiles $\hat{\rho}^{4}$ and $\hat{\rho}^{5}$, which 
contain one sample of dopaminergic neurons instead of two.
  Again we can plot the density profiles of the remaining dopaminergic neuron
  in each of the refitted models ($\hat{\rho}^{4}_{5}$ and $\hat{\rho}^{5}_{4}$, Figures \ref{figureDensitiesMutilated_4_5}(b) and \ref{figureDensitiesMutilated_4_5}(c) respectively), as well as the sum of the densities of A9 dopaminergic neurons and A10 dopaminergic neurons in the original model (i.e. $\rho_4+\rho_5$, Figure \ref{figureDensitiesMutilated_4_5}(c)). The three plots show the highest 
 fraction of their signal in the midbrain (hence the sections through mibrain in all three rows
 of Figure \ref{figureDensitiesMutilated_4_5}). generally speaking they look quite similar.
 Again this supports 
 the conjectures \ref{conjectureSum12} and \ref{conjectureSum12}, in the case of $t_1=4$ and $t=5$.\\

\begin{figure}
\includegraphics[width=\textwidth,keepaspectratio]{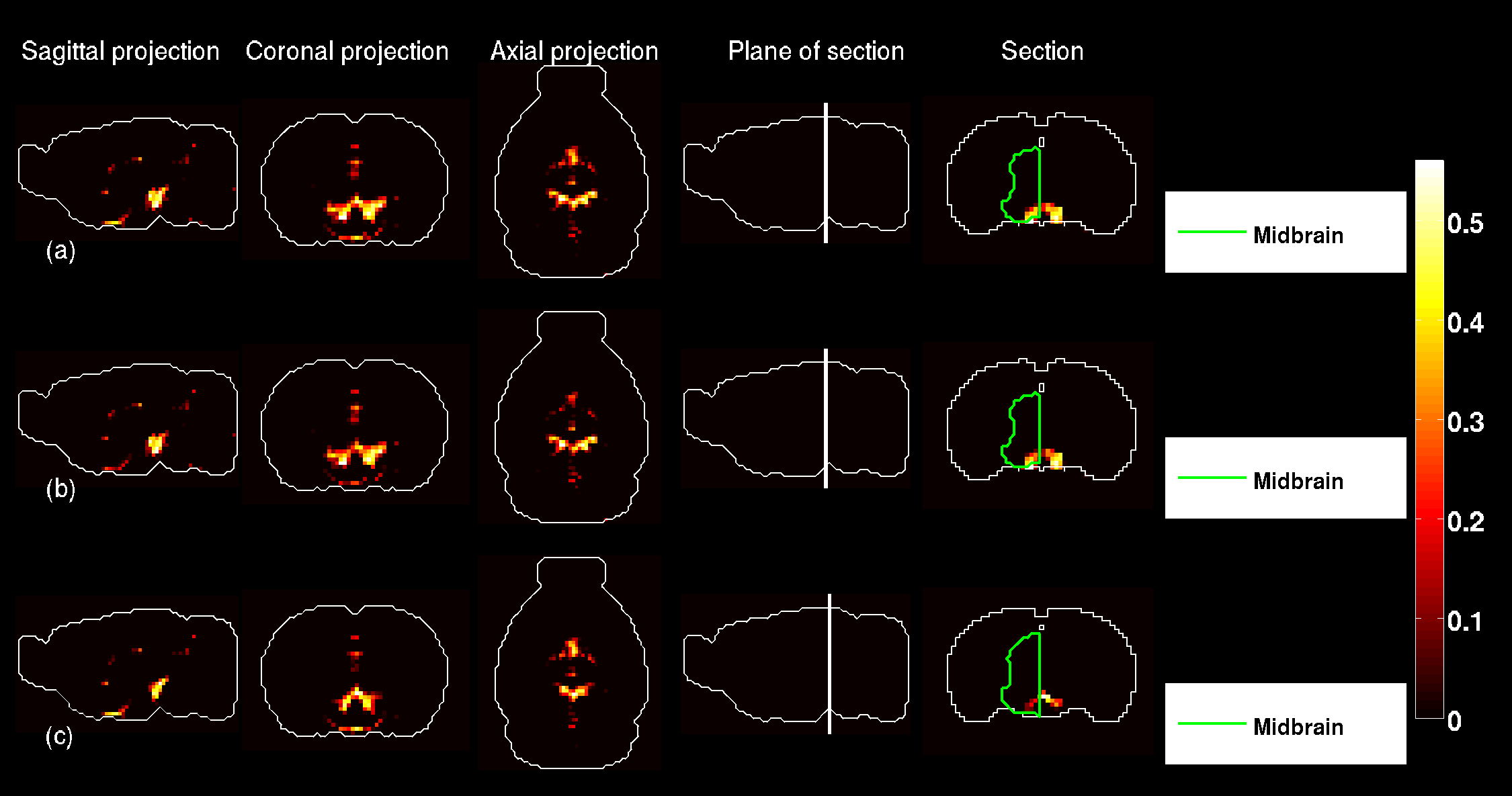}
\caption{{\bf{A9 dopaminergic neurons and A10 dopaminergic neurons are labeled by indices $t_1=4$ and $t_2=5$ in the origional fitting panel}}. (a) Heat map of the sum of estimated brain-wide densities in the original model, $\rho_{t_1}+\rho_{t_2}$. (b) Heat map of $\hat{\rho}^{t_2}_{t_1}$, the density of cell type $t_1=4$ when cell type $t_2=5$ is omitted from the fitting panel. (c) Heat map of $\hat{\rho}^{t_1}_{t_2}$, the density of cell type $t_2=5$ when cell type $t_1=4$ is omitted from the fitting panel. The three profiles are very close to each other by eye, consistently with the conjectures of Equations \ref{conjectureSum12} and \ref{conjectureSum12}.}
\label{figureDensitiesMutilated_4_5}
\end{figure}

Computing the correlation matrices $\hat{\Gamma}^{4}$ and $\hat{\Gamma}^{5}$
 and reading off the entries  
 The correlations conjectured in Equations \ref{sumDefCorr1} and \ref{sumDefCorr2} 
 are less close to 1 than in the case of medium spiny neurons, but still
\begin{equation}
\hat{\Gamma}^{4}_{5}= 0.87,\;\;\;\;\hat{\Gamma}^{5}_{4}=0.96,
\end{equation}
 and one can check that in each of the two refitted models,
 the sum of profiles $\rho_4+\rho_5$ has higher correlation
 with the remaining dopaminergic neurons than any 
 cell type in the original fitting panel:
\begin{equation}
\hat{\Gamma}^{t_1}(t_2,t_2) = {\mathrm{max}}_{1\leq t\leq T} \hat{\Gamma}^{t_1}(t_2,t),\;\;\;\hat{\Gamma}^{t_2}(t_1,t_1) = {\mathrm{max}}_{1\leq t\leq T} \hat{\Gamma}^{t_2}(t_1,t).
\end{equation}

 As in the case of medium spiny neurons, plotting the matrices
 $\hat{\Gamma}^{4}$ and $\hat{\Gamma}^{5}$ as heat maps (Figures
 \ref{projectionsPairfit_4_5}b and Figures
 \ref{projectionsPairfit_4_5}c) shows that the entries are roughly
 symmetric, and that the two matrices resemble the type-by-type matrix
 of $\Gamma$ of pairwise correlations in the original model (Figure
 \ref{projectionsPairfit_4_5}a ), apart from the ``crosses'' of zeroes
 (at row 4 and column 15 on Figure \ref{projectionsPairfit_4_5}b, row
 16 and column 16 on Figure \ref{projectionsPairfit_4_5}c). Moreover,
 the diagonal coefficients of $\hat{\Gamma}^{4}$ and
 $\hat{\Gamma}^{5}$ are all close to 1 (their sorted values are
 plotted on Figure \ref{diagCorrels_4_5}). Again we conclude 
 that refitting the model to a fitting panel containing 
 one dopaminergic neuron instead of two leads 
  to stable density profile, with the 
  remaining dopaminergic neuron inheriting most 
 of the sum of the two original density profiles $\rho_4$ and $\rho_5$.  
 
\begin{SCfigure}
\includegraphics[width=0.6\textwidth,keepaspectratio]{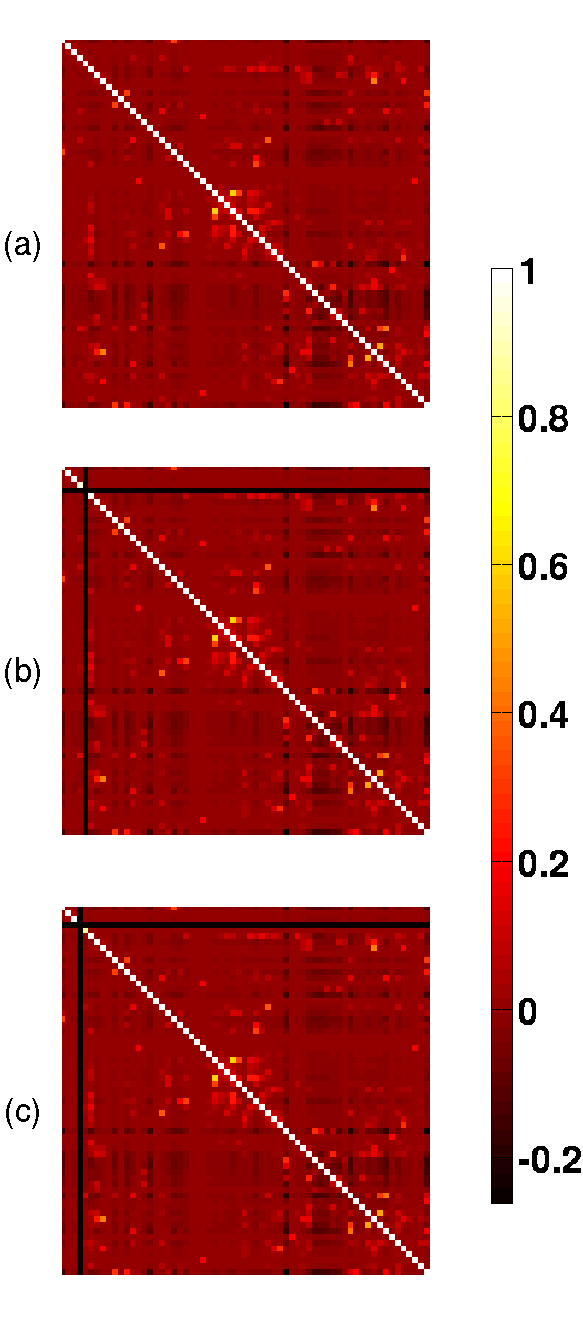}
\caption{{\bf{Heat maps of the type-by-type matrices of pairwise
      correlations between density profiles, for $t_1= 4$ and $t_2=5$ }}. (a) The matrix $\Gamma$ computed using density
  profiles in the original model, or $\Gamma$, Equation
  \ref{GammaOriginal}. (b) Using density profiles $\hat{\rho}^{4}$
  from the model using only A10 dopaminergic neurons, labeled $t_2=5$, and
  densities from the original model, except at column 5 where the sum
  $\rho^{4}+\rho^{5}$ is used (see Equation \ref{sumDefCorr1}). (c) Using density profiles $\hat{\rho}^{5}$
  from the model using only A10 dopaminergic neurons, labeled $t_1=4$, and
  densities from the original model, except at column 4 where the sum
  $\rho^{4}+\rho^{5}$ is used (see Equation \ref{sumDefCorr2}). The three matrices
 look similar (apart from the zeroes at row and column indices 4 and 5 on (b) and (c) respectively). This suggests that the results of the model are stable upon using 
 either sample of dopaminergic neurons, instead of both of them.}
\label{projectionsPairfit_4_5}
\end{SCfigure}

\begin{figure}
\includegraphics[width=\textwidth,keepaspectratio]{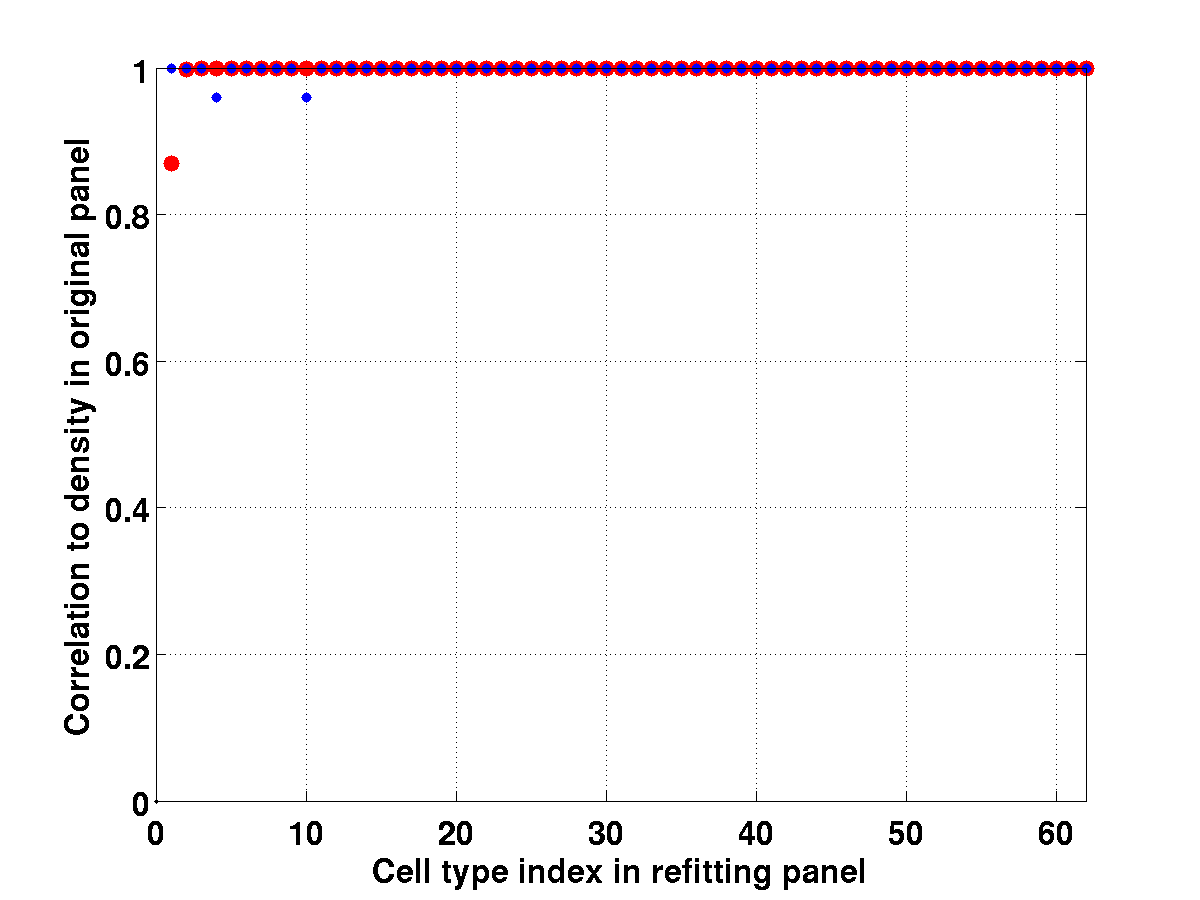}
\caption{Sorted diagonal elements of the type-by-type correlation
  matrix $\hat{\Gamma}^{4}$ (in red), and $\hat{\Gamma}^{5}$ (in blue,
  plotted in the same order as the diagonal coefficients of
  $\hat{\Gamma}^{4}$).}
\label{diagCorrels_4_5}
\end{figure}

\item{{\bf{The pair of pyramidal neurons labeled by indices
 $t_1=3$ and $t_2=3$.}}} We computed the refitted densities $\hat{\rho}^{2}$ 
 and $\hat{\rho}^3$ and repeated the above analysis. Figures 
 \ref{projectionsPairfit_2_3}, \ref{diagCorrels_2_3} and
\ref{figureDensitiesMutilated_2_3} confirm conjectures 
 \ref{conjectureSum12} and \ref{conjectureSum12}. 
 However, they are included mostly for 
 completeness, as the estimated densities of these two pyramidal 
 neurons are very low in the original model (only 52 voxels , about 1 in 1000, 
have a positive density in the sum $\rho_2+\rho_3$). They stay low 
 in the refitted model ($\hat{\rho}^2_3$ has only  49 voxels with non-zero density
  and $\hat{\rho}^3_2$ only 44). The high diagonal  correlation coefficients
 of Figure \ref{diagCorrels_2_3} show that the rest of the estimated 
 densities hardly change when refitting.

\begin{figure}
\includegraphics[width=\textwidth,keepaspectratio]{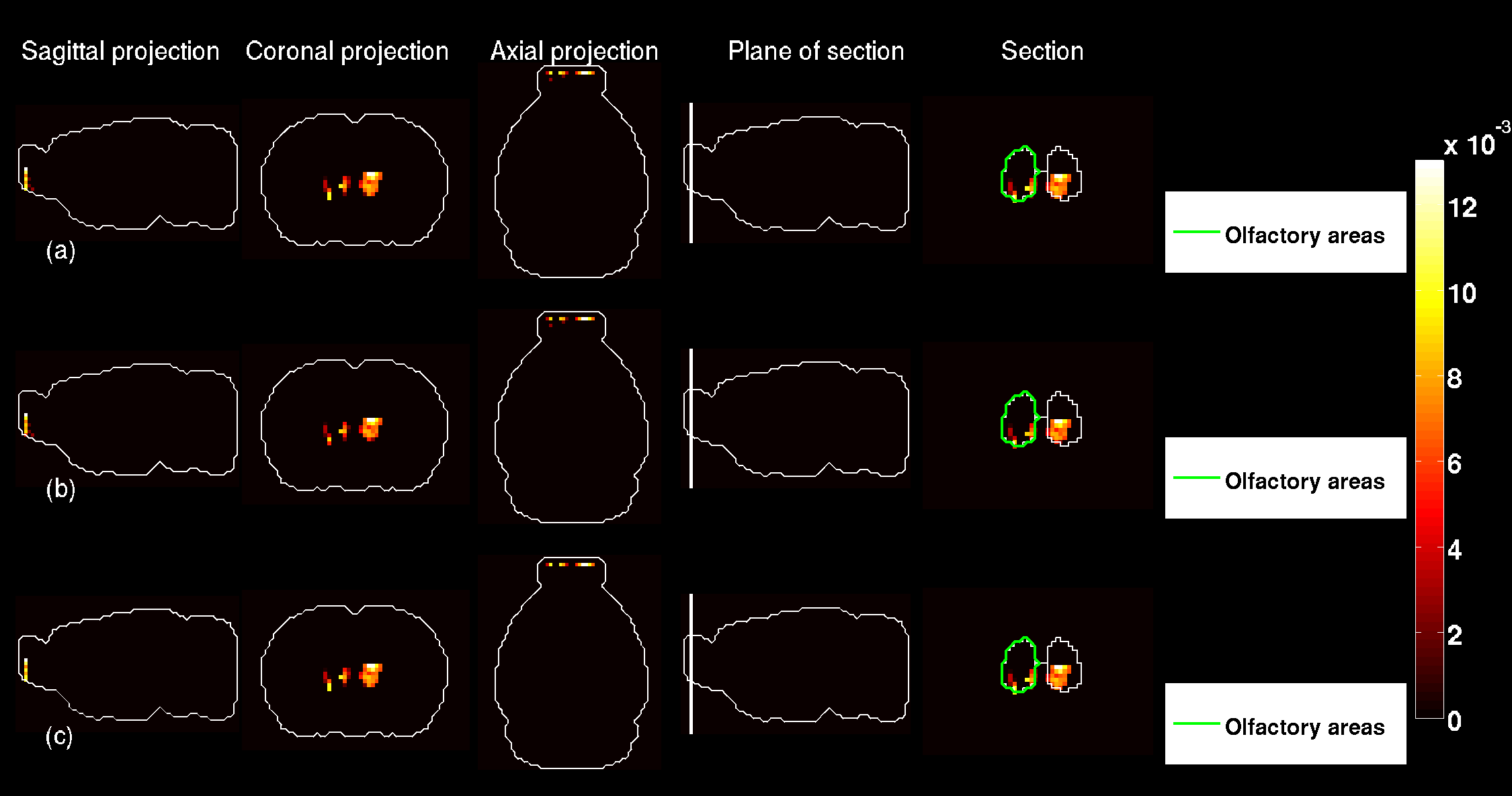}
\caption{{\bf{The pair of pyramidal neurons labeled by indices $t_1=2$ and $t_2=3$ in the origional fitting panel}}. (a) Heat map of the sum of estimated brain-wide densities in the original model, $\rho_{t_1}+\rho_{t_2}$. (b) Heat map of $\hat{\rho}^{t_2}_{t_1}$, the density of cell type $t_1=2$ when cell type $t_2=3$ is omitted from the fitting panel; it is 
 much closer to the sum $\rho_{t_1}+\rho_{t_2}$ than to the rather sparse profile $\rho_{t_1}$. (c) Heat map of $\hat{\rho}^{t_1}_{t_2}$, the density of cell type $t_2=3$ when cell type $t_1=2$ is omitted from the fitting panel. The three profiles are very close to each other by eye, consistently with the conjectures of Equations \ref{conjectureSum12} and \ref{conjectureSum12}.}
\label{figureDensitiesMutilated_2_3}
\end{figure}

\begin{SCfigure}
\includegraphics[width=0.6\textwidth,keepaspectratio]{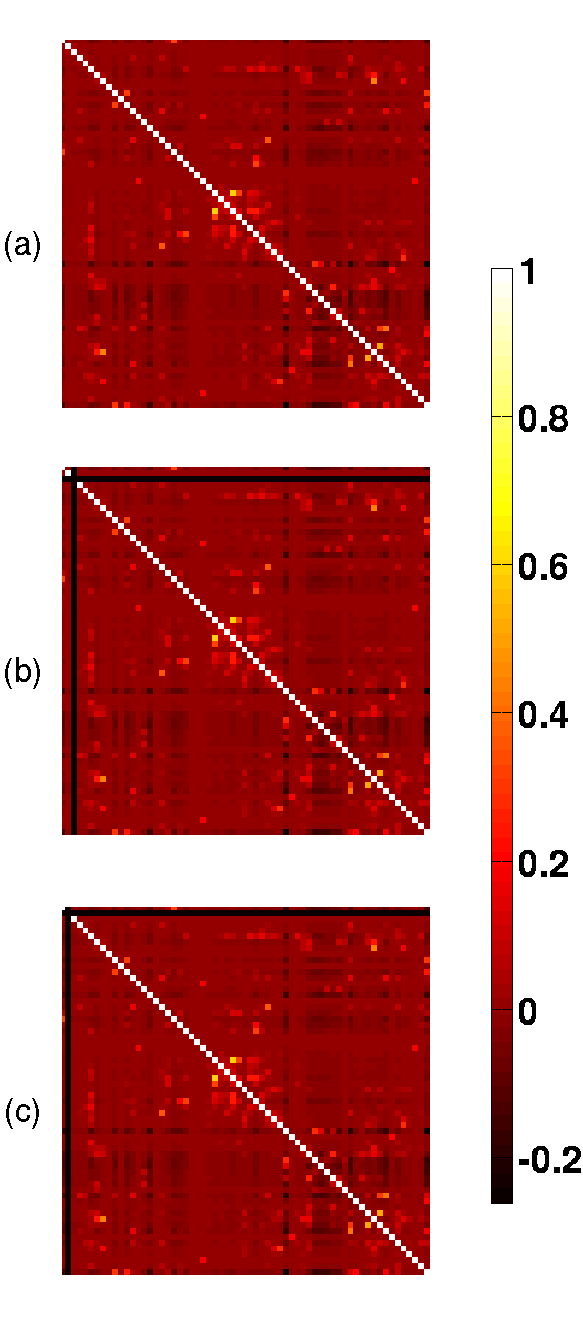}
\caption{{\bf{Heat maps of the type-by-type matrices of pairwise
      correlations between density profiles, for $t_1= 2$ and $t_2=3$ }}. (a) The matrix $\Gamma$ computed using density
  profiles in the original model, or $\Gamma$, Equation
  \ref{GammaOriginal}. (b) Using density profiles $\hat{\rho}^{2}$
  from the model using only the cell type labeled $t_2$ (instead of $t_1$ and $t_2$), and
  densities from the original model, except at column 3 where the sum
  $\rho^{2}+\rho^{3}$ is used (see Equation \ref{sumDefCorr1}). (c) Using density profiles $\hat{\rho}^{3}$
  from the model using only the cell type labeled labeled $t_1=2$ (instead of $t_1$ and $t_2$), and
  densities from the original model, except at column 2 where the sum
  $\rho^{2}+\rho^{3}$ is used (see Equation \ref{sumDefCorr2}). The three matrices
 look similar (apart from the zeroes at row and column indices 2 and 3 on (b) and (c) respectively). This suggests that the results of the model are stable upon using 
 either of the cell types labels $(2,3)$, instead of both of them.}
\label{projectionsPairfit_2_3}
\end{SCfigure}

\begin{figure}
\includegraphics[width=\textwidth,keepaspectratio]{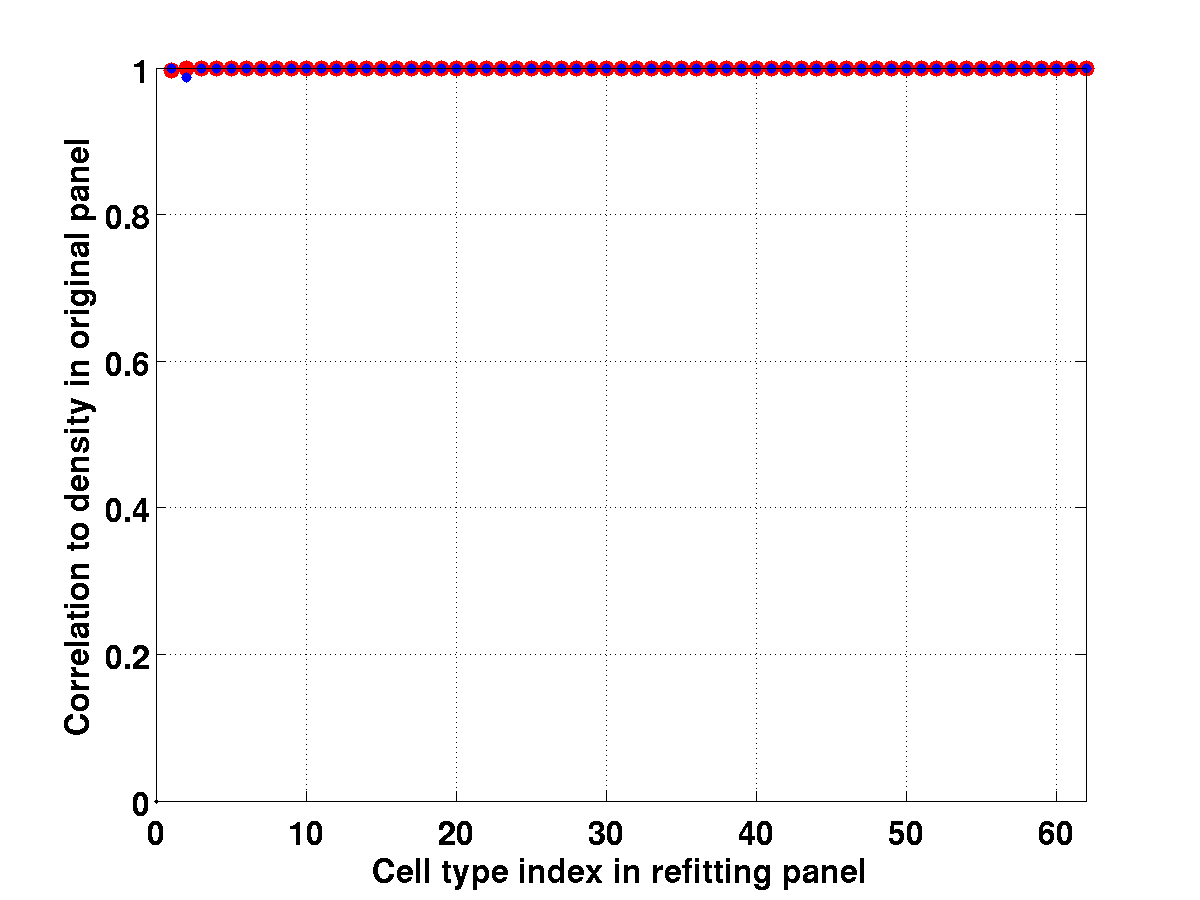}
\caption{Sorted diagonal elements of the type-by-type correlation
  matrix $\hat{\Gamma}^{2}$ (in red), and $\hat{\Gamma}^{3}$ (in blue,
  plotted in the same order as the diagonal coefficients of
  $\hat{\Gamma}^{2}$).}
\label{diagCorrels_2_3}
\end{figure}
\end{itemize}

\clearpage

\subsection{The set of pyramidal neurons}
 The previous subsection dealt with pairs of indices identified 
 computationally for their strong similarity to each other, as well
 as dissimilarity with the rest of the fitting panel (they happened 
 to have similar names in the taxonomy of cell types). 
 In this section we refit the model based on a choice made using 
 the names of the cell-type-specific samples: we group together 
 the largest set of cell types with similar names (pyramidal neurons),
 and replace them with a composite cell type made of the 
 average transcription profiles of all pyramidal neurons 
 in the data set.\\

 We combined all the $T_{pyr}=18$ transcriptomic profiles 
 of pyramidal neurons into their average:\\
 \begin{equation}
 \CPyr( g ) = \frac{1}{T_{\mathrm{pyr}}}\sum_{i=1}^{T_{\mathrm{pyr}}}C(t_i,g),
\label{CPyr}
 \end{equation}
where $t_i$ are the indices of the cell types labeled 
 as pyramidal neurons.
  We refitted the model to a set of $T'=47$ cell types consisting of 
 this composite pyramidal cell type with the data of the non-pyramidal 
 cell types concatenated with the data in $\CPyr$.\\
 
 More precisely, let us rewrite the new fitting panel as a matrix $\CNew$ of size $T- T_{\mathrm{pyr}} + 1$ by concatenating 
 $\CPyr$ and the rows of $C$ corresponding to non-pyramidal indices.
 Let $n_1,\dots, n_{T'}$ be the indices of the non-pyramidal cell
 types (row indices of $C$), where $T'= T -T_{\mathrm{pyr}}$ is the number of non-pyramidal
 cell types (i.e. for all $i$ in $[1..T']$, $C(n_i,.)$ is the transcription
 profile of a non-pyramidal cell):
 
\begin{align}
\CNew(1,.) &= \CPyr( g )\\
\forall i \in [1..T'],\;\;\CNew(i+1, . ) &= C(n_i,.).
\label{CNew}
\end{align}

The brain-wide density profile is the refitted profile, denoted by 
 $\rhoNew$, are the solutions of the usual optimization problems at 
 each voxel:\\
\begin{equation}
\forall v \in [1..V]\;\;\left(\rhoNew_t( v )\right)_{1\leq t \leq T'}= {\mathrm{argmin}}_{\nu\in {\mathbf{R}}_+^{T'+1}}\sum_{g=1}^G \left( E(v,g)-\sum_{t=1}^{T'+1}\CNew (t,g) \nu(t) \right)^2.
\label{modelPyr}
\end{equation}
 There are fewer degrees of freedom in the optimization problem \ref{modelPyr} than in the original 
 model, so the fitting is expected to be less accurate, 
 but one can ask how correlated the density $\rhoNew_1$ of the composite pyramidal 
 neuron is to the sum of the density profiles of all the pyramidal neurons in the original model.

 Let us denote by $\rhoPyr$ the sum of the density profiles
 of all the pyramidal cells in the original model:
\begin{equation}
\rhoPyr = \sum_{i=1}^{T_\mathrm{pyr}}\rho_{t_i}.
\end{equation}
The natural matrix of correlations $\GammaNew$ to compute is the 
 following:
\begin{align}
\GammaNew(t,u)&= \mathcal{C}(\rhoNew_t,\rho_{n_u}) \;\;\forall t >1,\; \forall u > 1,\\
\GammaNew( 1, 1)&=  \mathcal{C}(\rhoNew_t, \rhoPyr),\\
\GammaNew( 1, u)&=  \mathcal{C}(\rhoNew_1, \rho_{n_u})\;\;\forall u > 1,\\
\GammaNew( u, 1)&=  \mathcal{C}(\rhoNew_u, \rhoPyr)\;\;\forall u > 1,
\label{GammaNew}
\end{align}
which is expected to be close to the following $T'+1$ by $T'+1$ matrix
 based on the profiles $\rho$ of the original model only:
\begin{align}
\GammaOld(1,1) &= \mathcal{C}(\rhoPyr,\rhoPyr)\\
\forall i,j \in [1..T']\;\; \Gamma(i+1,j+1) &= \mathcal{C}(\rho_{n_i}, \rho_{n_j}),\\
\GammaOld(1,i+1) = \GammaOld( i+1, 1 ) &= \mathcal{C}( \rhoPyr,\rho_{n_i} ).
\label{GammaOld}
\end{align}
 The correlation $\GammaNew(1,1)= 0.8875$ is indeed close to $1$, moreover
 it is the maximum entry in the first row (the next-highest value in the first row is 0.3621) and in the first  
 column of $\GammaNew$ (the next-highest value in the first column is -0.0033):
 \begin{equation}
 \mathrm{argmax}_{i\in [1..T'+1]}\GammaNew(.,1) = 1,\;\;\;\mathrm{argmax}_{i\in [1..T'+1]}\GammaNew(1,.) = 1.
 \end{equation}
 The plots of $\rhoPyr$ and $\rhoNew_1$ show that indeed the two profiles are mostly 
 concentrated in the cerebral cortex. The most visible difference comes from the 
 hippocampus, wich is more highlighted in $\rhoPyr$ than in the density $\rhoNew_1$ of the composite pyramidal 
 cell type.

% CODE: pyramidal_refit.m
% pyramidalRefit = pyramidal_refit( Ref, E, C, fitVoxelsToTypes, colsToUseInAllen, colsToUseInTypes );

\begin{figure}
\includegraphics[width=\textwidth,keepaspectratio]{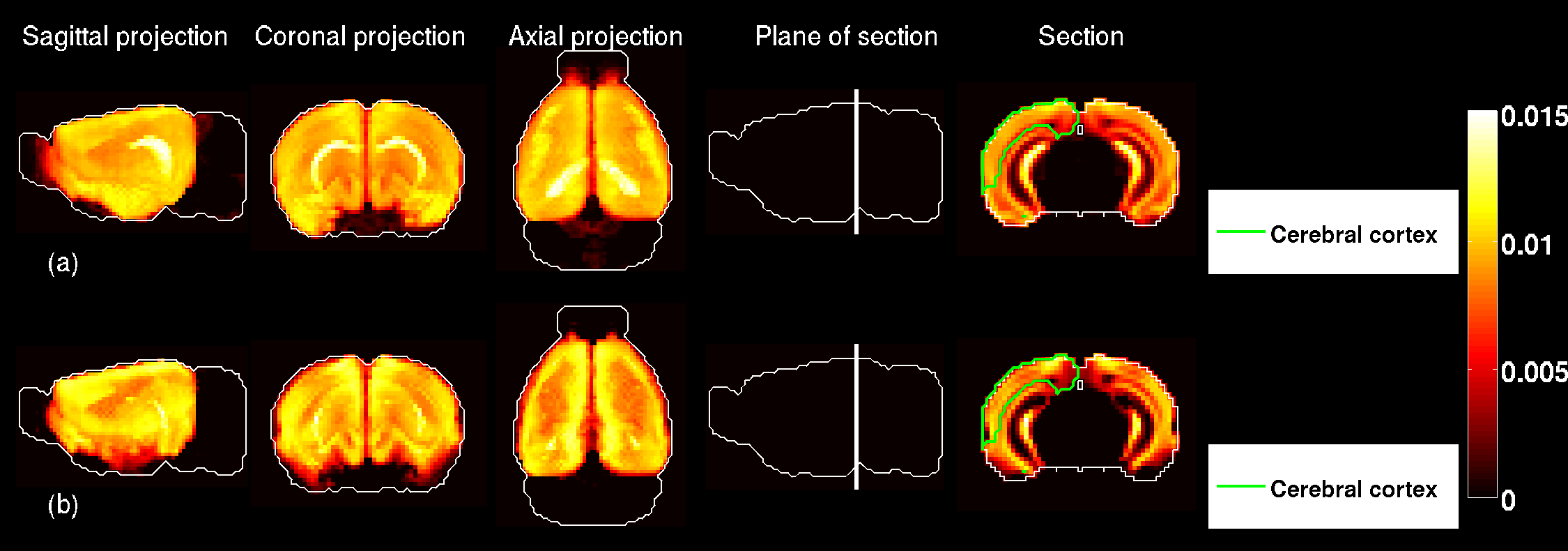}
\caption{(a) Heat map of $\rhoPyr$, the sum of  brain-wide densities of pyramidal neurons in the
 original model. (5) Heat map of the profile $\rhoNew_1$, the estimated brain-wide 
 density of the composite pyramidal cell defined in Equation \ref{CPyr}. The correlation between 
these two brain-wide profiles is $\GammaNew(1,1)= 0.8875$.}
\label{figurePyramidalDensities}
\end{figure}

\begin{figure}
\includegraphics[width=\textwidth,keepaspectratio]{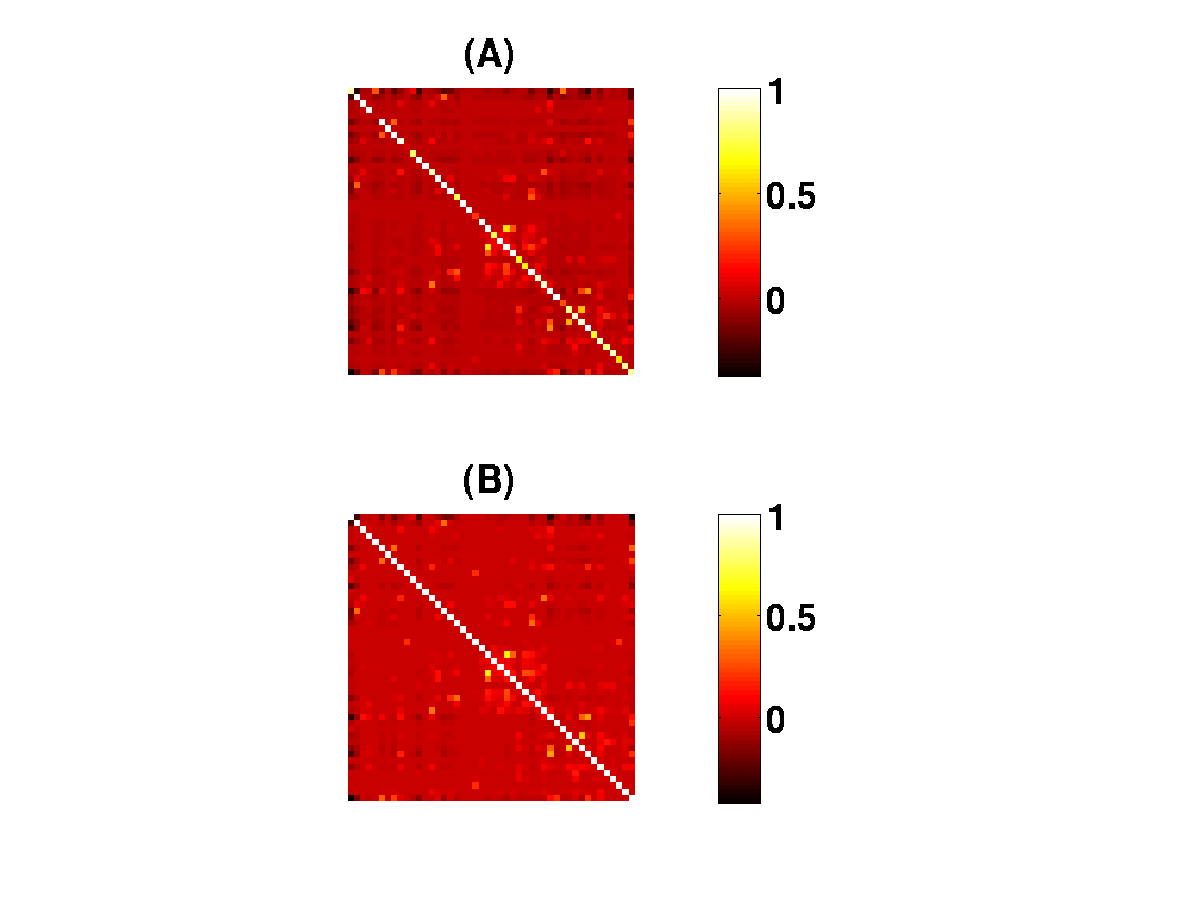}
\caption{(a) Heat map of the type-by-type correlation matrix $\GammaNew$ defined in Equation \ref{GammaNew}. 
 (b) Heat map of the type-by-type correlation matrix $\GammaOld$ defined in Equation \ref{GammaOld}.}
\label{correlsGridPyrAvg}
\end{figure}

\begin{figure}
\includegraphics[width=\textwidth,keepaspectratio]{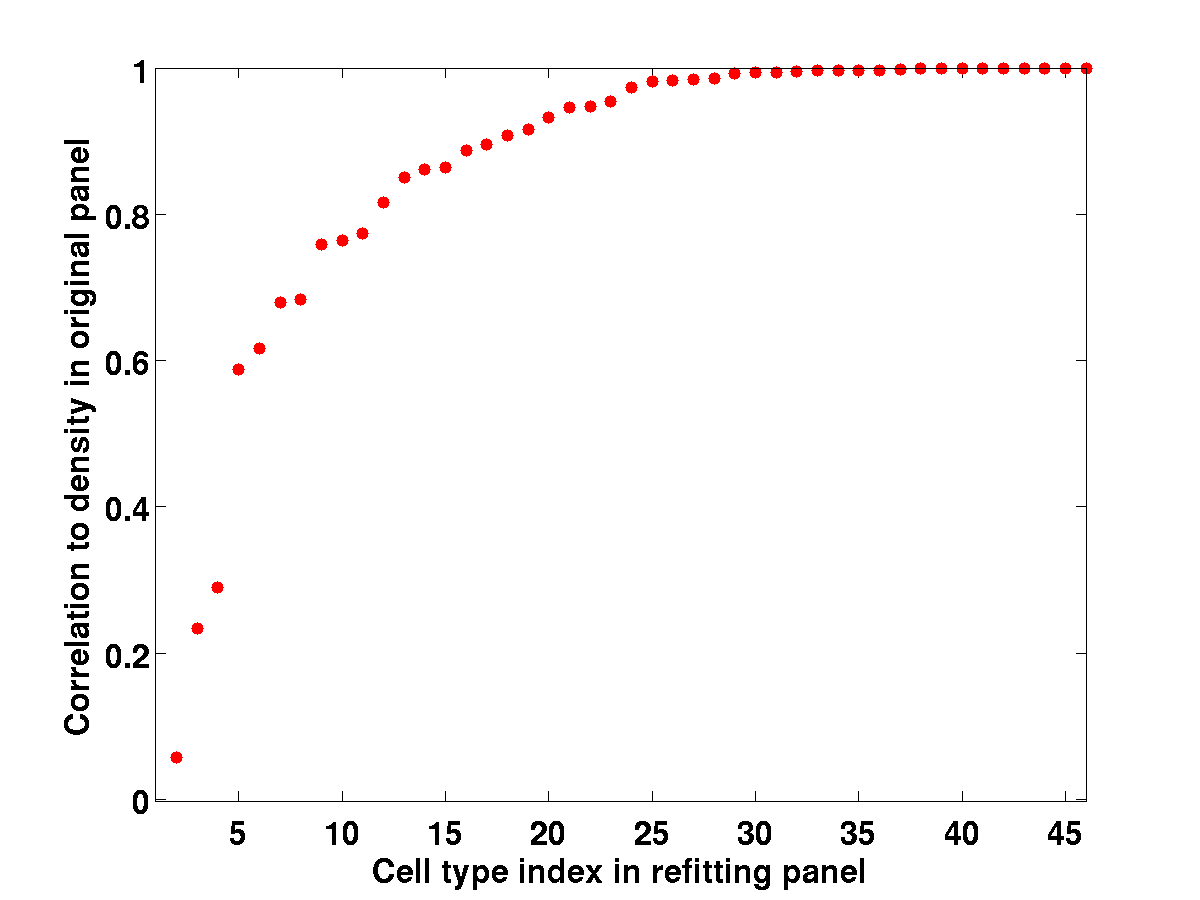}
\caption{Sorted diagonal elements of the type-by-type correlation matrices $\GammaNew$ (a) and $\GammaOld$ (b).}
\label{diagCorrelsPyramidalStudy}
\end{figure}

 Figures \ref{correlsGridPyrAvg} and \ref{diagCorrelsPyramidalStudy} show some similarity 
 between $\GammaNew$ and $\GammaOld$, moreover the diagonal correlations
 in $\GammaNew$ are larger than 0.8 for 35 out of 48 cell types in the new fitting panel 
 defined in Equation \ref{CNew}.

%{\bf{Hippocampus hiatus?}}
%Plotting the projections of the estimated profile
% of $\CPyr$ and that of the sum of the profiles of all the pyramidal neurons,
% it appears that the refitted profile has inherited 
% from most of the sum of the densities of the pyramidal neurons,
% except in the hippocampus and in the amygdala.

%  We refitted the model by leaving the hippocampal 
% and amygdalar samples as independent rows, 
% and computed the composite pyramidal profile using only 
% the 16 pyramidal neurons that are
%  extracted from the cerebral cortex.\\

\subsection{Background value in microarray data}

It can be deduced by examination of the ISH data set (the voxel-by-gene matrix $E$ representing the 
 Allen Atlas of the adult mouse brain)
 and the microarray data set (the type-by-gene matrix $C$) that the background intensities
 are lower in the Allen Atlas than in the cell-type-specific microarray 
 data. Out of the $G=2,131$ genes analyzed in the intersection 
 of the two datasets:\\
1. no gene has zero signal in any of the 
 cell types, moreover, the minimum entry
 of the type-by-gene matrix $C$ is a relatively large
 fraction of the average value of the average of all the entries:\\
 \begin{equation}
 m_C := {\mathrm{min}}_{1\leq t \leq T, 1\leq g \leq G}C( t, g ),
\label{minimumC}
 \end{equation}
 \begin{equation}
 \frac{m_C}{\frac{1}{GT}\sum_{1\leq t \leq T}
\sum_{1\leq g \leq G} C(t,g)} =  46.5\%.
\label{minimumC}
 \end{equation}
2. Each gene has zero ISH signal in at least 0.38 percent of the voxels,
 and the average proportion of voxels with zero signal is larger than 4 percent   
 \begin{equation}
 m_E := {\mathrm{min}}_{1\leq v \leq V, 1\leq g \leq G}E( v, g ) = 0,
\label{minimumE}
\end{equation}
 \begin{equation}
 \;\;\;\frac{1}{VG}\sum_{1\leq g \leq G}\left|\{v \in [1..V], E(v,g)=0\}\right| = 4.16\%. 
 \end{equation}

\begin{figure}
\includegraphics[width=\textwidth,keepaspectratio]{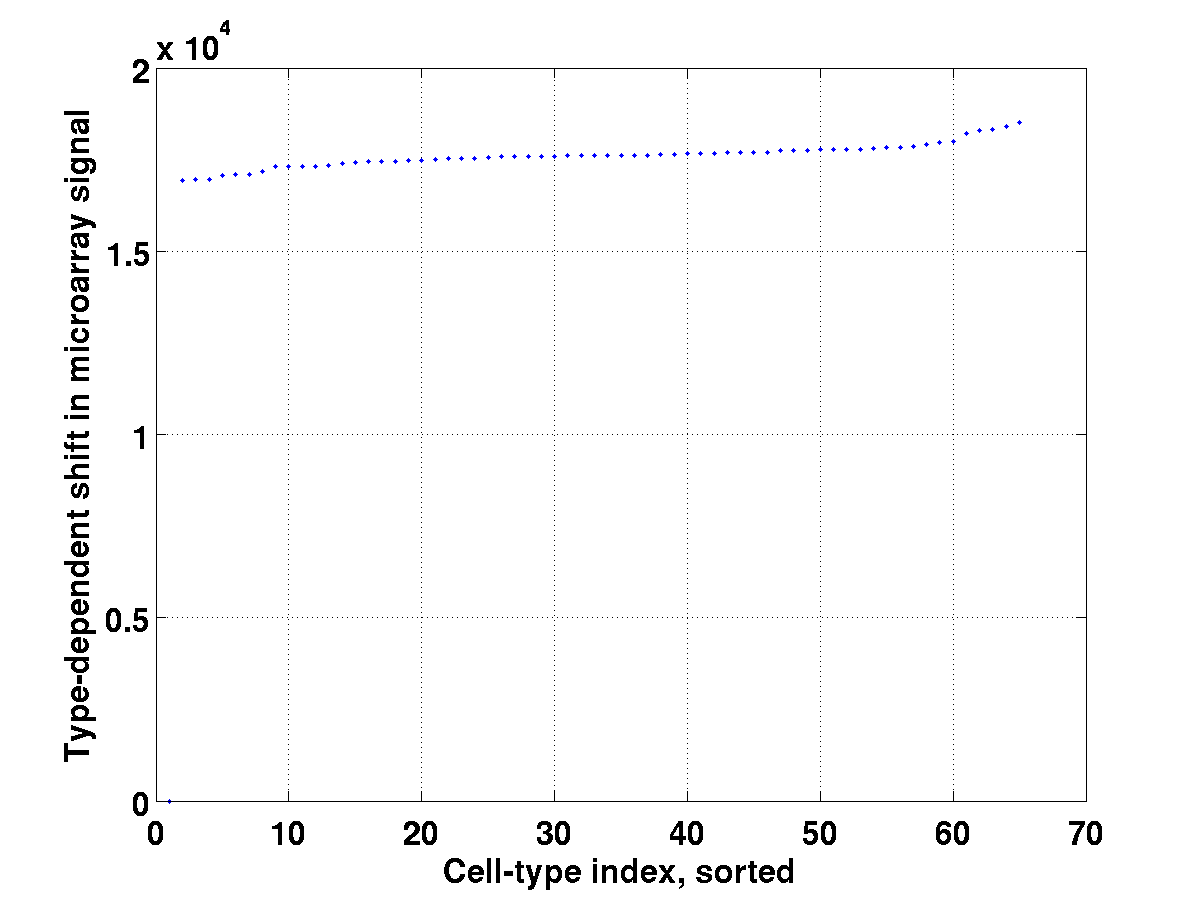}
\caption{Sorted values of the type-dependent shift term $\sum_{g'}C(t,g')$ in 
 Equation \ref{CtrueInverted}. All the values are within 5.2 percent of the average value across cell types.}
\label{cellTypeShift}
\end{figure}

 Let us denote the true expression
 signal of gene labeled $g$ in cell type labeled $t$ by $\Ctrue(t,g)$.
 The cross-hybridization between genes $g$ and $g'$ can be modeled 
 by a factor $\mathcal{T}_{gg'}$ relating $C$ to the true signal:\\
 \begin{equation}
C(t,g) = \sum_{g'}\Ctrue(t,g')\mathcal{T}_{g'g}.
\label{multiplicativeCrossHyb}
\end{equation}
A zero cross-hybridization would correspond to a matrix $\mathcal{T}$
 equal to the identity matrix in gene space. The simplest choice 
 for a non-zero cross-hybridization assumes that a uniform 
 fraction $\alpha$ of the true signal of each gene
 is smeared across all other genes: 
\begin{equation}
\mathcal{T}_{gg'}:= \left(1-\alpha\right)\delta_{gg'} + \alpha U_{gg'},
\label{TUniform}
\end{equation}
where uniform cross-hybridization $U$ is the rank-one matrix whose entries 
are all equal to one: 
\begin{equation}
\forall g,g',\; U_{gg'} = 1.
\end{equation}
Equation \ref{multiplicativeCrossHyb} can be used as follows to express
 $\Ctrue$ in terms of $C$:
\begin{align}
\Ctrue &= \frac{1}{1-\alpha}C\left( I_G+\frac{\alpha}{1-\alpha} U\right)^{-1} = \frac{1}{1-\alpha}C \sum_{k\geq 0}
 \left(-\frac{\alpha}{1-\alpha}\right)^kU^k\\
  & = \frac{1}{1-\alpha}C\left(I_G+U\sum_{k\geq1}\left(-\frac{\alpha}{1-\alpha}\right)^k\right) =\frac{1}{1-\alpha} C\left(I_G -\alpha U\right),
\end{align}
where we have used the fact that $U$ is a projector, and 
 as such has all its positive powers equal to $U$.
 We can therefore rewrite the entries of $\Ctrue$ (up to a collective multiplicative 
 factor that does not modify the relative values of entries in $C$) as the 
 sum of the corresponding entry in $C$ and a negative term that does only 
 depend on the type index $t$, and not on the gene $g$:
\begin{equation}
\Ctrue(t,g) = \frac{1}{1-\alpha}\left( C(t,g) - \alpha \sum_{g'}C(t,g')\right).
\label{CtrueInverted}
\end{equation} 
 Plotting the $t$-dependent shift-term in the above equation as a function 
 of $t$, one observes (Figure \ref{cellTypeShift}) that it is roughly constant across cell types.\\

% CODE:
% crossHybridizationStudy = cross_hybridization_study( Ref, EUsed, CUsed );

% {\bf{Insert the comment about different technologies here.}}\\ 

 %Having lower residual terms in the refitted model 
 %than in the original model fitted to the full 
 %matrix $C$ is consistent with the assumption that 
 %the true signal differs from the matrix $C$ by a uniform term.
 
Consider the contribution of two cell types labeled $t_1$ and $t_2$,
 and two genes labeled $g_1$ and $g_2$, such that $g_1$ is expressed
 only in $t_1$ and $g_2$ is expressed only in $t_2$. 
 This situation corresponds to a diagonal submatrix matrix of size 2 in the
  true signal. 
 Cross-hybridization between $g_1$ and $g_2$ adds non-diagonal elements
 to this matrix, and if they are large enough the determinant
 of the matrix can become zero, resulting in a degenerate 
  solution to the system $E(v,.) = C\,x$, even though 
 the system  $E(v,.) = \Ctrue\,x$ is a linear 
 system with positive solutions.\\

 Moreover, suppose a perfect solution $\rho^{\mathrm{perfect}}$ exists at some voxel labeled $v$
 for the fitting to a true signal:\\
 \begin{equation}
 E(v,. ) = \sum_{t=1}^T\rho^{\mathrm{perfect}}_t(v) \Ctrue(t,: ) + 0.
 \end{equation} 
 If the model is fitted to a microarray signal that differs from 
  $\Ctrue$ by a type-by-gene matrix with uniform entries all equal to some value $c$, 
 one can write a test-vector $\phi_t(v) \in \mathbf{R}_+^T$  as the sum of the 
 vector $\rho^{\mathrm{perfect}}_t(v)$ and a term $\delta\phi_t(v) \in \mathbf{R}^T$, and optimize over the 
 values of $\delta\phi$ (whose optimal value goes to zero when $c$ goes to 0):\\
 \begin{equation}
 {\mathrm{Err_c}}(\phi) := \sum_{g=1}^G\left( E(v,g ) - \sum_{t=1}^T(\phi_t(v) ( \Ctrue(t,g) +c ) \right)^2. 
 \end{equation}
 %\begin{equation}
 %{\mathrm{argmin}}_{\phi \in {\mathbf{R}}_+^T}{\mathrm{Err_c}} =: \phi^{\mathrm{true}}+ \delta\phi.
 %\end{equation}
 \begin{equation}
 \phi_t(v) :=  \rho^{\mathrm{perfect}}_t(v) \delta\phi_t(v)
 \end{equation}

 \begin{align}
{\mathrm{Err_c}}(\phi) &= \sum_{g=1}^G\left( \sum_{t=1}^T(\Ctrue(t,g) +c)\left(\rho^{\mathrm{perfect}}_t(v)+ \delta\phi_t(v)\right) - \sum_{t=1}^T\Ctrue(t,g)\rho^{\mathrm{perfect}}_t(v) \right)^2\\
&= \sum_{g=1}^G\left( \sum_{t=1}^T( c \rho^{\mathrm{perfect}}_t(v) +  \Ctrue(t,g) \delta\phi_t(v)+ c \delta\phi_t(v) ) \right)^2 \\
\label{heuristic}
\end{align}
  At $c$=0, each term in the sum over genes is identically zero, but one can see that the dependence on the 
 gene index $g$ cannot be maintained when $c$ is increased, hence an increased
 value of the residual term.\\

% CODE for numerical study of residual terms
% refitProfiles = rossHybridizationStudy;
% compareResiduals = compare_residuals( Ref, EUsed, CUsed, fitVoxelsToTypes, refitProfiles );

 The parameter $\alpha$ of Equation \ref{TUniform} is not fixed by the previous analysis,
 and any value can be tried provided it does not give rise to
 any negative entry in $\Ctrue$ defined by \ref{CtrueInverted}.
 We can therefore simulate a worst-case scenario, 
 by subtracting the lowest entry of $C$ from all the other entries (which
 corresponds to the approximation where the $t$-dependence of 
the second term in \ref{CtrueInverted} is neglected, i.e. where the dots on 
 Figure \ref{cellTypeShift} are considered to be on a horizontal line).
 We therefore took the following type-by-gene matrix $\COffset$ as the new fitting panel:
 \begin{equation}
 \COffset(t,g) := C(t,g) - m_C.
\label{COffset}
 \end{equation}
 We solved the following 
 optimization problem at each voxel, giving rise
 to a new estimate called $\rhoOffset$ for 
 the brain-wide density of each cell type:
\begin{equation}
\left({\rhoOffset}_t( v )\right)_{1\leq t \leq T}= {\mathrm{argmin}}_{\nu\in {\mathbf{R}}_+^T}\sum_{g=1}^G \left( E(v,g)-\sum_{t=1}^T \COffset(t,g) \nu(t) \right)^2.
\label{voxelByVoxel}
\end{equation}

% {\bf{Insert the comment about {\emph{Gabra6}} here.}} 

%{\bf{Plots: for the 6 main first, with a number indicating 
% deviation. Then run as a loop and single out the bad guys. Are they?}}\\

The brain-wide correlation profiles between each cell types and the Allen Atlas
 do not change under the transformation $C\longrightarrow \COffset$ because the uniform  shift 
 in all the entries of the matrix $C$ is compensated by the shift in the
entries of the average row $\overline{C}$. For each cell type, labeled $t$, we can plot 
 projections and sections of the new estimated profile derived ${\rhoOffset}_t$ from $\Ctrue$,
 next to the original plot, for visual comparison. The top region
 by density is the one through which the section is taken, which shows that 
 the top region is conserved {\bf{except for cell types labeled ??}}.\\

\begin{figure}
\includegraphics[width=\textwidth,keepaspectratio]{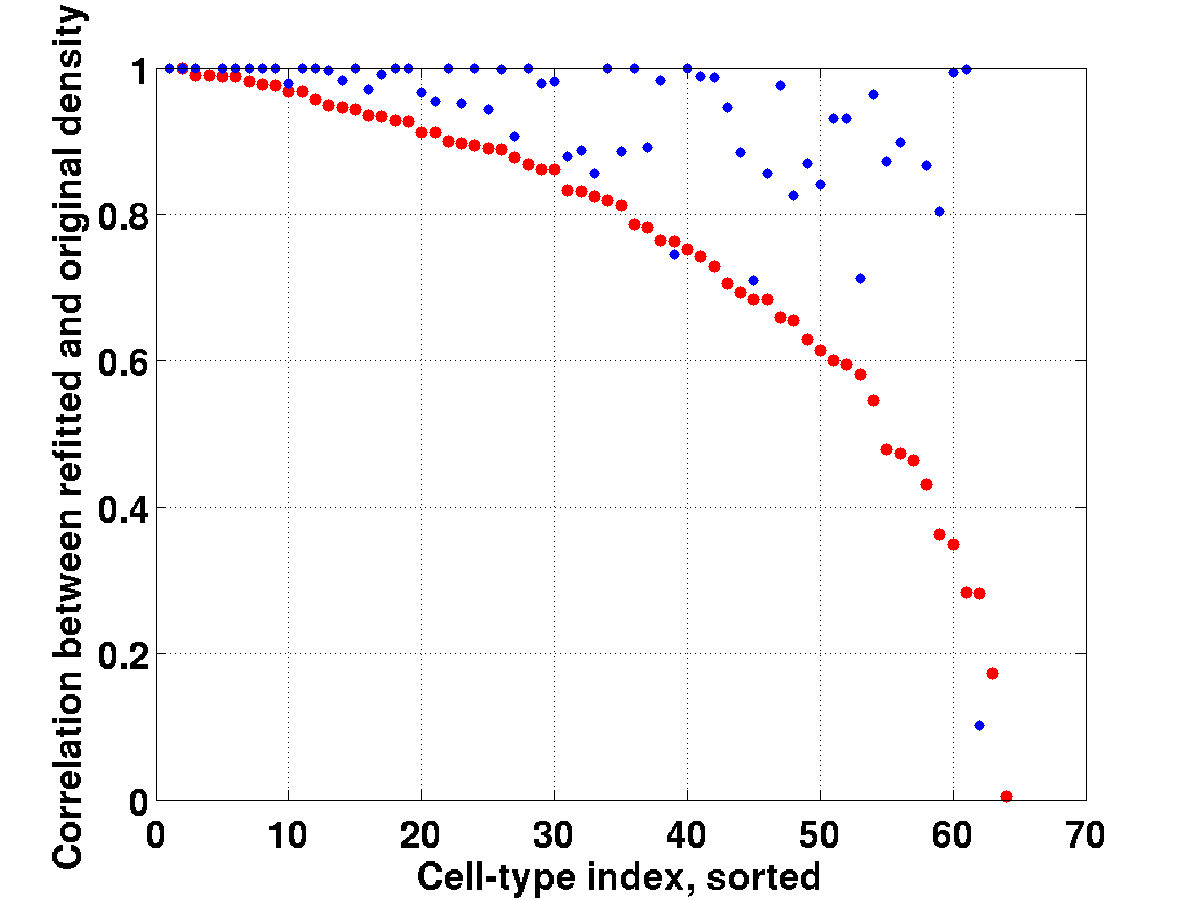}
\caption{Sorted values of the correlation coefficient $\mathcal{C}(
  \rho_t,{\rhoOffset}_t)$, defined in Equation \ref{correlationOffset}
  for all values of the cell-type label $t$ (in red) (see the tables
  for assignment of each value to a definite cell type, the average value is $75.3\% $).  The
  correlations between anatomic fractions in the ARA (Equation
  \ref{correlationOffsetARA} are plotted in blue (the average value is $92.8\%$, the cell types are
  ordered in the same way as for the red dots)}
\label{crossHybridCorrelations}
\end{figure}

 Moreover, we can compute the following correlation coefficient
 between the new correlation profile and the one obtained in the original linear model for the same
 cell type,
 as defined in Equation \ref{corrProfiles} for any pair of brain-wide profiles:
\begin{equation}
\mathcal{C}( \rho_t,{\rhoOffset}_t) = \frac{\sum_{v=1}^V\left(\rho_t(v)-\overline{\rho_t}\right)
 \left({\rhoOffset}_t(v)-\overline{{\rhoOffset}_t}\right)}{\sqrt{\left(\sum_{v=1}^V\left(\rho_t(v)-\overline{\rho_t}\right)^2\right)
\left(\sum_{v=1}^V\left({\rhoOffset}_t(v)-\overline{{\rhoOffset}_t}\right)^2\right)}},
\label{correlationOffset}
\end{equation}
It turns out that 35 of the $T=64$ cell types in the data set 
give rise so correlation values larger than 80 percent.
 Another correlation can be computed between the fractions 
 of the density binned by the regions of the ARA (defined in Equation \ref{fittingRegion}).
\begin{equation}
\mathcal{C^{\mathrm{ARA}}}( \rho_t,{\rhoOffset}_t) = \frac{\sum_{r=1}^R\left({\overline{\rho}}(r,t)-\langle\overline{\rho(.,t)}\rangle\right)
 \left({\rhoOffset}(r,t)- \langle\overline{\rhoOffset(.,t)}\rangle\right)}{\sqrt{\left(\sum_{r=1}^R\left({\overline{\rho}}(r,t)-
  \langle\overline{\rho(.,t)}\rangle \right)^2\right)
\left(\sum_{r=1}^R\left({\rhoOffset}(r,t)-\langle\overline{\rhoOffset(.,t)}\rangle\right)^2\right)}},
\label{correlationOffsetARA}
\end{equation} 
where $R=13$ is the number of regions in the coarsest version of the ARA, and the fractions of densities 
 in regions are defined as in Equation \ref{fittingRegion}:

\begin{equation}
{\overline{\rho}}( r, t ) = \frac{1}{|\sum_{v \in {\mathtt{{Brain\;Annotation}}}}\rho_t(v) |}
                                  \sum_{v \in V_r}\rho_t(v),
\label{fittingRegionRe} 
\end{equation}
\begin{equation}
{\overline{\rhoOffset}}( r, t ) = \frac{1}{|\sum_{v \in {\mathtt{{Brain\;Annotation}}}}{\rhoOffset}_t(v) |}
                                  \sum_{v \in V_r}{\rhoOffset}_t(v),
\label{fittingRegionReRe} 
\end{equation}
and used the values to compute the average of residual fraction across regions 
 in the ARA:
\begin{equation}
\langle\overline{\rho(.,t)}\rangle = \frac{1}{R}\sum_{r=1}^R{\overline{\rho}}( r, t ),\;\;\;
\langle\overline{\rhoOffset(.,t)}\rangle = \frac{1}{R}\sum_{r=1}^R{\overline{\rhoOffset}}( r, t ).
\end{equation}
The values of $\mathcal{C^{\mathrm{ARA}}}( \rho_t,{\rhoOffset}_t)$ are plotted in Figure \ref{crossHybridCorrelations} (56 of them are larger than 80 percent).

% CODE FOR CORRELATIONS
% crossHybridCorrelations = cross_hybrid_correlations( Ref, fitVoxelsToTypes, fitCorrected )

% Moreover, we computed the deviation factor as
% a fraction of the original density:
% \begin{equation}
% \frac{\delta \rho_t}{\rho_t}= \frac{\sqrt{\sum_{v=1}^V \left( \rhoNew_t(v)-\rho_t(v)\right)^2}}
% {\sqrt{\sum_{v=1}^V \rho_t(v)^2}}.
% \label{deviation}
% \end{equation} 

% CODE FOR FIGURES: plot_refit_figures( Ref, fitVoxelsToTypes, fitCorrected );
% CODE FOR TABLES: cell_types_table_suppl( Ref, 1:64 );
 
\begin{figure}
\includegraphics[width=1\textwidth,keepaspectratio]{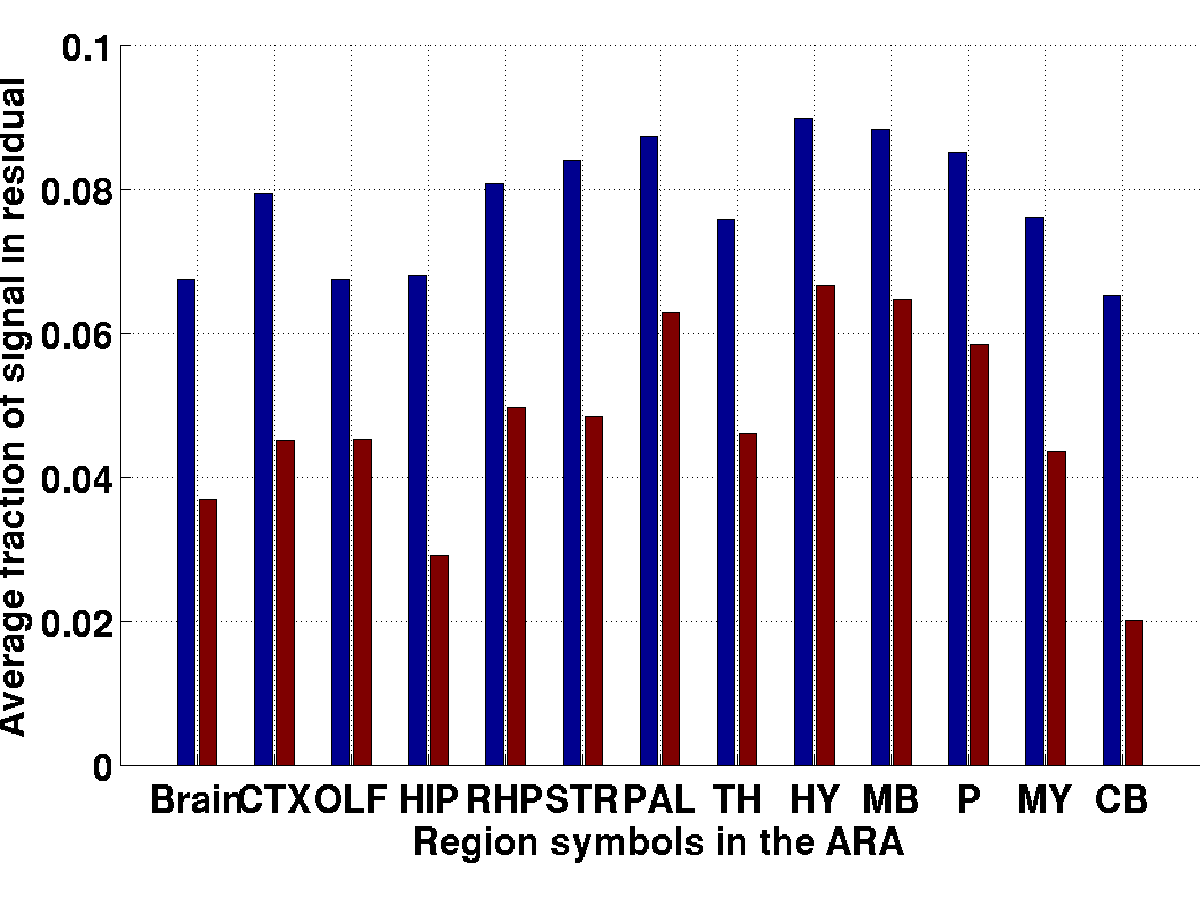}
\caption{Fraction of absolute values of the residual averaged over the regions
 in the coarsest version of the ARA (for the original model in blue, and for the refitted 
 model in red).}
\label{residualOriAndRefit}
 \end{figure}

The residual terms can be compared in order to estimate whether 
the fit between the Allen Atlas and the cell types is improved by 
 subtracting a cross-hybridization term to the microarray data, as 
 it should from the heuristic Equation \ref{heuristic} (even though the offset matrix 
 $\COffset$ is not expected to give rise to a perfect fit due to the imperfections
 of our cross-hybridization model, it must be closer to the signal 
 measured by ISH data as it brings some entries of the cell-type-specific matrix 
 closer to zero).
We computed the average fraction of the signal that 
 is contained in the absolute value of the residual,
\begin{equation}
{\mathrm{ResFracLocal}}( v ) = \left|1 - \frac{\sum_t \rho_t(v) C(t,g )}{E(v,g)}\right|
\label{resFracLocal}
\end{equation}
across each of the 
 13 regions in the coarsest partition of the ARA:
\begin{align}
{\mathrm{ResFrac}}( r )& = \frac{1}{\sum_{\in V_r } 1}\sum_{v\in V_r }\left|1 - \frac{\sum_t \rho_t(v) C(t,g )}{E(v,g)}\right|\\
{\mathrm{ResFrac}}_{\mathrm{refit}}( r ) &= \frac{1}{\sum_{\in V_r } 1}\sum_{v\in V_r }\left|1 - \frac{\sum_t \rho_t(v) ( C(t,g ) - m_C)}{E(v,g)}\right|
\end{align}
where $r$ labels regions in the ARA (see Figure \ref{residualOriAndRefit}).

%meanQuotientOri =   -0.0780
%meanQuotientRe =  -0.0440

Indeed the average residual fractions in the refitted model are consistently lower
 than in the original model fitted to the full microarray data (see Figure \ref{residualOriAndRefit}), which suggests   
 a better correspondence to the ISH data. For a plot of the voxel-by voxel values, see
 Figure \ref{resFracLocal}. The average value across the brain goes down when 
 going from the original model to the refitted model, and the highest values
 are localized in the same voxels in both models. The high values in the 
 most rostral voxels can be attributed to the loss of sections from the olfactory bulbs  
 in the ISH data, as these sections are the most fragile. The other cluster of 
 voxels with high residual, which is much more caudal, could be related to  
 registration errors.

 \begin{figure}
\includegraphics[width=1\textwidth,keepaspectratio]{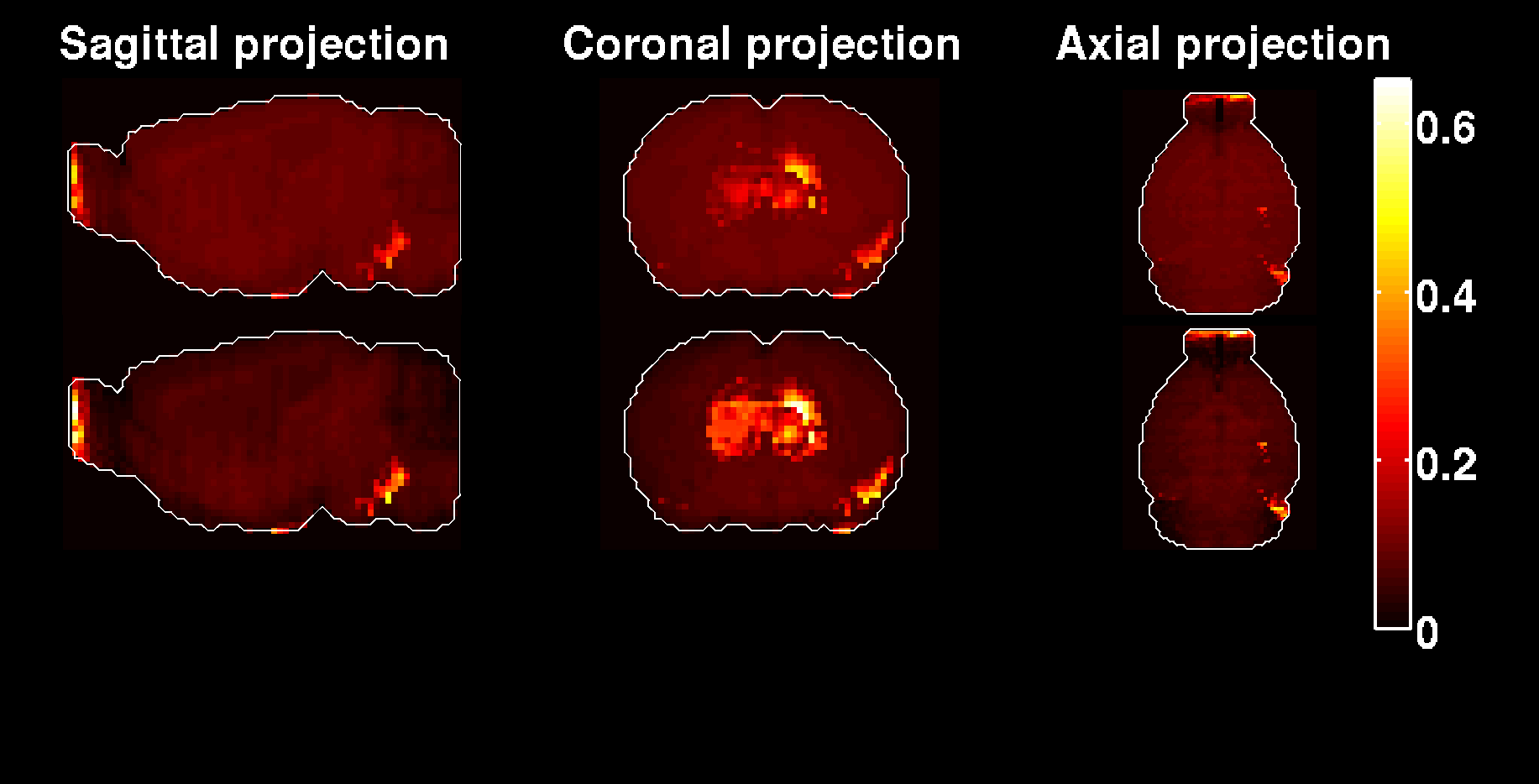}
\caption{Maximal-intensity projections of the absolute values of the residual at each voxel (original 
 model in the first row, refitted model in the second row).
 The maximum values are close to the boundary of the brain in both cases. The average value is
 (for the original model in the first row, average value 7.8\%, for the refitted model in the second row, average vale 4.4\%).}
\label{reFracLocal}
 \end{figure}

 % good guys and bad guys in numerical studies of residual

 % conclusion in terms of qualitative output (ranking of brain regions)

\clearpage
\section{Tables of plots of estimated densities, original and corrected for cross-hybridization }

\begin{table}
\begin{tabular}{|m{0.06\textwidth}|m{0.06\textwidth}|m{\widthParamForTable\textwidth}|}
\hline
\textbf{Index}&\textbf{Cell type}&\textbf{Heat maps of densities, original and refitted}\\\hline
1&\tiny{Purkinje Cells}&\includegraphics[width=0.85\textwidth,keepaspectratio]{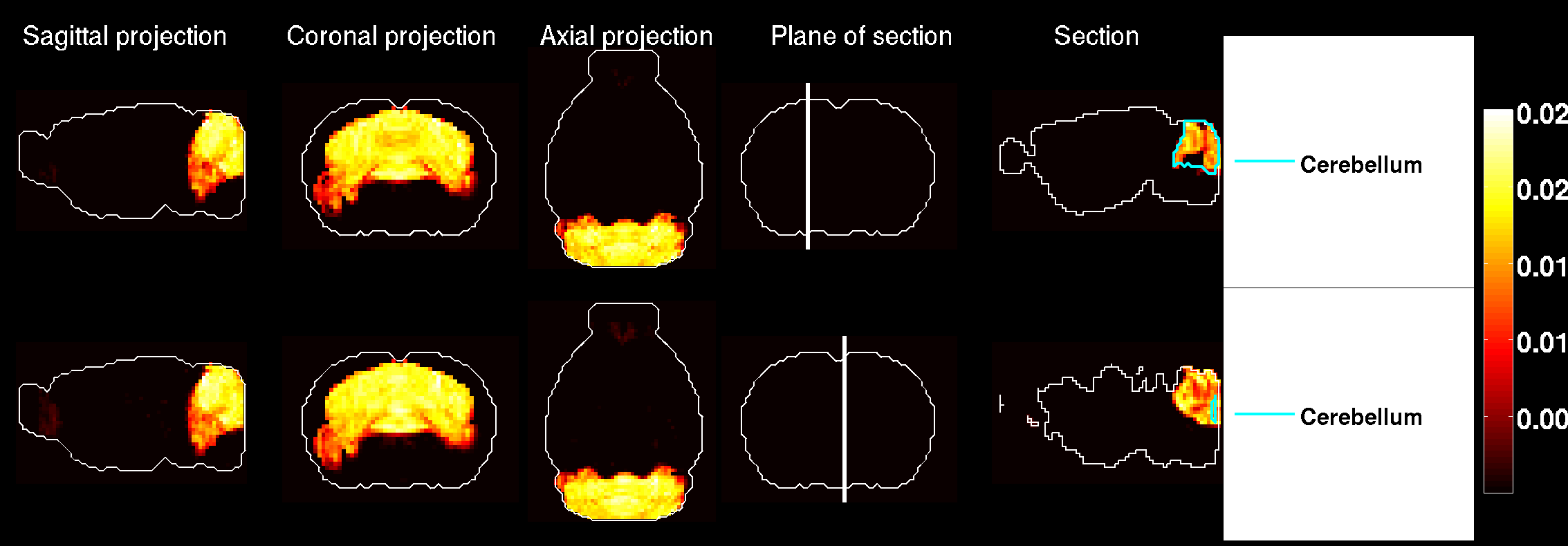}\\\hline
2&\tiny{Pyramidal Neurons}&\includegraphics[width=0.85\textwidth,keepaspectratio]{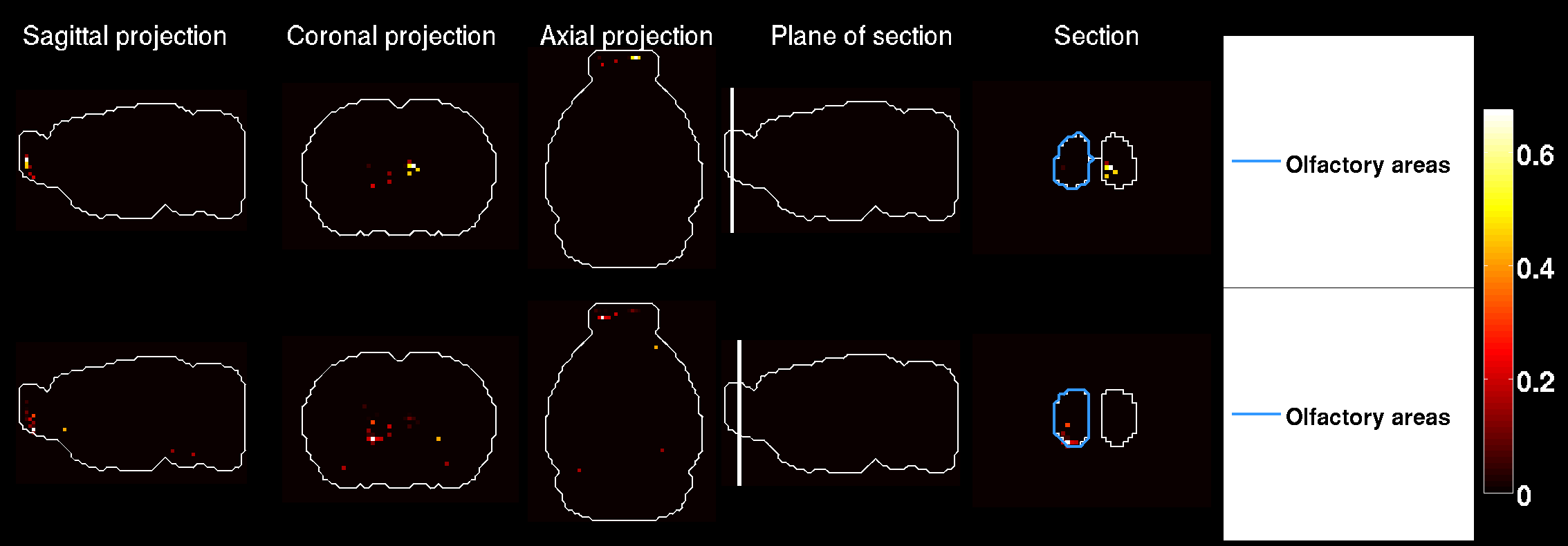}\\\hline
3&\tiny{Pyramidal Neurons}&\includegraphics[width=0.85\textwidth,keepaspectratio]{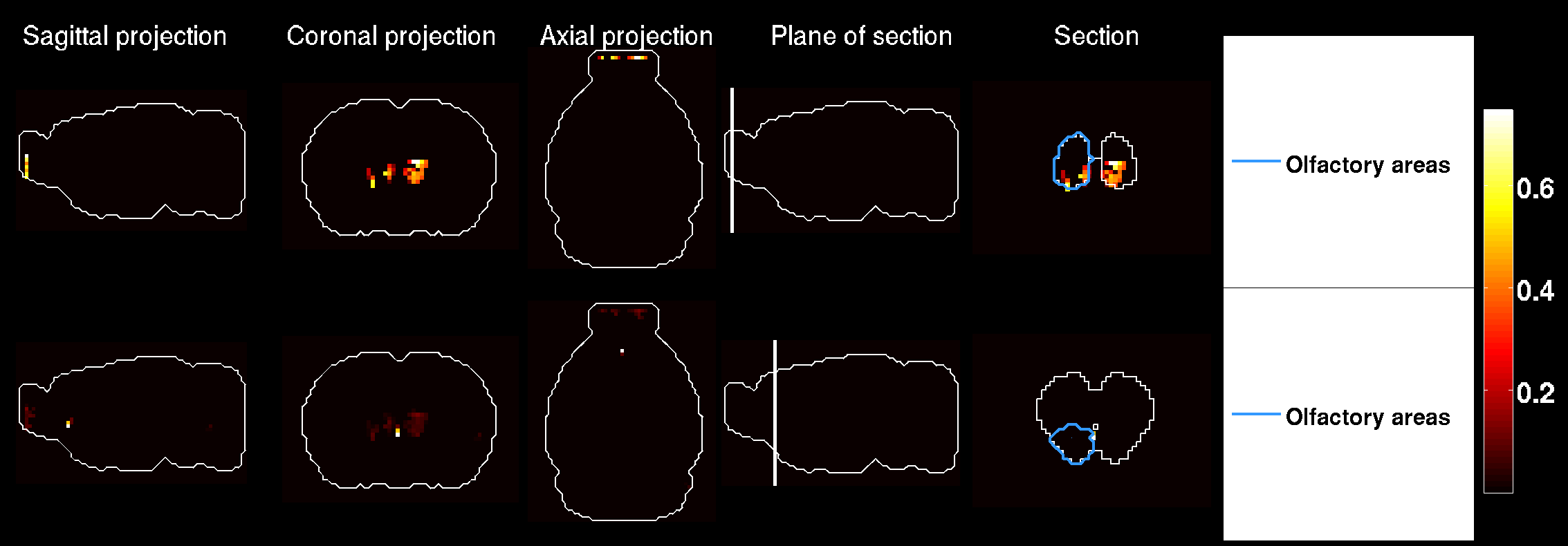}\\\hline
\end{tabular}
 
\caption{Brain-wide density profiles of \numTypesPerReTable cell types, in the 
 original linear model (first row of each figure), and in the model fitted to microarray data
 incorporating the maximum uniform correction compatible with positive entries (second row of each figure).}
\label{tableReFittings1}
\end{table}

\begin{table}
\begin{tabular}{|m{0.06\textwidth}|m{0.06\textwidth}|m{\widthParamForTable\textwidth}|}
\hline
\textbf{Index}&\textbf{Cell type}&\textbf{Heat maps of densities, original and refitted}\\\hline
4&\tiny{A9 Dopaminergic Neurons}&\includegraphics[width=0.85\textwidth,keepaspectratio]{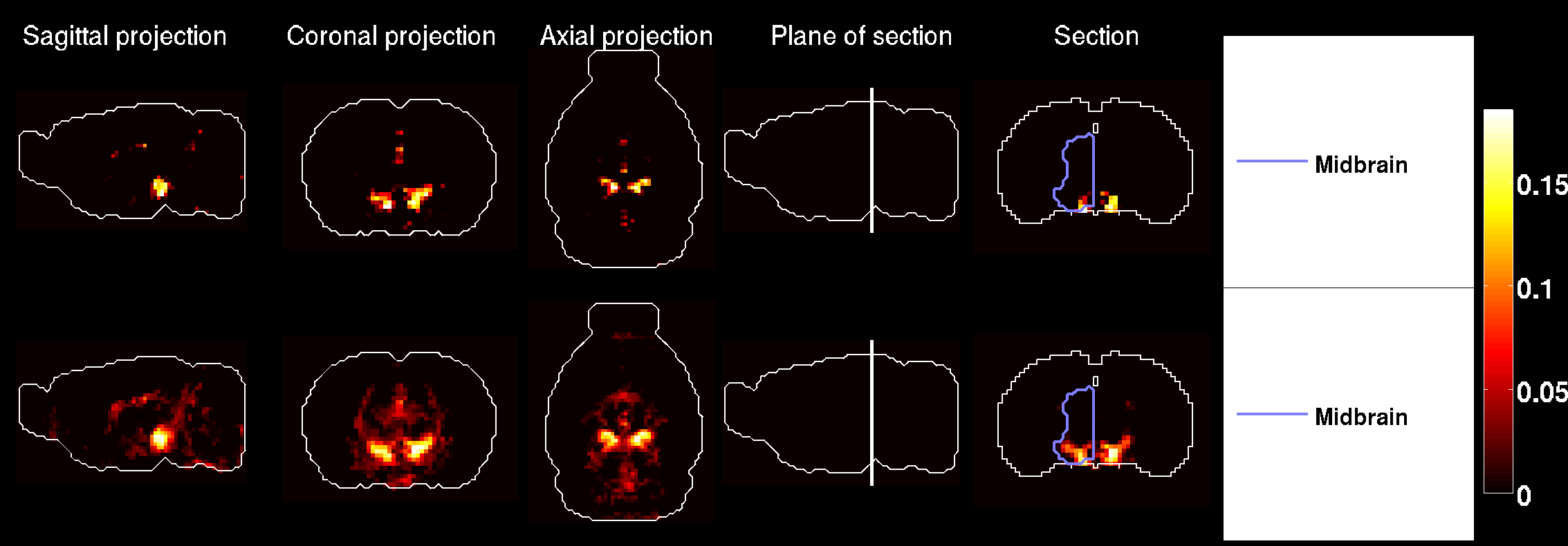}\\\\\hline
5&\tiny{A10 Dopaminergic Neurons}&\includegraphics[width=0.85\textwidth,keepaspectratio]{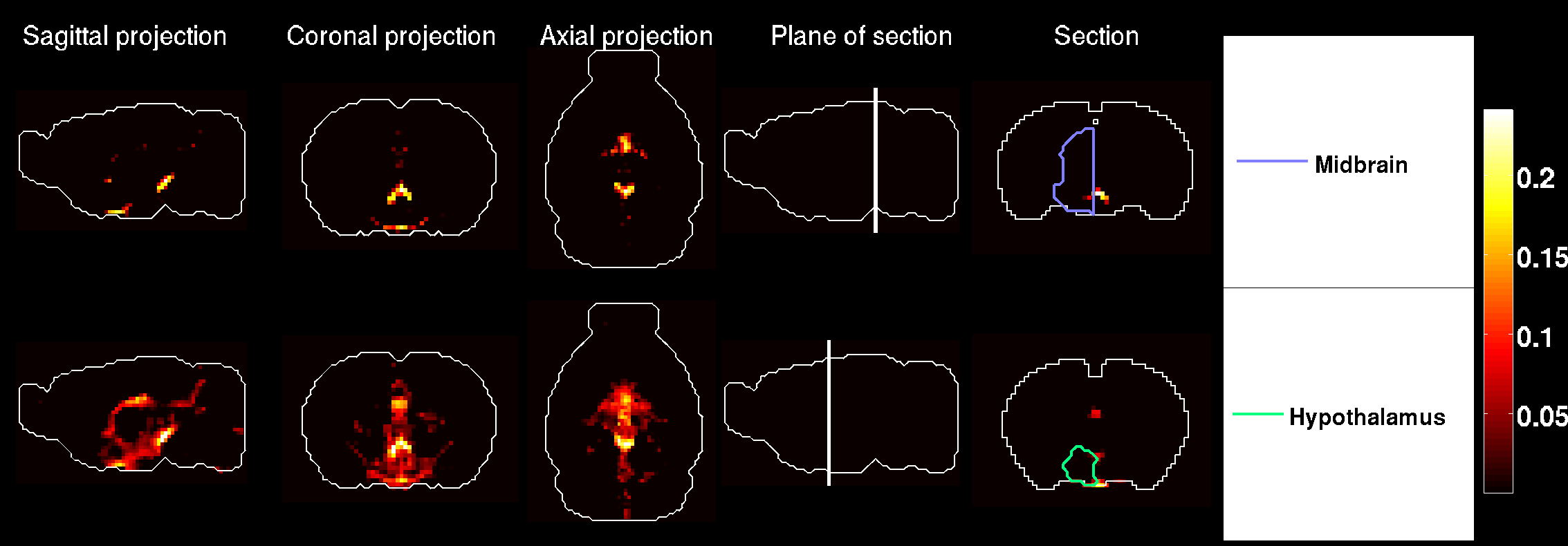}\\\\\hline
6&\tiny{Pyramidal Neurons}&\includegraphics[width=0.85\textwidth,keepaspectratio]{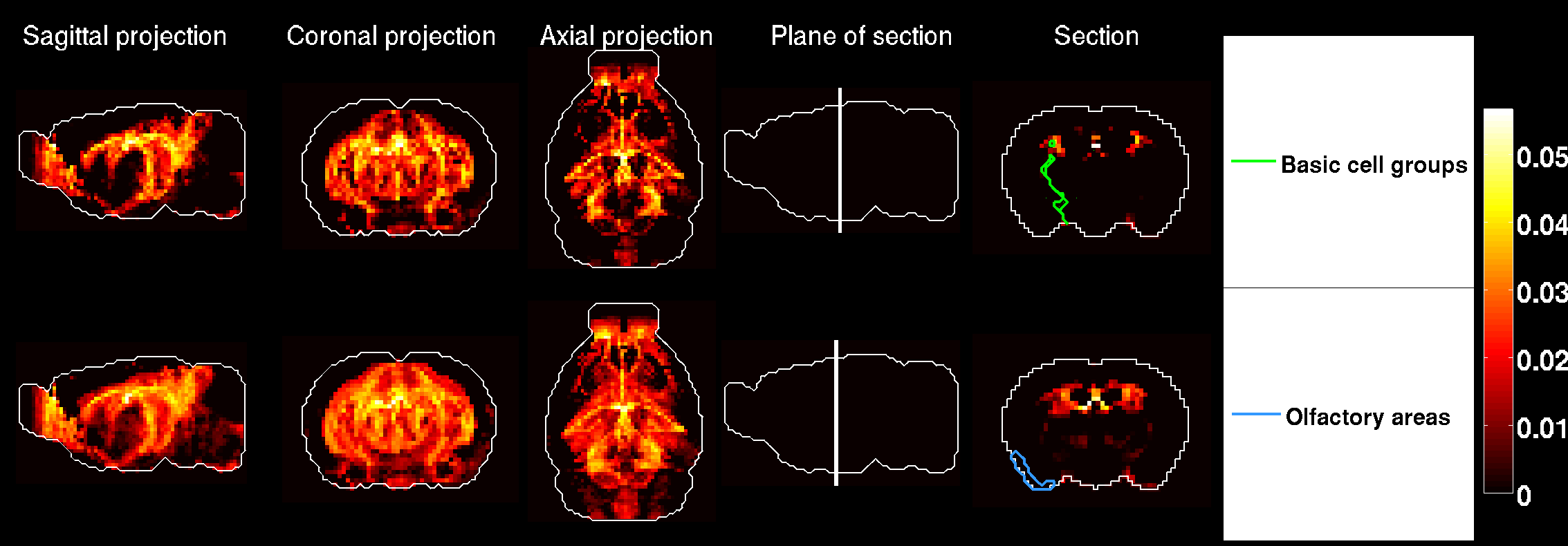}\\\\\hline
\end{tabular}
 
\caption{Brain-wide density profiles of \numTypesPerReTable cell types, in the 
 original linear model (first row of each figure), and in the model fitted to microarray data
 incorporating the maximum uniform correction compatible with positive entries (second row of each figure).}
\label{tableReFittings2}
\end{table}

\begin{table}
\begin{tabular}{|m{0.06\textwidth}|m{0.06\textwidth}|m{\widthParamForTable\textwidth}|}
\hline
\textbf{Index}&\textbf{Cell type}&\textbf{Heat maps of densities, original and refitted}\\\hline
7&\tiny{Pyramidal Neurons}&\includegraphics[width=0.85\textwidth,keepaspectratio]{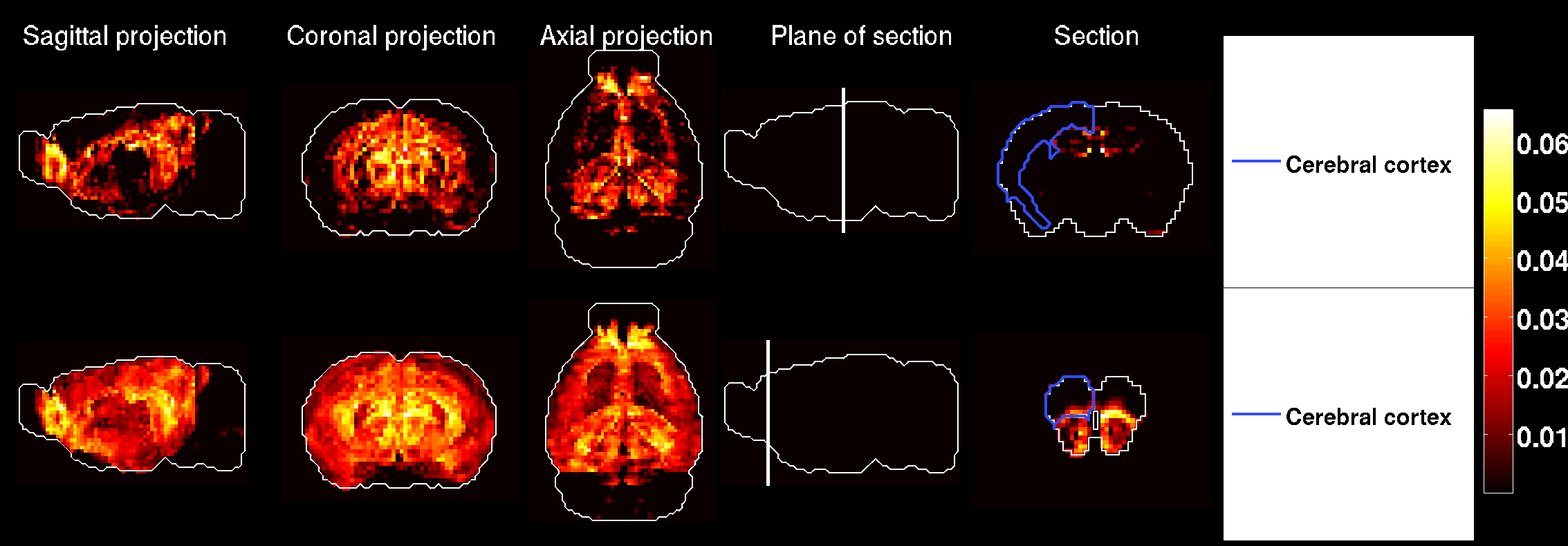}\\\hline
8&\tiny{Pyramidal Neurons}&\includegraphics[width=0.85\textwidth,keepaspectratio]{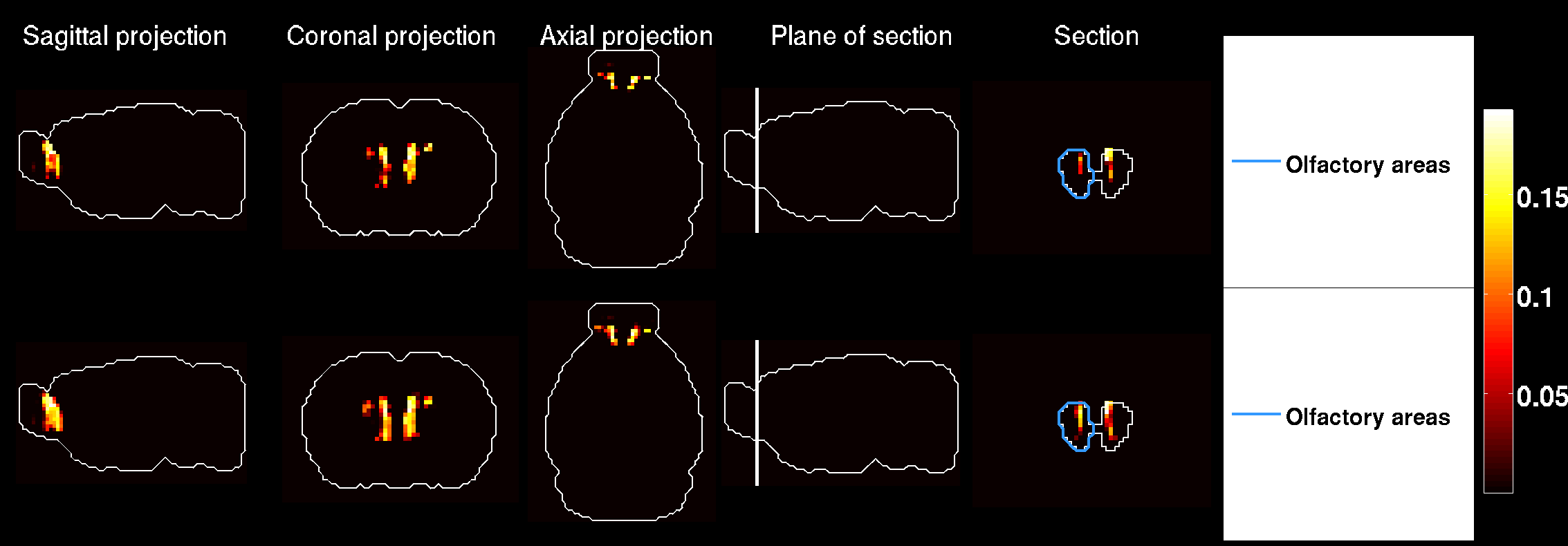}\\\hline
9&\tiny{Mixed Neurons}&\includegraphics[width=0.85\textwidth,keepaspectratio]{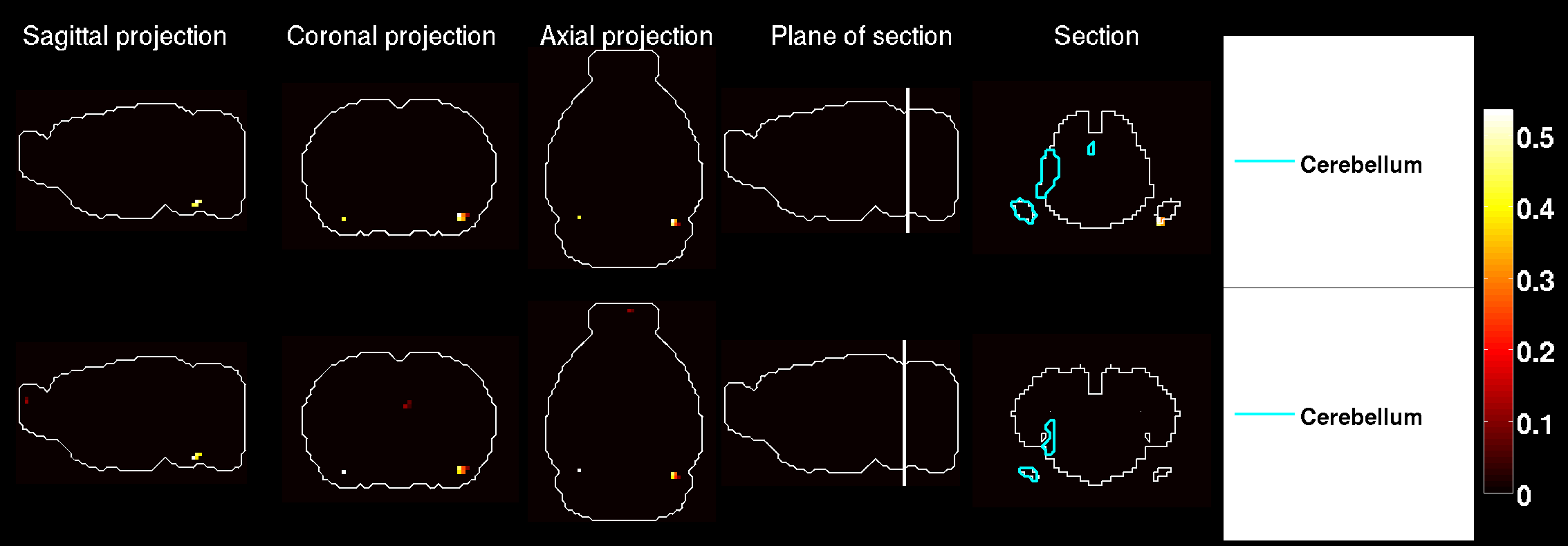}\\\hline
\end{tabular}
 
\caption{Brain-wide density profiles of \numTypesPerReTable cell types, in the 
 original linear model (first row of each figure), and in the model fitted to microarray data
 incorporating the maximum uniform correction compatible with positive entries (second row of each figure).}
\label{tableReFittings3}
\end{table}

\begin{table}
\begin{tabular}{|m{0.06\textwidth}|m{0.06\textwidth}|m{\widthParamForTable\textwidth}|}
\hline
\textbf{Index}&\textbf{Cell type}&\textbf{Heat maps of densities, original and refitted}\\\hline
10&\tiny{Motor Neurons, Midbrain Cholinergic Neurons}&\includegraphics[width=0.85\textwidth,keepaspectratio]{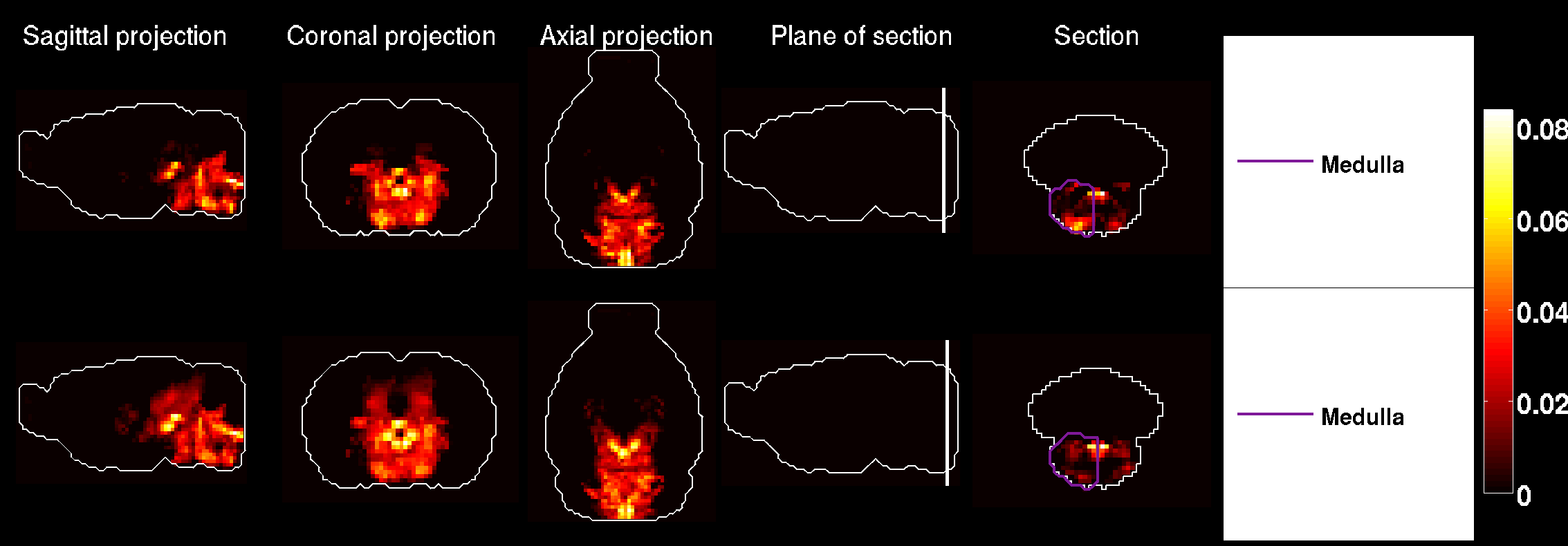}\\\hline
11&\tiny{Cholinergic Projection Neurons}&\includegraphics[width=0.85\textwidth,keepaspectratio]{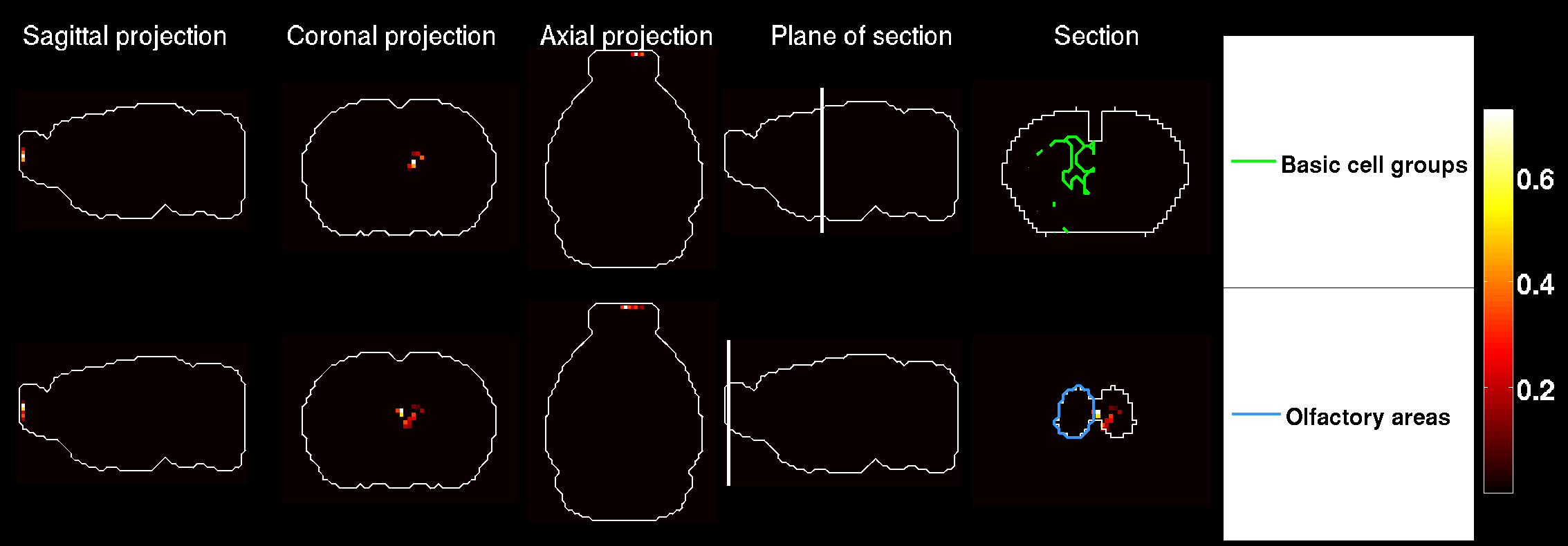}\\\hline
12&\tiny{Motor Neurons, Cholinergic Interneurons}&\includegraphics[width=0.85\textwidth,keepaspectratio]{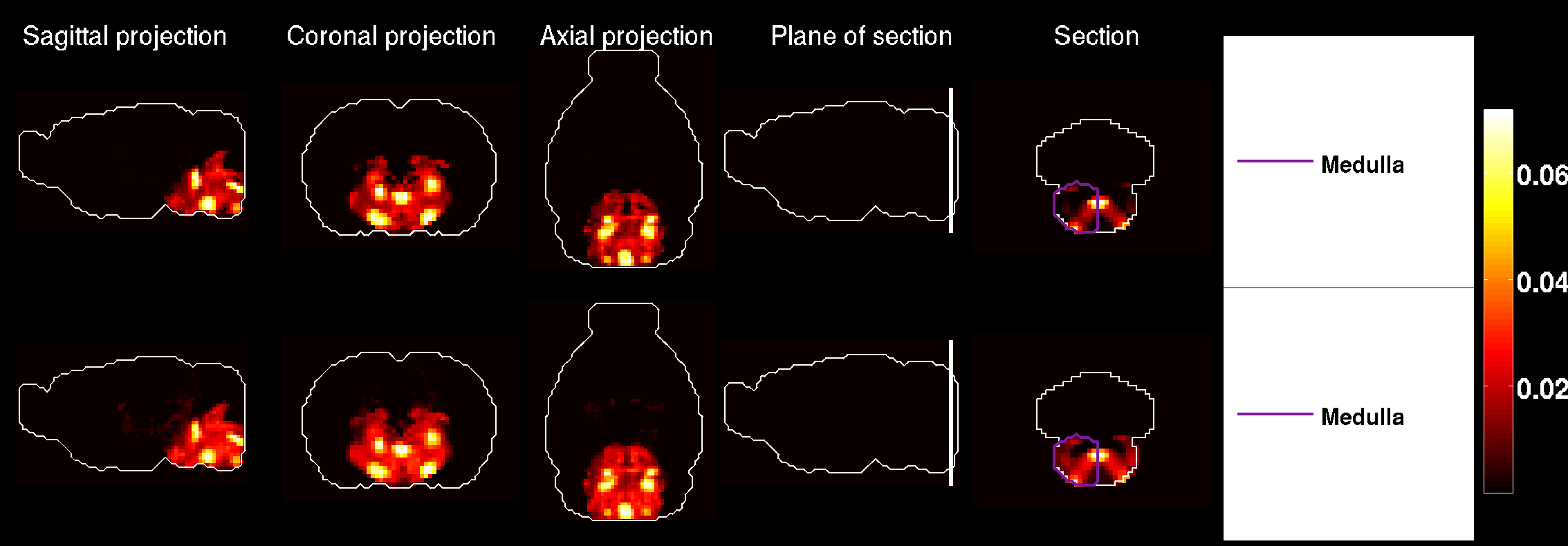}\\\hline
\end{tabular}
 
\caption{Brain-wide density profiles of \numTypesPerReTable cell types, in the 
 original linear model (first row of each figure), and in the model fitted to microarray data
 incorporating the maximum uniform correction compatible with positive entries (second row of each figure).}
\label{tableReFittings4}
\end{table}

\begin{table}
\begin{tabular}{|m{0.06\textwidth}|m{0.06\textwidth}|m{\widthParamForTable\textwidth}|}
\hline
\textbf{Index}&\textbf{Cell type}&\textbf{Heat maps of densities, original and refitted}\\\hline
13&\tiny{Cholinergic Neurons}&\includegraphics[width=0.85\textwidth,keepaspectratio]{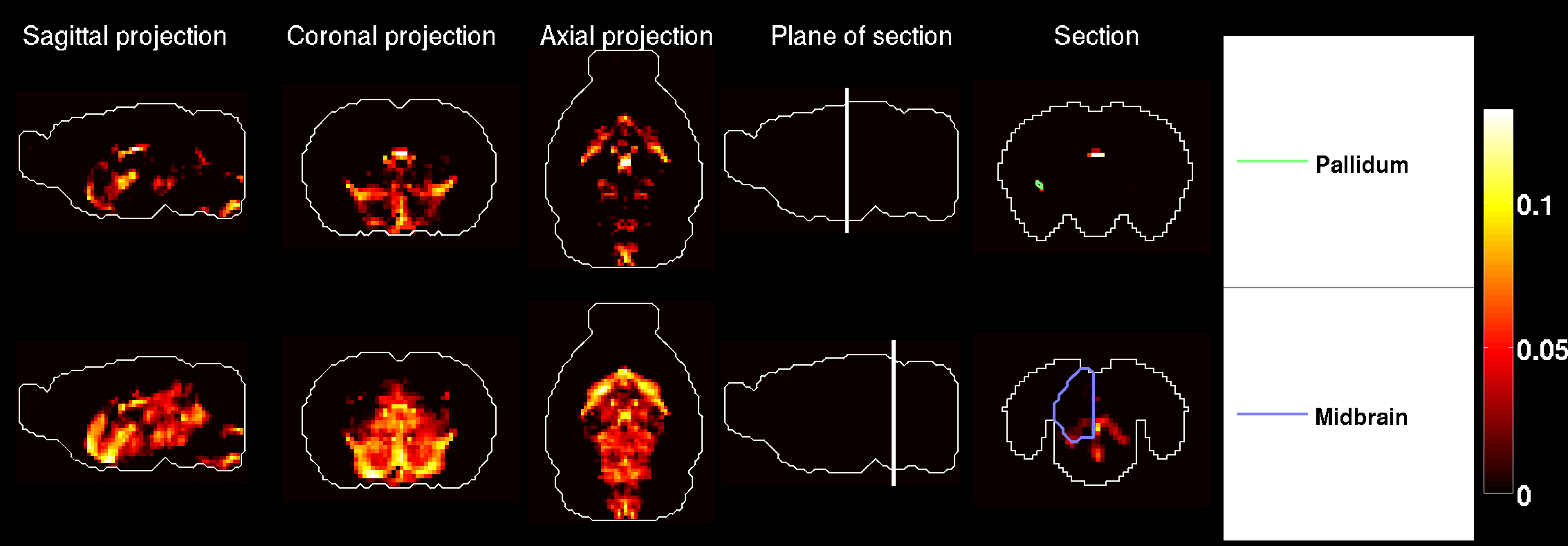}\\\hline
14&\tiny{Interneurons}&\includegraphics[width=0.85\textwidth,keepaspectratio]{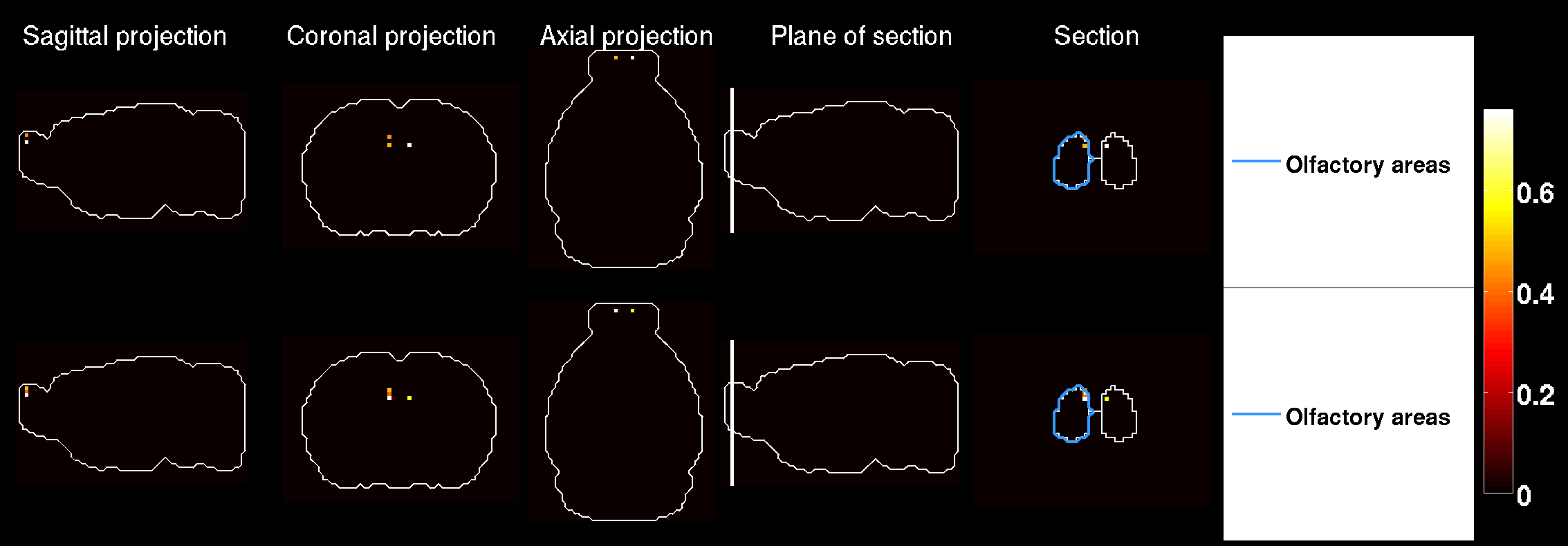}\\\hline
15&\tiny{Drd1+ Medium Spiny Neurons}&\includegraphics[width=0.85\textwidth,keepaspectratio]{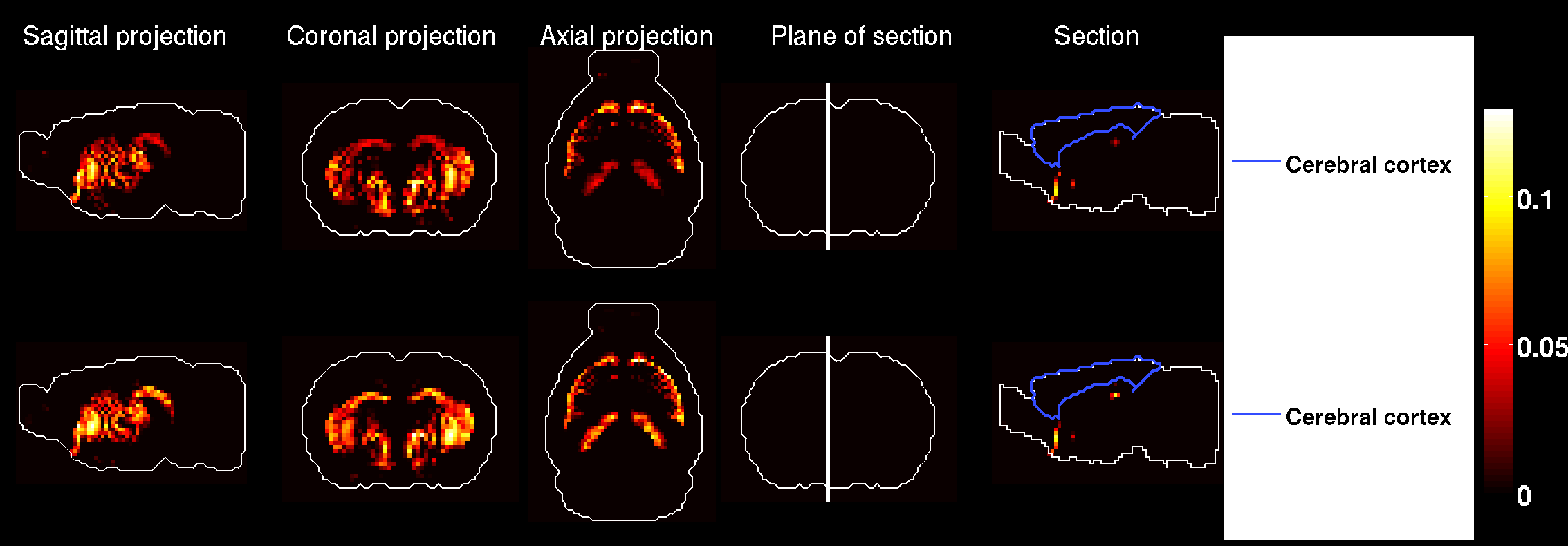}\\\hline
\end{tabular}
 
\caption{Brain-wide density profiles of \numTypesPerReTable cell types, in the 
 original linear model (first row of each figure), and in the model fitted to microarray data
 incorporating the maximum uniform correction compatible with positive entries (second row of each figure).}
\label{tableReFittings5}
\end{table}

\clearpage
\begin{table}
\begin{tabular}{|m{0.06\textwidth}|m{0.06\textwidth}|m{\widthParamForTable\textwidth}|}
\hline
\textbf{Index}&\textbf{Cell type}&\textbf{Heat maps of densities, original and refitted}\\\hline
16&\tiny{Drd2+ Medium Spiny Neurons}&\includegraphics[width=0.85\textwidth,keepaspectratio]{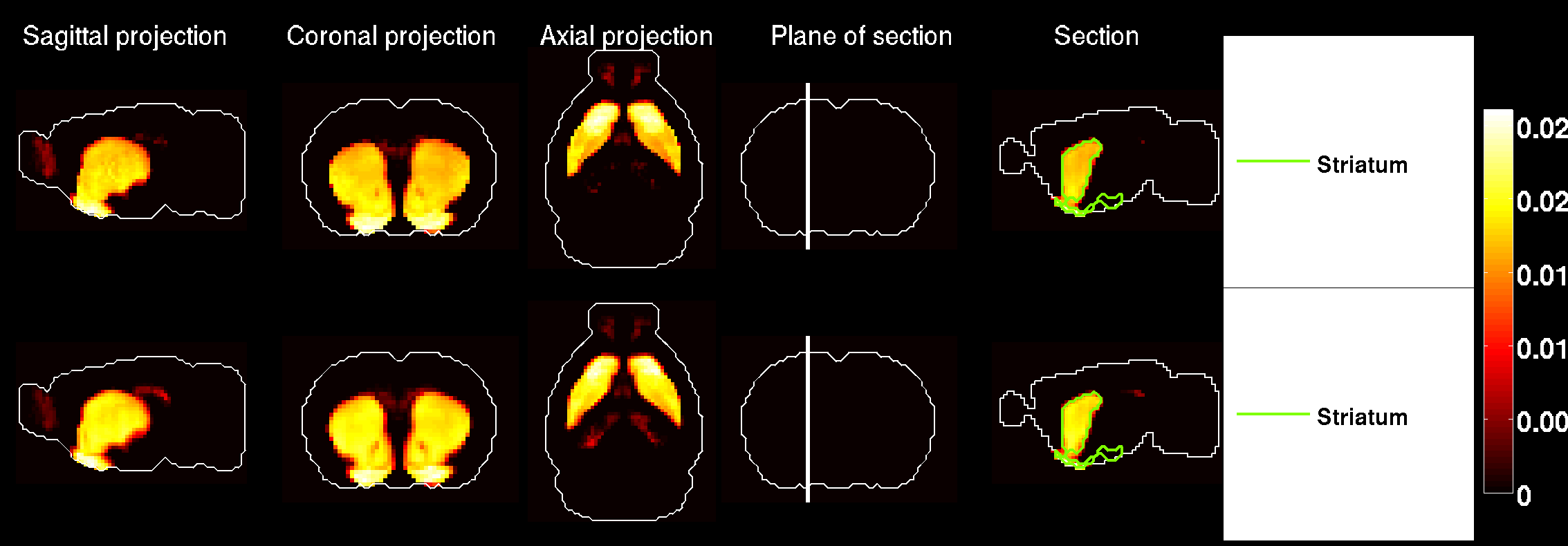}\\\\\hline
17&\tiny{Golgi Cells}&\includegraphics[width=0.85\textwidth,keepaspectratio]{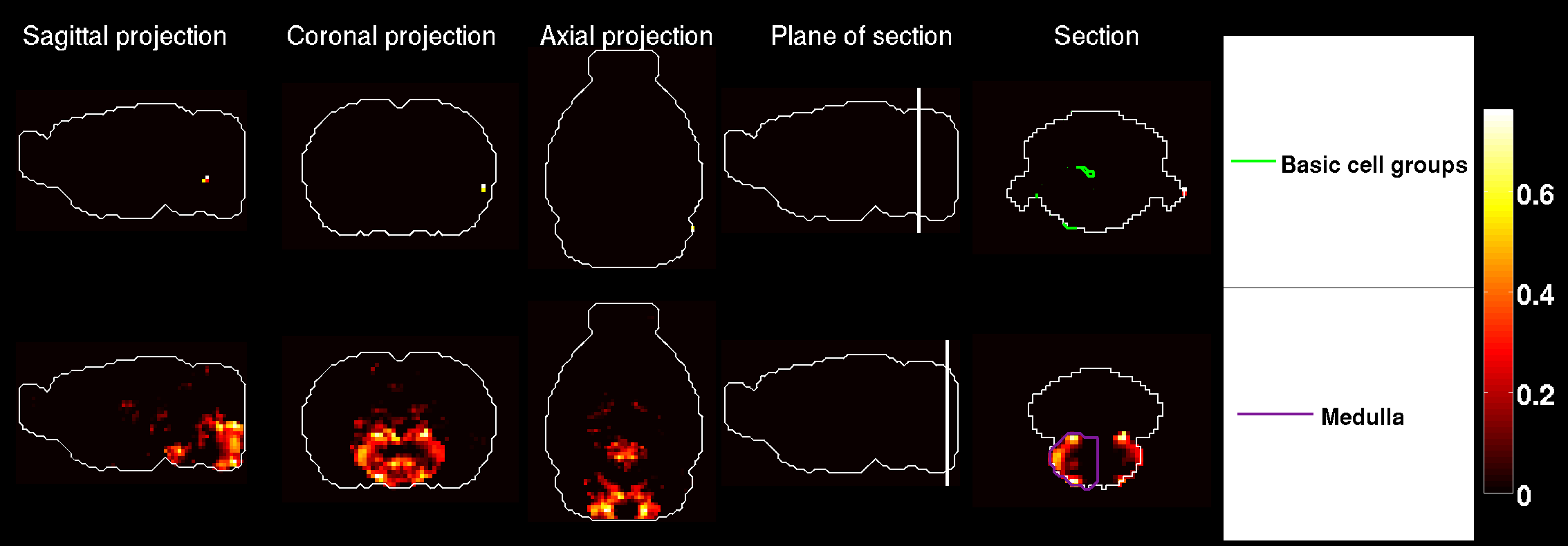}\\\\\hline
18&\tiny{Unipolar Brush cells (some Bergman Glia)}&\includegraphics[width=0.85\textwidth,keepaspectratio]{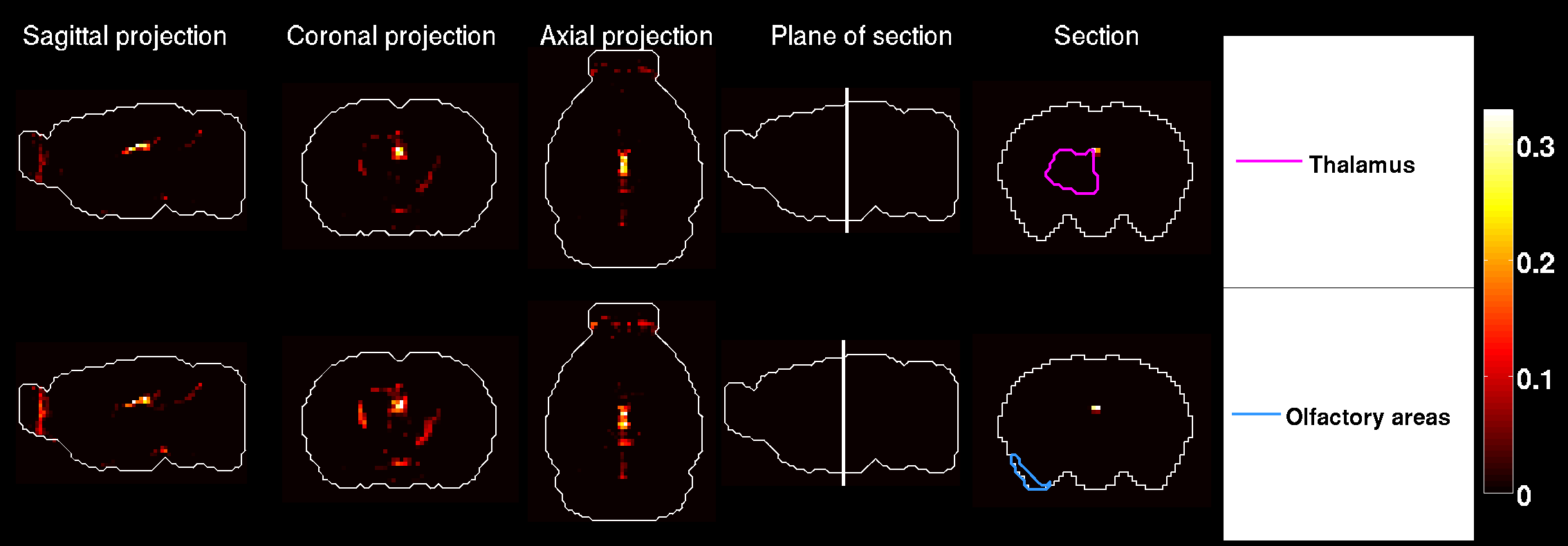}\\\\\hline
\end{tabular}
 
\caption{Brain-wide density profiles of \numTypesPerReTable cell types, in the 
 original linear model (first row of each figure), and in the model fitted to microarray data
 incorporating the maximum uniform correction compatible with positive entries (second row of each figure).}
\label{tableReFittings7}
\end{table}

\begin{table}
\begin{tabular}{|m{0.06\textwidth}|m{0.06\textwidth}|m{\widthParamForTable\textwidth}|}
\hline
\textbf{Index}&\textbf{Cell type}&\textbf{Heat maps of densities, original and refitted}\\\hline
19&\tiny{Stellate Basket Cells}&\includegraphics[width=0.85\textwidth,keepaspectratio]{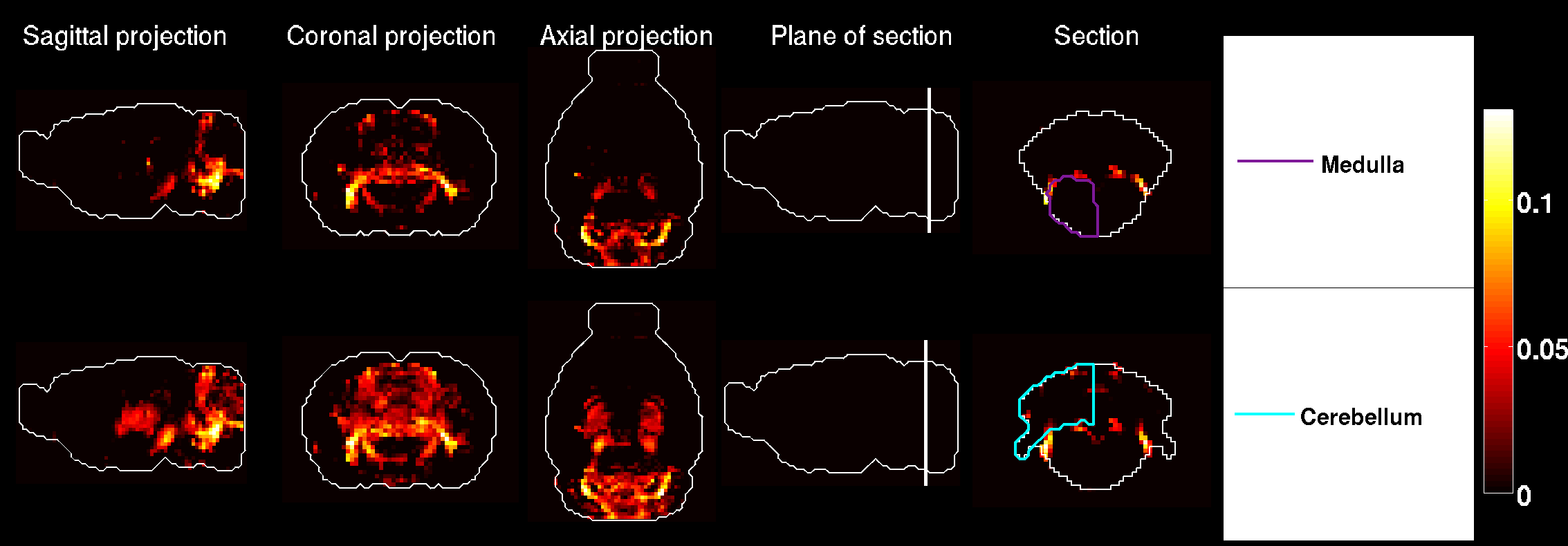}\\\hline
20&\tiny{Granule Cells}&\includegraphics[width=0.85\textwidth,keepaspectratio]{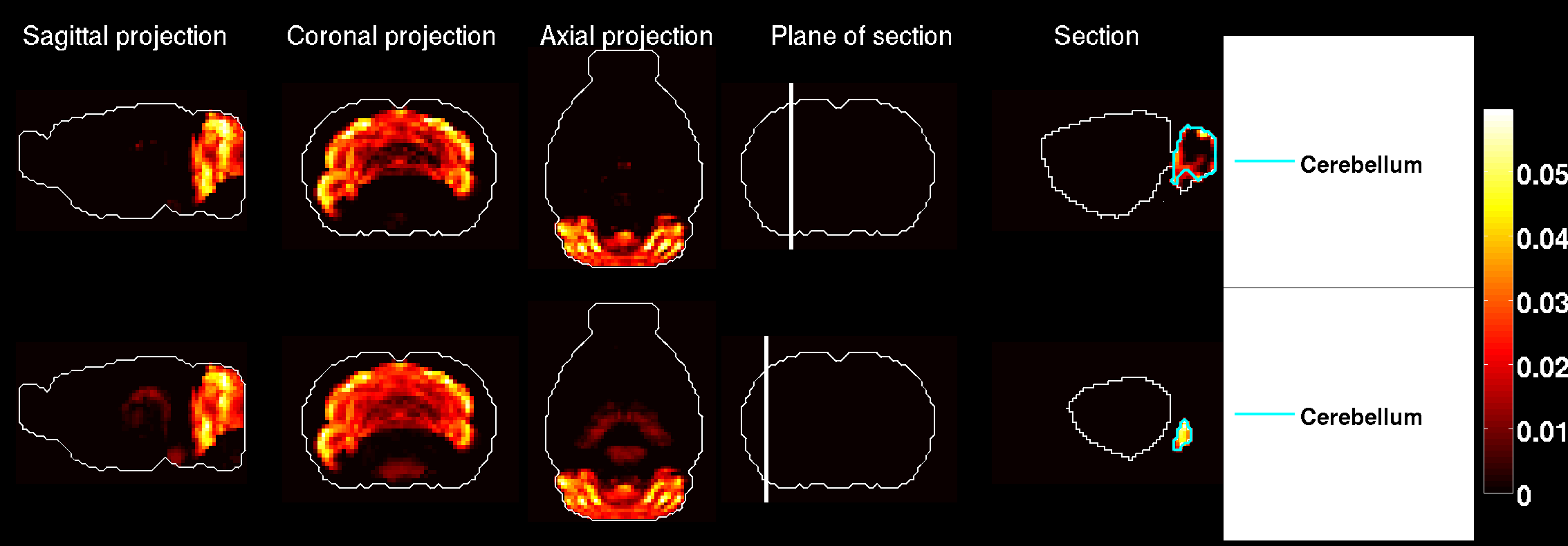}\\\hline
21&\tiny{Mature Oligodendrocytes}&\includegraphics[width=0.85\textwidth,keepaspectratio]{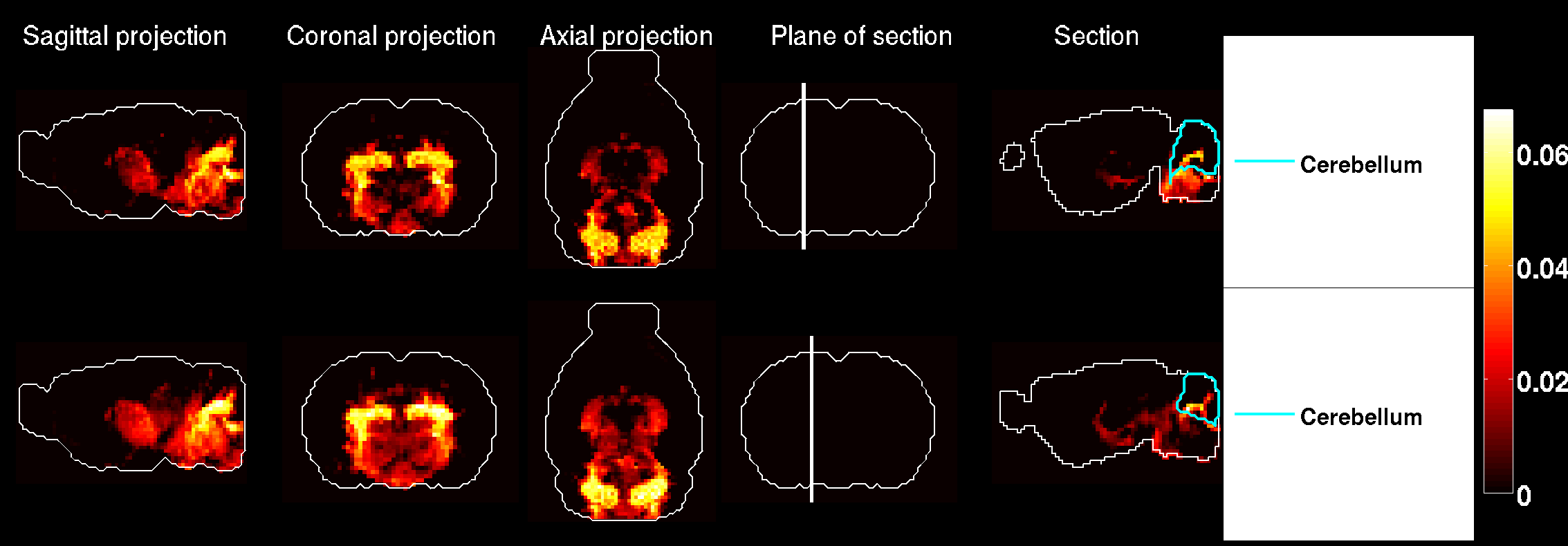}\\\hline
\end{tabular}
 
\caption{Brain-wide density profiles of \numTypesPerReTable cell types, in the 
 original linear model (first row of each figure), and in the model fitted to microarray data
 incorporating the maximum uniform correction compatible with positive entries (second row of each figure).}
\label{tableReFittings7}
\end{table}

\begin{table}
\begin{tabular}{|m{0.06\textwidth}|m{0.06\textwidth}|m{\widthParamForTable\textwidth}|}
\hline
\textbf{Index}&\textbf{Cell type}&\textbf{Heat maps of densities, original and refitted}\\\hline
22&\tiny{Mature Oligodendrocytes}&\includegraphics[width=0.85\textwidth,keepaspectratio]{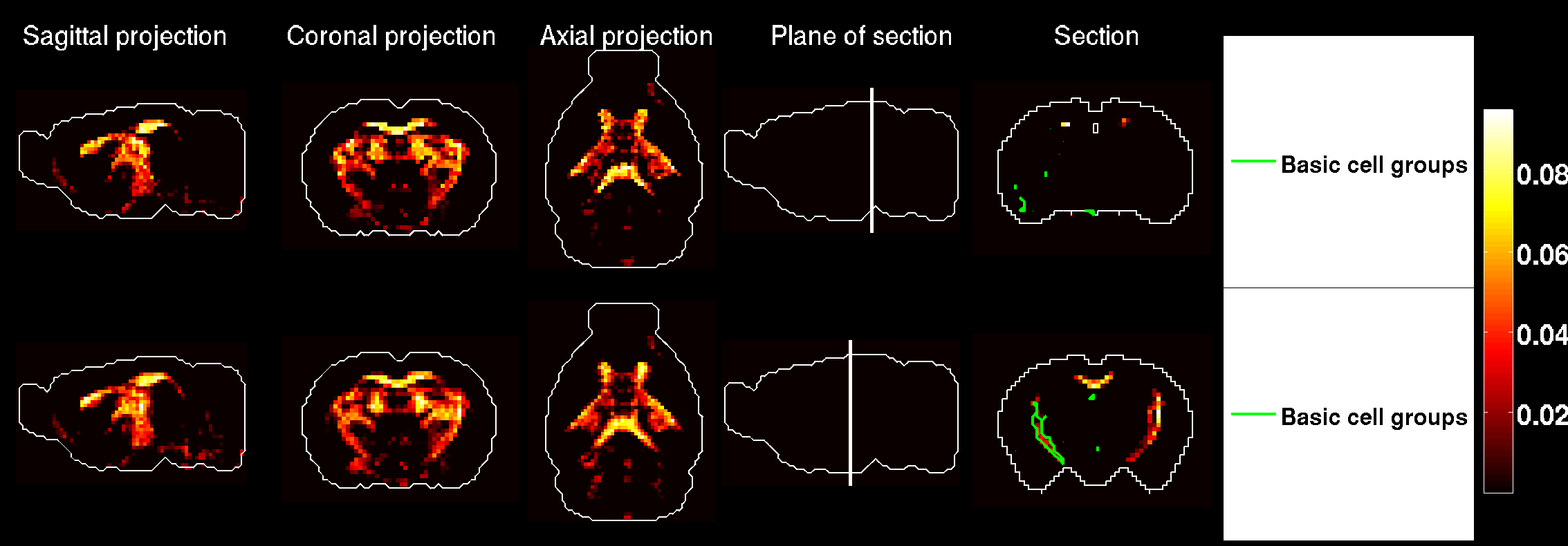}\\\hline
23&\tiny{Mixed Oligodendrocytes}&\includegraphics[width=0.85\textwidth,keepaspectratio]{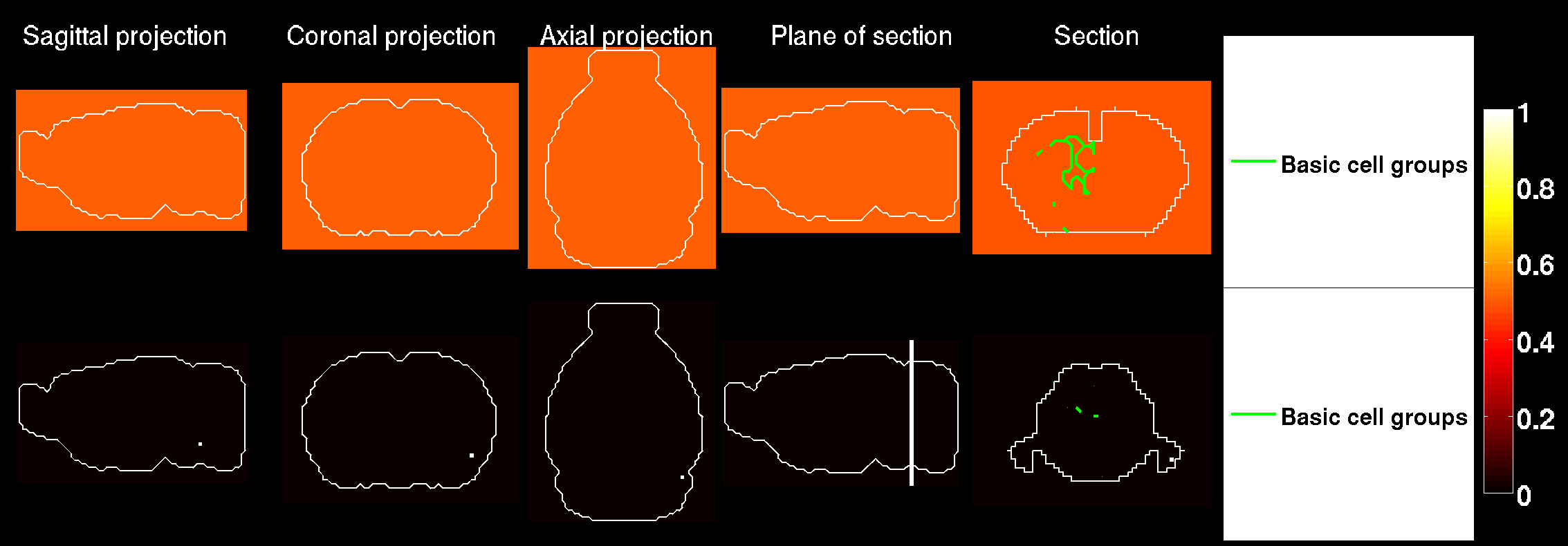}\\\hline
24&\tiny{Mixed Oligodendrocytes}&\includegraphics[width=0.85\textwidth,keepaspectratio]{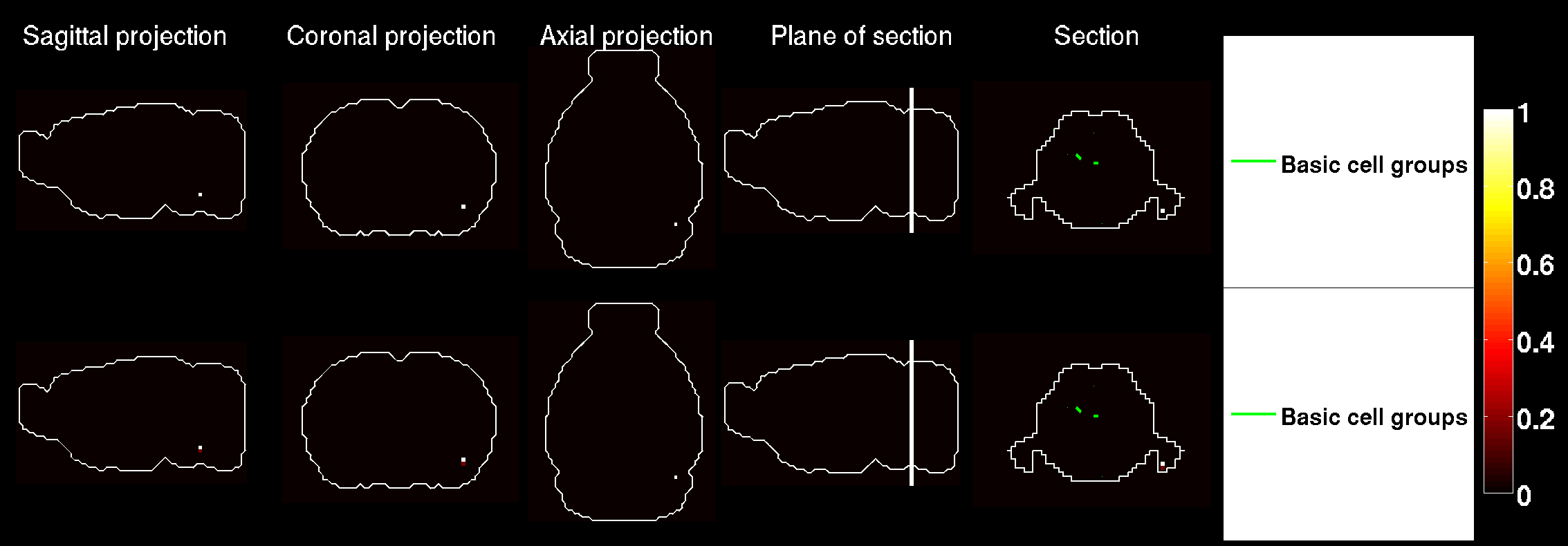}\\\hline
\end{tabular}
 
\caption{Brain-wide density profiles of \numTypesPerReTable cell types, in the 
 original linear model (first row of each figure), and in the model fitted to microarray data
 incorporating the maximum uniform correction compatible with positive entries (second row of each figure).}
\label{tableReFittings8}
\end{table}

\begin{table}
\begin{tabular}{|m{0.06\textwidth}|m{0.06\textwidth}|m{\widthParamForTable\textwidth}|}
\hline
\textbf{Index}&\textbf{Cell type}&\textbf{Heat maps of densities, original and refitted}\\\hline
25&\tiny{Purkinje Cells}&\includegraphics[width=0.85\textwidth,keepaspectratio]{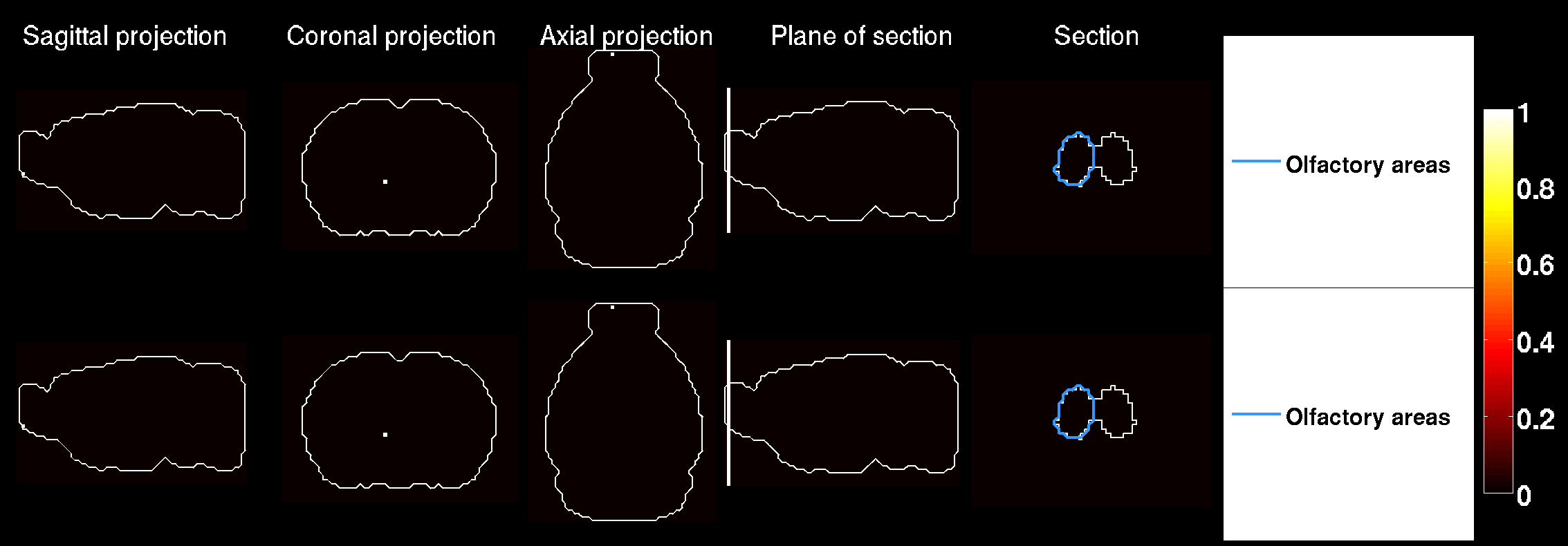}\\\hline
26&\tiny{Neurons}&\includegraphics[width=0.85\textwidth,keepaspectratio]{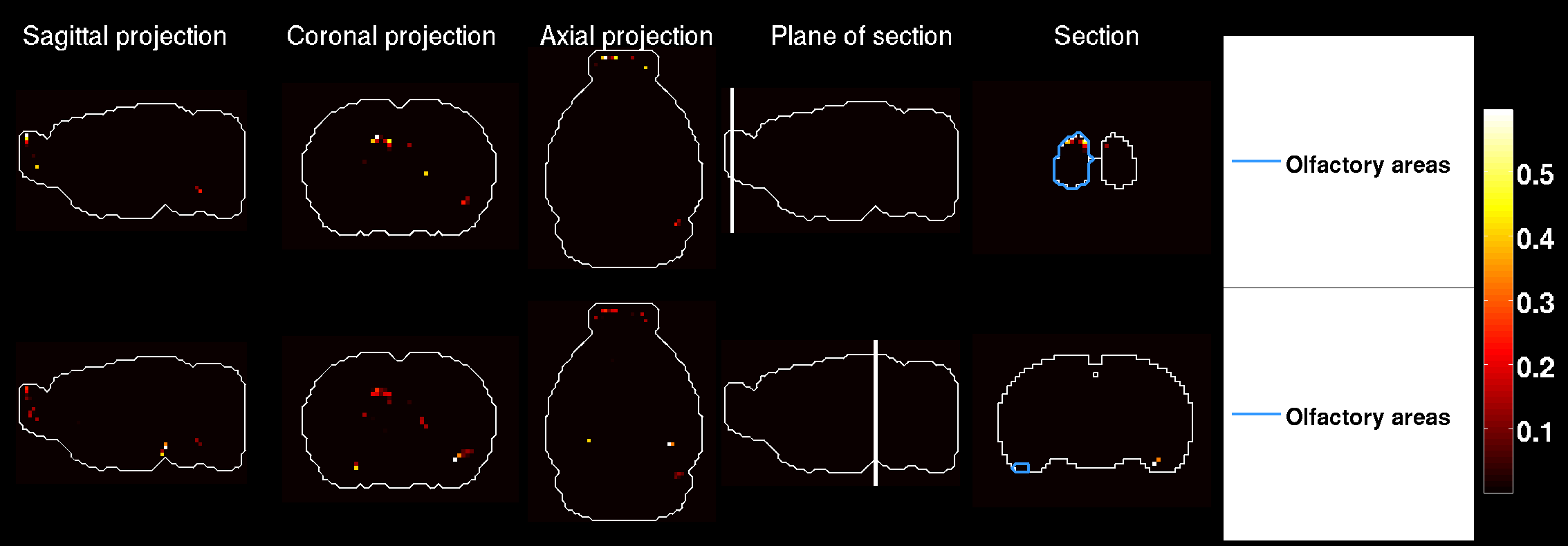}\\\hline
27&\tiny{Bergman Glia}&\includegraphics[width=0.85\textwidth,keepaspectratio]{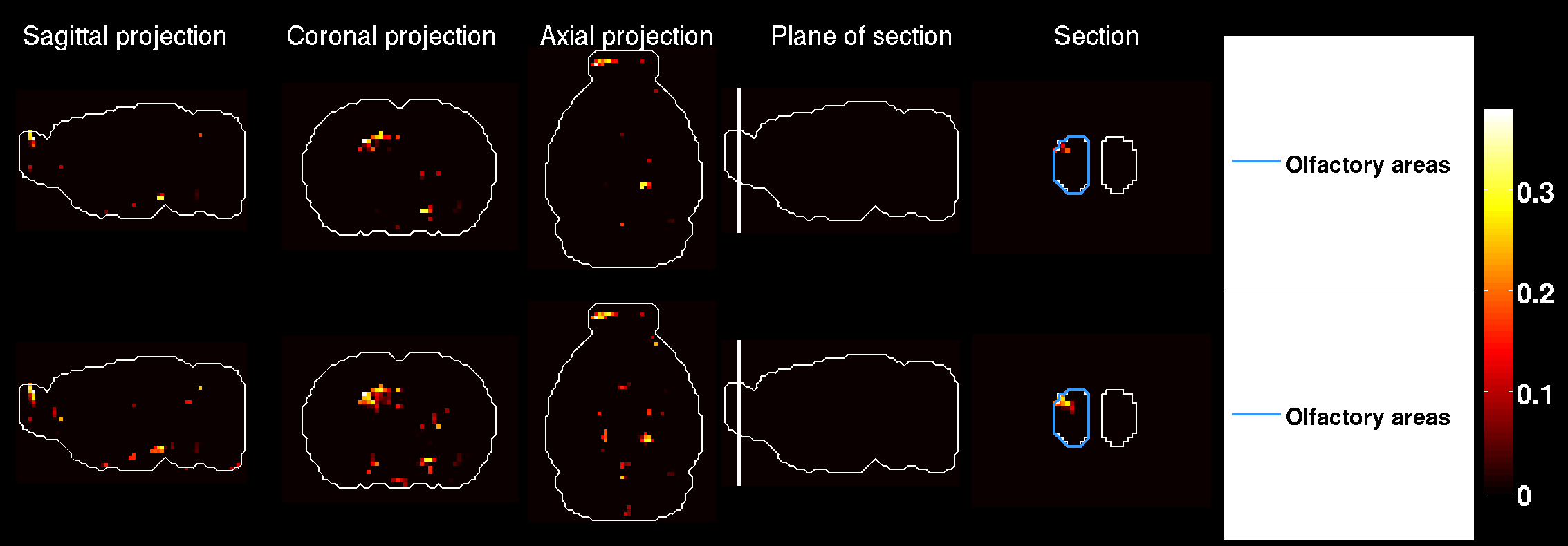}\\\hline
\end{tabular}
 
\caption{Brain-wide density profiles of \numTypesPerReTable cell types, in the 
 original linear model (first row of each figure), and in the model fitted to microarray data
 incorporating the maximum uniform correction compatible with positive entries (second row of each figure).}
\label{tableReFittings9}
\end{table}

\begin{table}
\begin{tabular}{|m{0.06\textwidth}|m{0.06\textwidth}|m{\widthParamForTable\textwidth}|}
\hline
\textbf{Index}&\textbf{Cell type}&\textbf{Heat maps of densities, original and refitted}\\\hline
28&\tiny{Astroglia}&\includegraphics[width=0.85\textwidth,keepaspectratio]{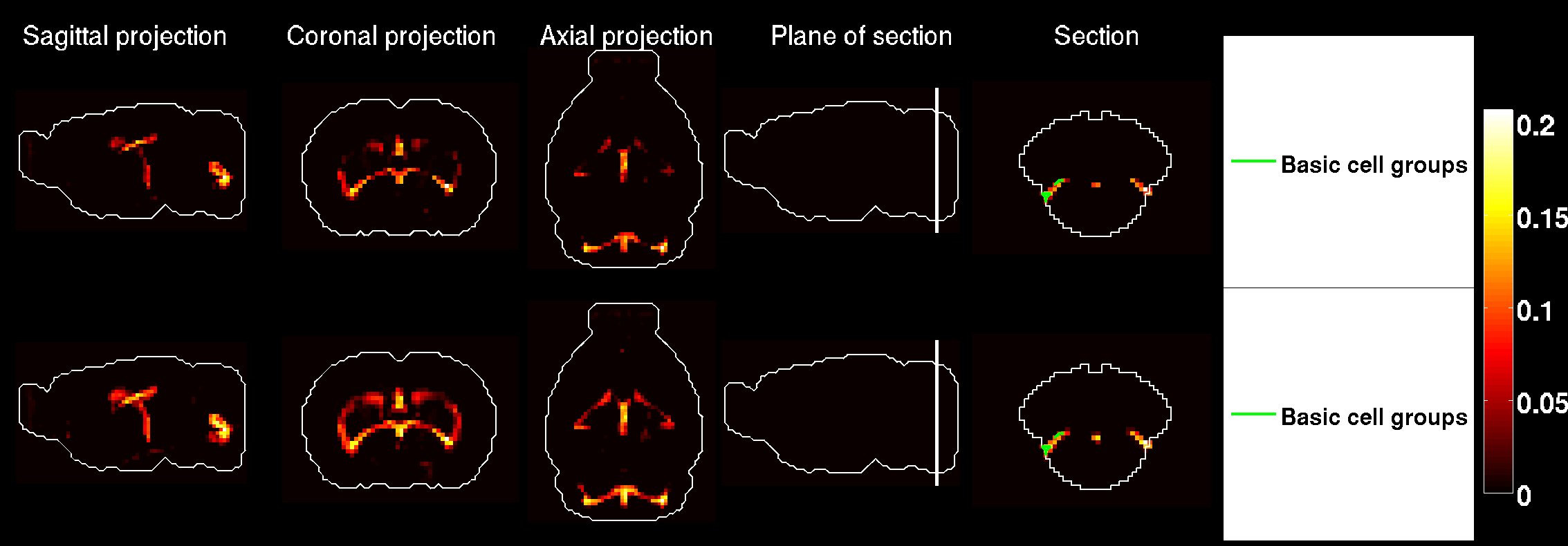}\\\hline
29&\tiny{Astroglia}&\includegraphics[width=0.85\textwidth,keepaspectratio]{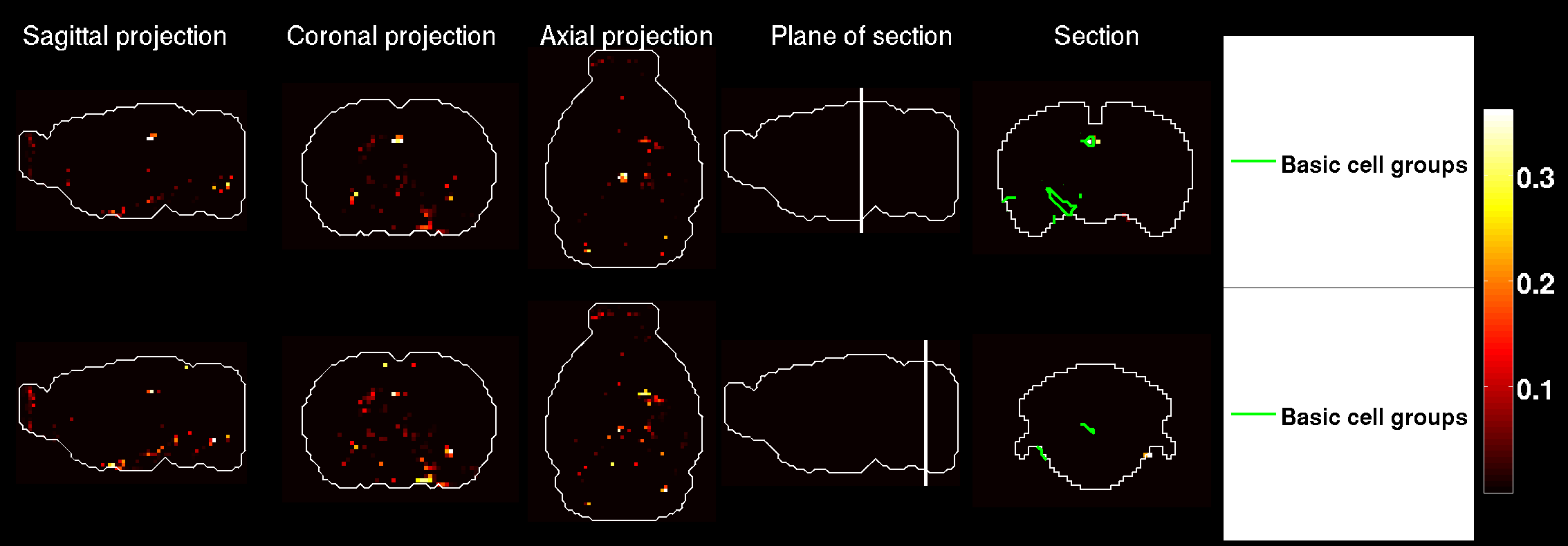}\\\hline
30&\tiny{Astrocytes}&\includegraphics[width=0.85\textwidth,keepaspectratio]{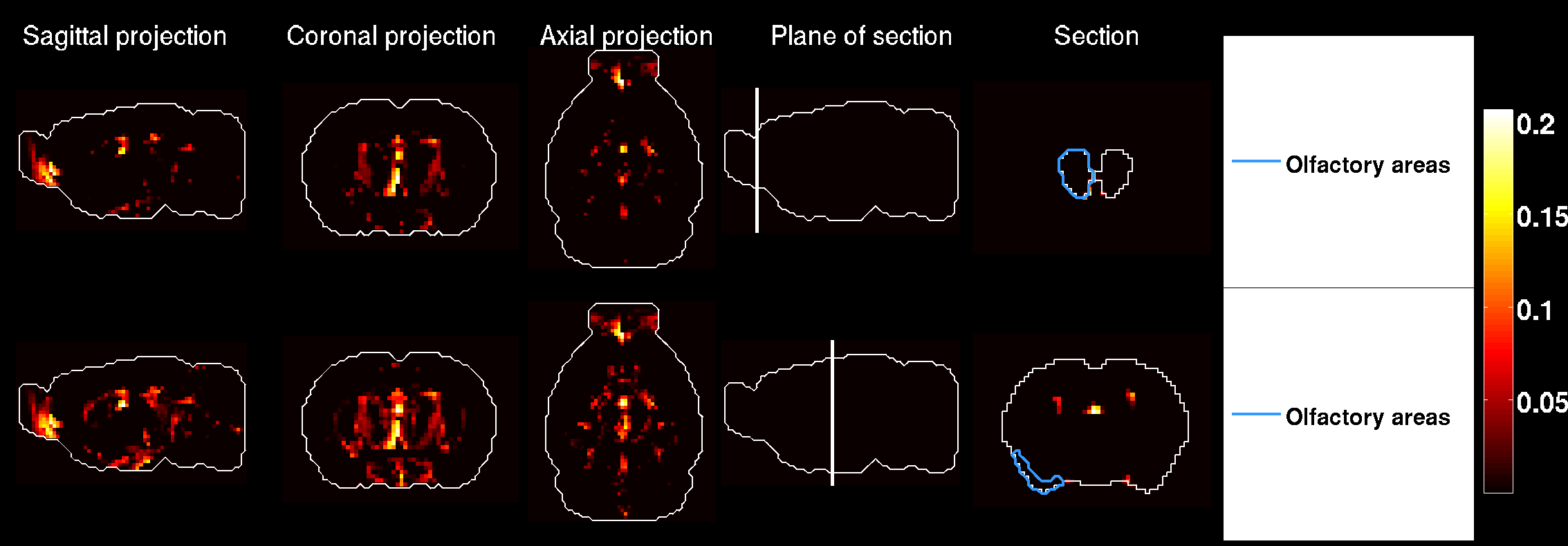}\\\hline
\end{tabular}
 
\caption{Brain-wide density profiles of \numTypesPerReTable cell types, in the 
 original linear model (first row of each figure), and in the model fitted to microarray data
 incorporating the maximum uniform correction compatible with positive entries (second row of each figure).}
\label{tableReFittings10}
\end{table}

\begin{table}
\begin{tabular}{|m{0.06\textwidth}|m{0.06\textwidth}|m{\widthParamForTable\textwidth}|}
\hline
\textbf{Index}&\textbf{Cell type}&\textbf{Heat maps of densities, original and refitted}\\\hline
31&\tiny{Astrocytes}&\includegraphics[width=0.85\textwidth,keepaspectratio]{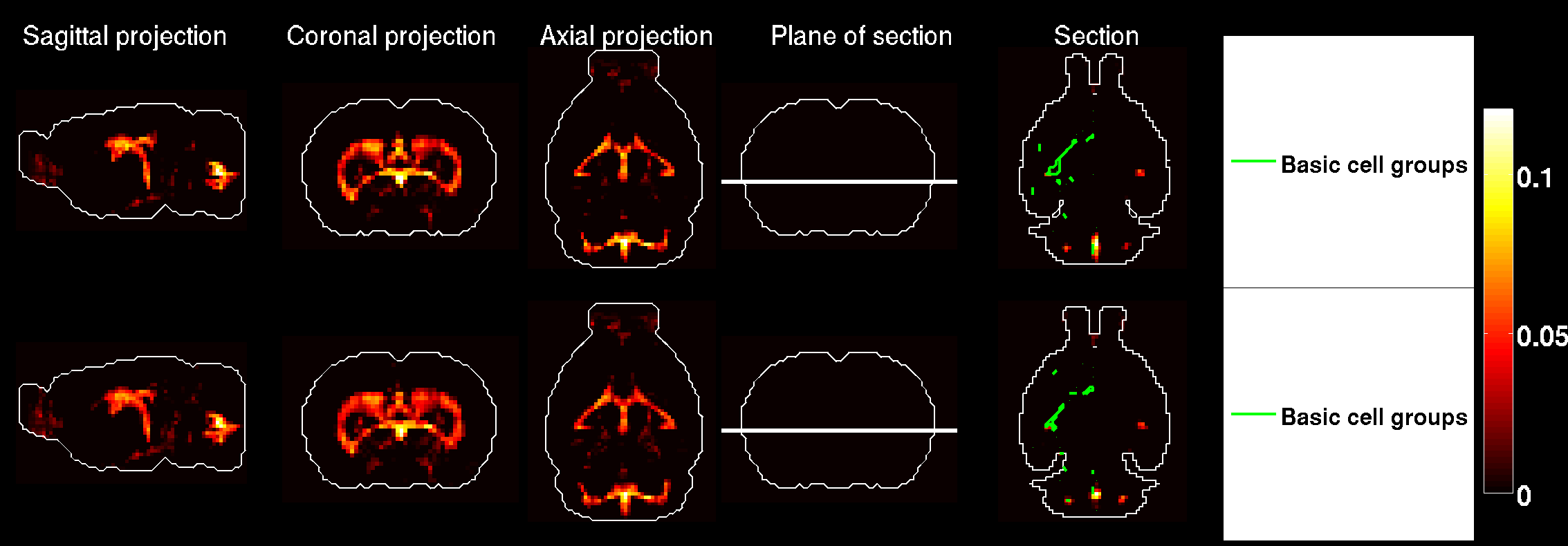}\\\hline
32&\tiny{Astrocytes}&\includegraphics[width=0.85\textwidth,keepaspectratio]{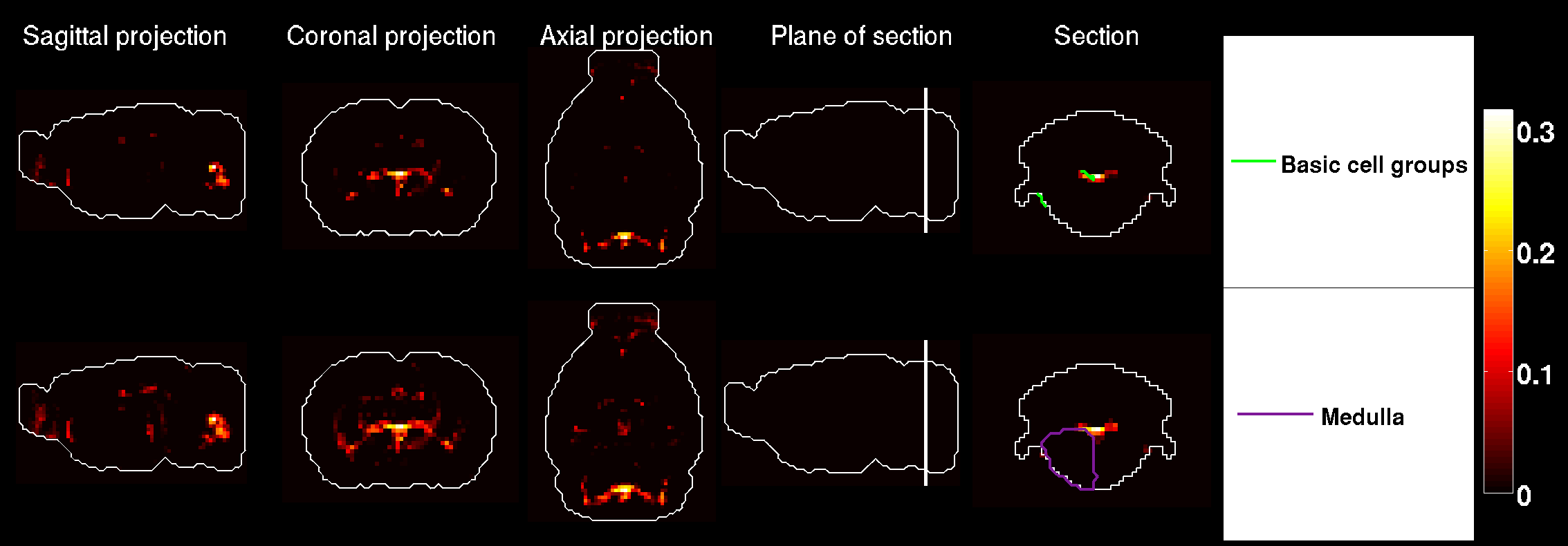}\\\hline
33&\tiny{Mixed Neurons}&\includegraphics[width=0.85\textwidth,keepaspectratio]{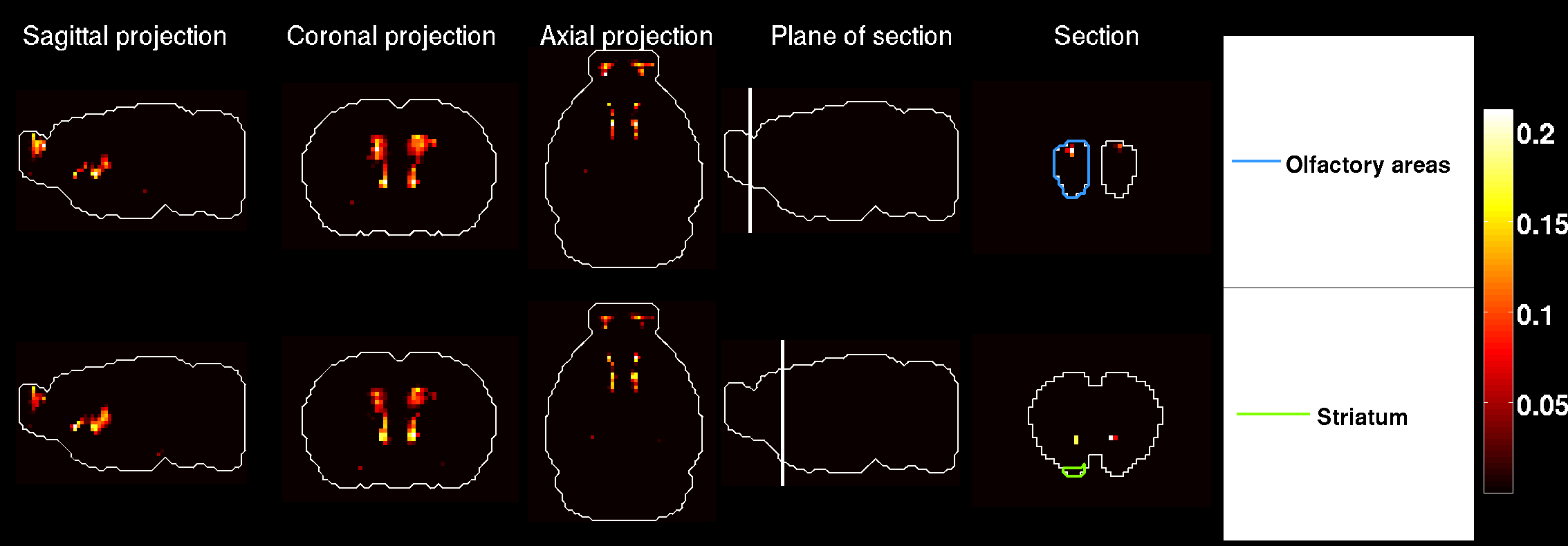}\\\hline
\end{tabular}
 
\caption{Brain-wide density profiles of \numTypesPerReTable cell types, in the 
 original linear model (first row of each figure), and in the model fitted to microarray data
 incorporating the maximum uniform correction compatible with positive entries (second row of each figure).}
\label{tableReFittings11}
\end{table}

\begin{table}
\begin{tabular}{|m{0.06\textwidth}|m{0.06\textwidth}|m{\widthParamForTable\textwidth}|}
\hline
\textbf{Index}&\textbf{Cell type}&\textbf{Heat maps of densities, original and refitted}\\\hline
34&\tiny{Mixed Neurons}&\includegraphics[width=0.85\textwidth,keepaspectratio]{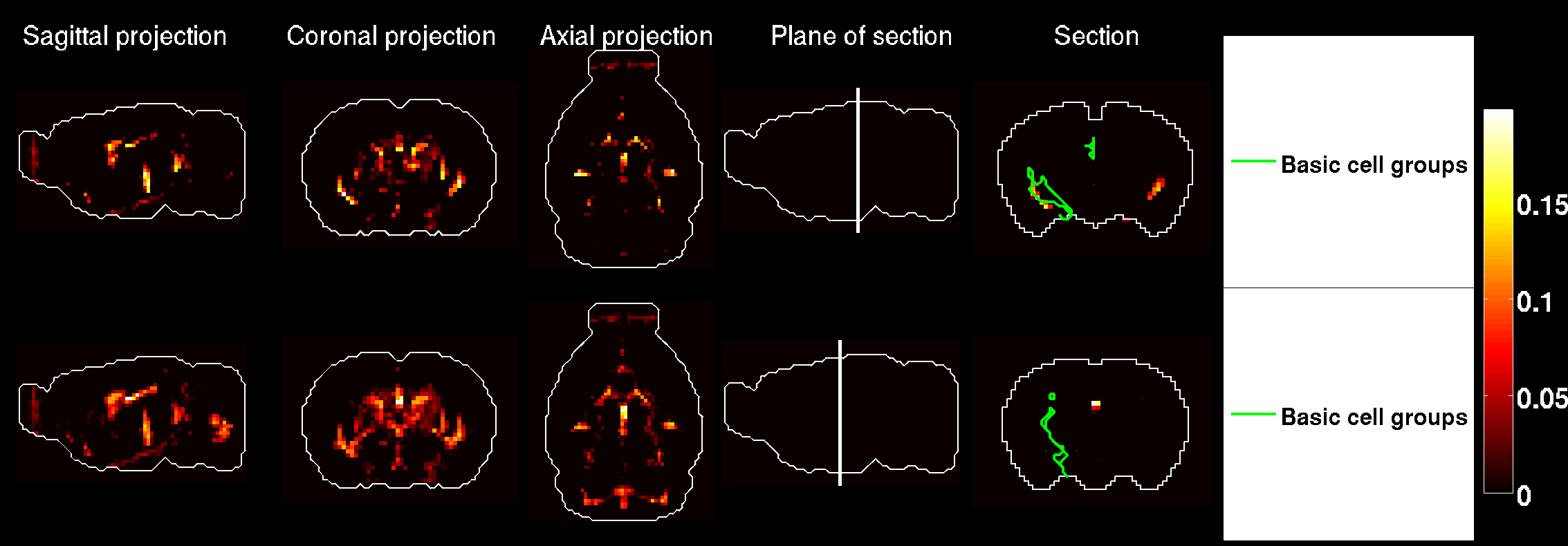}\\\hline
35&\tiny{Mature Oligodendrocytes}&\includegraphics[width=0.85\textwidth,keepaspectratio]{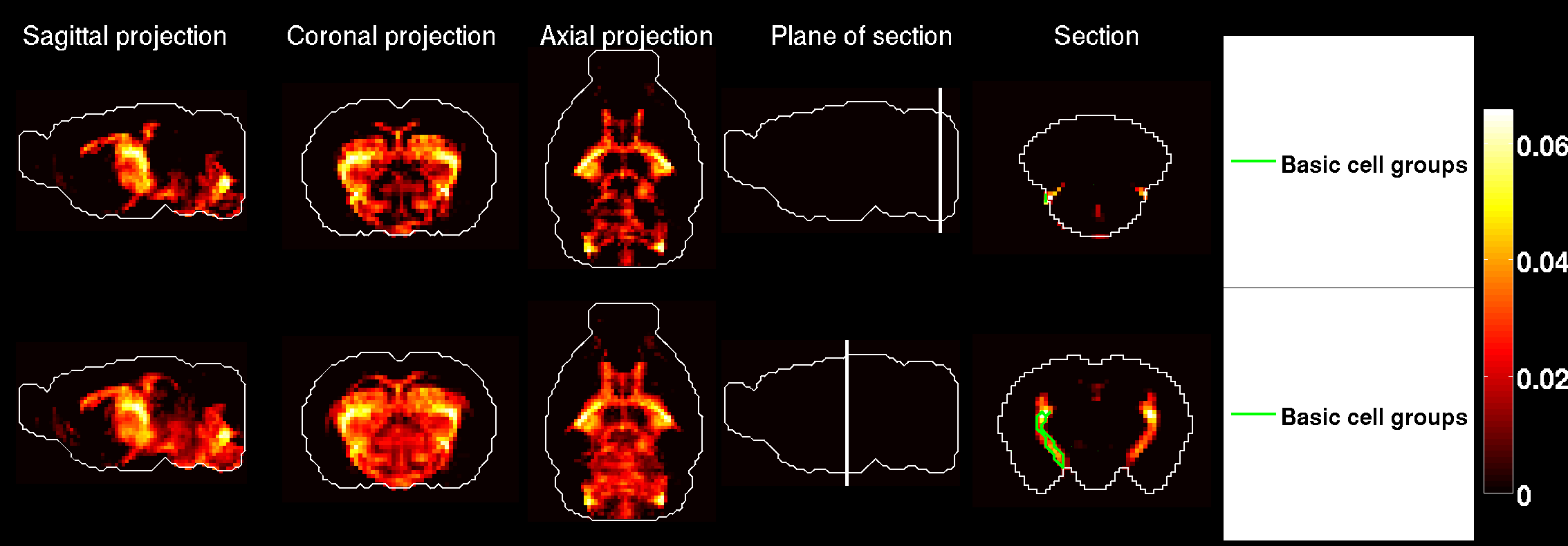}\\\hline
36&\tiny{Oligodendrocytes}&\includegraphics[width=0.85\textwidth,keepaspectratio]{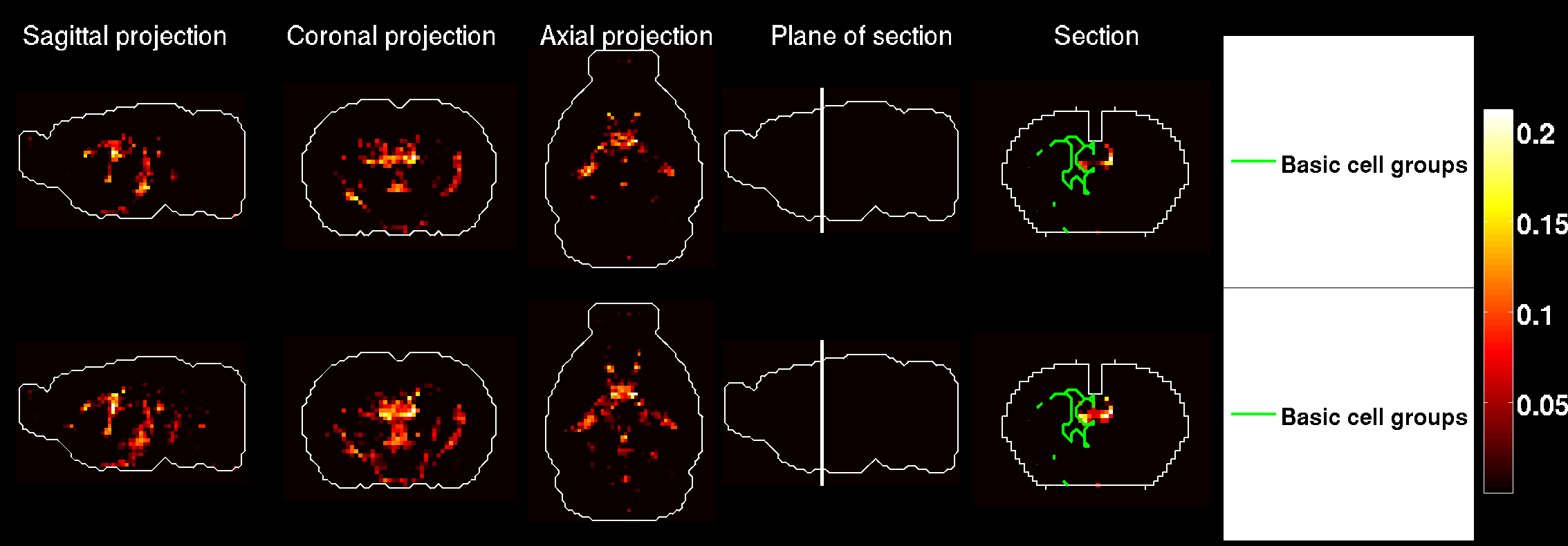}\\\hline
\end{tabular}
 
\caption{Brain-wide density profiles of \numTypesPerReTable cell types, in the 
 original linear model (first row of each figure), and in the model fitted to microarray data
 incorporating the maximum uniform correction compatible with positive entries (second row of each figure).}
\label{tableReFittings12}
\end{table}

\begin{table}
\begin{tabular}{|m{0.06\textwidth}|m{0.06\textwidth}|m{\widthParamForTable\textwidth}|}
\hline
\textbf{Index}&\textbf{Cell type}&\textbf{Heat maps of densities, original and refitted}\\\hline
37&\tiny{Oligodendrocyte Precursors}&\includegraphics[width=0.85\textwidth,keepaspectratio]{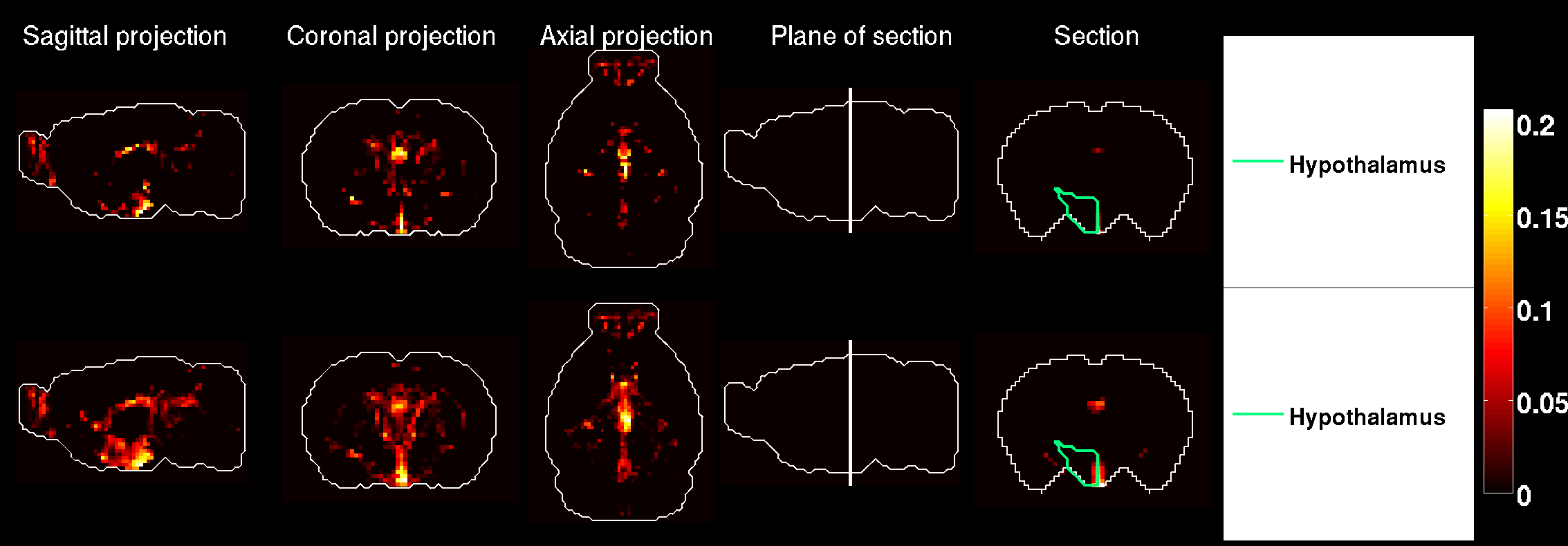}\\\hline
38&\tiny{Pyramidal Neurons, Callosally projecting, P3}&\includegraphics[width=0.85\textwidth,keepaspectratio]{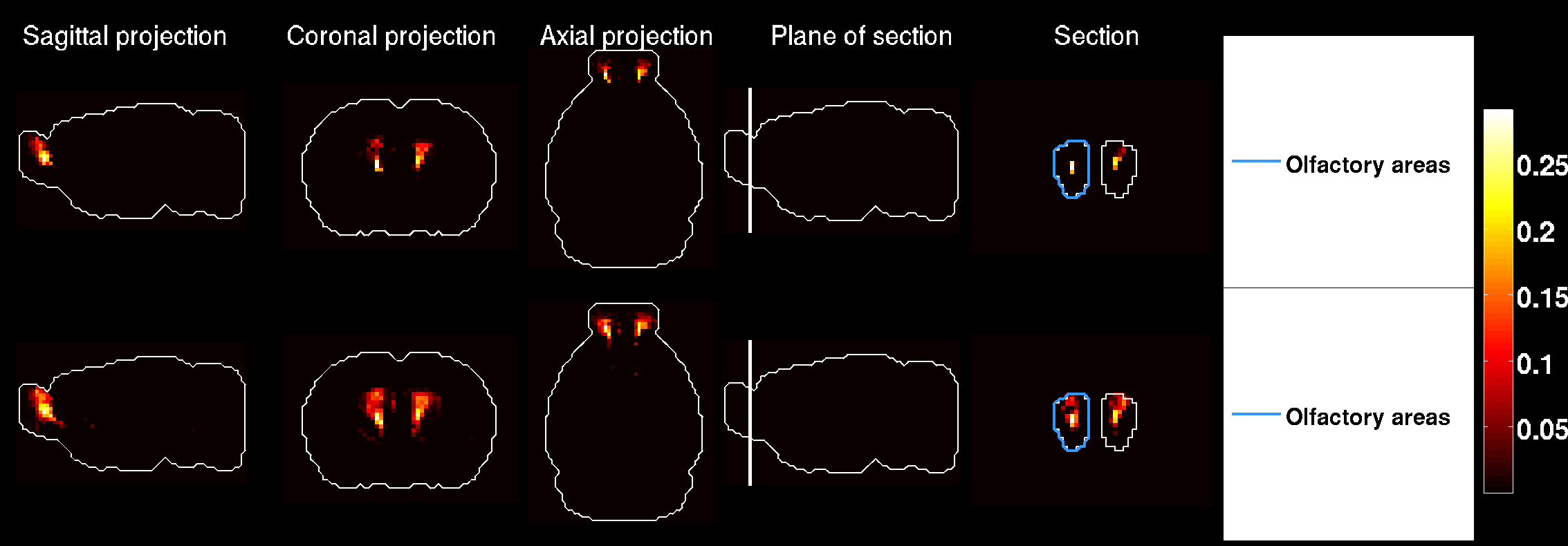}\\\hline
39&\tiny{Pyramidal Neurons, Callosally projecting, P6}&\includegraphics[width=0.85\textwidth,keepaspectratio]{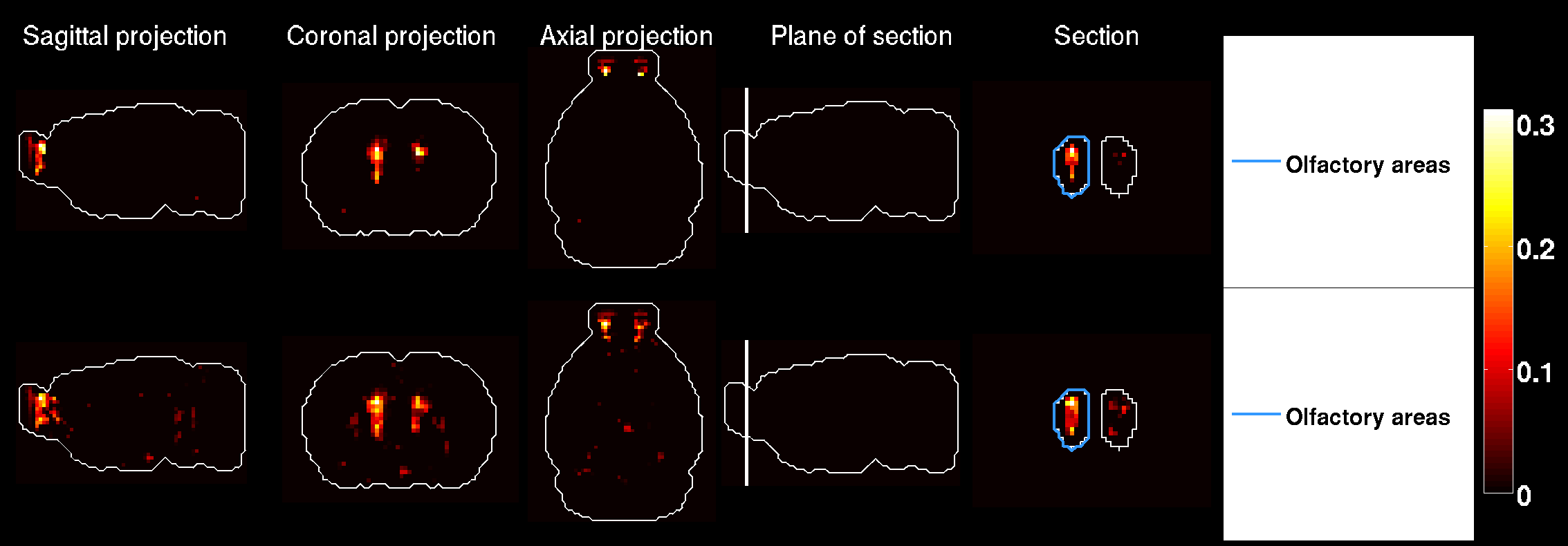}\\\hline
\end{tabular}
 
\caption{Brain-wide density profiles of \numTypesPerReTable cell types, in the 
 original linear model (first row of each figure), and in the model fitted to microarray data
 incorporating the maximum uniform correction compatible with positive entries (second row of each figure).}
\label{tableReFittings13}
\end{table}

\begin{table}
\begin{tabular}{|m{0.06\textwidth}|m{0.06\textwidth}|m{\widthParamForTable\textwidth}|}
\hline
\textbf{Index}&\textbf{Cell type}&\textbf{Heat maps of densities, original and refitted}\\\hline
40&\tiny{Pyramidal Neurons, Callosally projecting, P14}&\includegraphics[width=0.85\textwidth,keepaspectratio]{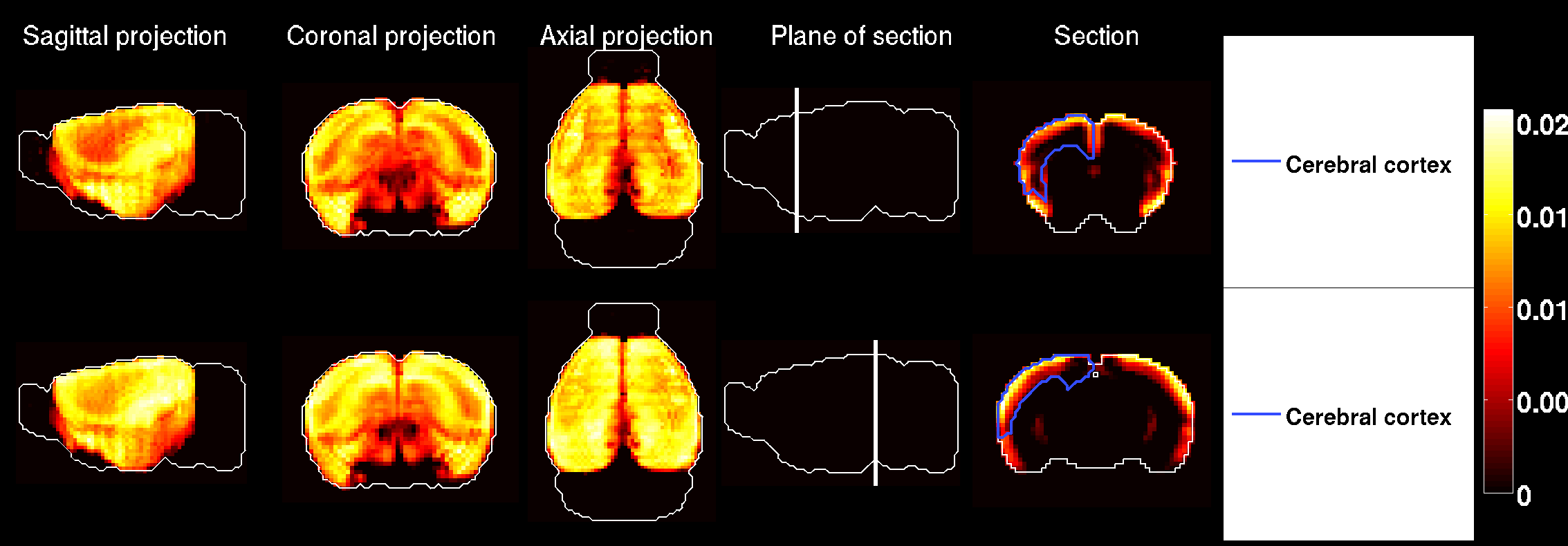}\\\hline
41&\tiny{Pyramidal Neurons, Corticospinal, P3}&\includegraphics[width=0.85\textwidth,keepaspectratio]{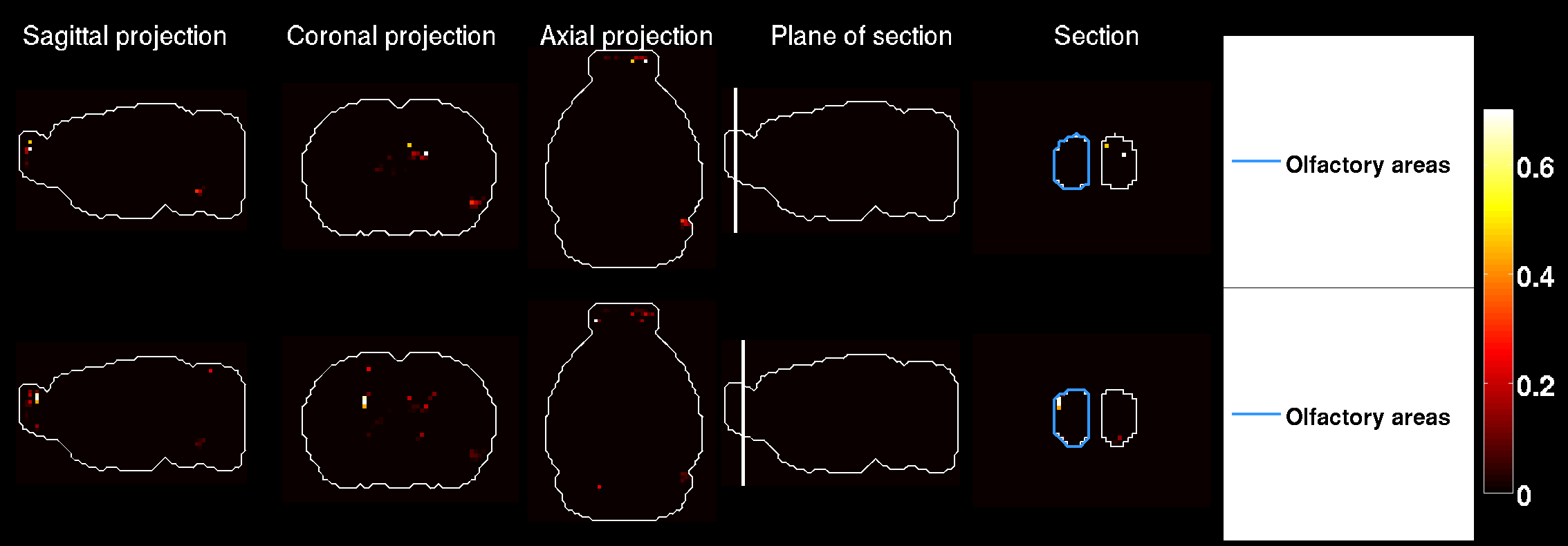}\\\hline
42&\tiny{Pyramidal Neurons, Corticospinal, P6}&\includegraphics[width=0.85\textwidth,keepaspectratio]{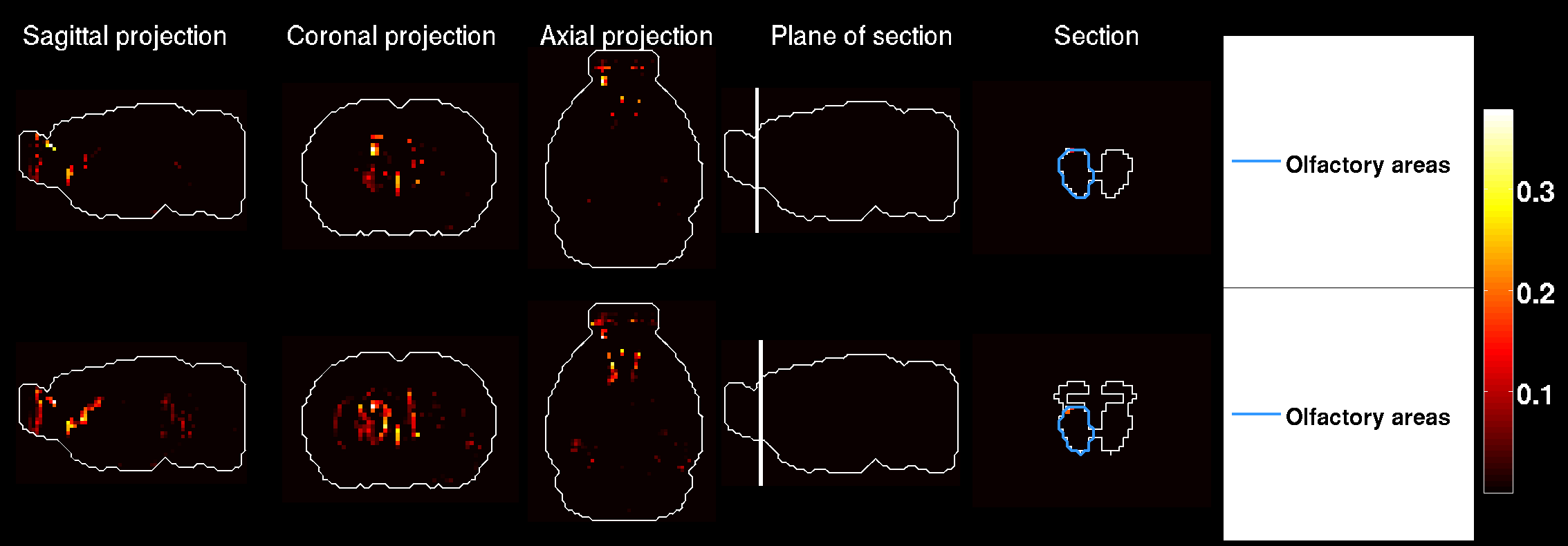}\\\hline
\end{tabular}
 
\caption{Brain-wide density profiles of \numTypesPerReTable cell types, in the 
 original linear model (first row of each figure), and in the model fitted to microarray data
 incorporating the maximum uniform correction compatible with positive entries (second row of each figure).}
\label{tableReFittings14}
\end{table}

\begin{table}
\begin{tabular}{|m{0.06\textwidth}|m{0.06\textwidth}|m{\widthParamForTable\textwidth}|}
\hline
\textbf{Index}&\textbf{Cell type}&\textbf{Heat maps of densities, original and refitted}\\\hline
43&\tiny{Pyramidal Neurons, Corticospinal, P14}&\includegraphics[width=0.85\textwidth,keepaspectratio]{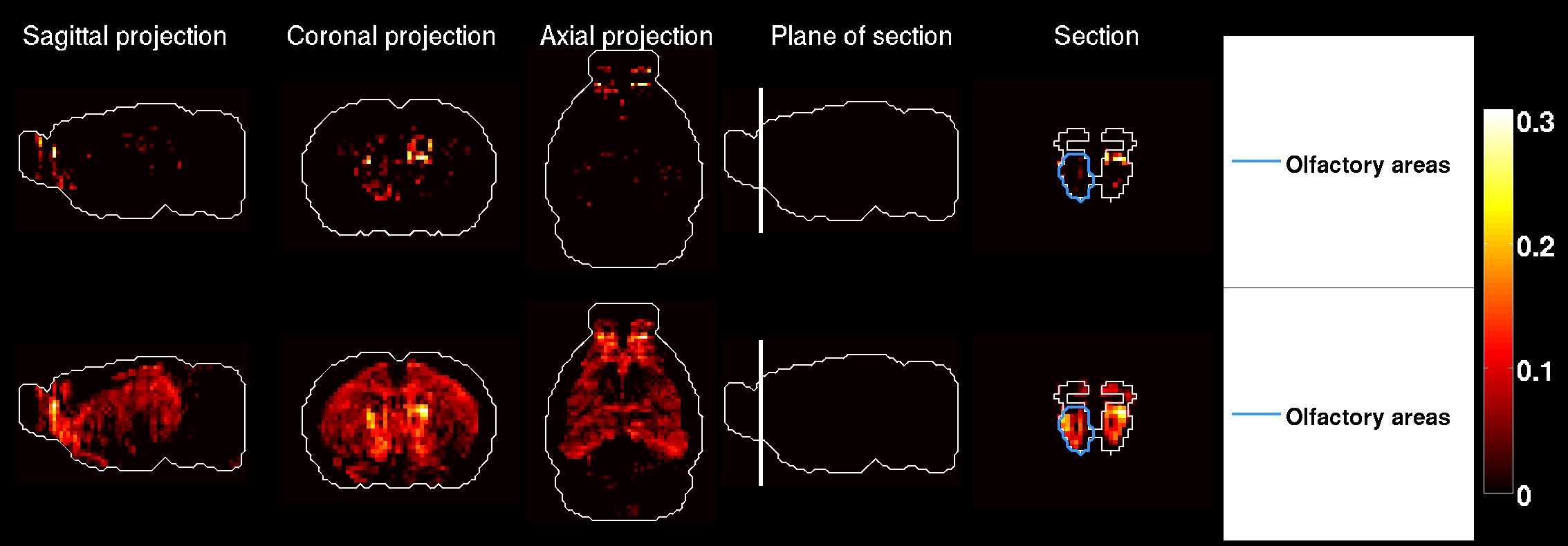}\\\hline
44&\tiny{Pyramidal Neurons, Corticotectal, P14}&\includegraphics[width=0.85\textwidth,keepaspectratio]{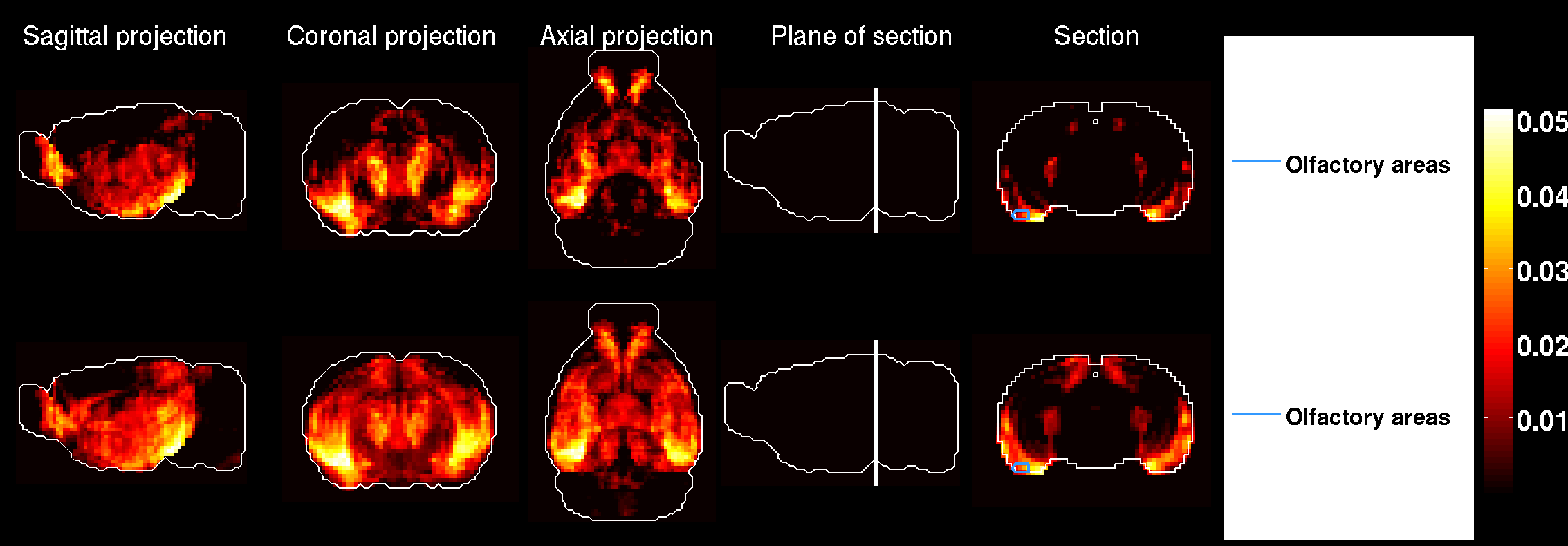}\\\hline
45&\tiny{Pyramidal Neurons}&\includegraphics[width=0.85\textwidth,keepaspectratio]{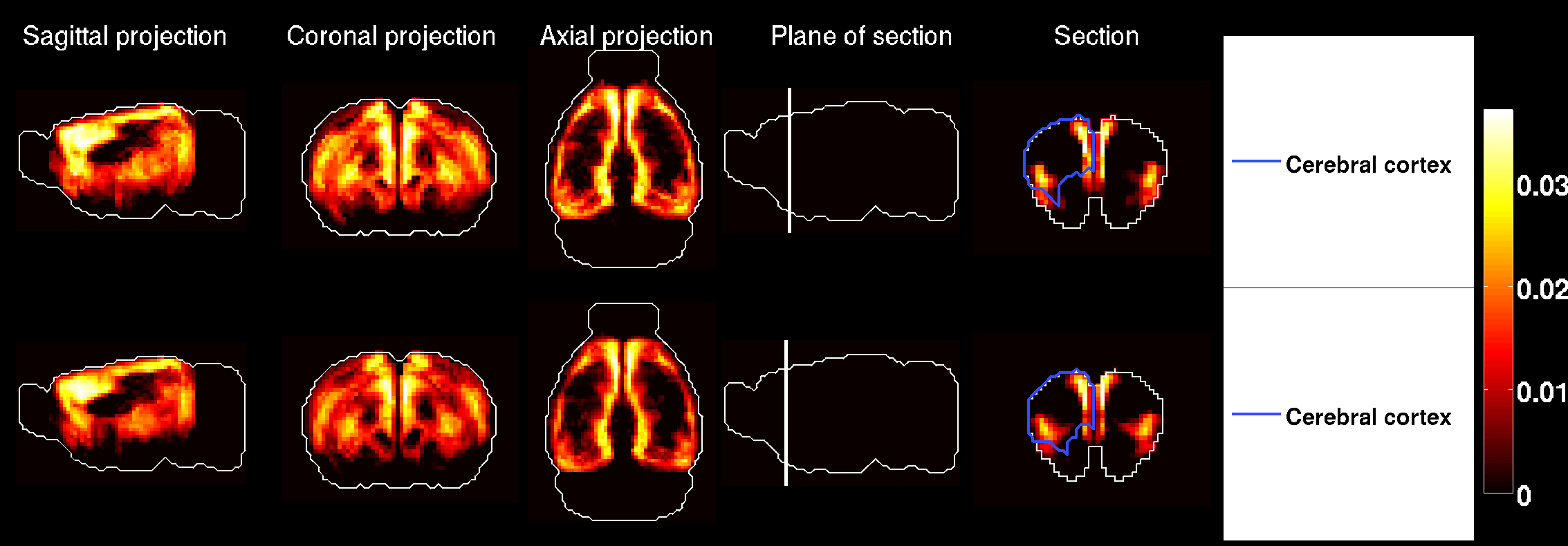}\\\hline
\end{tabular}
 
\caption{Brain-wide density profiles of \numTypesPerReTable cell types, in the 
 original linear model (first row of each figure), and in the model fitted to microarray data
 incorporating the maximum uniform correction compatible with positive entries (second row of each figure).}
\label{tableReFittings15}
\end{table}

\begin{table}
\begin{tabular}{|m{0.06\textwidth}|m{0.06\textwidth}|m{\widthParamForTable\textwidth}|}
\hline
\textbf{Index}&\textbf{Cell type}&\textbf{Heat maps of densities, original and refitted}\\\hline
46&\tiny{Pyramidal Neurons}&\includegraphics[width=0.85\textwidth,keepaspectratio]{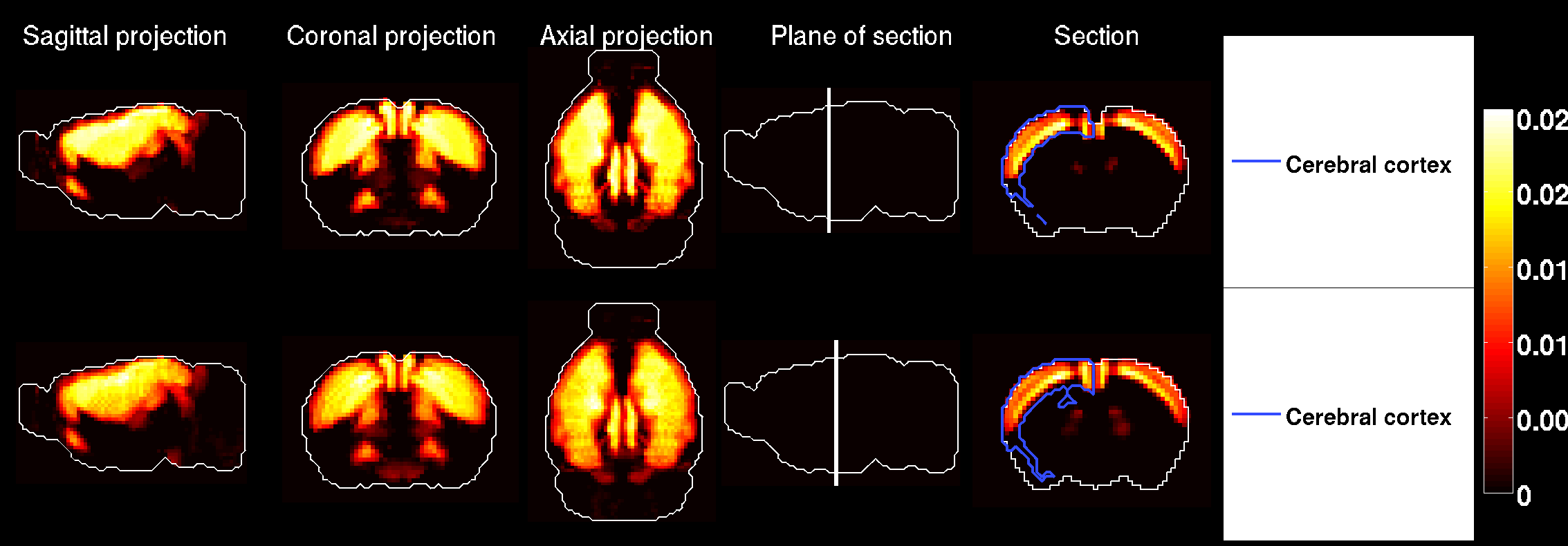}\\\hline
47&\tiny{Pyramidal Neurons}&\includegraphics[width=0.85\textwidth,keepaspectratio]{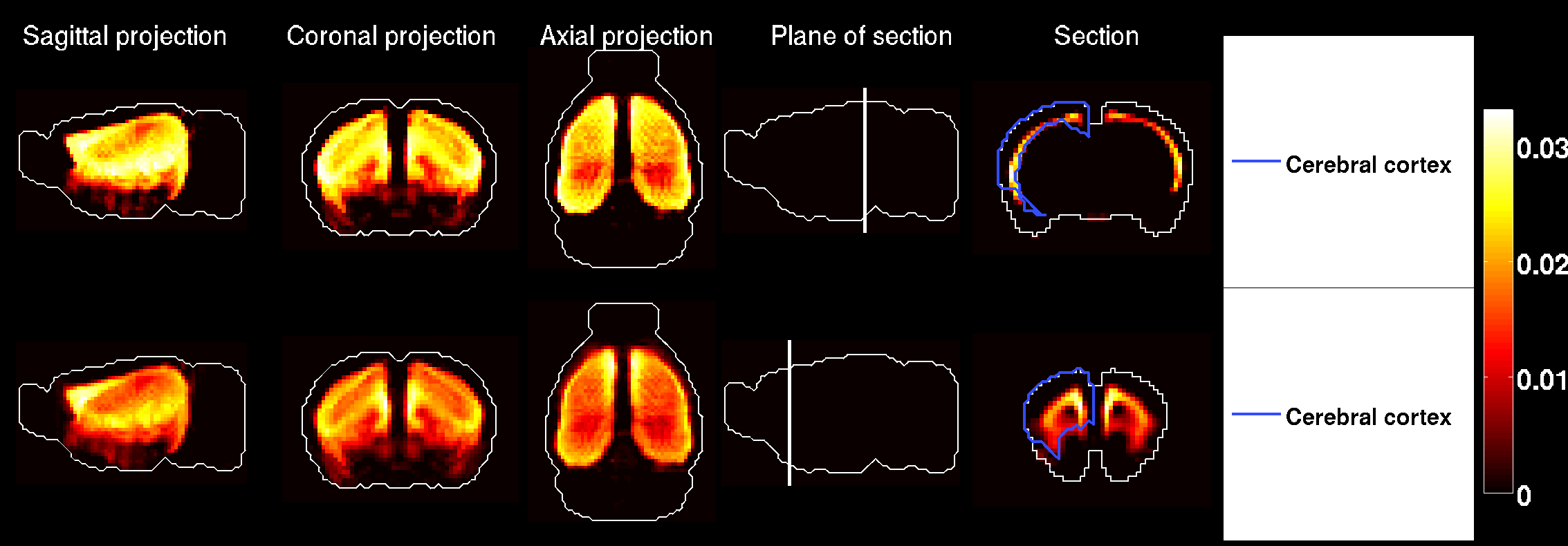}\\\hline
48&\tiny{Pyramidal Neurons}&\includegraphics[width=0.85\textwidth,keepaspectratio]{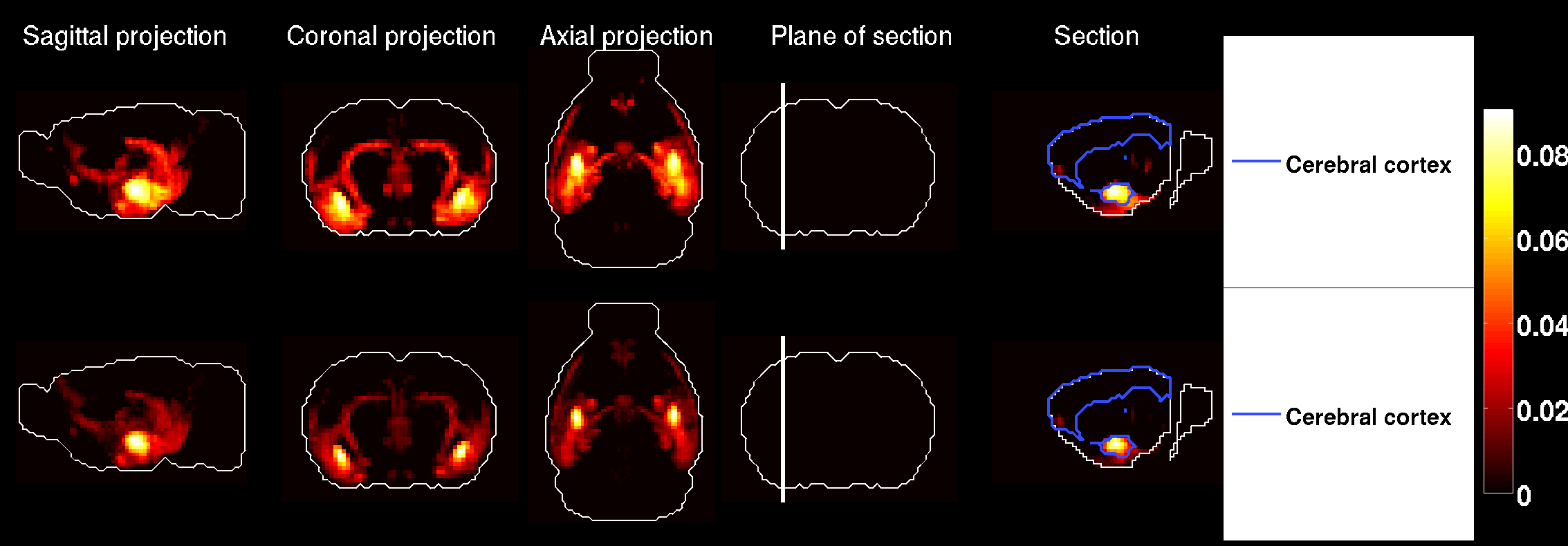}\\\hline
\end{tabular}
 
\caption{Brain-wide density profiles of \numTypesPerReTable cell types, in the 
 original linear model (first row of each figure), and in the model fitted to microarray data
 incorporating the maximum uniform correction compatible with positive entries (second row of each figure).}
\label{tableReFittings16}
\end{table}

\begin{table}
\begin{tabular}{|m{0.06\textwidth}|m{0.06\textwidth}|m{\widthParamForTable\textwidth}|}
\hline
\textbf{Index}&\textbf{Cell type}&\textbf{Heat maps of densities, original and refitted}\\\hline
49&\tiny{Pyramidal Neurons}&\includegraphics[width=0.85\textwidth,keepaspectratio]{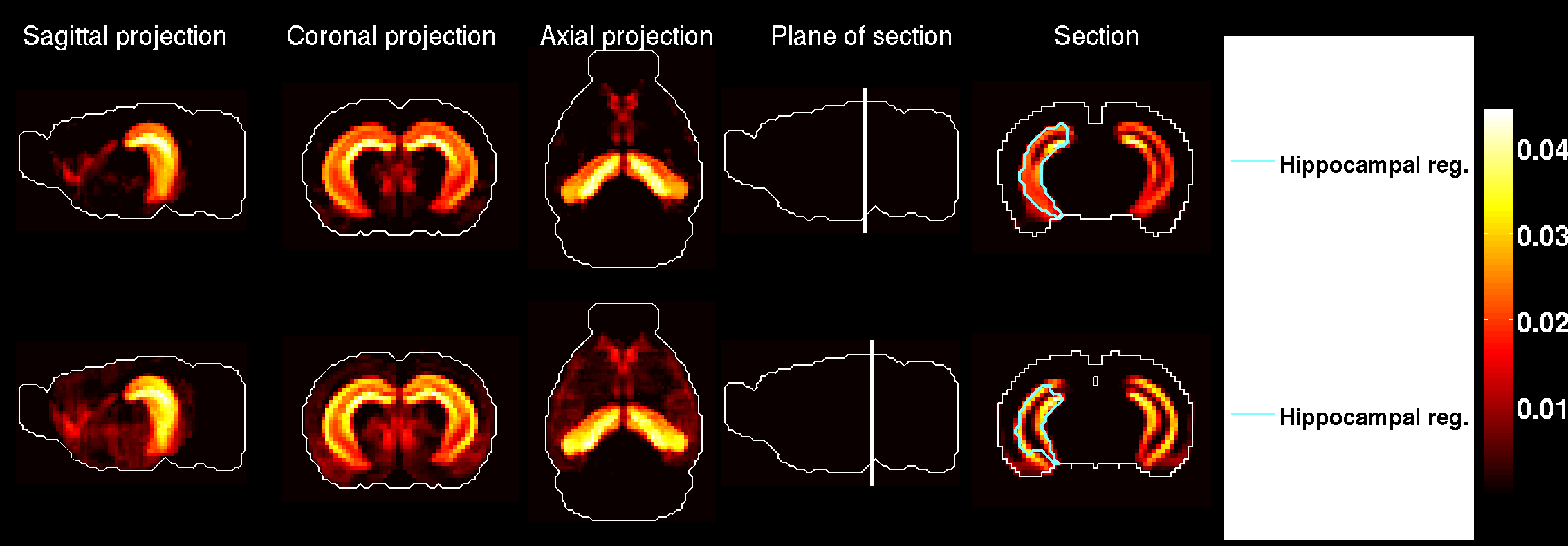}\\\hline
50&\tiny{Pyramidal Neurons}&\includegraphics[width=0.85\textwidth,keepaspectratio]{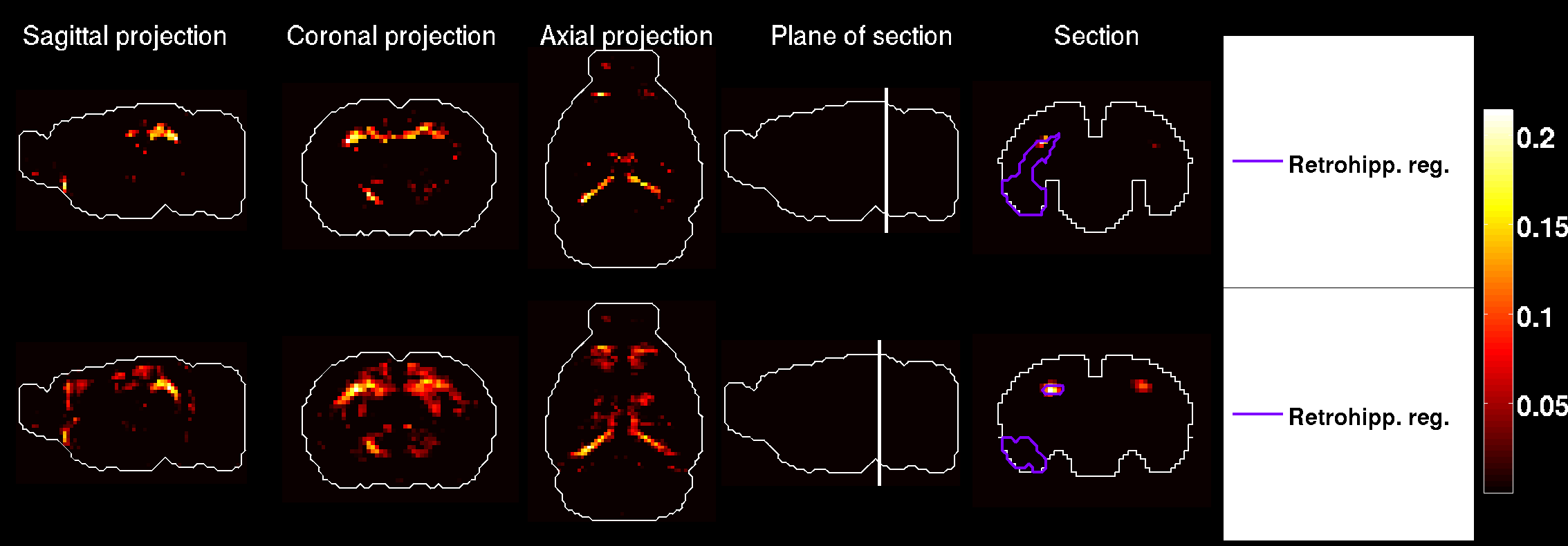}\\\hline
51&\tiny{Tyrosine Hydroxylase Expressing}&\includegraphics[width=0.85\textwidth,keepaspectratio]{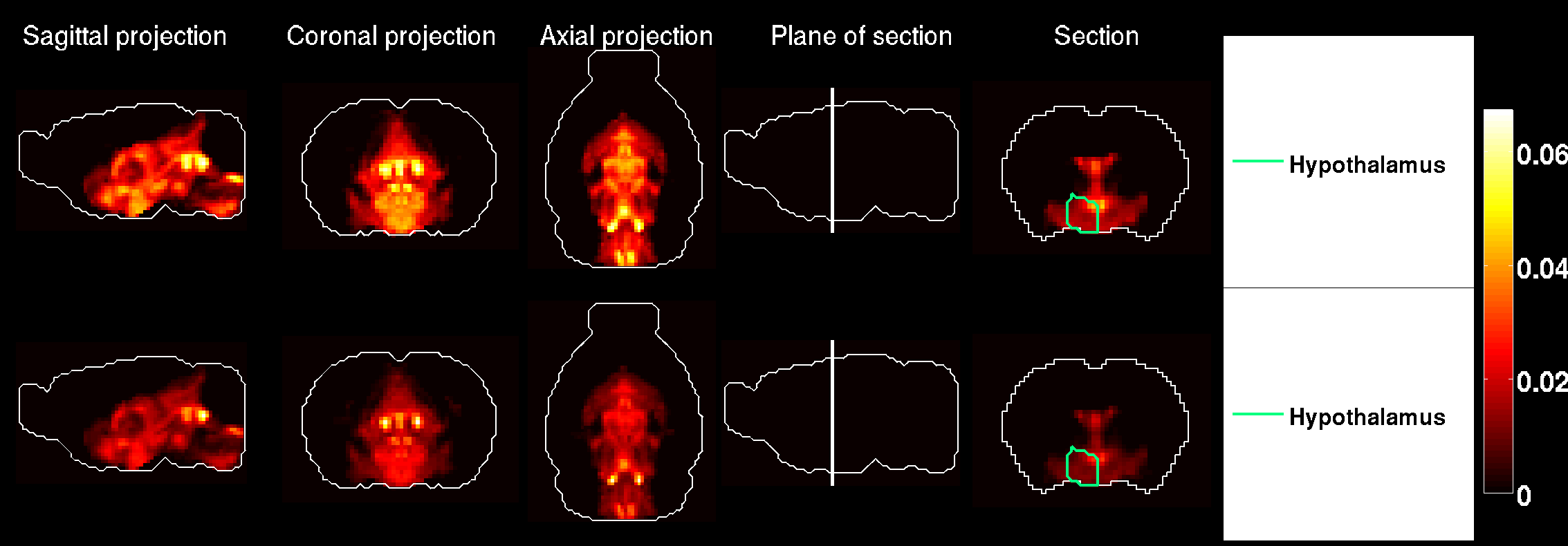}\\\hline
\end{tabular}
 
\caption{Brain-wide density profiles of \numTypesPerReTable  cell types, in the 
 original linear model (first row of each figure), and in the model fitted to microarray data
 incorporating the maximum uniform correction compatible with positive entries (second row of each figure).}
\label{tableReFittings17}
\end{table}

\begin{table}
\begin{tabular}{|m{0.06\textwidth}|m{0.06\textwidth}|m{\widthParamForTable\textwidth}|}
\hline
\textbf{Index}&\textbf{Cell type}&\textbf{Heat maps of densities, original and refitted}\\\hline
52&\tiny{Purkinje Cells}&\includegraphics[width=0.85\textwidth,keepaspectratio]{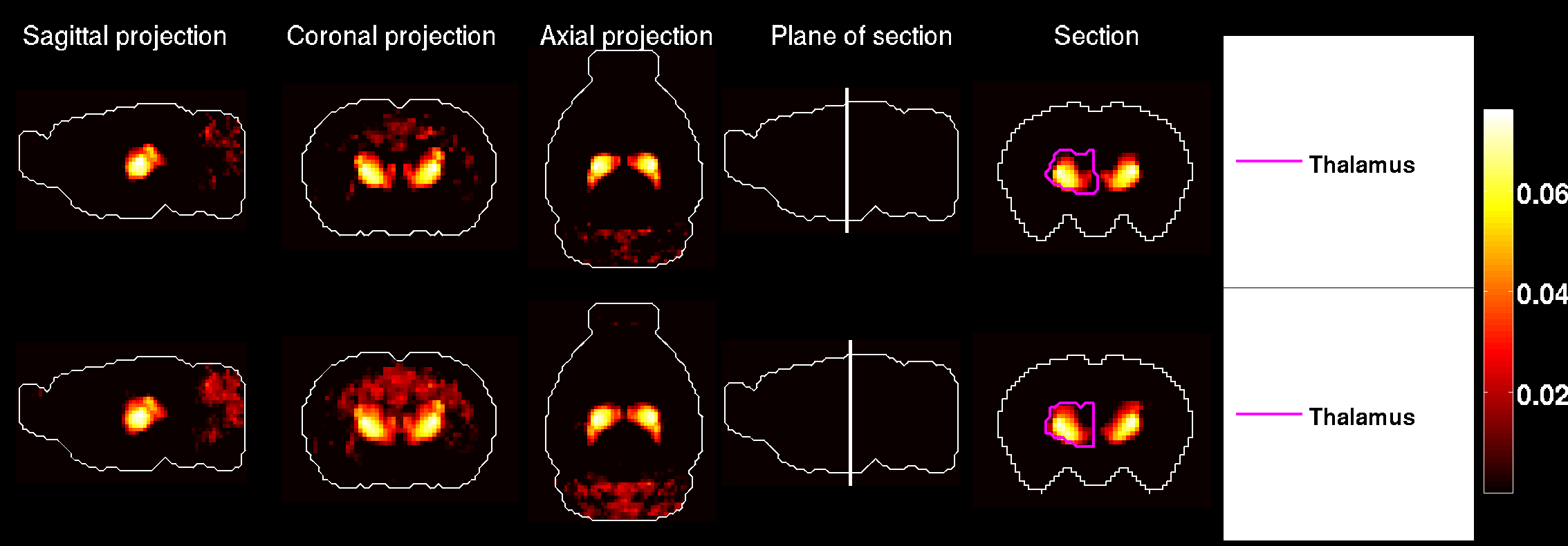}\\\hline
53&\tiny{Glutamatergic Neuron (not well defined)}&\includegraphics[width=0.85\textwidth,keepaspectratio]{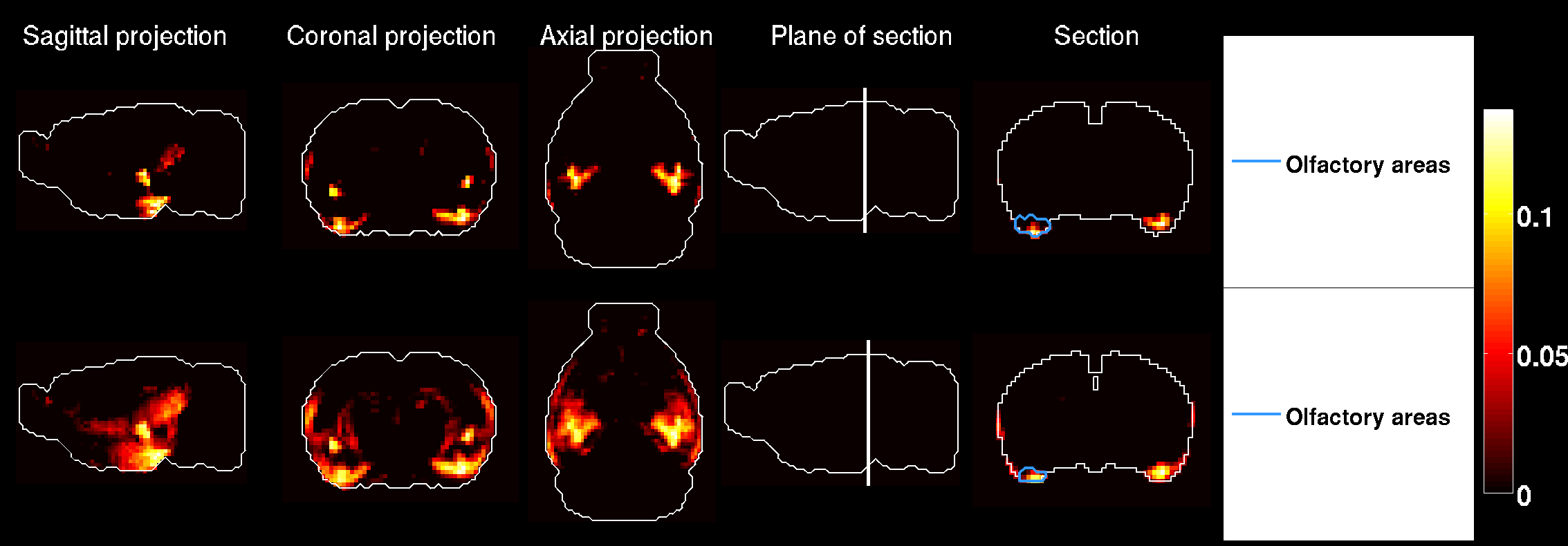}\\\hline
54&\tiny{GABAergic Interneurons, VIP+}&\includegraphics[width=0.85\textwidth,keepaspectratio]{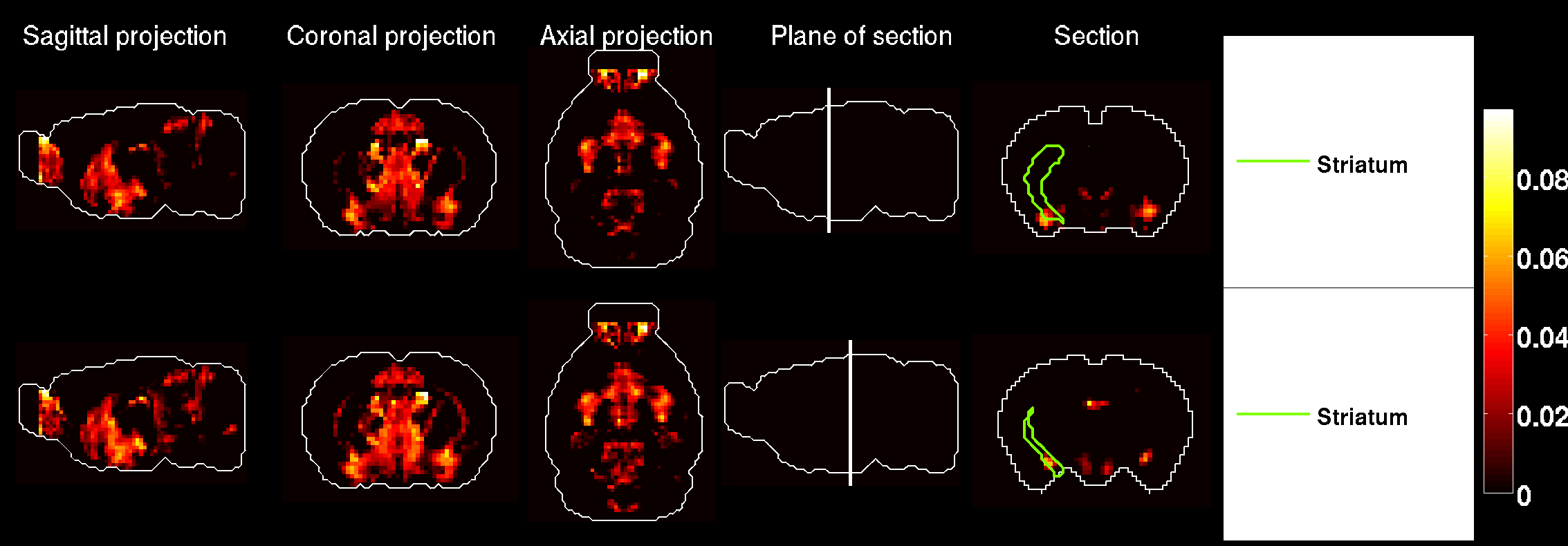}\\\hline
\end{tabular}
 
\caption{Brain-wide density profiles of \numTypesPerReTable  cell types, in the 
 original linear model (first row of each figure), and in the model fitted to microarray data
 incorporating the maximum uniform correction compatible with positive entries (second row of each figure).}
\label{tableReFittings18}
\end{table}

\begin{table}
\begin{tabular}{|m{0.06\textwidth}|m{0.06\textwidth}|m{\widthParamForTable\textwidth}|}
\hline
\textbf{Index}&\textbf{Cell type}&\textbf{Heat maps of densities, original and refitted}\\\hline
55&\tiny{GABAergic Interneurons, VIP+}&\includegraphics[width=0.85\textwidth,keepaspectratio]{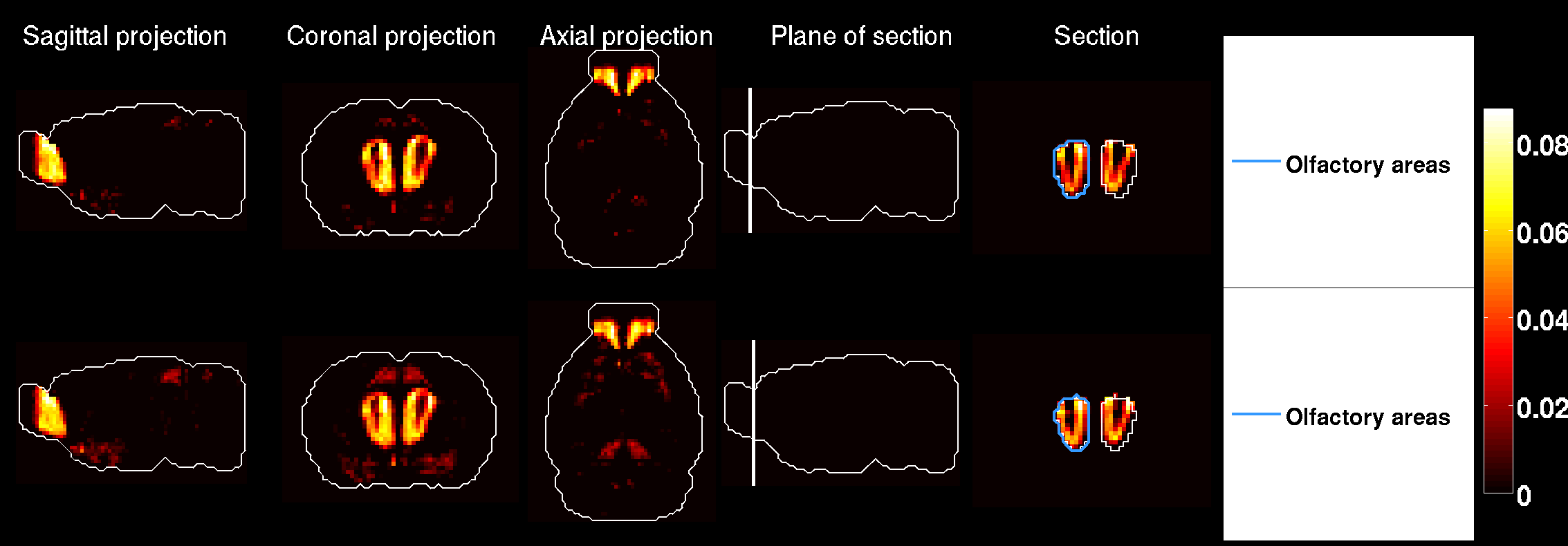}\\\hline
56&\tiny{GABAergic Interneurons, SST+}&\includegraphics[width=0.85\textwidth,keepaspectratio]{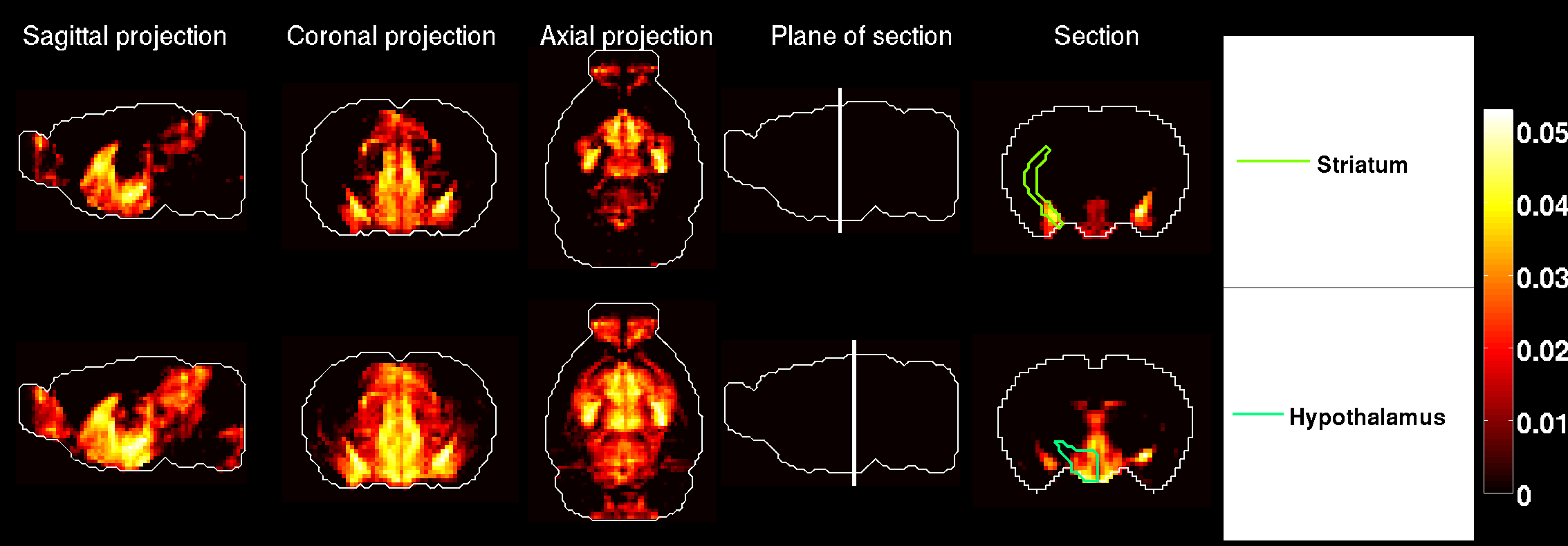}\\\hline
57&\tiny{GABAergic Interneurons, SST+}&\includegraphics[width=0.85\textwidth,keepaspectratio]{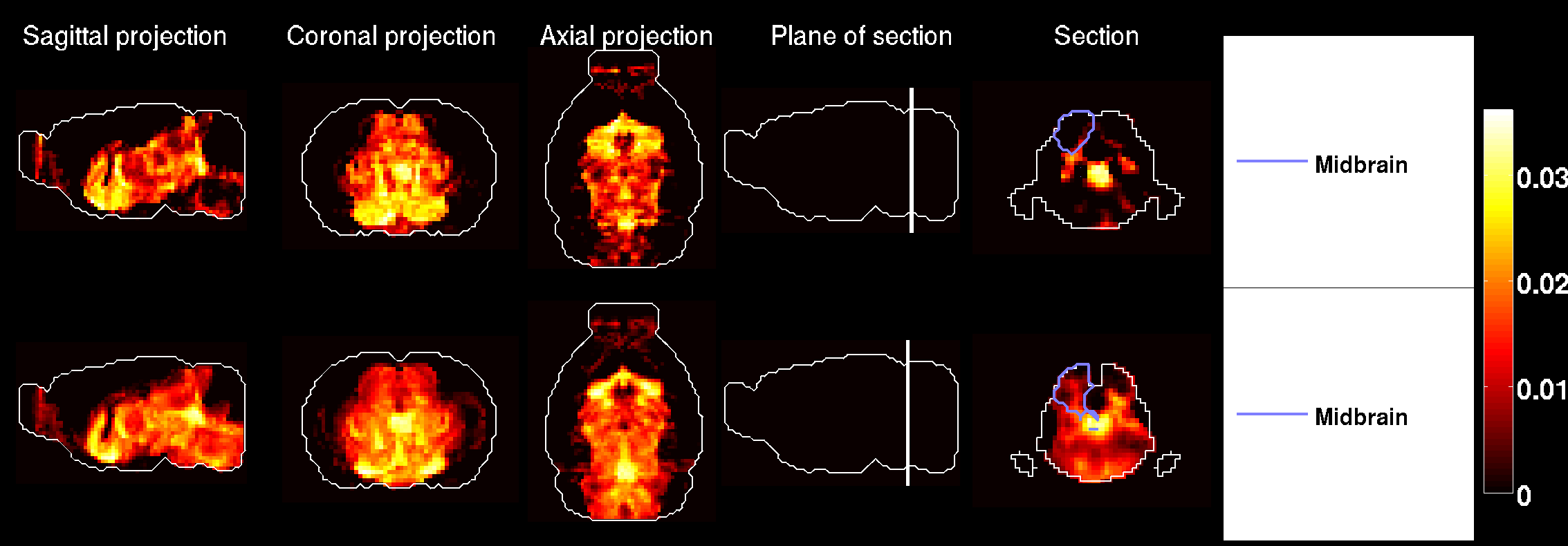}\\\hline
\end{tabular}
 
\caption{Brain-wide density profiles of \numTypesPerReTable  cell types, in the 
 original linear model (first row of each figure), and in the model fitted to microarray data
 incorporating the maximum uniform correction compatible with positive entries (second row of each figure).}
\label{tableReFittings19}
\end{table}

\begin{table}
\begin{tabular}{|m{0.06\textwidth}|m{0.06\textwidth}|m{\widthParamForTable\textwidth}|}
\hline
\textbf{Index}&\textbf{Cell type}&\textbf{Heat maps of densities, original and refitted}\\\hline
58&\tiny{GABAergic Interneurons, PV+}&\includegraphics[width=0.85\textwidth,keepaspectratio]{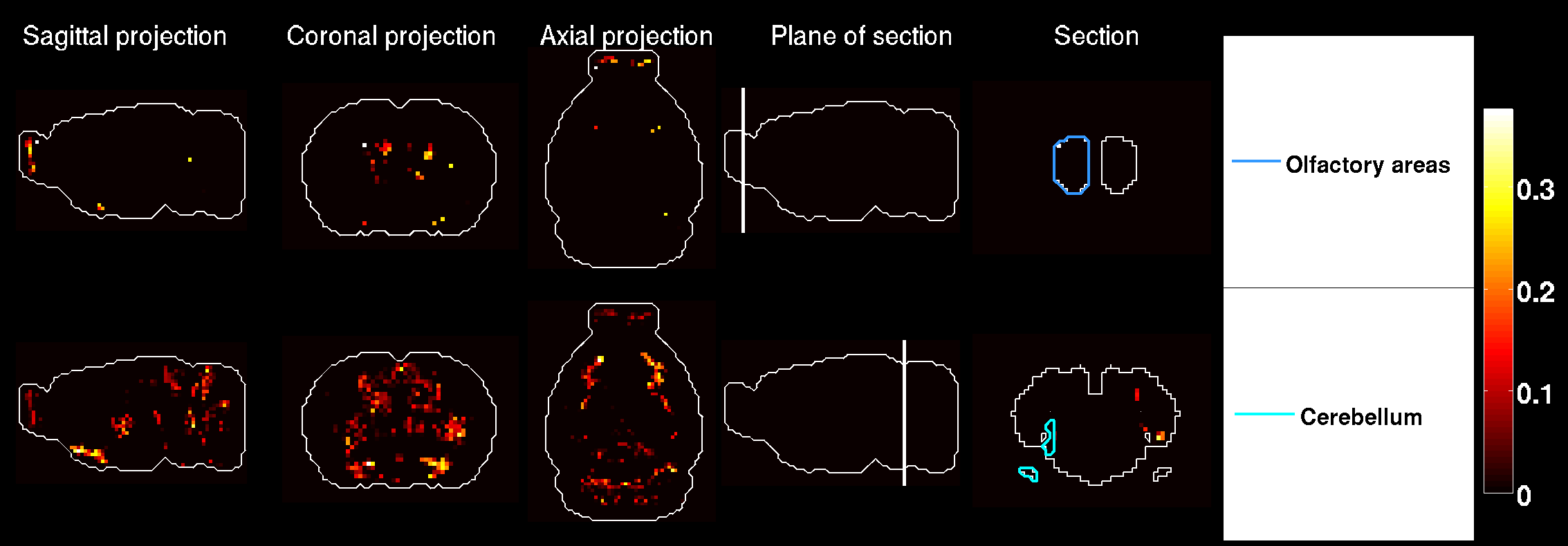}\\\hline
59&\tiny{GABAergic Interneurons, PV+}&\includegraphics[width=0.85\textwidth,keepaspectratio]{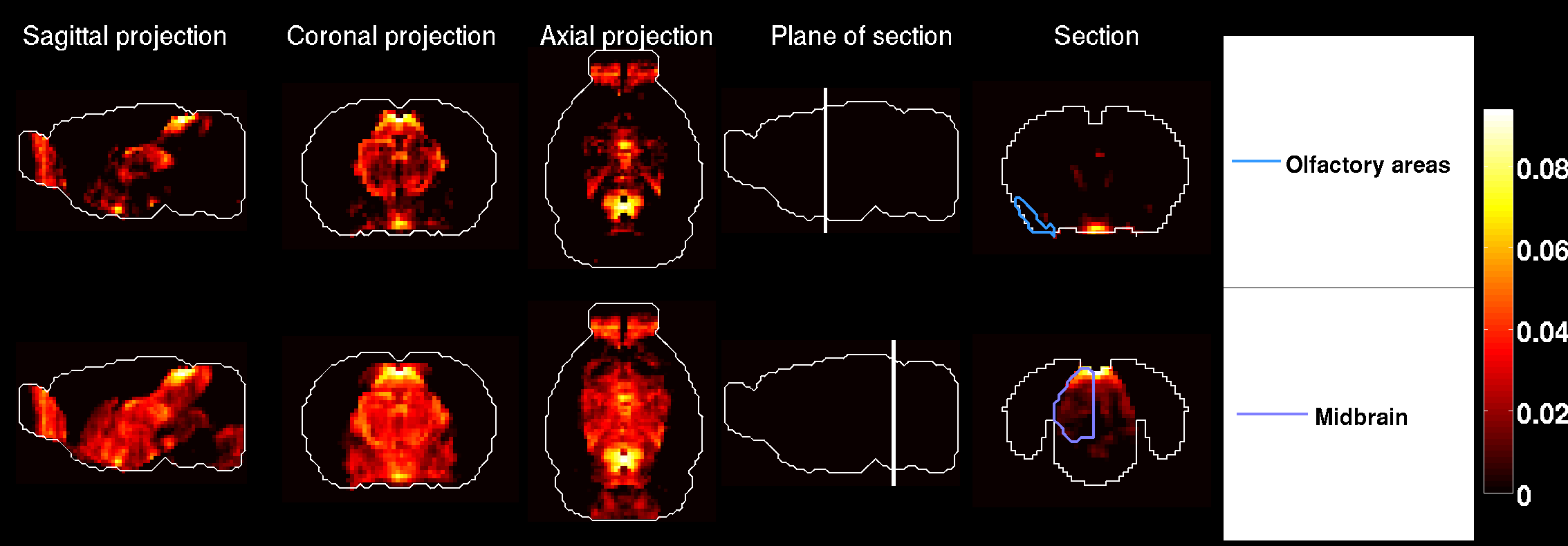}\\\hline
60&\tiny{GABAergic Interneurons, PV+, P7}&\includegraphics[width=0.85\textwidth,keepaspectratio]{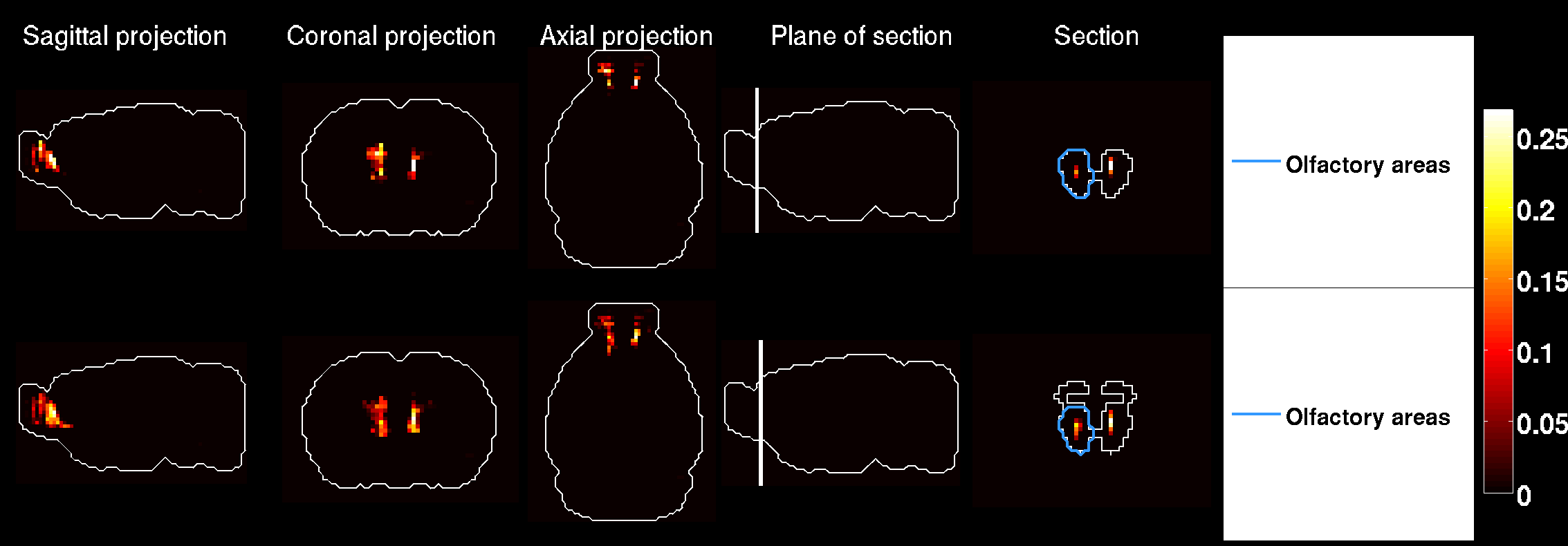}\\\hline
\end{tabular}
 
\caption{Brain-wide density profiles of \numTypesPerReTable  cell types, in the 
 original linear model (first row of each figure), and in the model fitted to microarray data
 incorporating the maximum uniform correction compatible with positive entries (second row of each figure).}
\label{tableReFittings20}
\end{table}

\begin{table}
\begin{tabular}{|m{0.06\textwidth}|m{0.06\textwidth}|m{\widthParamForTable\textwidth}|}
\hline
\textbf{Index}&\textbf{Cell type}&\textbf{Heat maps of densities, original and refitted}\\\hline
61&\tiny{GABAergic Interneurons, PV+, P10}&\includegraphics[width=0.85\textwidth,keepaspectratio]{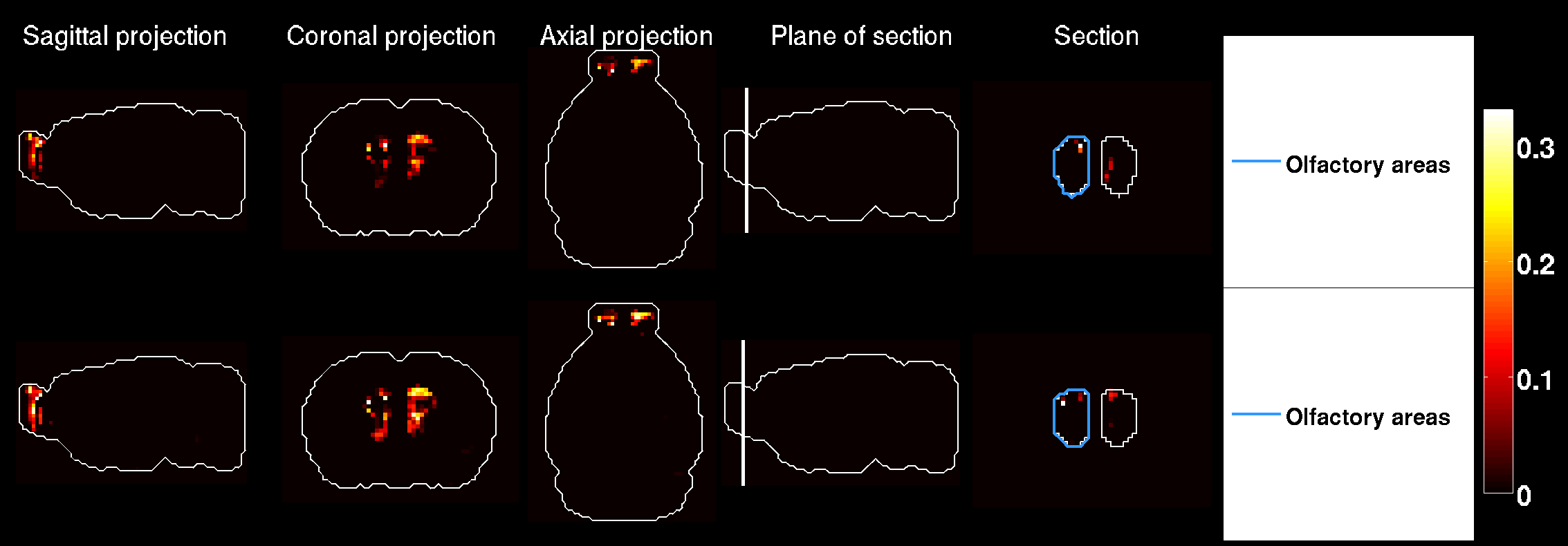}\\\hline
62&\tiny{GABAergic Interneurons, PV+, P13-P15}&\includegraphics[width=0.85\textwidth,keepaspectratio]{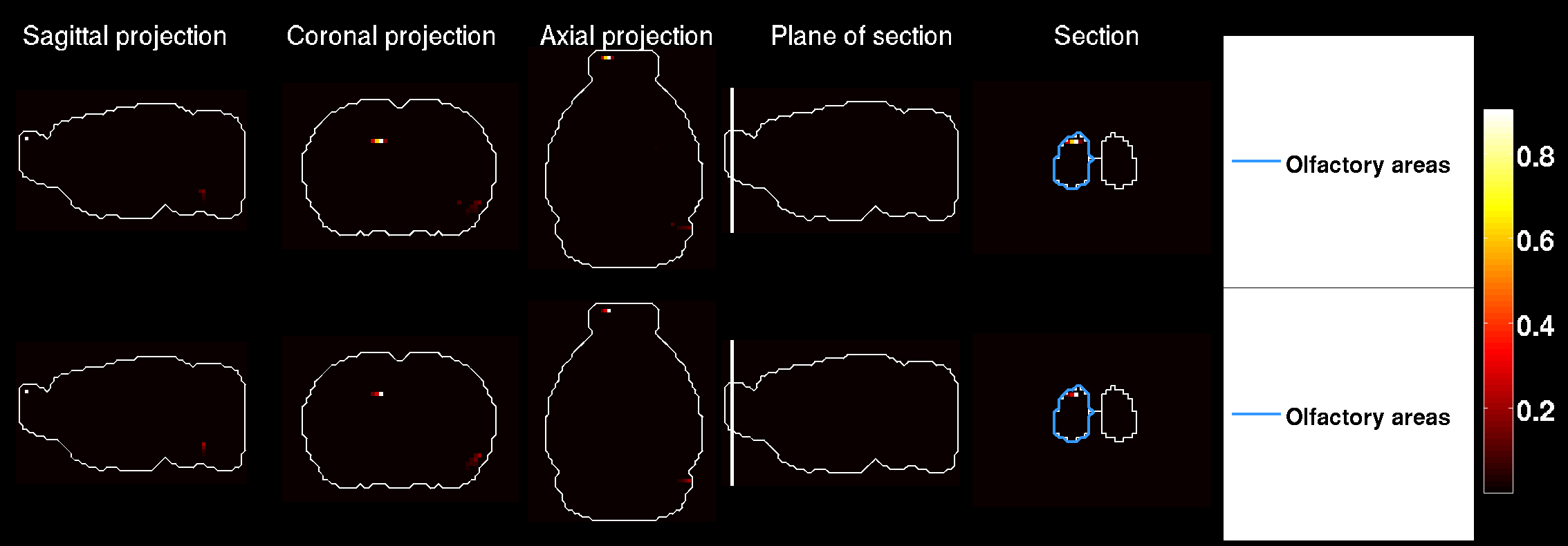}\\\hline
63&\tiny{GABAergic Interneurons, PV+, P25}&\includegraphics[width=0.85\textwidth,keepaspectratio]{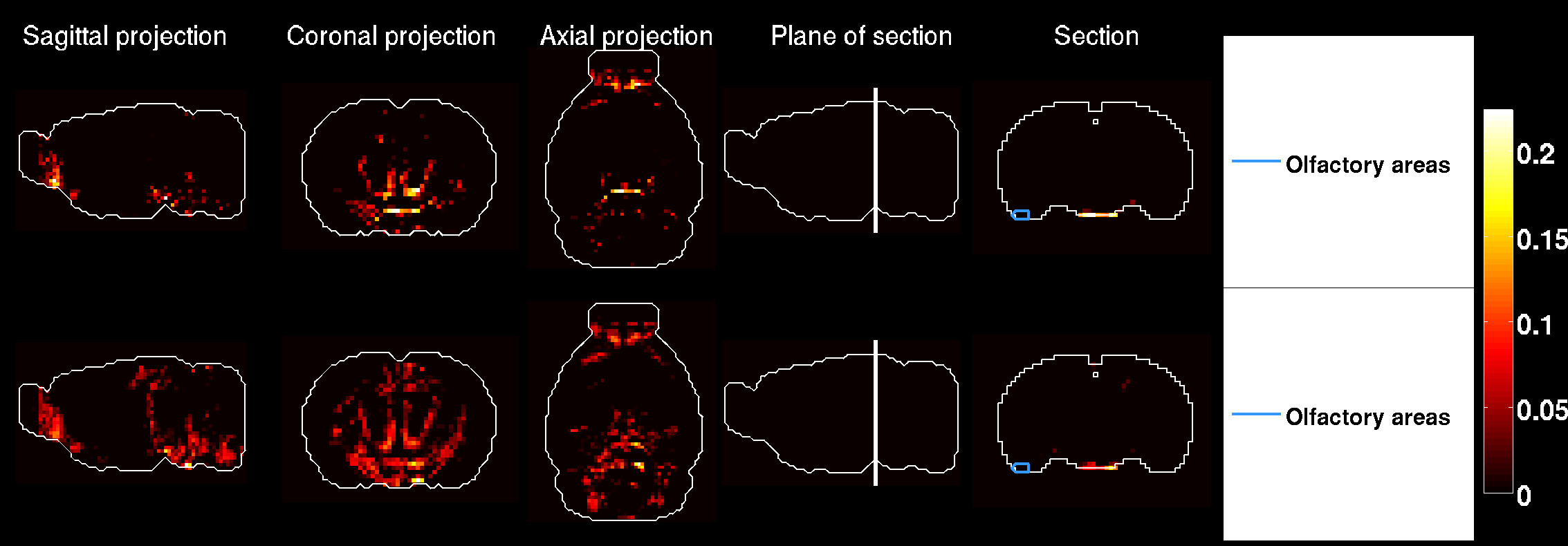}\\\hline
\end{tabular}
 
\caption{Brain-wide density profiles of \numTypesPerReTable  cell types, in the 
 original linear model (first row of each figure), and in the model fitted to microarray data
 incorporating the maximum uniform correction compatible with positive entries (second row of each figure).}
\label{tableReFittings21}
\end{table}

\begin{table}
\begin{tabular}{|m{0.06\textwidth}|m{0.06\textwidth}|m{\widthParamForTable\textwidth}|}
\hline
\textbf{Index}&\textbf{Cell type}&\textbf{Heat maps of densities, original and refitted}\\\hline
64&\tiny{GABAergic Interneurons, PV+}&\includegraphics[width=0.85\textwidth,keepaspectratio]{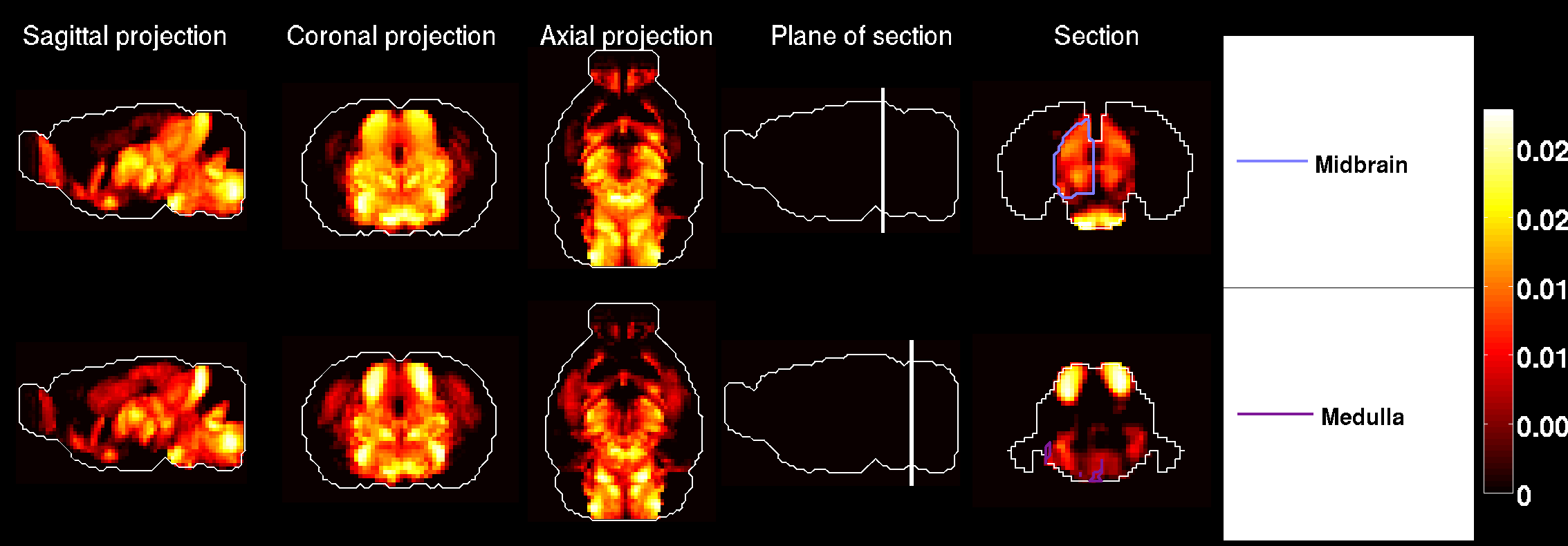}\\\hline
\end{tabular}
 
\caption{Brain-wide density profiles of \numTypesPerReTable  cell types, in the 
 original linear model (first row of each figure), and in the model fitted to microarray data
 incorporating the maximum uniform correction compatible with positive entries (second row of each figure).}
\label{tableReFittings22}
\end{table}

\clearpage
\section{Conclusions}

Some of the neuroanatomical patterns discovered 
 both in the correlation and density volumes confirm prior
 biological knowledge in the sense that the top region 
 by correlation or density coincides with the region 
 from which the corresponding cell-type-specific sample was extracted. 
 This is the case in the cerebral cortex, the hippocampal 
 region, the striatum, the ventral midbrain, the medulla 
 and the cerebellum.\\

 The anatomical analysis conducted here is supervised 
 in the sense that the brain regions are always taken to be
 one of the sets of voxels carrying the same label
 under some version of the voxelized Allen Reference Atlas.
 In the case of the two samples extracted from the amygdala (indices
 48 and 53), our method cannot return the label 'amygdala' as 
 a top region by correlation and density, since the amygdala
 is split between several regions. 
 This kind of issue is bound to occur for any choice of atlas,
 since there is no universal agreement on the nomenclature 
 of brain regions \cite{concordance}. Clustering methods could 
 be applied to the correlation and density profiles,
 following the approach taken in  \cite{methodsPaper} for the
 analysis of the voxel-by-gene matrix of expression energies.\\

 The rankings of brain regions yield some surprising results that can be linked to
 the relative paucity of cell types in the study, compared to the
 whole diversity of cells in the mouse brain, and to the fact that the
 cell samples are not distributed uniformly across the brain.
 It can be noted that there is more solidarity between the GABAergic interneurons 
 (indices 54--64), than between their top regions (by correlation or density),
 and their anatomical origin.\\

 A richer microarray data set with a more uniform sampling of brain regions will
 modify the numerical results. In particular, it will be interesting to see
 if positive densities in the hypothalamus and in the olfactory areas
 can be estimated for samples extracted from these regions.\\

\section{Acknowledgments}
 We thank Benjamin Okaty, Sacha Nelson and Ken Sugino for collating
 and transferring the microarray data, and for help with the 
 anatomical analysis of results. We thank Jason W. Bohland, Hemant Bokil, Lydia Ng
 and Domenico Orlando for discussions.
This research is supported by the NIH-NIDA Grant 1R21DA027644-01,
{\emph{Computational analysis of co-expression networks in the mouse
    and human brain}}.

\clearpage

\clearpage
\section{Tables of brain-wide correlations}
\begin{table}
\begin{tabular}{|m{0.06\textwidth}|m{0.13\textwidth}|m{\widthParamTable\textwidth}|}
\hline
\textbf{Index}&\textbf{Description}&\textbf{Brain-wide heat maps of correlations}\\\hline
1&\small{Purkinje Cells}&\includegraphics[width=\widthParamTable\textwidth]{correlsFig1.png}\\
2&\small{Pyramidal Neurons}&\includegraphics[width=\widthParamTable\textwidth]{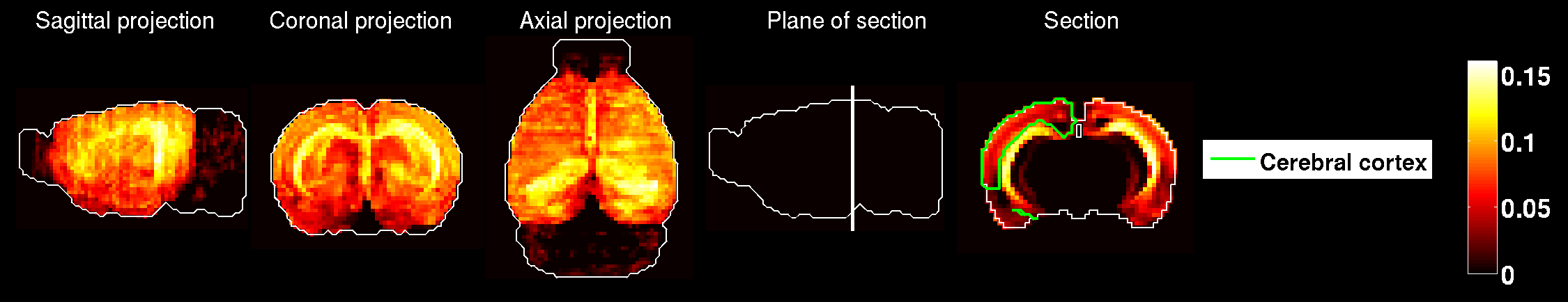}\\
3&\small{Pyramidal Neurons}&\includegraphics[width=\widthParamTable\textwidth]{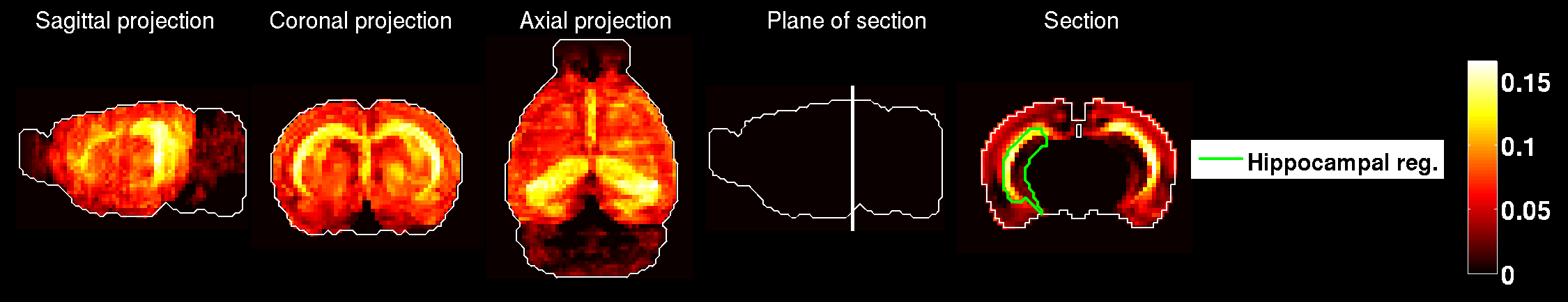}\\
4&\small{A9 Dopaminergic Neurons}&\includegraphics[width=\widthParamTable\textwidth]{correlsFig4.png}\\
5&\small{A10 Dopaminergic Neurons}&\includegraphics[width=\widthParamTable\textwidth]{correlsFig5.png}\\
6&\small{Pyramidal Neurons}&\includegraphics[width=\widthParamTable\textwidth]{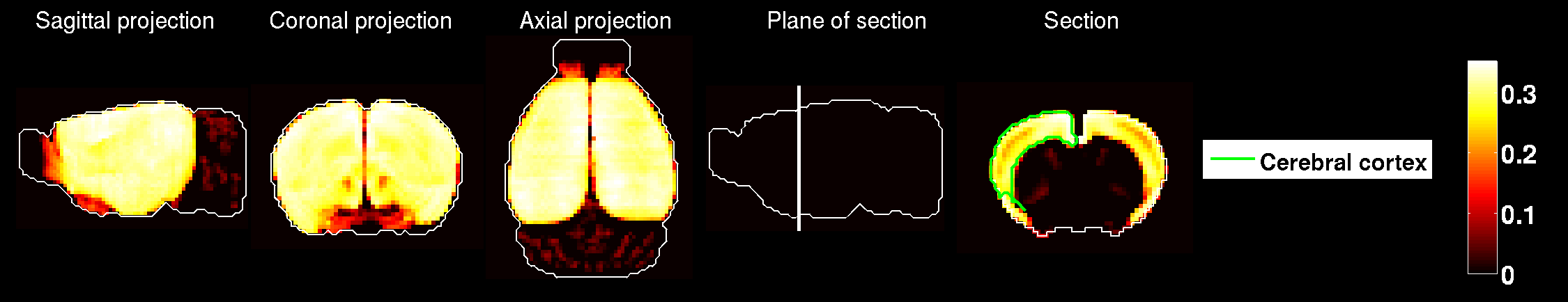}\\\hline
\end{tabular}
\caption{Brain-wide correlation profiles between \numTypesPerTable  cell types and the Allen Atlas.}
\label{tableCorrels1}
\end{table}
%\newpage

\begin{table}
\begin{tabular}{|m{0.06\textwidth}|m{0.13\textwidth}|m{\widthParamTable\textwidth}|}
\hline
\textbf{Index}&\textbf{Description}&\textbf{Brain-wide heat maps of correlations}\\\hline
7&\small{Pyramidal Neurons}&\includegraphics[width=\widthParamTable\textwidth]{correlsFig7.png}\\
8&\small{Pyramidal Neurons}&\includegraphics[width=\widthParamTable\textwidth]{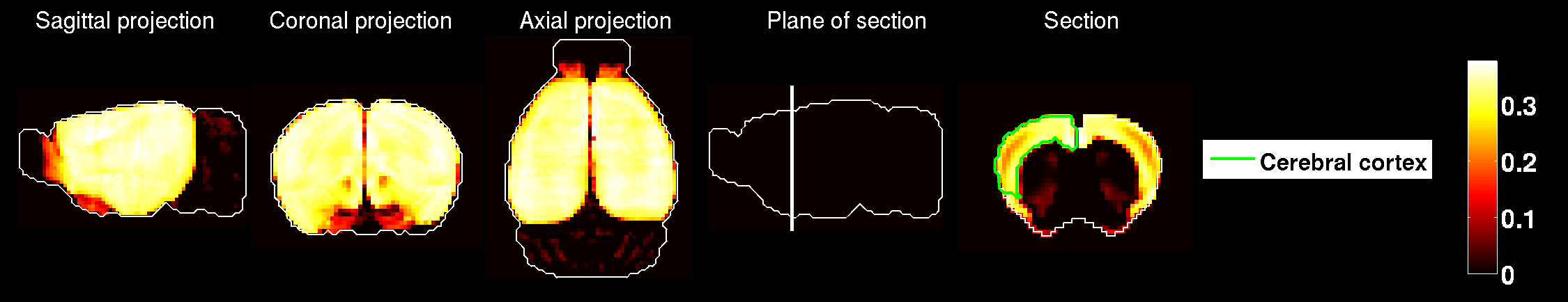}\\
9&\small{Mixed Neurons}&\includegraphics[width=\widthParamTable\textwidth]{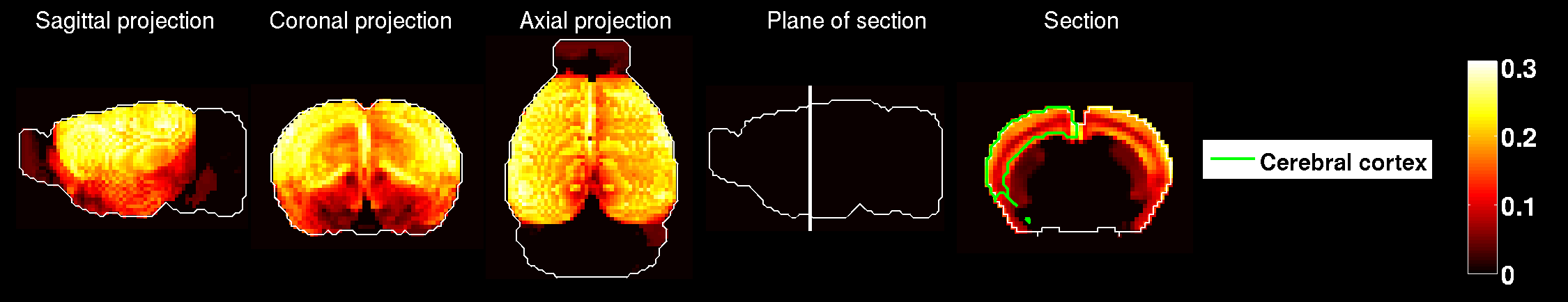}\\
10&\small{Motor Neurons, Midbrain Cholinergic Neurons}&\includegraphics[width=\widthParamTable\textwidth]{correlsFig10.png}\\
11&\small{Cholinergic Projection Neurons}&\includegraphics[width=\widthParamTable\textwidth]{correlsFig11.png}\\
12&\small{Motor Neurons, Cholinergic Interneurons}&\includegraphics[width=\widthParamTable\textwidth]{correlsFig12.png}\\\hline
\end{tabular}
\caption{\label{tableCorrels2}Brain-wide correlation profiles between \numTypesPerTable cell types and the Allen Atlas.}
\end{table}
%\newpage

\begin{table}
\begin{tabular}{|m{0.06\textwidth}|m{0.13\textwidth}|m{\widthParamTable\textwidth}|}
\hline
\textbf{Index}&\textbf{Description}&\textbf{Brain-wide heat maps of correlations}\\\hline
13&\small{Cholinergic Neurons}&\includegraphics[width=\widthParamTable\textwidth]{correlsFig13.png}\\
14&\small{Interneurons}&\includegraphics[width=\widthParamTable\textwidth]{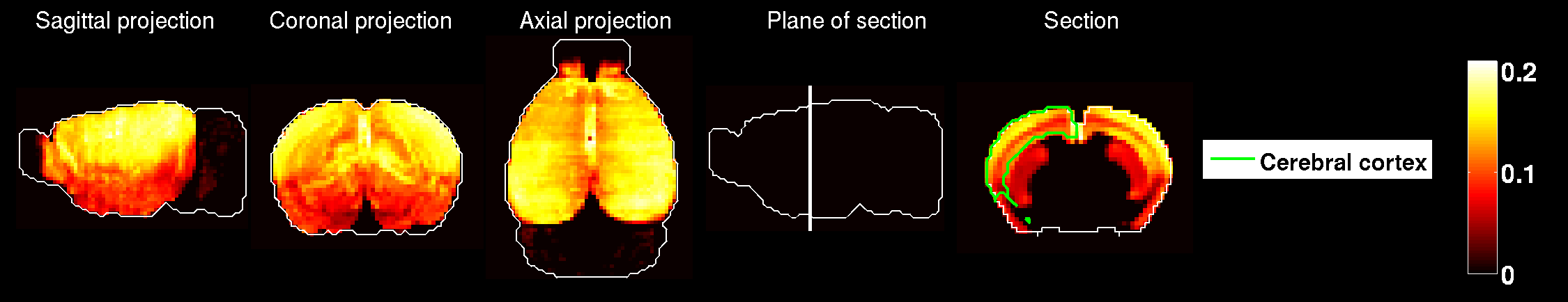}\\
15&\small{Drd1+ Medium Spiny Neurons}&\includegraphics[width=\widthParamTable\textwidth]{correlsFig15.png}\\
16&\small{Drd2+ Medium Spiny Neurons}&\includegraphics[width=\widthParamTable\textwidth]{correlsFig16.png}\\
17&\small{Golgi Cells}&\includegraphics[width=\widthParamTable\textwidth]{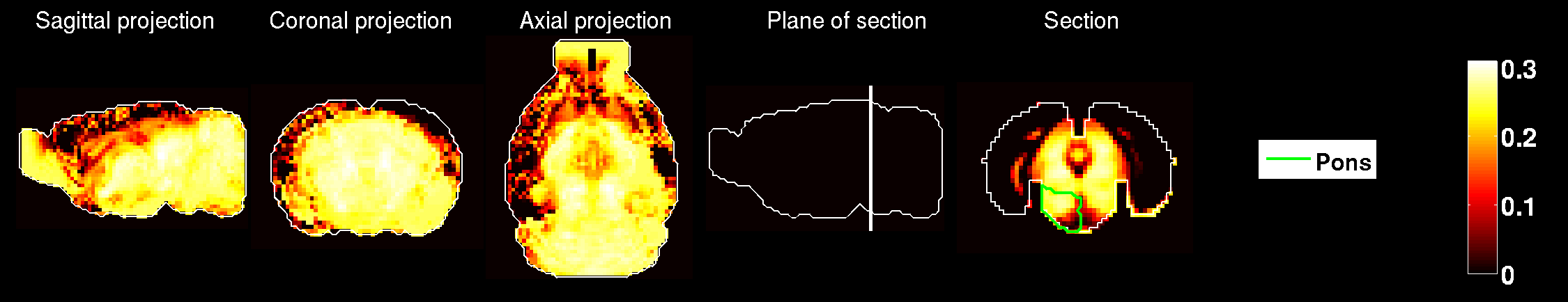}\\
18&\small{Unipolar Brush cells (some Bergman Glia)}&\includegraphics[width=\widthParamTable\textwidth]{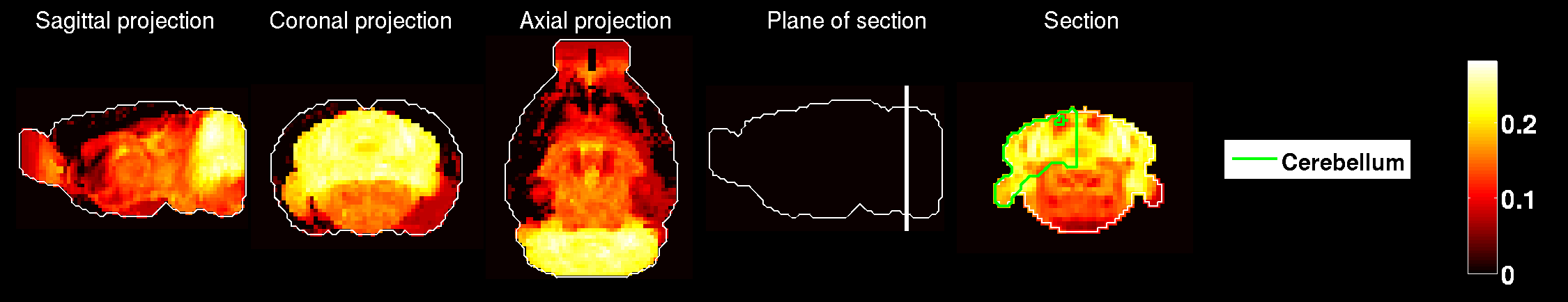}\\\hline
\end{tabular}
\caption{\label{tableCorrels3}Brain-wide correlation profiles between \numTypesPerTable cell types and the Allen Atlas.}
\end{table}
%\newpage

\begin{table}
\begin{tabular}{|m{0.06\textwidth}|m{0.13\textwidth}|m{\widthParamTable\textwidth}|}
\hline
\textbf{Index}&\textbf{Description}&\textbf{Brain-wide heat maps of correlations}\\\hline
19&\small{Stellate Basket Cells}&\includegraphics[width=\widthParamTable\textwidth]{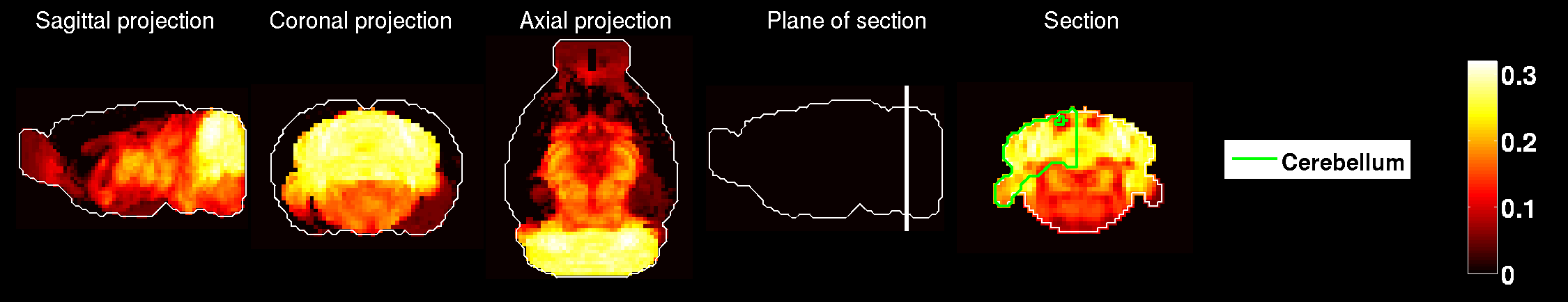}\\
20&\small{Granule Cells}&\includegraphics[width=\widthParamTable\textwidth]{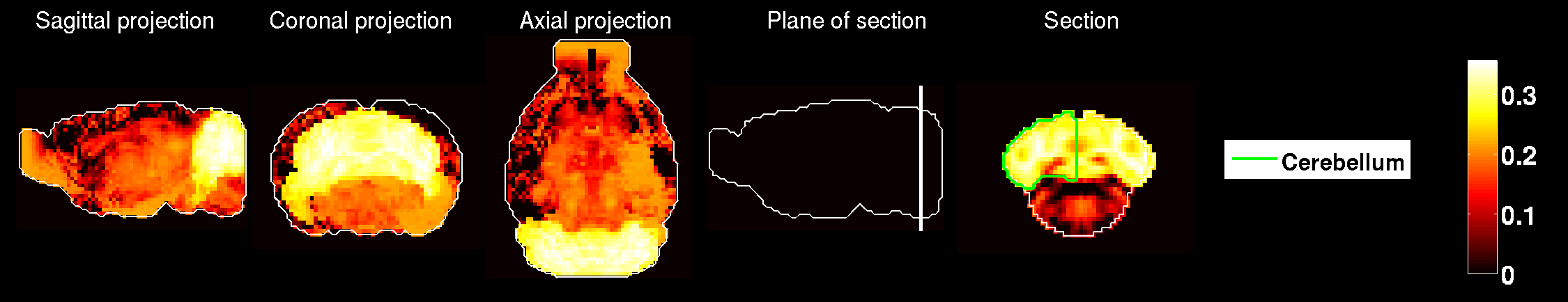}\\
21&\small{Mature Oligodendrocytes}&\includegraphics[width=\widthParamTable\textwidth]{correlsFig21.png}\\
22&\small{Mature Oligodendrocytes}&\includegraphics[width=\widthParamTable\textwidth]{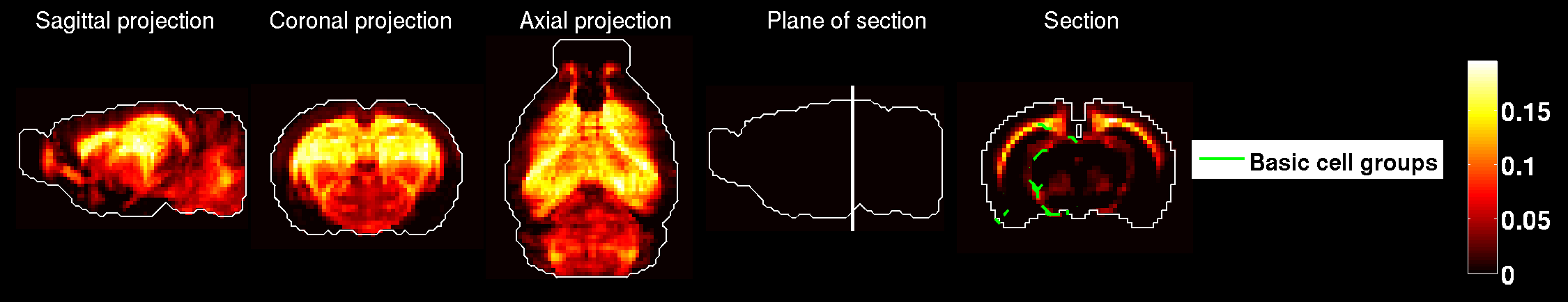}\\
23&\small{Mixed Oligodendrocytes}&\includegraphics[width=\widthParamTable\textwidth]{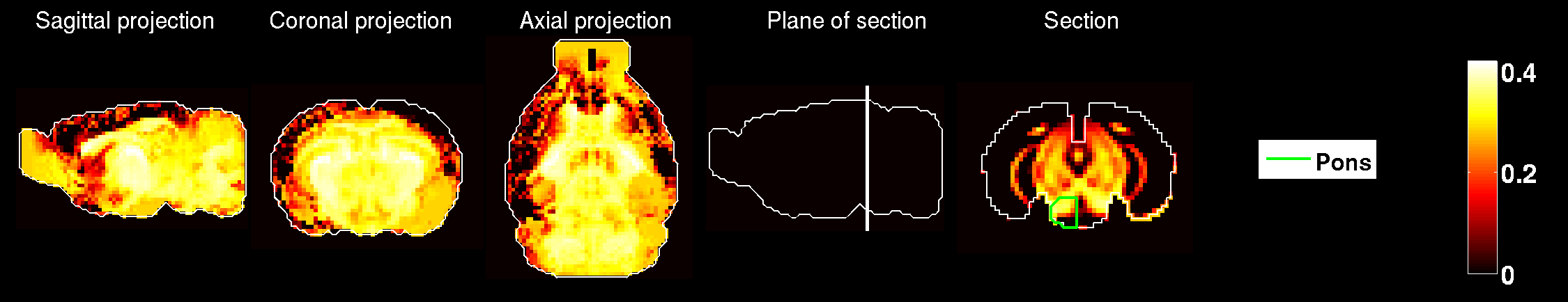}\\
24&\small{Mixed Oligodendrocytes}&\includegraphics[width=\widthParamTable\textwidth]{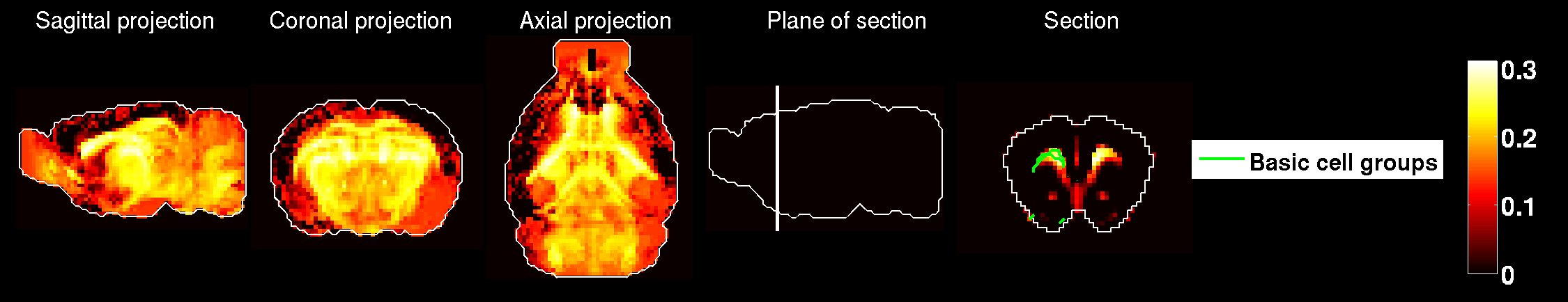}\\\hline
\end{tabular}
\caption{\label{tableCorrels4}Brain-wide correlation profiles between \numTypesPerTable cell types and the Allen Atlas.}
\end{table}
%\newpage

\begin{table}
\begin{tabular}{|m{0.06\textwidth}|m{0.13\textwidth}|m{\widthParamTable\textwidth}|}
\hline
\textbf{Index}&\textbf{Description}&\textbf{Brain-wide heat maps of correlations}\\\hline
25&\small{Purkinje Cells}&\includegraphics[width=\widthParamTable\textwidth]{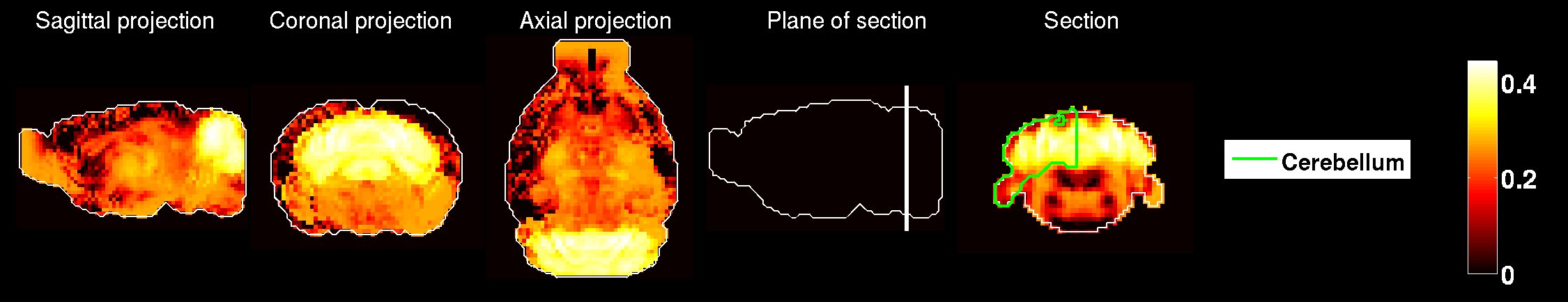}\\
26&\small{Neurons}&\includegraphics[width=\widthParamTable\textwidth]{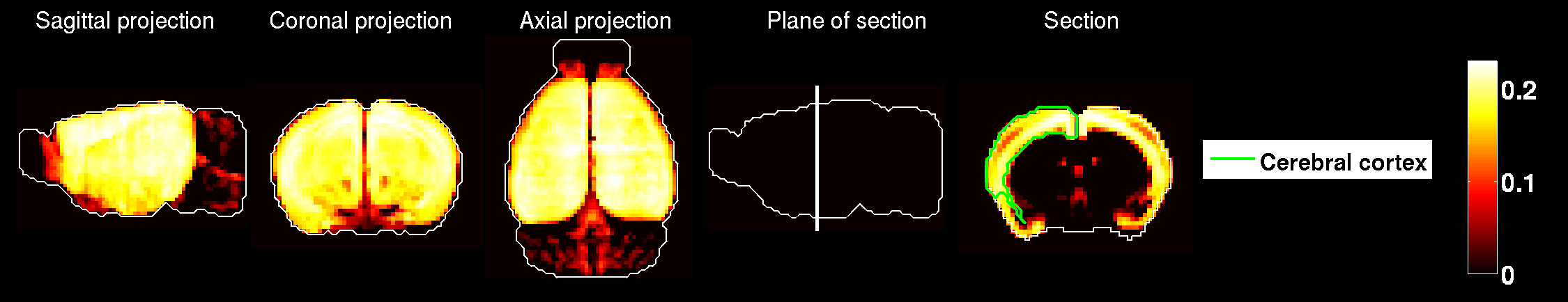}\\
27&\small{Bergman Glia}&\includegraphics[width=\widthParamTable\textwidth]{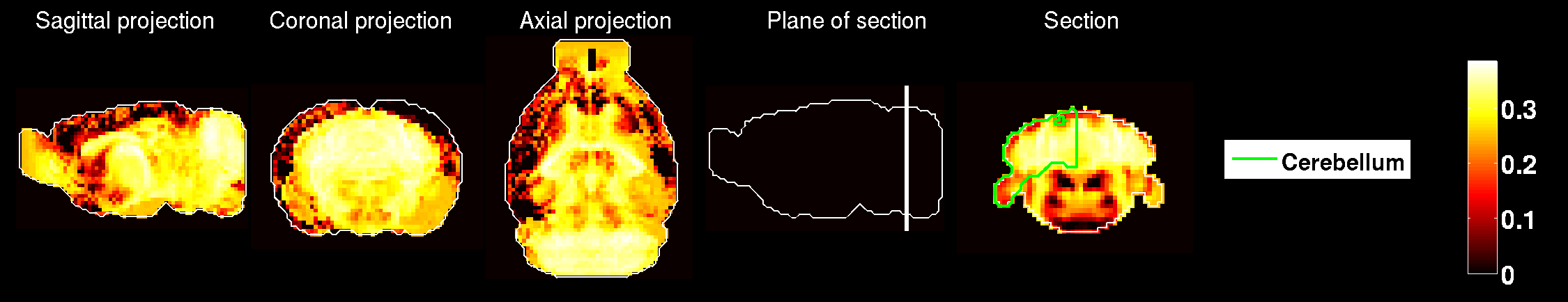}\\
28&\small{Astroglia}&\includegraphics[width=\widthParamTable\textwidth]{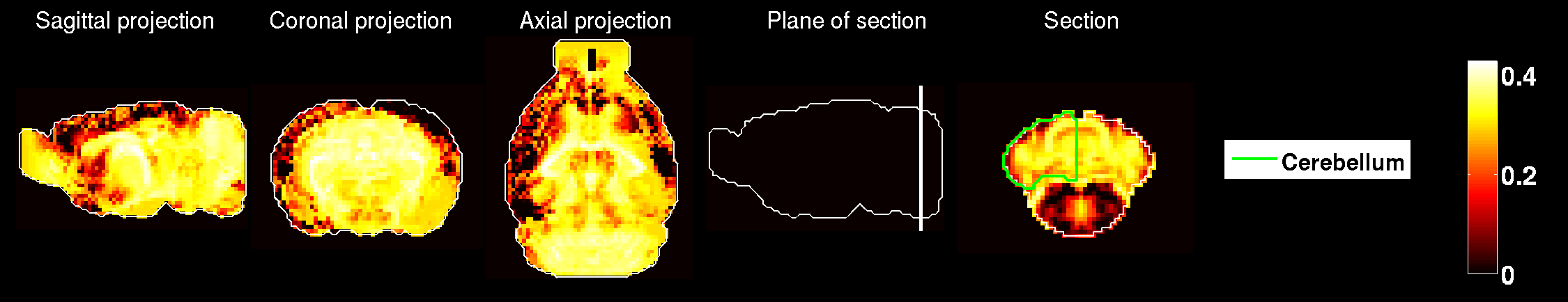}\\
29&\small{Astroglia}&\includegraphics[width=\widthParamTable\textwidth]{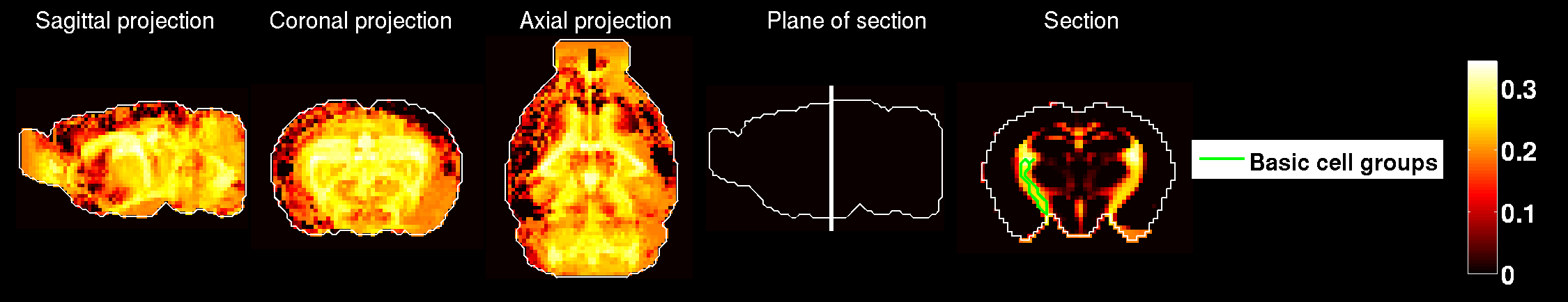}\\
30&\small{Astrocytes}&\includegraphics[width=\widthParamTable\textwidth]{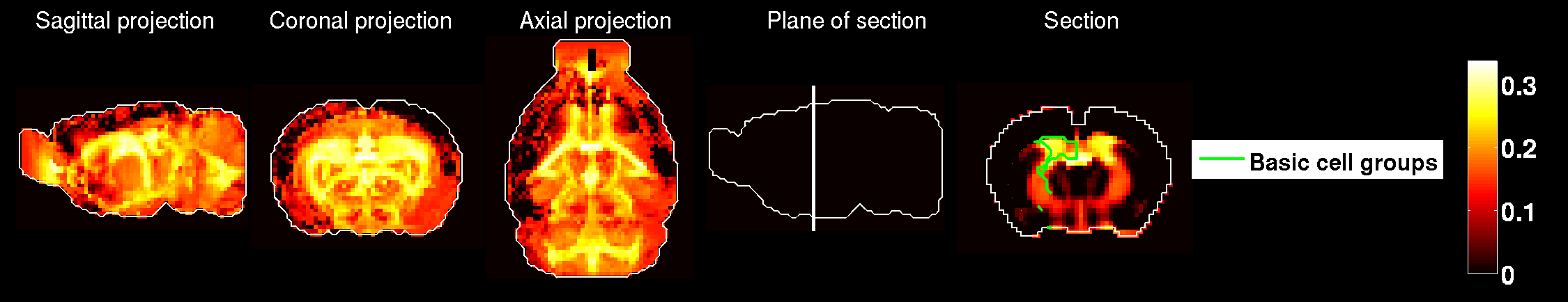}\\\hline
\end{tabular}
\caption{\label{tableCorrels5}Brain-wide correlation profiles between \numTypesPerTable cell types and the Allen Atlas.}
\end{table}
%\newpage

\begin{table}
\begin{tabular}{|m{0.06\textwidth}|m{0.13\textwidth}|m{\widthParamTable\textwidth}|}
\hline
\textbf{Index}&\textbf{Description}&\textbf{Brain-wide heat maps of correlations}\\\hline
31&\small{Astrocytes}&\includegraphics[width=\widthParamTable\textwidth]{correlsFig31.png}\\
32&\small{Astrocytes}&\includegraphics[width=\widthParamTable\textwidth]{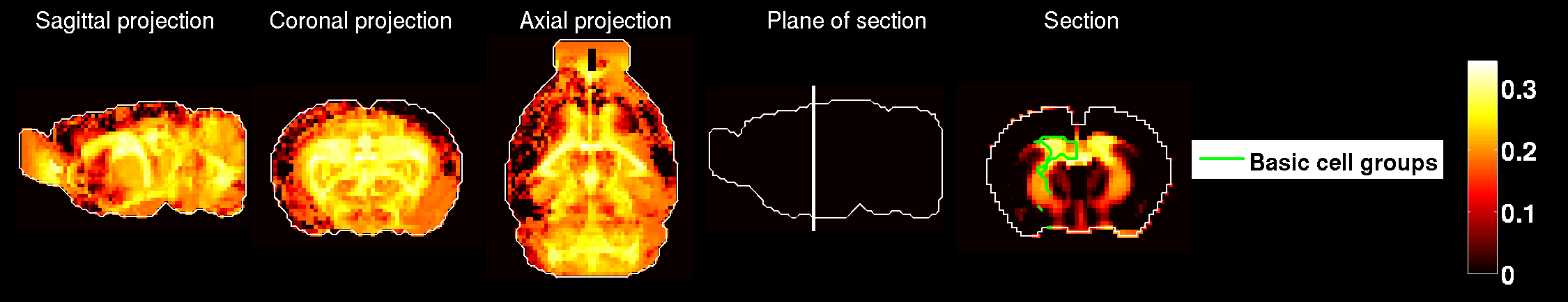}\\
33&\small{Mixed Neurons}&\includegraphics[width=\widthParamTable\textwidth]{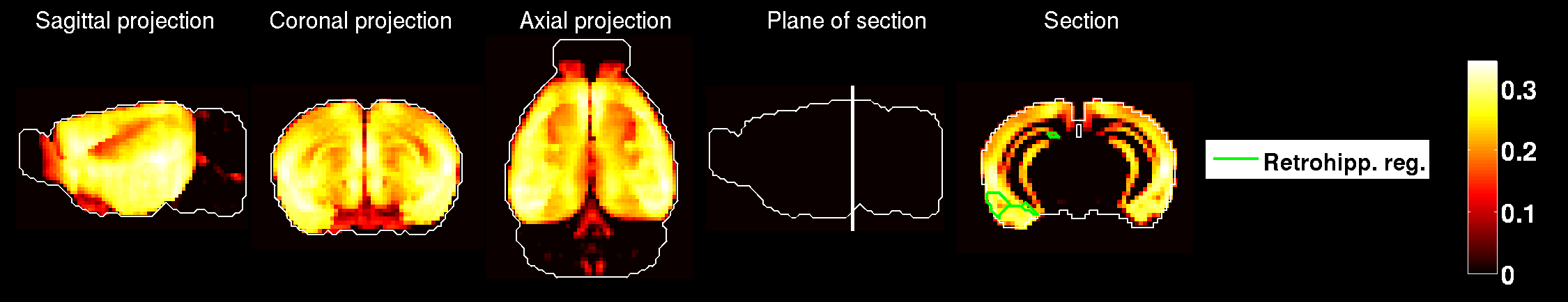}\\
34&\small{Mixed Neurons}&\includegraphics[width=\widthParamTable\textwidth]{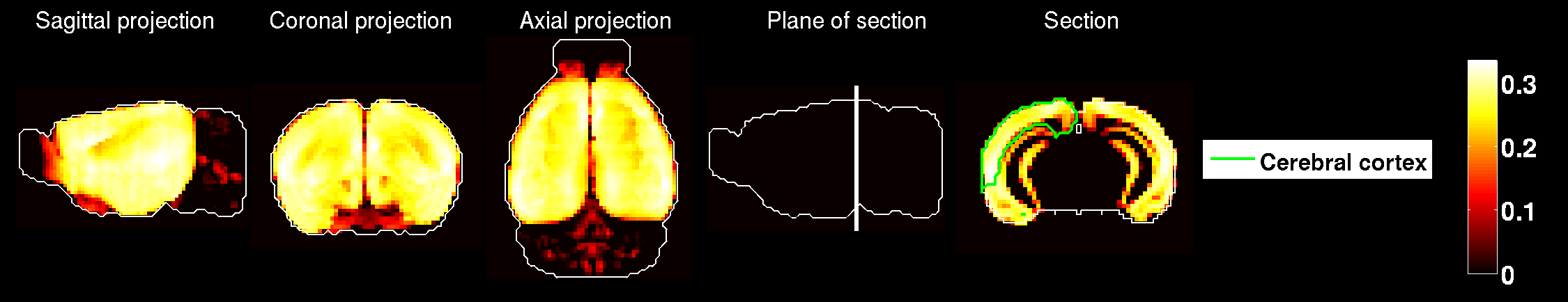}\\
35&\small{Mature Oligodendrocytes}&\includegraphics[width=\widthParamTable\textwidth]{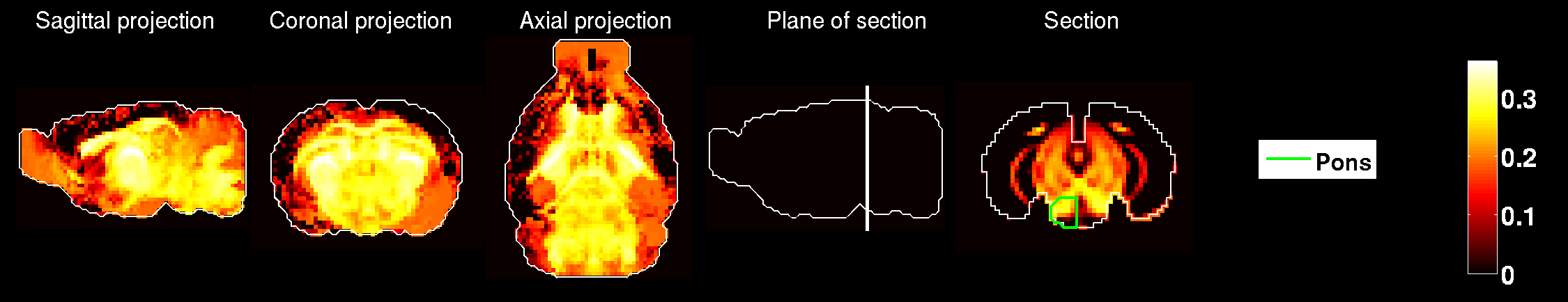}\\
36&\small{Oligodendrocytes}&\includegraphics[width=\widthParamTable\textwidth]{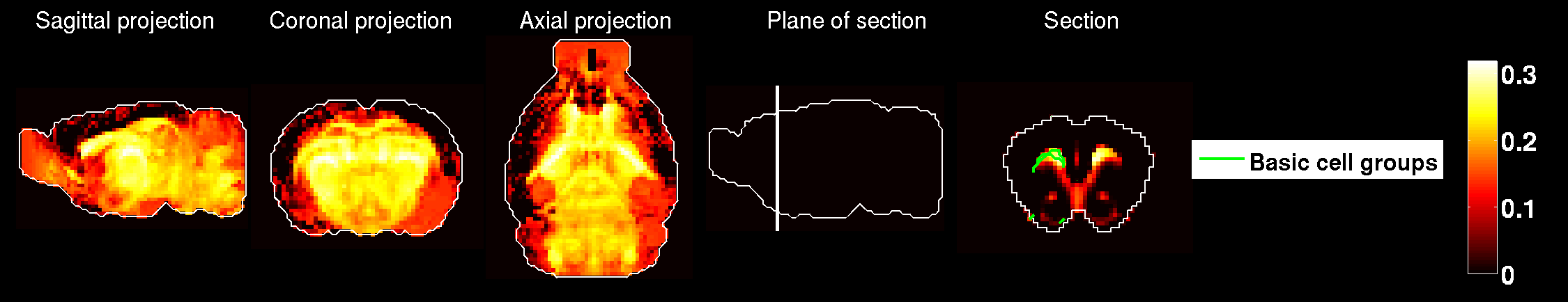}\\\hline
\end{tabular}
\caption{\label{tableCorrels6}Brain-wide correlation profiles between \numTypesPerTable cell types and the Allen Atlas.}
\end{table}
%\newpage

\begin{table}
\begin{tabular}{|m{0.06\textwidth}|m{0.13\textwidth}|m{\widthParamTable\textwidth}|}
\hline
\textbf{Index}&\textbf{Description}&\textbf{Brain-wide heat maps of correlations}\\\hline
37&\small{Oligodendrocyte Precursors}&\includegraphics[width=\widthParamTable\textwidth]{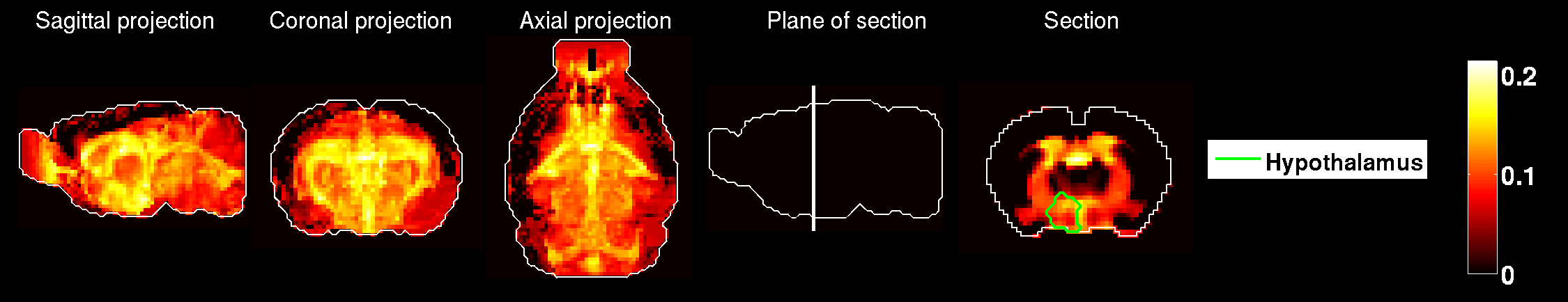}\\
38&\small{Pyramidal Neurons, Callosally projecting, P3}&\includegraphics[width=\widthParamTable\textwidth]{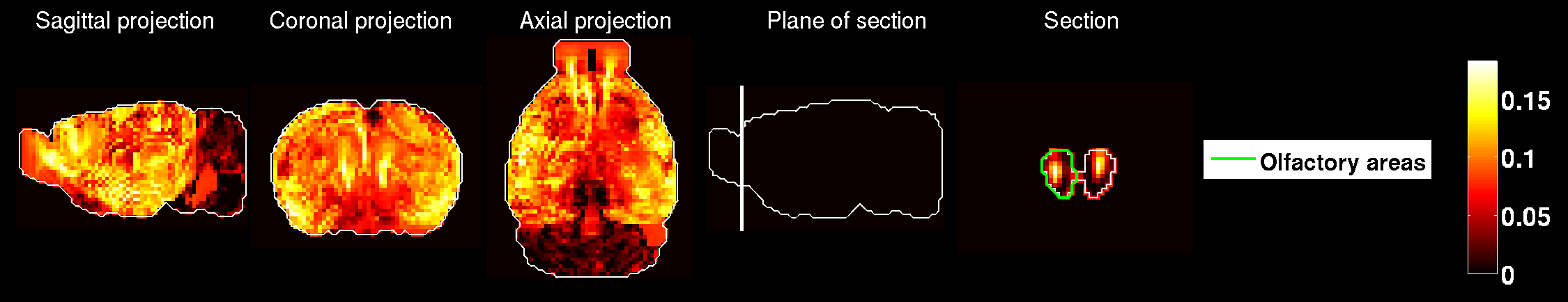}\\
39&\small{Pyramidal Neurons, Callosally projecting, P6}&\includegraphics[width=\widthParamTable\textwidth]{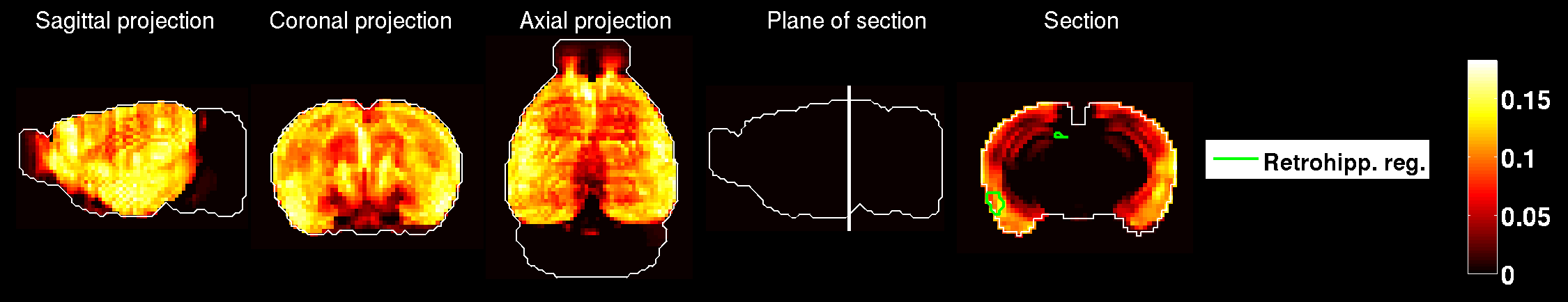}\\
40&\small{Pyramidal Neurons, Callosally projecting, P14}&\includegraphics[width=\widthParamTable\textwidth]{correlsFig40.png}\\
41&\small{Pyramidal Neurons, Corticospinal, P3}&\includegraphics[width=\widthParamTable\textwidth]{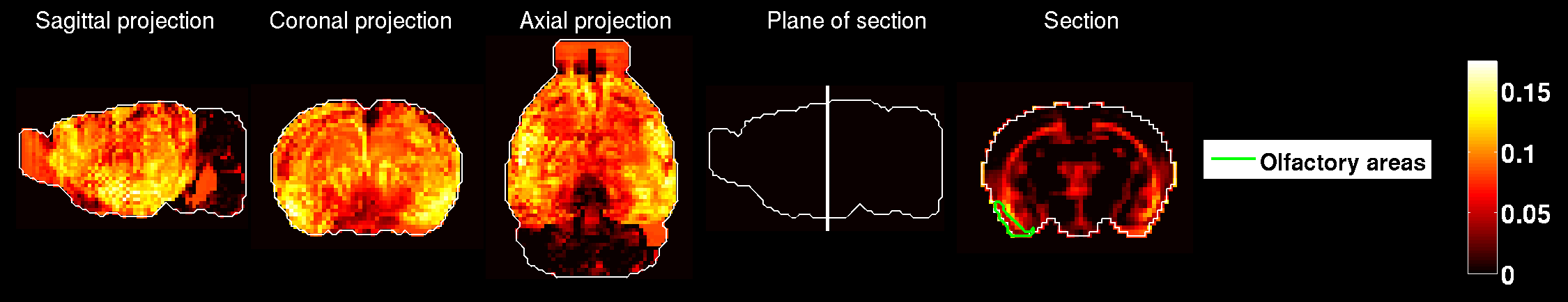}\\
42&\small{Pyramidal Neurons, Corticospinal, P6}&\includegraphics[width=\widthParamTable\textwidth]{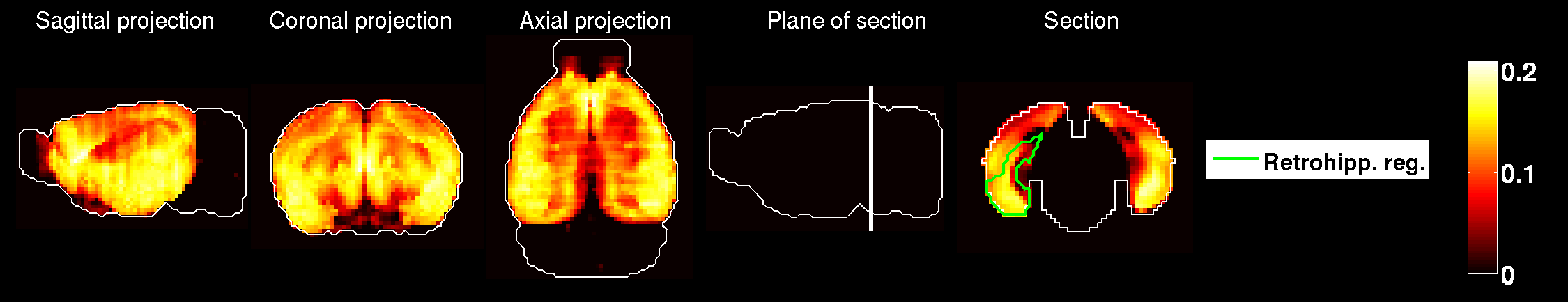}\\\hline
\end{tabular}
\caption{\label{tableCorrels7}Brain-wide correlation profiles between \numTypesPerTable cell types and the Allen Atlas.}
\end{table}
%\newpage

\begin{table}
\begin{tabular}{|m{0.06\textwidth}|m{0.13\textwidth}|m{\widthParamTable\textwidth}|}
\hline
\textbf{Index}&\textbf{Description}&\textbf{Brain-wide heat maps of correlations}\\\hline
43&\small{Pyramidal Neurons, Corticospinal, P14}&\includegraphics[width=\widthParamTable\textwidth]{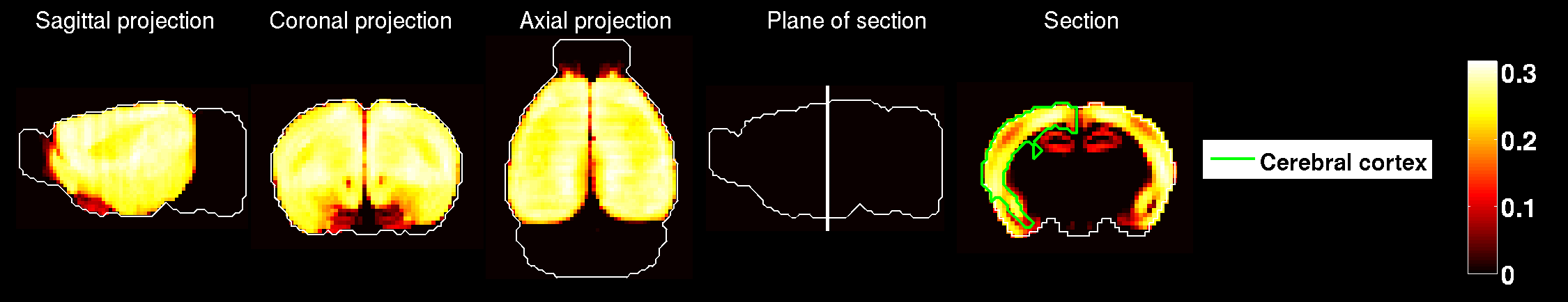}\\
44&\small{Pyramidal Neurons, Corticotectal, P14}&\includegraphics[width=\widthParamTable\textwidth]{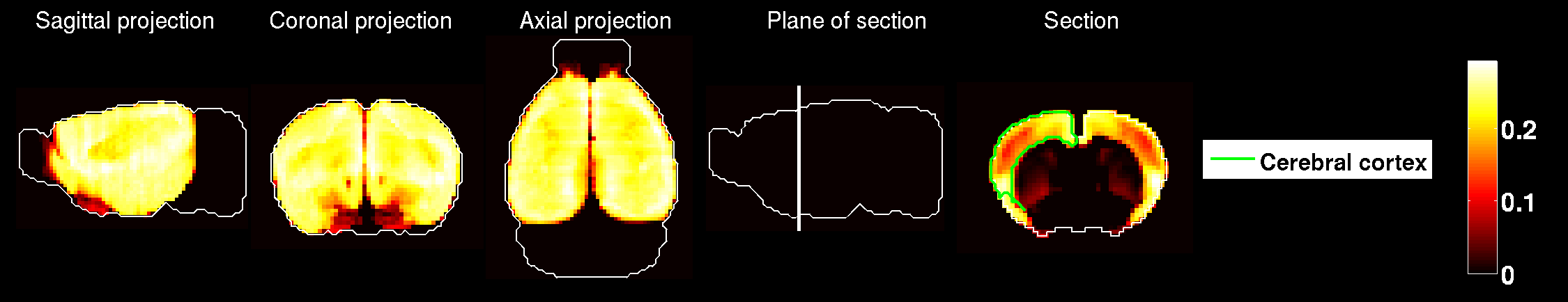}\\
45&\small{Pyramidal Neurons}&\includegraphics[width=\widthParamTable\textwidth]{correlsFig45.png}\\
46&\small{Pyramidal Neurons}&\includegraphics[width=\widthParamTable\textwidth]{correlsFig46.png}\\
47&\small{Pyramidal Neurons}&\includegraphics[width=\widthParamTable\textwidth]{correlsFig47.png}\\
48&\small{Pyramidal Neurons}&\includegraphics[width=\widthParamTable\textwidth]{correlsFig48.png}\\\hline
\end{tabular}
\caption{\label{tableCorrels8}Brain-wide correlation profiles between \numTypesPerTable cell types and the Allen Atlas.}
\end{table}
%\newpage

\begin{table}
\begin{tabular}{|m{0.06\textwidth}|m{0.13\textwidth}|m{\widthParamTable\textwidth}|}
\hline
\textbf{Index}&\textbf{Description}&\textbf{Brain-wide heat maps of correlations}\\\hline
49&\small{Pyramidal Neurons}&\includegraphics[width=\widthParamTable\textwidth]{correlsFig49.png}\\
50&\small{Pyramidal Neurons}&\includegraphics[width=\widthParamTable\textwidth]{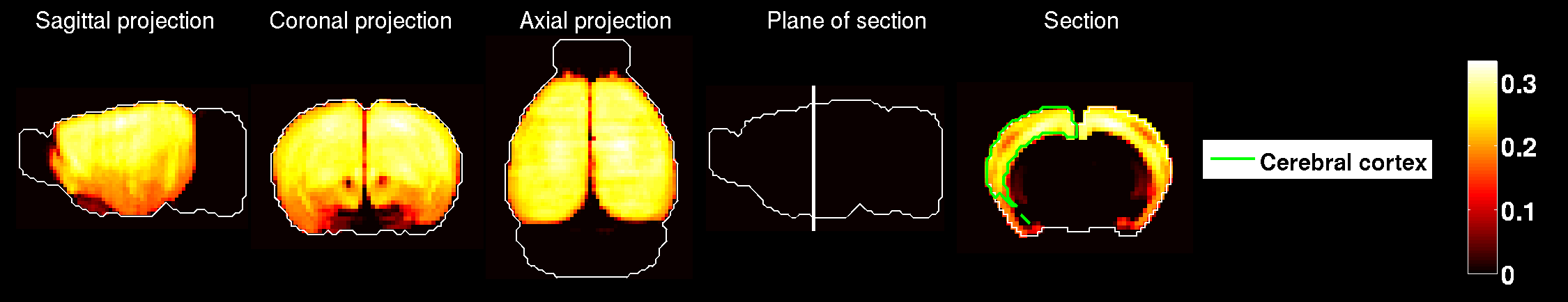}\\
51&\small{Tyrosine Hydroxylase Expressing}&\includegraphics[width=\widthParamTable\textwidth]{correlsFig51.png}\\
52&\small{Purkinje Cells}&\includegraphics[width=\widthParamTable\textwidth]{correlsFig52.png}\\
53&\small{Glutamatergic Neuron (not well defined)}&\includegraphics[width=\widthParamTable\textwidth]{correlsFig53.png}\\
54&\small{GABAergic Interneurons, VIP+}&\includegraphics[width=\widthParamTable\textwidth]{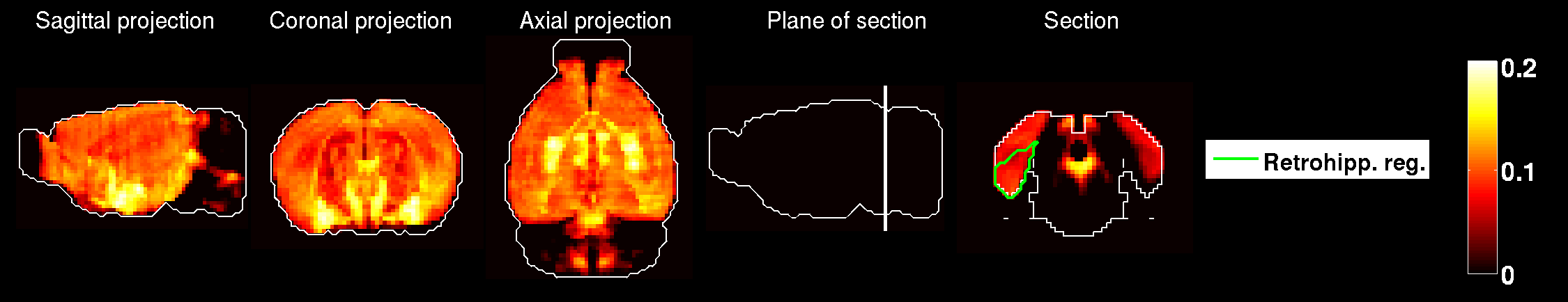}\\\hline
\end{tabular}
\caption{\label{tableCorrels9}Brain-wide correlation profiles between \numTypesPerTable cell types and the Allen Atlas.}
\end{table}
%\newpage

\begin{table}
\begin{tabular}{|m{0.06\textwidth}|m{0.13\textwidth}|m{\widthParamTable\textwidth}|}
\hline
\textbf{Index}&\textbf{Description}&\textbf{Brain-wide heat maps of correlations}\\\hline
55&\small{GABAergic Interneurons, VIP+}&\includegraphics[width=\widthParamTable\textwidth]{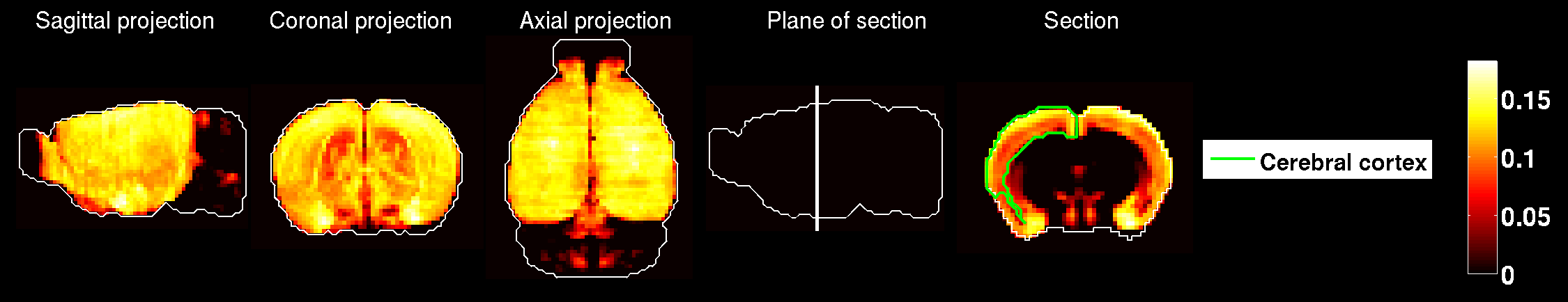}\\
56&\small{GABAergic Interneurons, SST+}&\includegraphics[width=\widthParamTable\textwidth]{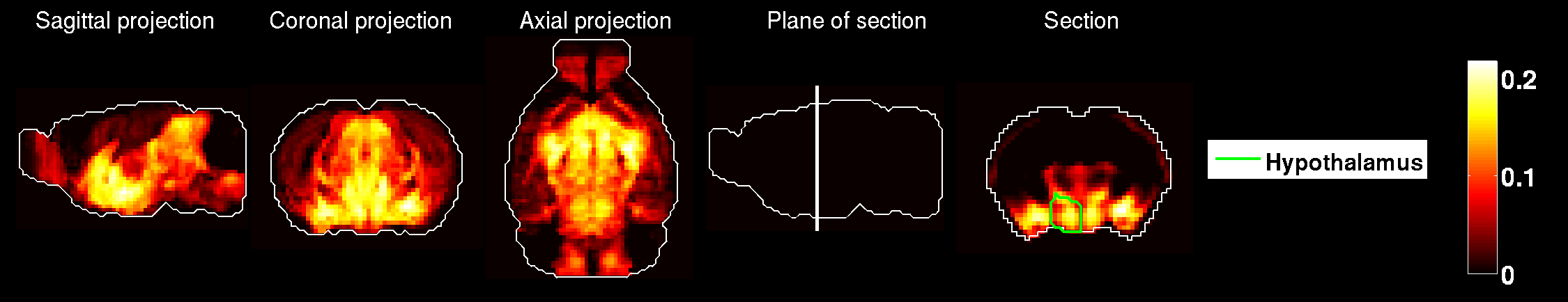}\\
57&\small{GABAergic Interneurons, SST+}&\includegraphics[width=\widthParamTable\textwidth]{correlsFig57.png}\\
58&\small{GABAergic Interneurons, PV+}&\includegraphics[width=\widthParamTable\textwidth]{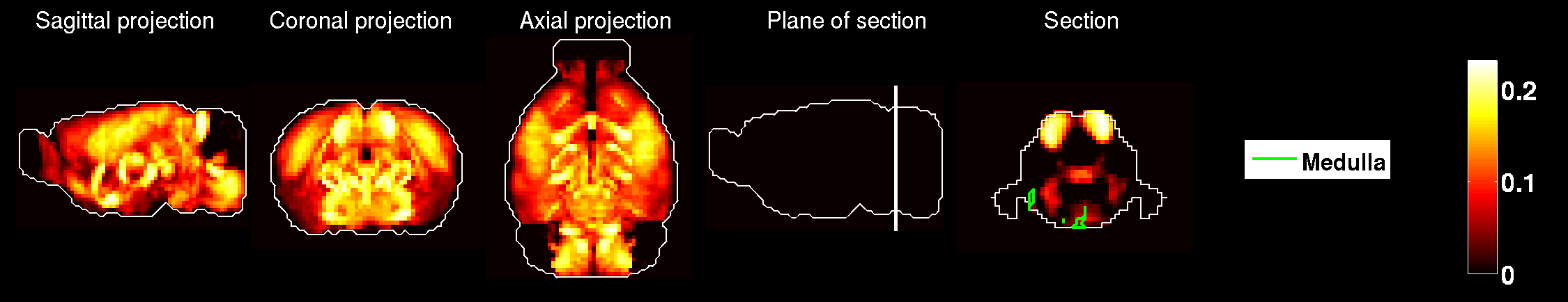}\\
59&\small{GABAergic Interneurons, PV+}&\includegraphics[width=\widthParamTable\textwidth]{correlsFig59.png}\\
60&\small{GABAergic Interneurons, PV+, P7}&\includegraphics[width=\widthParamTable\textwidth]{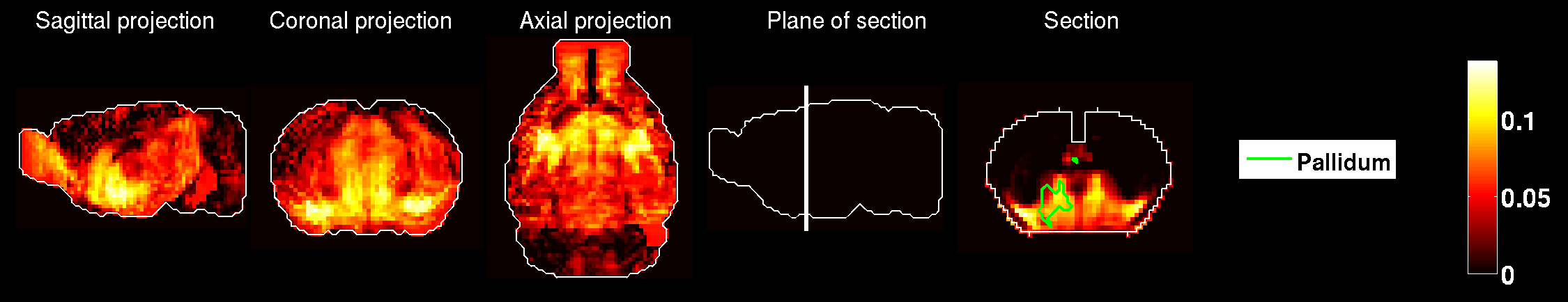}\\\hline
\end{tabular}
\caption{\label{tableCorrels10}Brain-wide correlation profiles between \numTypesPerTable cell types and the Allen Atlas.}
\end{table}
%\newpage

\begin{table}
\begin{tabular}{|m{0.06\textwidth}|m{0.13\textwidth}|m{\widthParamTable\textwidth}|}
\hline
\textbf{Index}&\textbf{Description}&\textbf{Brain-wide heat maps of correlations}\\\hline
61&\small{GABAergic Interneurons, PV+, P10}&\includegraphics[width=\widthParamTable\textwidth]{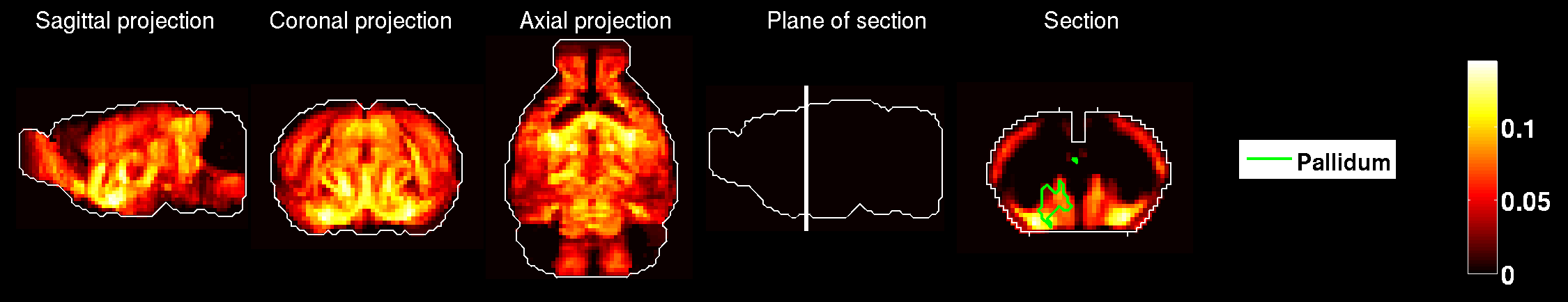}\\
62&\small{GABAergic Interneurons, PV+, P13-P15}&\includegraphics[width=\widthParamTable\textwidth]{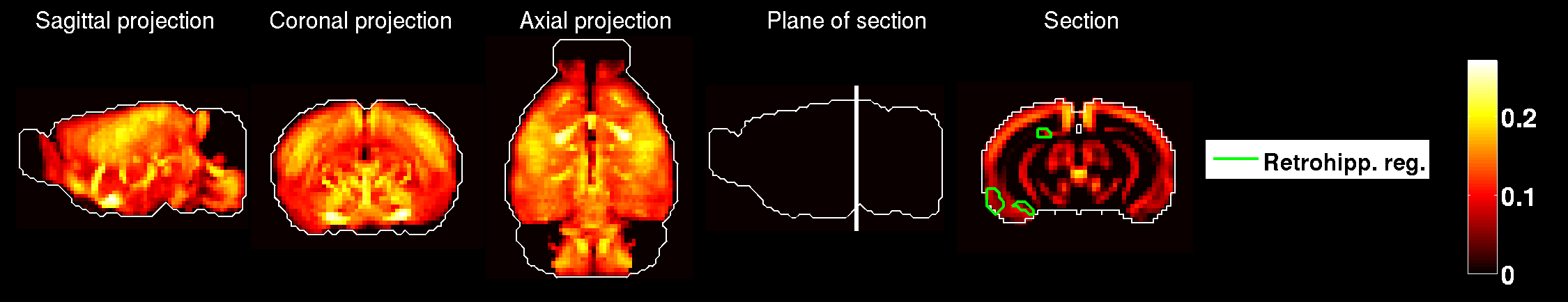}\\
63&\small{GABAergic Interneurons, PV+, P25}&\includegraphics[width=\widthParamTable\textwidth]{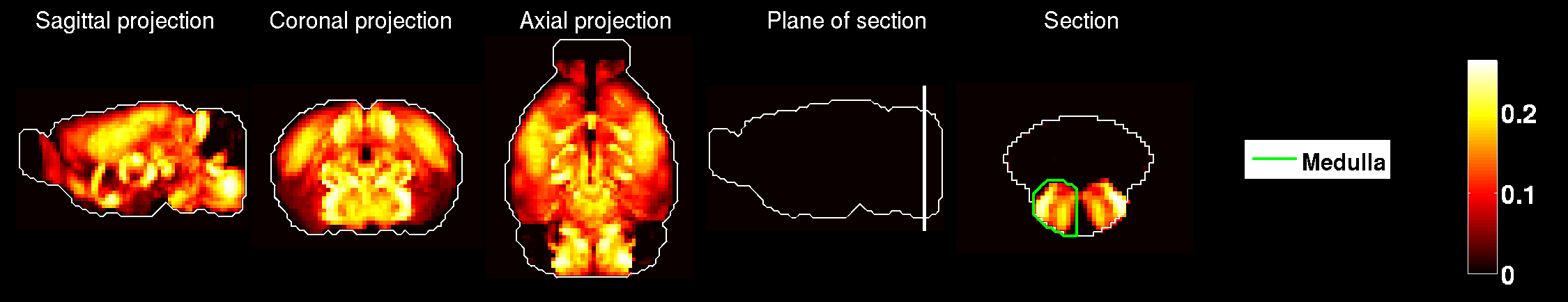}\\
64&\small{GABAergic Interneurons, PV+}&\includegraphics[width=\widthParamTable\textwidth]{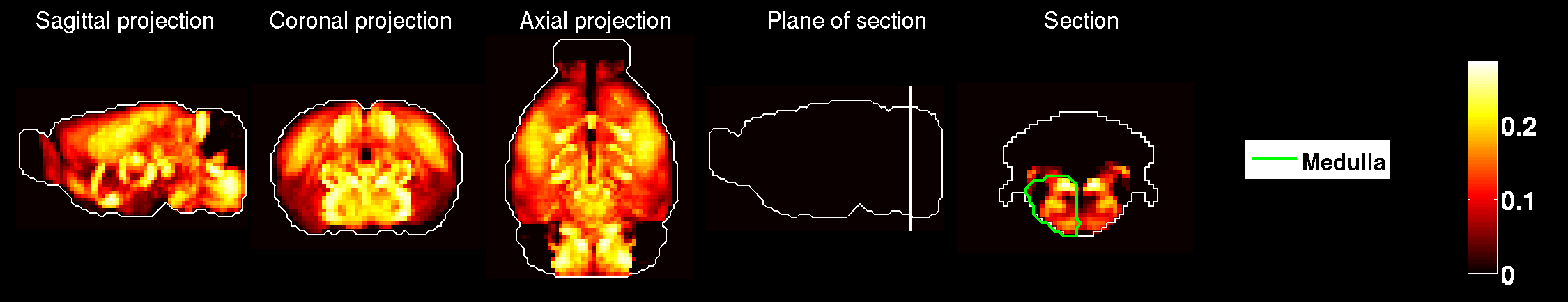}\\\hline
\end{tabular}
\caption{\label{tableCorrels11}Brain-wide correlation profiles between 4 cell types and the Allen Atlas.}
\end{table}

\clearpage
\section{Tables of estimated brain-wide densities}

\begin{table}
\begin{tabular}{|m{0.06\textwidth}|m{0.13\textwidth}|m{\widthParamTable\textwidth}|}
\hline
\textbf{Index}&\textbf{Description}&\textbf{Brain-wide heat maps of density profiles}\\\hline
1&\small{Purkinje Cells}&\includegraphics[width=\widthParamTable\textwidth]{fittingsFig1.png}\\
2&\small{Pyramidal Neurons}&\includegraphics[width=\widthParamTable\textwidth]{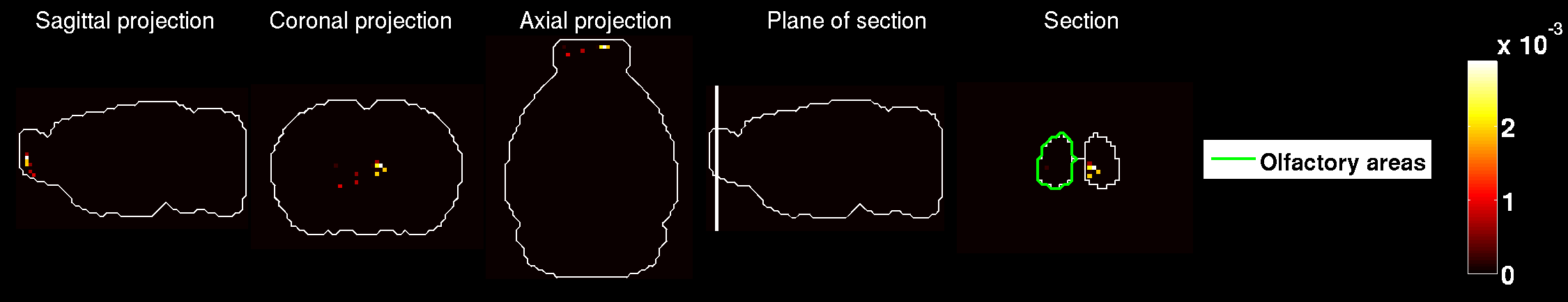}\\
3&\small{Pyramidal Neurons}&\includegraphics[width=\widthParamTable\textwidth]{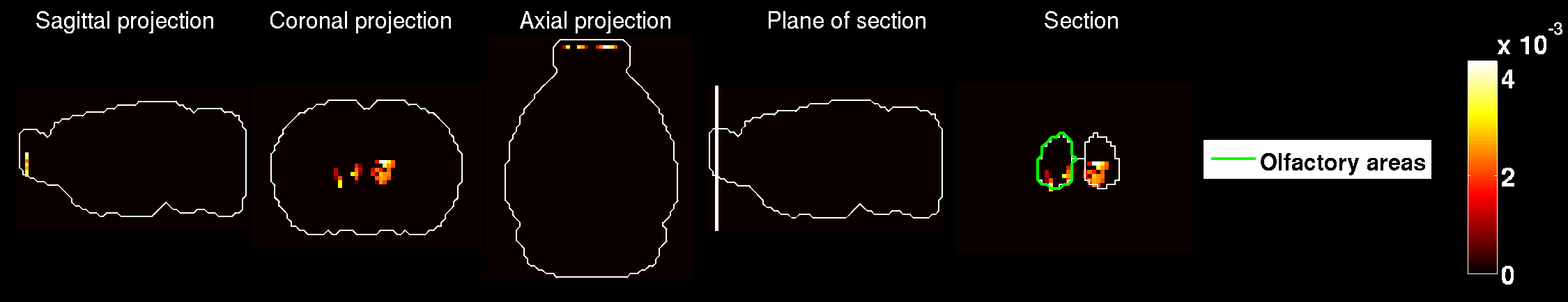}\\
4&\small{A9 Dopaminergic Neurons}&\includegraphics[width=\widthParamTable\textwidth]{fittingsFig4.png}\\
5&\small{A10 Dopaminergic Neurons}&\includegraphics[width=\widthParamTable\textwidth]{fittingsFig5.png}\\
6&\small{Pyramidal Neurons}&\includegraphics[width=\widthParamTable\textwidth]{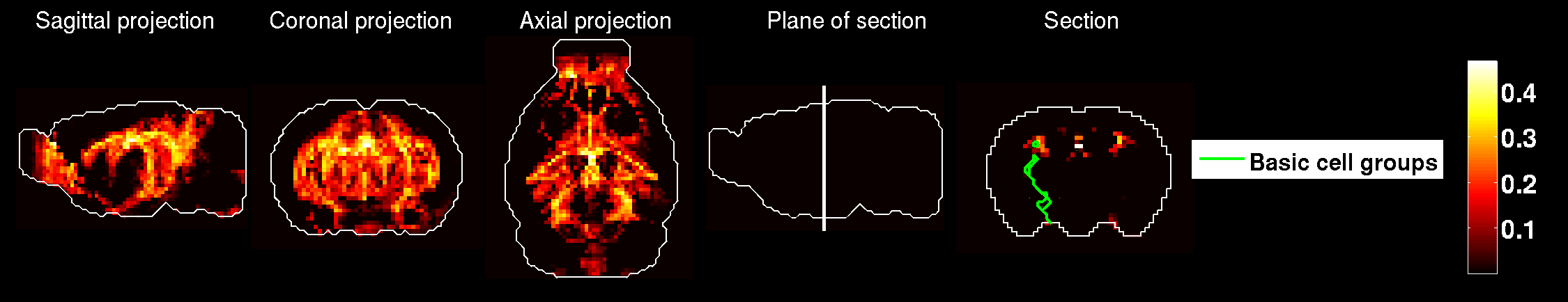}\\\hline
\end{tabular}
\caption{Brain-wide density profiles of \numTypesPerTable cell types.}
\label{tableFittings1}
\end{table}
%\newpage

\begin{table}
\begin{tabular}{|m{0.06\textwidth}|m{0.13\textwidth}|m{\widthParamTable\textwidth}|}
\hline
\textbf{Index}&\textbf{Description}&\textbf{Brain-wide heat maps of density profiles}\\\hline
7&\small{Pyramidal Neurons}&\includegraphics[width=\widthParamTable\textwidth]{fittingsFig7.png}\\
8&\small{Pyramidal Neurons}&\includegraphics[width=\widthParamTable\textwidth]{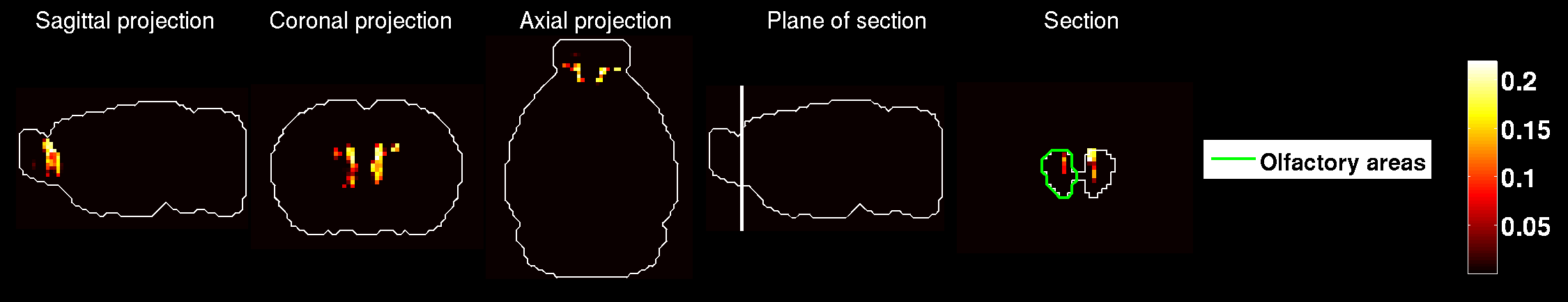}\\
9&\small{Mixed Neurons}&\includegraphics[width=\widthParamTable\textwidth]{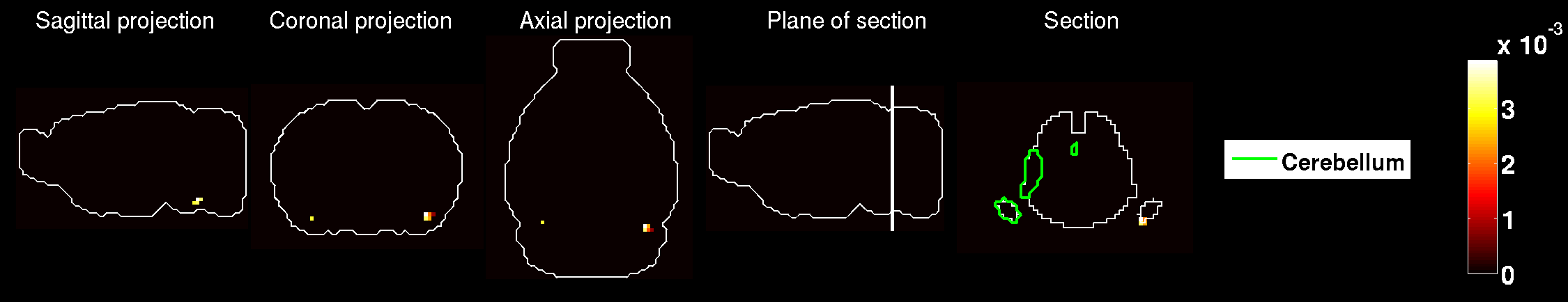}\\
10&\small{Motor Neurons, Midbrain Cholinergic Neurons}&\includegraphics[width=\widthParamTable\textwidth]{fittingsFig10.png}\\
11&\small{Cholinergic Projection Neurons}&\includegraphics[width=\widthParamTable\textwidth]{fittingsFig11.png}\\
12&\small{Motor Neurons, Cholinergic Interneurons}&\includegraphics[width=\widthParamTable\textwidth]{fittingsFig12.png}\\\hline
\end{tabular}
\caption{Brain-wide density profiles of \numTypesPerTable cell types.}
\label{tableFittings2}
\end{table}
%\newpage

\begin{table}
\begin{tabular}{|m{0.06\textwidth}|m{0.13\textwidth}|m{\widthParamTable\textwidth}|}
\hline
\textbf{Index}&\textbf{Description}&\textbf{Brain-wide heat maps of density profiles}\\\hline
13&\small{Cholinergic Neurons}&\includegraphics[width=\widthParamTable\textwidth]{fittingsFig13.png}\\
14&\small{Interneurons}&\includegraphics[width=\widthParamTable\textwidth]{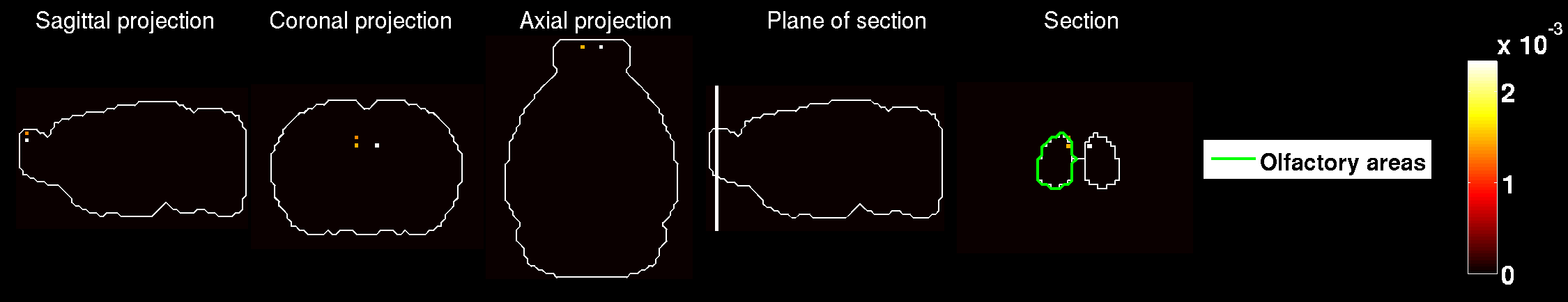}\\
15&\small{Drd1+ Medium Spiny Neurons}&\includegraphics[width=\widthParamTable\textwidth]{fittingsFig15.png}\\
16&\small{Drd2+ Medium Spiny Neurons}&\includegraphics[width=\widthParamTable\textwidth]{fittingsFig16.png}\\
17&\small{Golgi Cells}&\includegraphics[width=\widthParamTable\textwidth]{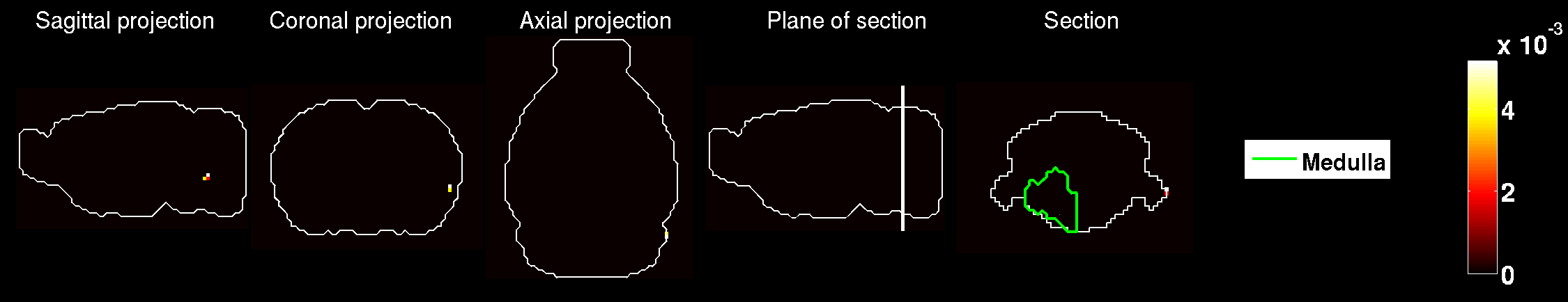}\\
18&\small{Unipolar Brush cells (some Bergman Glia)}&\includegraphics[width=\widthParamTable\textwidth]{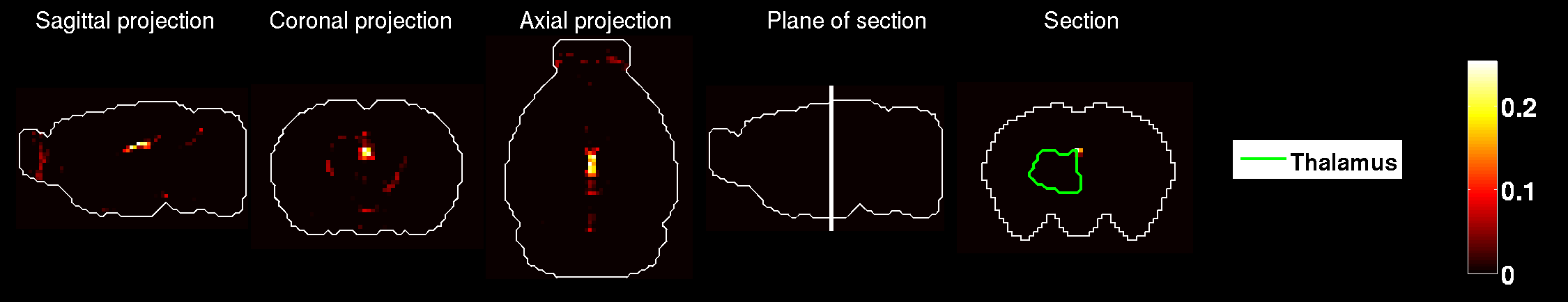}\\\hline
\end{tabular}
\caption{Brain-wide density profiles of \numTypesPerTable cell types.}
\label{tableFittings3}
\end{table}
%\newpage

\begin{table}
\begin{tabular}{|m{0.06\textwidth}|m{0.13\textwidth}|m{\widthParamTable\textwidth}|}
\hline
\textbf{Index}&\textbf{Description}&\textbf{Brain-wide heat maps of density profiles}\\\hline
19&\small{Stellate Basket Cells}&\includegraphics[width=\widthParamTable\textwidth]{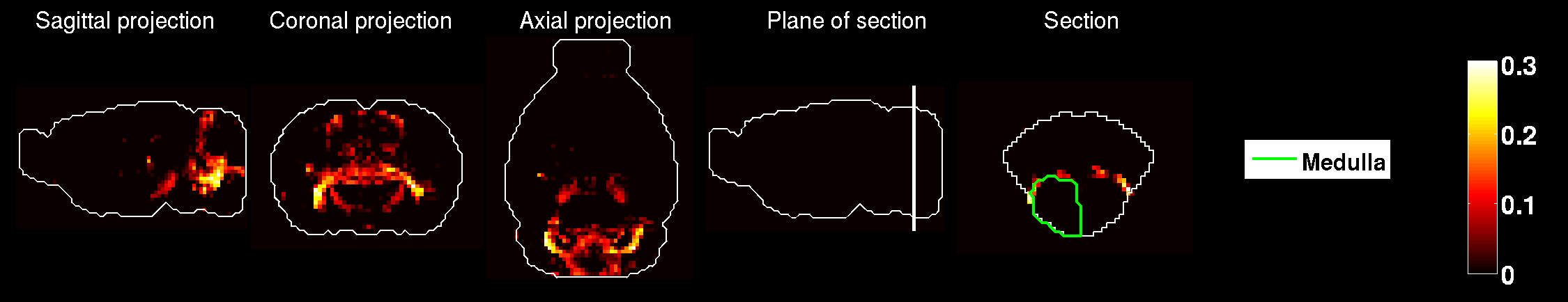}\\
20&\small{Granule Cells}&\includegraphics[width=\widthParamTable\textwidth]{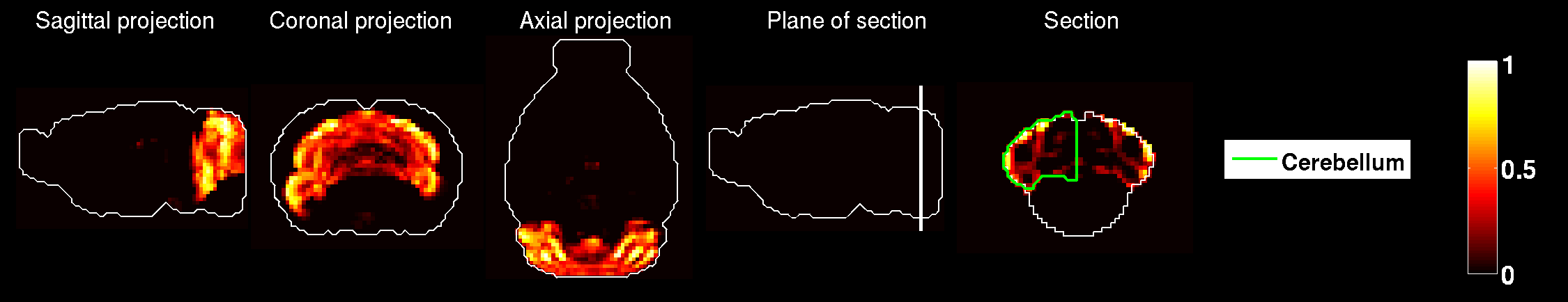}\\
21&\small{Mature Oligodendrocytes}&\includegraphics[width=\widthParamTable\textwidth]{fittingsFig21.png}\\
22&\small{Mature Oligodendrocytes}&\includegraphics[width=\widthParamTable\textwidth]{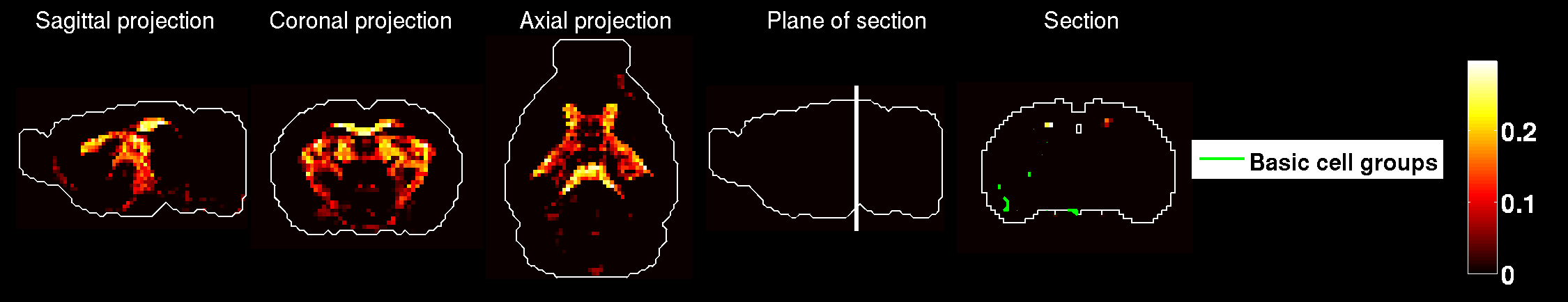}\\
%23&\small{Mixed Oligodendrocytes}&\includegraphics[width=\widthParamTable\textwidth]{fittingsFig23.png}\\
23&\small{Mixed Oligodendrocytes}& N/A \\
24&\small{Mixed Oligodendrocytes}&\includegraphics[width=\widthParamTable\textwidth]{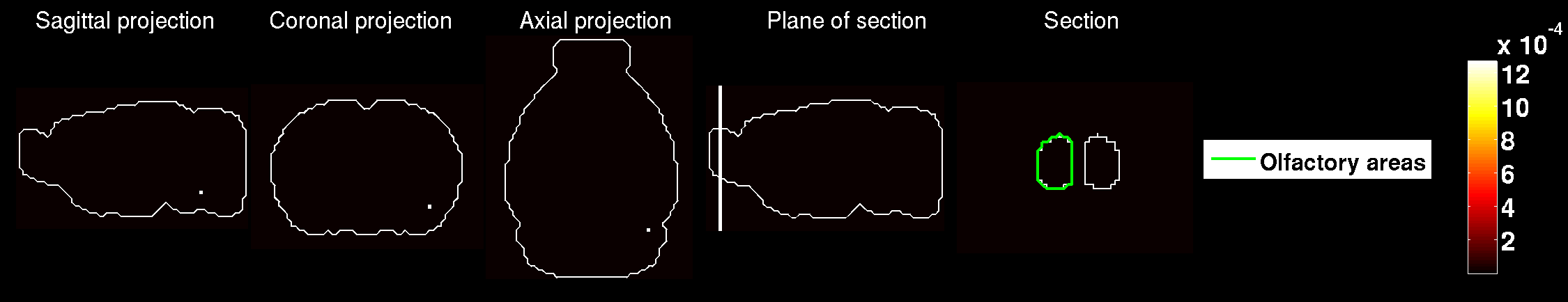}\\\hline
\end{tabular}
\caption{Brain-wide density profiles of \numTypesPerTable cell types.}
\label{tableFittings4}
\end{table}
%\newpage

\begin{table}
\begin{tabular}{|m{0.06\textwidth}|m{0.13\textwidth}|m{\widthParamTable\textwidth}|}
\hline
\textbf{Index}&\textbf{Description}&\textbf{Brain-wide heat maps of density profiles}\\\hline
25&\small{Purkinje Cells}&\includegraphics[width=\widthParamTable\textwidth]{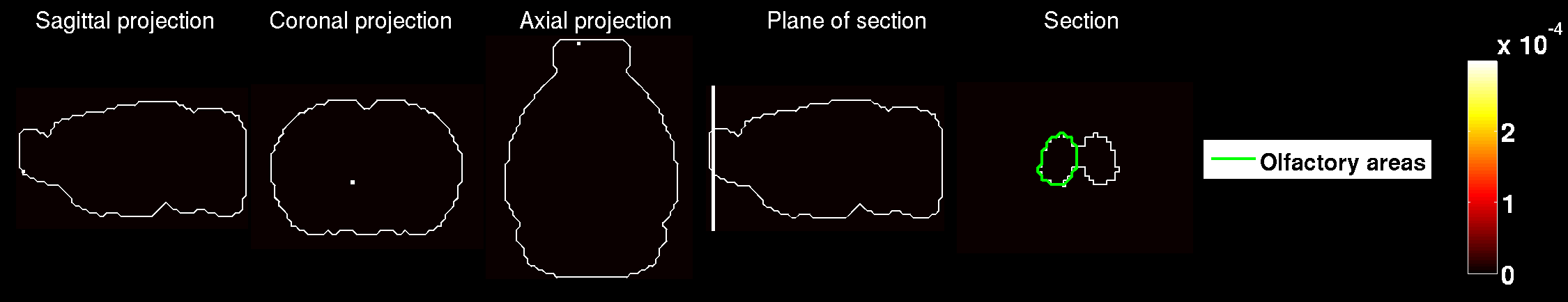}\\
26&\small{Neurons}&\includegraphics[width=\widthParamTable\textwidth]{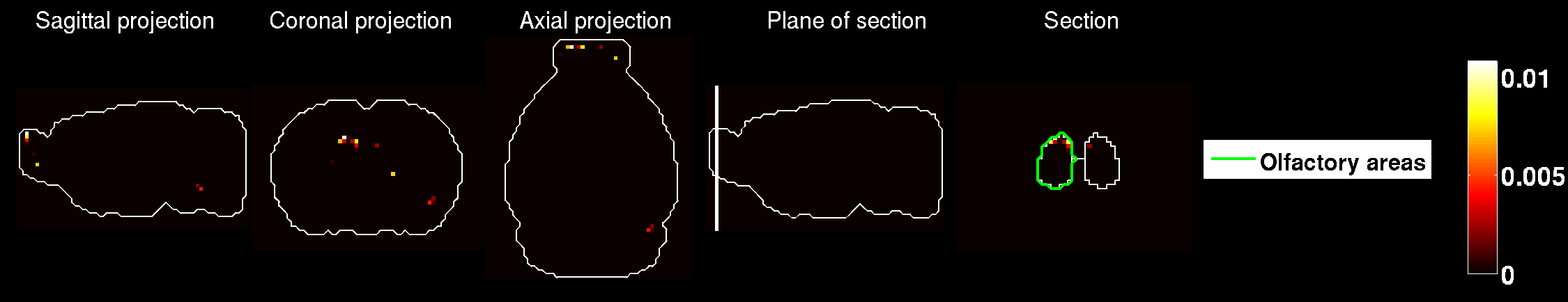}\\
27&\small{Bergman Glia}&\includegraphics[width=\widthParamTable\textwidth]{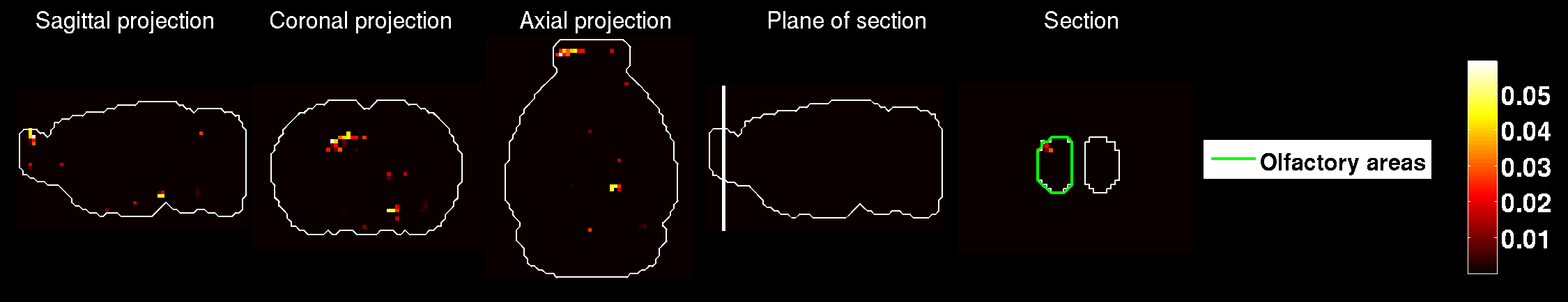}\\
28&\small{Astroglia}&\includegraphics[width=\widthParamTable\textwidth]{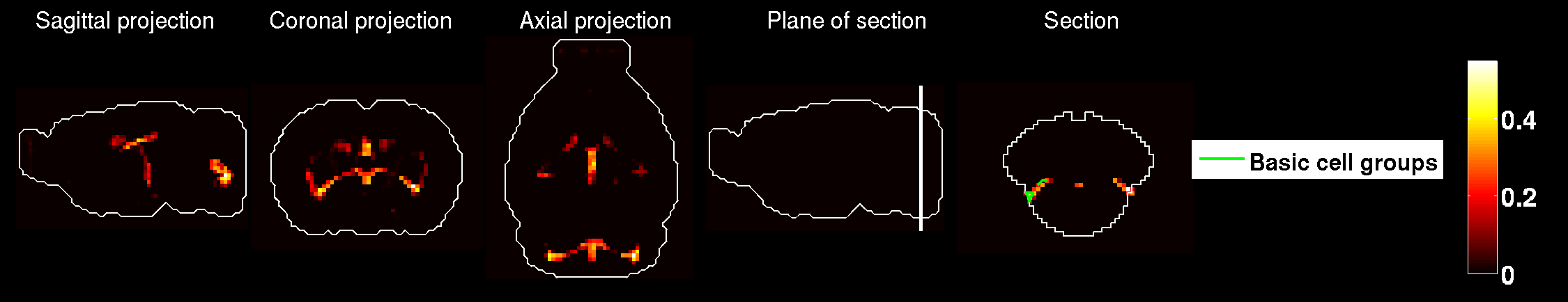}\\
29&\small{Astroglia}&\includegraphics[width=\widthParamTable\textwidth]{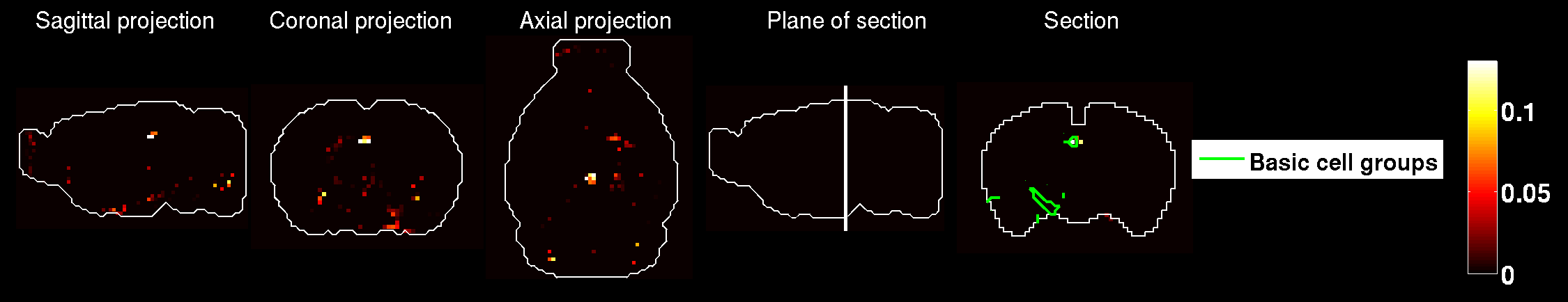}\\
30&\small{Astrocytes}&\includegraphics[width=\widthParamTable\textwidth]{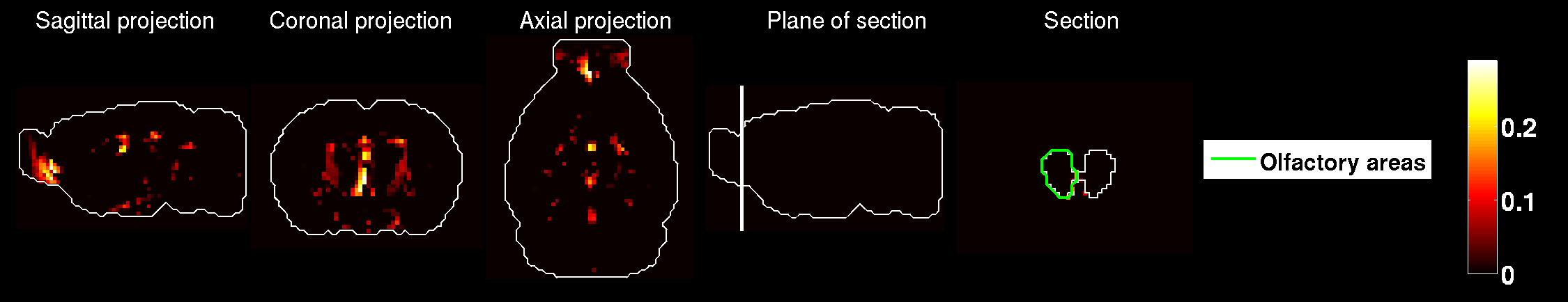}\\\hline
\end{tabular}
\caption{Brain-wide density profiles of \numTypesPerTable cell types.}
\label{tableFittings5}
\end{table}
%\newpage

\begin{table}
\begin{tabular}{|m{0.06\textwidth}|m{0.13\textwidth}|m{\widthParamTable\textwidth}|}
\hline
\textbf{Index}&\textbf{Description}&\textbf{Brain-wide heat maps of density profiles}\\\hline
31&\small{Astrocytes}&\includegraphics[width=\widthParamTable\textwidth]{fittingsFig31.png}\\
32&\small{Astrocytes}&\includegraphics[width=\widthParamTable\textwidth]{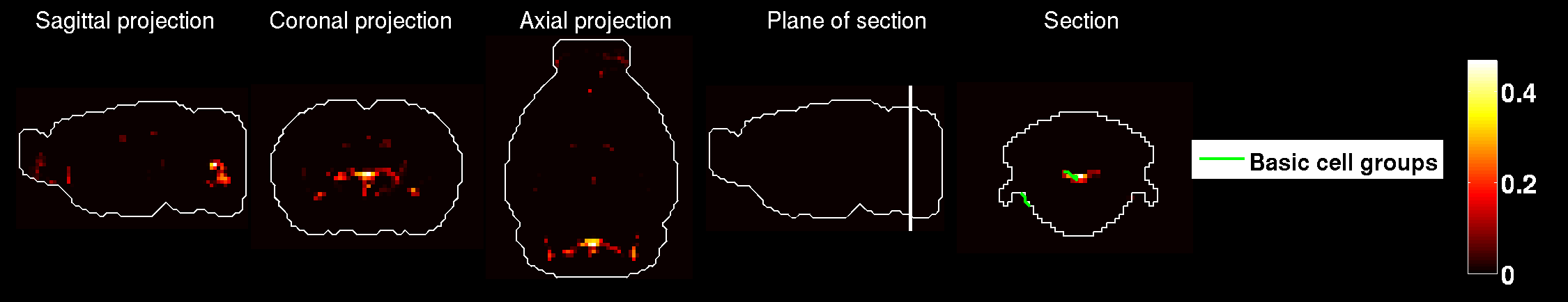}\\
33&\small{Mixed Neurons}&\includegraphics[width=\widthParamTable\textwidth]{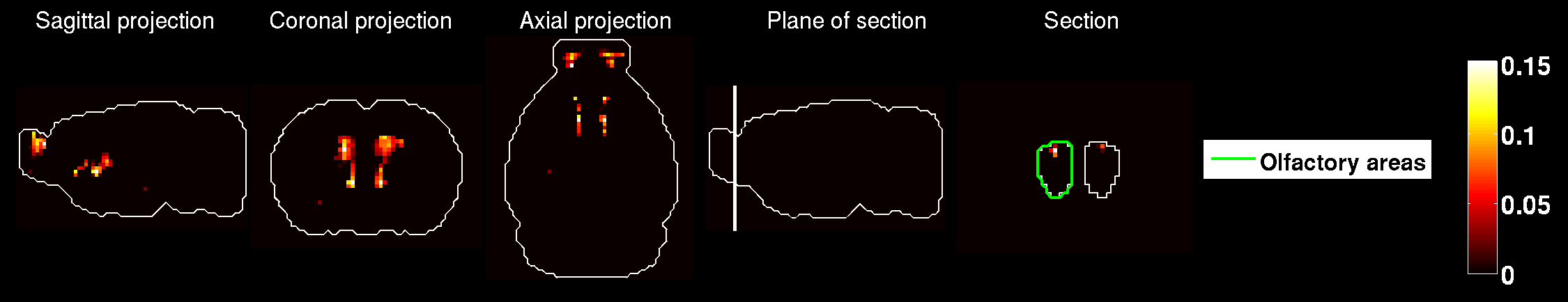}\\
34&\small{Mixed Neurons}&\includegraphics[width=\widthParamTable\textwidth]{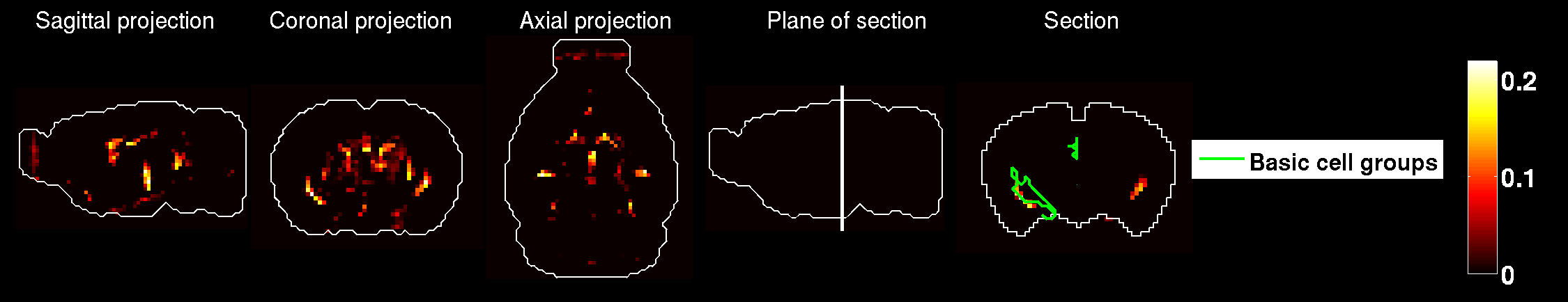}\\
35&\small{Mature Oligodendrocytes}&\includegraphics[width=\widthParamTable\textwidth]{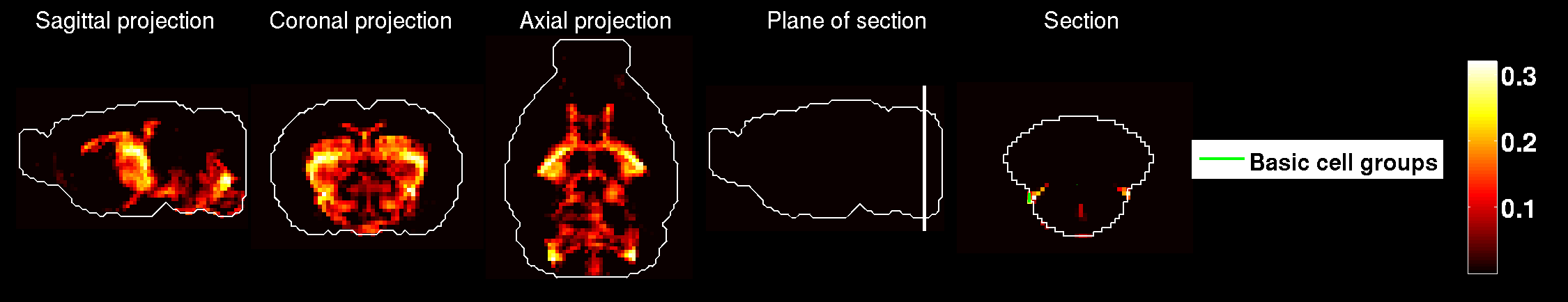}\\
36&\small{Oligodendrocytes}&\includegraphics[width=\widthParamTable\textwidth]{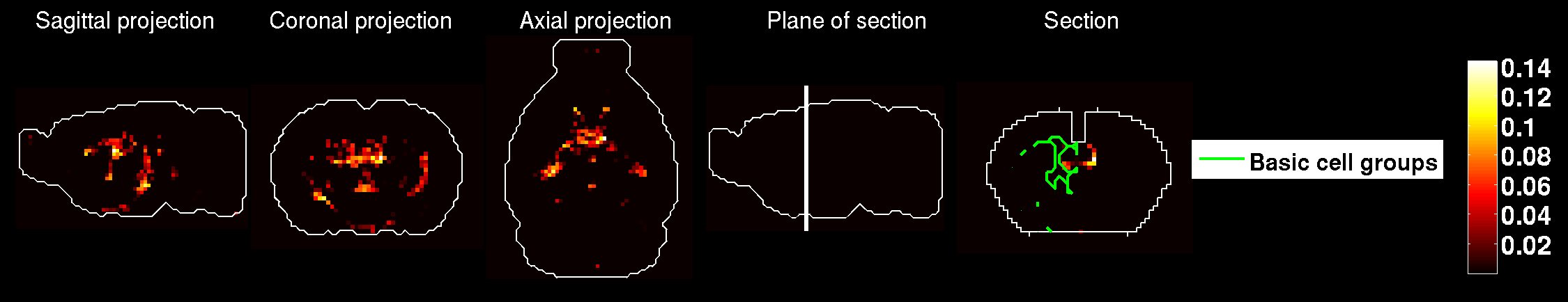}\\\hline
\end{tabular}
\caption{Brain-wide density profiles of \numTypesPerTable cell types.}
\label{tableFittings6}
\end{table}
%\newpage

\begin{table}
\begin{tabular}{|m{0.06\textwidth}|m{0.13\textwidth}|m{\widthParamTable\textwidth}|}
\hline
\textbf{Index}&\textbf{Description}&\textbf{Brain-wide heat maps of density profiles}\\\hline
37&\small{Oligodendrocyte Precursors}&\includegraphics[width=\widthParamTable\textwidth]{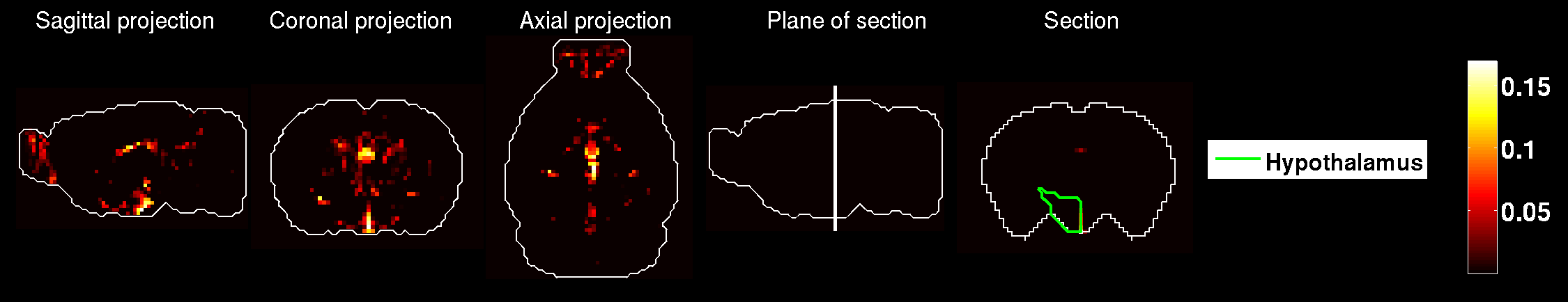}\\
38&\small{Pyramidal Neurons, Callosally projecting, P3}&\includegraphics[width=\widthParamTable\textwidth]{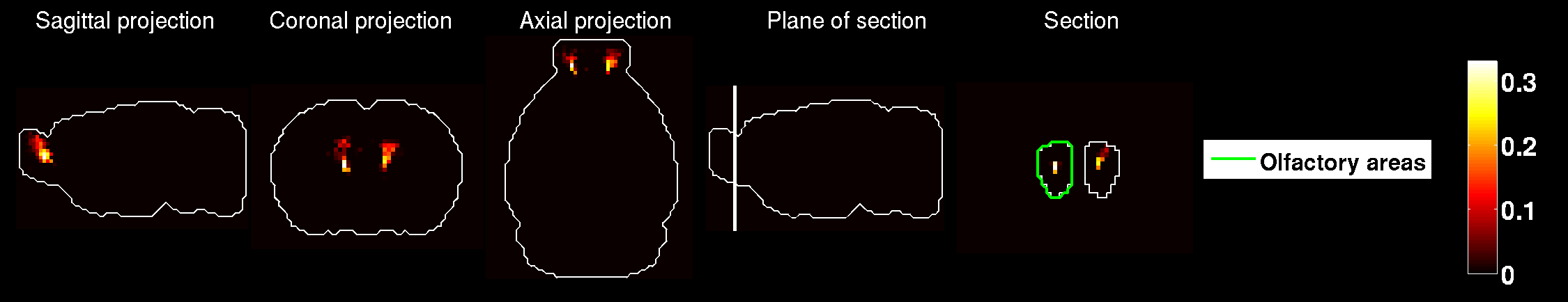}\\
39&\small{Pyramidal Neurons, Callosally projecting, P6}&\includegraphics[width=\widthParamTable\textwidth]{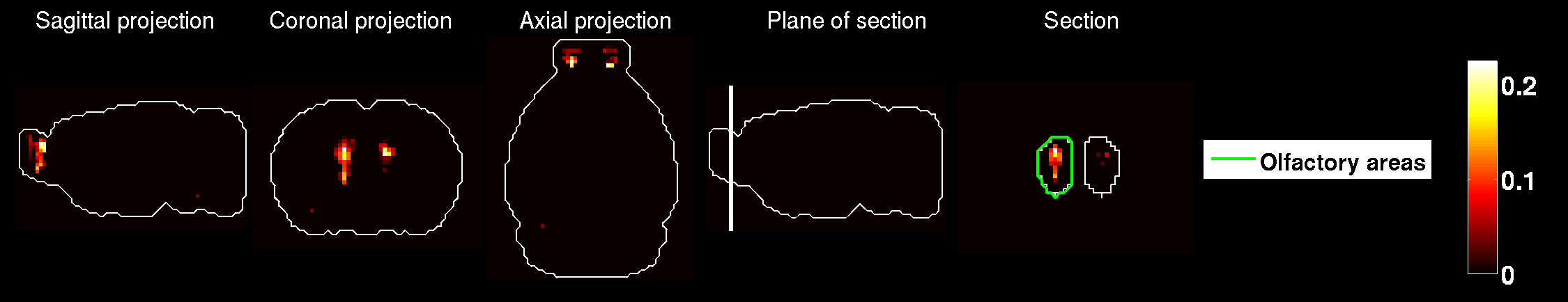}\\
40&\small{Pyramidal Neurons, Callosally projecting, P14}&\includegraphics[width=\widthParamTable\textwidth]{fittingsFig40.png}\\
41&\small{Pyramidal Neurons, Corticospinal, P3}&\includegraphics[width=\widthParamTable\textwidth]{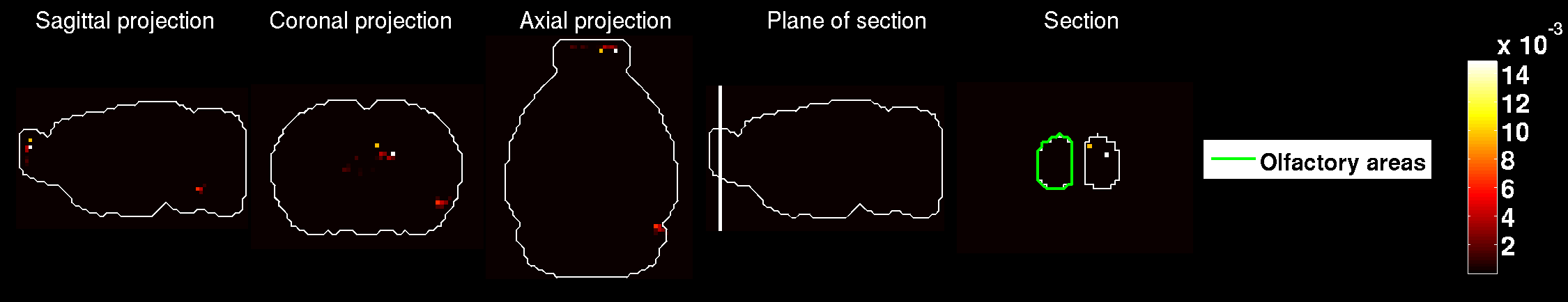}\\
42&\small{Pyramidal Neurons, Corticospinal, P6}&\includegraphics[width=\widthParamTable\textwidth]{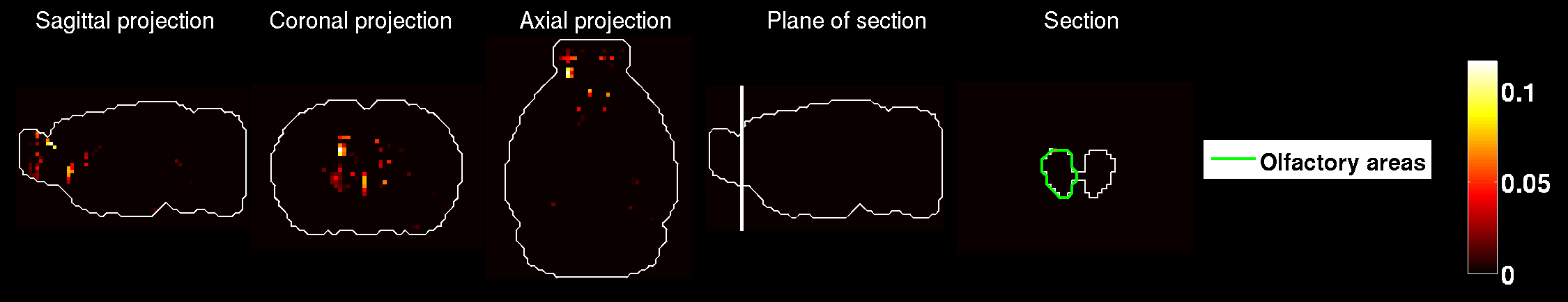}\\\hline
\end{tabular}
\caption{Brain-wide density profiles of \numTypesPerTable cell types.}
\label{tableFittings7}
\end{table}
%\newpage

\begin{table}
\begin{tabular}{|m{0.06\textwidth}|m{0.13\textwidth}|m{\widthParamTable\textwidth}|}
\hline
\textbf{Index}&\textbf{Description}&\textbf{Brain-wide heat maps of density profiles}\\\hline
43&\small{Pyramidal Neurons, Corticospinal, P14}&\includegraphics[width=\widthParamTable\textwidth]{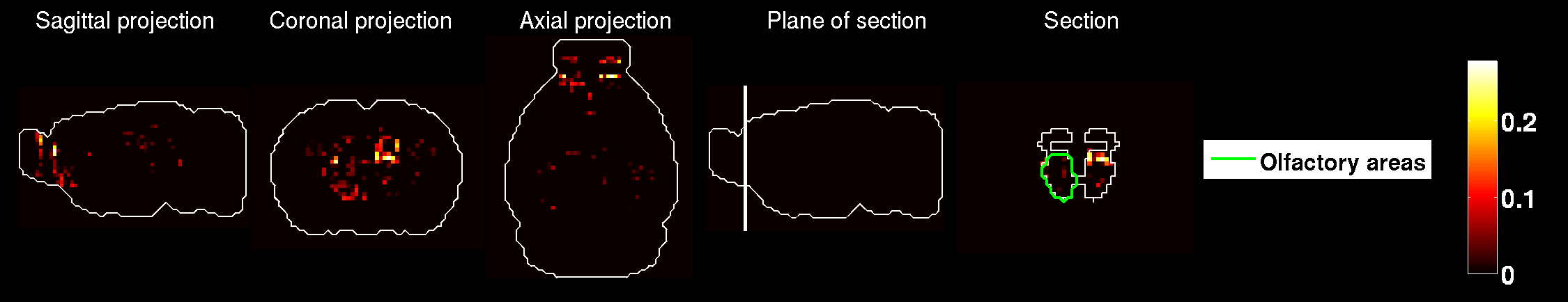}\\
44&\small{Pyramidal Neurons, Corticotectal, P14}&\includegraphics[width=\widthParamTable\textwidth]{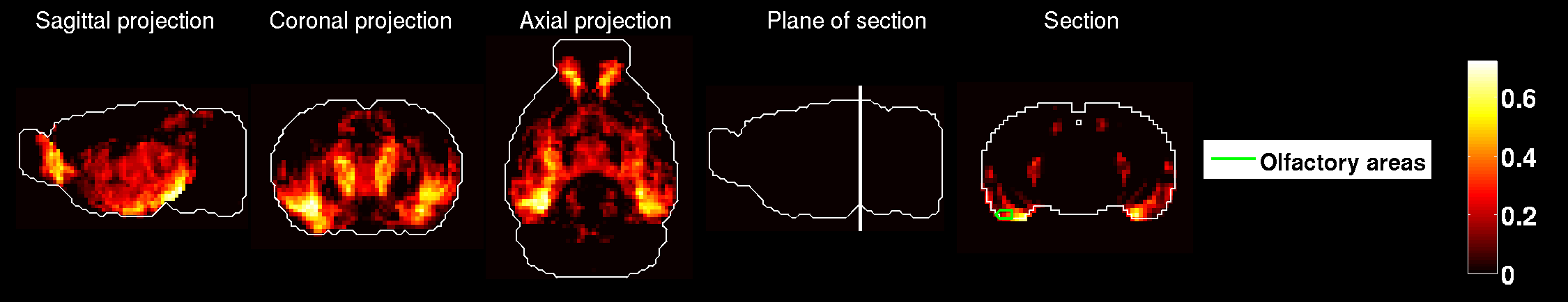}\\
45&\small{Pyramidal Neurons}&\includegraphics[width=\widthParamTable\textwidth]{fittingsFig45.png}\\
46&\small{Pyramidal Neurons}&\includegraphics[width=\widthParamTable\textwidth]{fittingsFig46.png}\\
47&\small{Pyramidal Neurons}&\includegraphics[width=\widthParamTable\textwidth]{fittingsFig47.png}\\
48&\small{Pyramidal Neurons}&\includegraphics[width=\widthParamTable\textwidth]{fittingsFig48.png}\\\hline
\end{tabular}
\caption{Brain-wide density profiles of \numTypesPerTable cell types.}
\label{tableFittings8}
\end{table}
%\newpage

\begin{table}
\begin{tabular}{|m{0.06\textwidth}|m{0.13\textwidth}|m{\widthParamTable\textwidth}|}
\hline
\textbf{Index}&\textbf{Description}&\textbf{Brain-wide heat maps of density profiles}\\\hline
49&\small{Pyramidal Neurons}&\includegraphics[width=\widthParamTable\textwidth]{fittingsFig49.png}\\
50&\small{Pyramidal Neurons}&\includegraphics[width=\widthParamTable\textwidth]{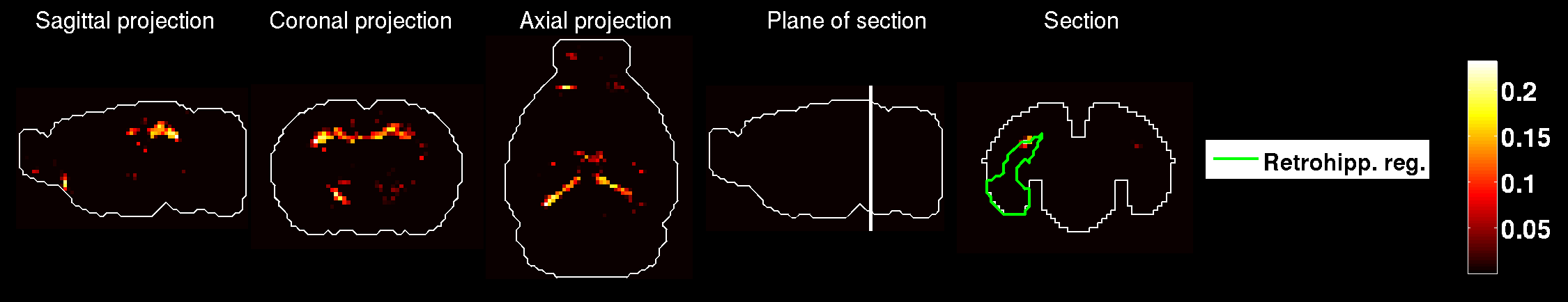}\\
51&\small{Tyrosine Hydroxylase Expressing}&\includegraphics[width=\widthParamTable\textwidth]{fittingsFig51.png}\\
52&\small{Purkinje Cells}&\includegraphics[width=\widthParamTable\textwidth]{fittingsFig52.png}\\
53&\small{Glutamatergic Neuron (not well defined)}&\includegraphics[width=\widthParamTable\textwidth]{fittingsFig53.png}\\
54&\small{GABAergic Interneurons, VIP+}&\includegraphics[width=\widthParamTable\textwidth]{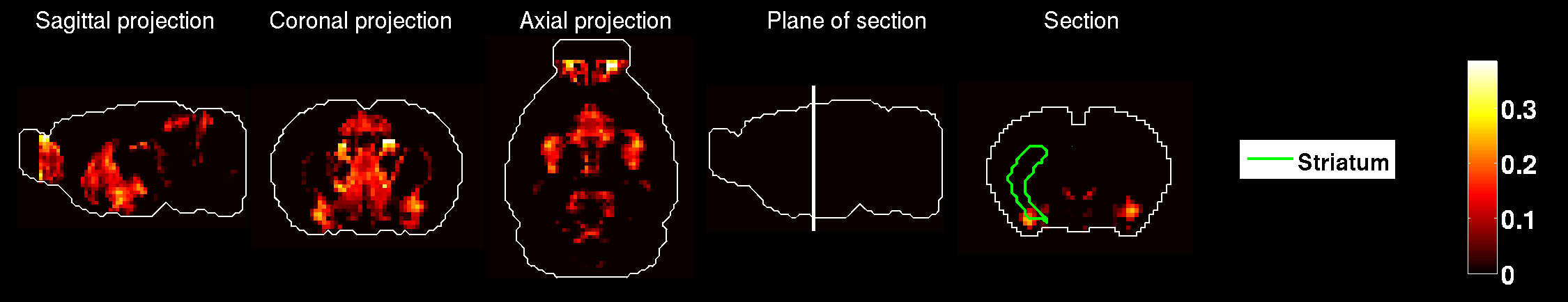}\\\hline
\end{tabular}
\caption{Brain-wide density profiles of \numTypesPerTable cell types.}
\label{tableFittings9}
\end{table}
%\newpage

\begin{table}
\begin{tabular}{|m{0.06\textwidth}|m{0.13\textwidth}|m{\widthParamTable\textwidth}|}
\hline
\textbf{Index}&\textbf{Description}&\textbf{Brain-wide heat maps of density profiles}\\\hline
55&\small{GABAergic Interneurons, VIP+}&\includegraphics[width=\widthParamTable\textwidth]{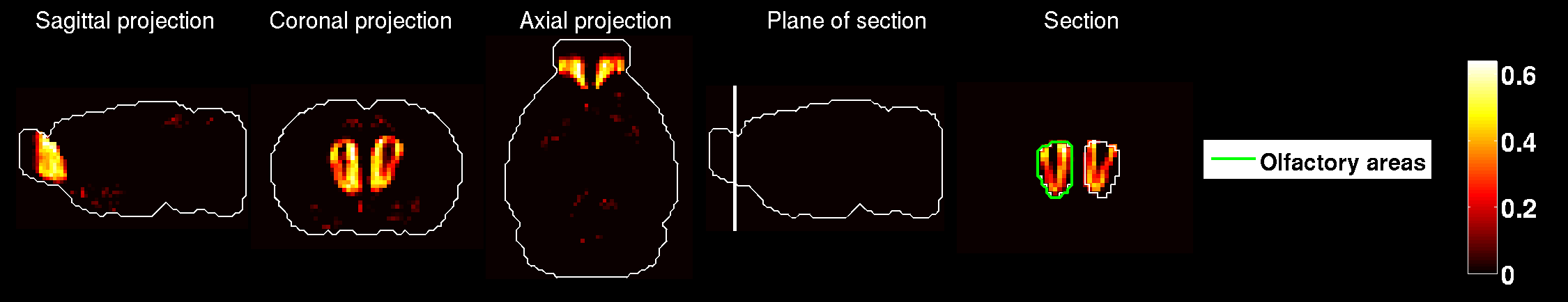}\\
56&\small{GABAergic Interneurons, SST+}&\includegraphics[width=\widthParamTable\textwidth]{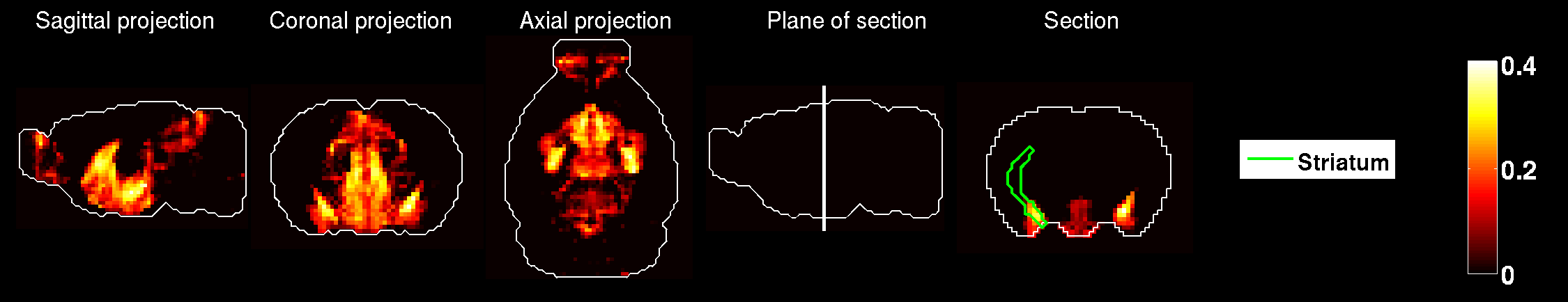}\\
57&\small{GABAergic Interneurons, SST+}&\includegraphics[width=\widthParamTable\textwidth]{fittingsFig57.png}\\
58&\small{GABAergic Interneurons, PV+}&\includegraphics[width=\widthParamTable\textwidth]{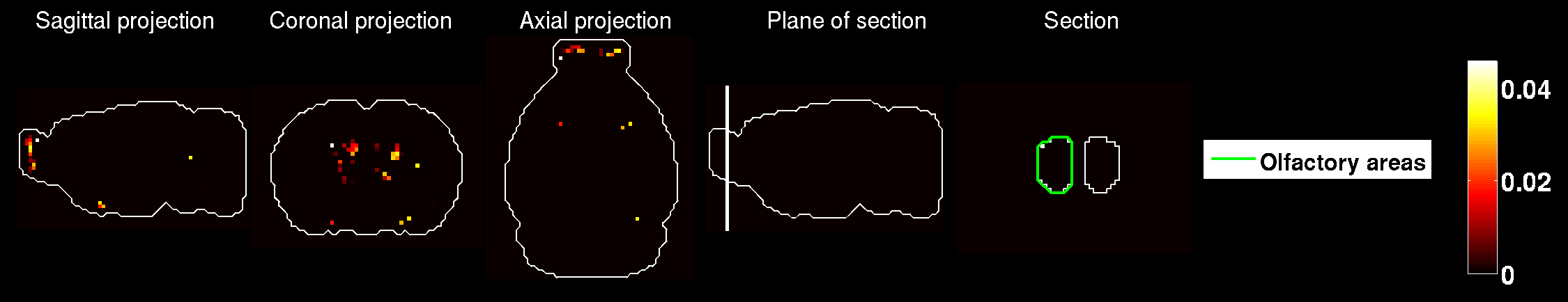}\\
59&\small{GABAergic Interneurons, PV+}&\includegraphics[width=\widthParamTable\textwidth]{fittingsFig59.png}\\
60&\small{GABAergic Interneurons, PV+, P7}&\includegraphics[width=\widthParamTable\textwidth]{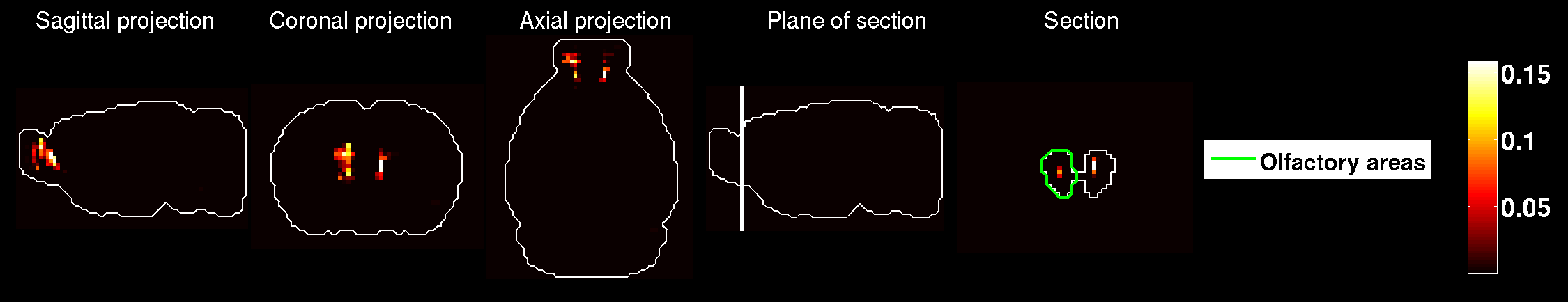}\\\hline
\end{tabular}
\caption{Brain-wide density profiles of \numTypesPerTable cell types.}
\label{tableFittings10}
\end{table}
%\newpage

\begin{table}
\begin{tabular}{|m{0.06\textwidth}|m{0.13\textwidth}|m{\widthParamTable\textwidth}|}
\hline
\textbf{Index}&\textbf{Description}&\textbf{Brain-wide heat maps of density profiles}\\\hline
61&\small{GABAergic Interneurons, PV+, P10}&\includegraphics[width=\widthParamTable\textwidth]{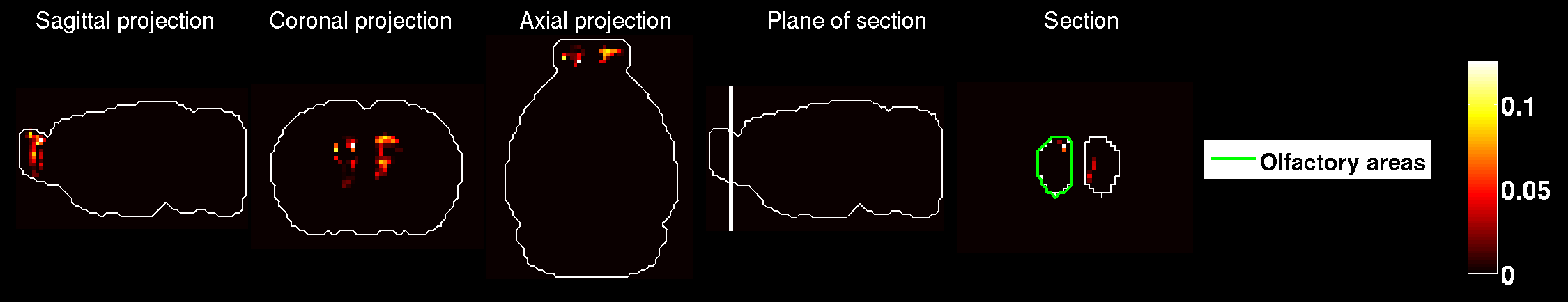}\\
62&\small{GABAergic Interneurons, PV+, P13-P15}&\includegraphics[width=\widthParamTable\textwidth]{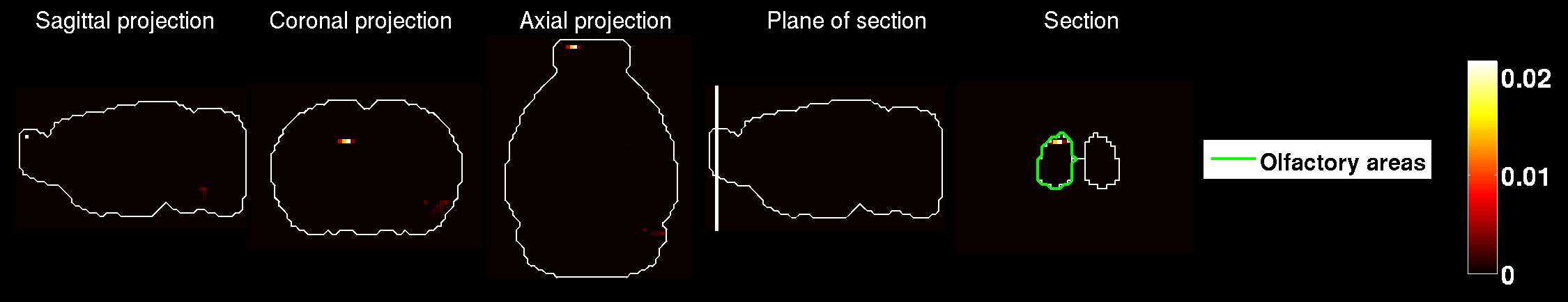}\\
63&\small{GABAergic Interneurons, PV+, P25}&\includegraphics[width=\widthParamTable\textwidth]{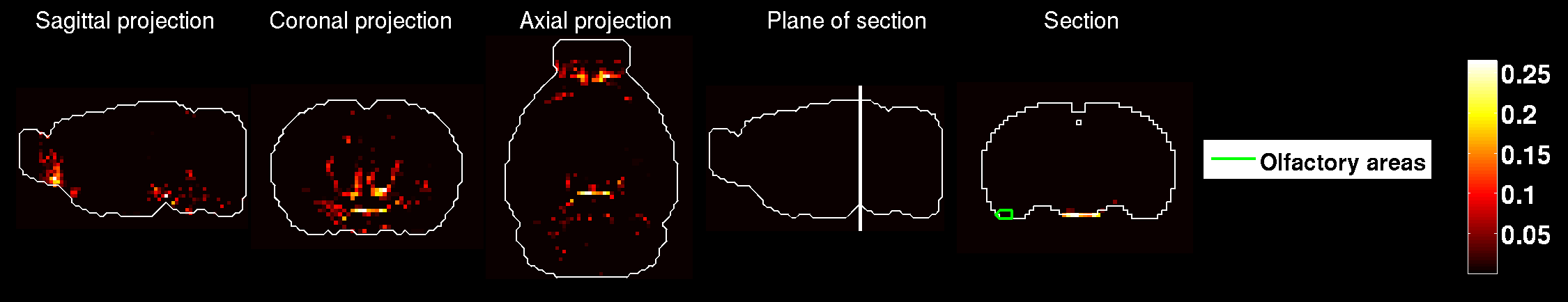}\\
64&\small{GABAergic Interneurons, PV+}&\includegraphics[width=\widthParamTable\textwidth]{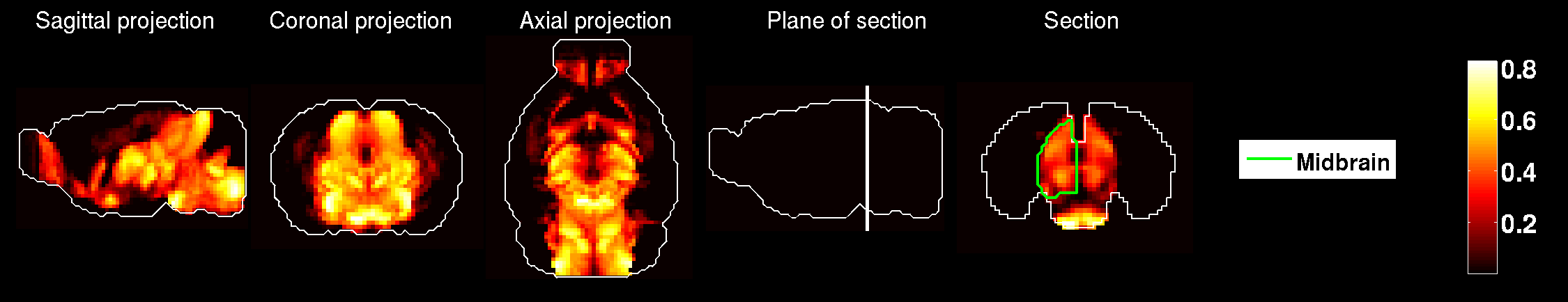}\\\hline
\end{tabular}
\caption{Brain-wide fitting profiles between 4 cell-types and the Allen Atlas.}
\label{tableFittings11}
\end{table}
%\newpage

\clearpage
\section{Tables of description of cell-type specific samples }
%metadataTables = metadata_tables( Ref )
\begin{table}
\centering
\begin{tabular}{|m{0.07\textwidth}|m{0.44\textwidth}|m{0.32\textwidth}|m{0.12\textwidth}|}
\hline
\textbf{Index}&\textbf{Description}&\textbf{Age of animal}&\textbf{Reference}\\\hline
1&Purkinje Cells&$\sim$P60&\cite{RossnerCells}\\\hline
2&Pyramidal Neurons&$\sim$P60&\cite{RossnerCells}\\\hline
3&Pyramidal Neurons&$\sim$P60&\cite{RossnerCells}\\\hline
4&A9 Dopaminergic Neurons&Adult (precise age not given)&\cite{ChungCells}\\\hline
5&A10 Dopaminergic Neurons&Adult (precise age not given)&\cite{ChungCells}\\\hline
6&Pyramidal Neurons&Adult (precise age not given)&\cite{DoyleCells}\\\hline
7&Pyramidal Neurons&Adult (precise age not given)&\cite{DoyleCells}\\\hline
8&Pyramidal Neurons&Adult (precise age not given)&\cite{DoyleCells}\\\hline
9&Mixed Neurons&Adult (precise age not given)&\cite{DoyleCells}\\\hline
10&Motor Neurons, Midbrain Cholinergic Neurons&Adult (precise age not given)&\cite{DoyleCells}\\\hline
11&Cholinergic Projection Neurons&Adult (precise age not given)&\cite{DoyleCells}\\\hline
12&Motor Neurons, Cholinergic Interneurons&Adult (precise age not given)&\cite{DoyleCells}\\\hline
13&Cholinergic Neurons&Adult (precise age not given)&\cite{DoyleCells}\\\hline
14&Interneurons&Adult (precise age not given)&\cite{DoyleCells}\\\hline
15&Drd1+ Medium Spiny Neurons&Adult (precise age not given)&\cite{DoyleCells}\\\hline
16&Drd2+ Medium Spiny Neurons&Adult (precise age not given)&\cite{DoyleCells}\\\hline
17&Golgi Cells&Adult (precise age not given)&\cite{DoyleCells}\\\hline
18&Unipolar Brush cells (some Bergman Glia)&Adult (precise age not given)&\cite{DoyleCells}\\\hline
19&Stellate Basket Cells&Adult (precise age not given)&\cite{DoyleCells}\\\hline
20&Granule Cells&Adult (precise age not given)&\cite{DoyleCells}\\\hline
21&Mature Oligodendrocytes&Adult (precise age not given)&\cite{DoyleCells}\\\hline
22&Mature Oligodendrocytes&Adult (precise age not given)&\cite{DoyleCells}\\\hline
23&Mixed Oligodendrocytes&Adult (precise age not given)&\cite{DoyleCells}\\\hline
24&Mixed Oligodendrocytes&Adult (precise age not given)&\cite{DoyleCells}\\\hline
25&Purkinje Cells&Adult (precise age not given)&\cite{DoyleCells}\\\hline
26&Neurons&Adult (precise age not given)&\cite{DoyleCells}\\\hline
27&Bergman Glia&Adult (precise age not given)&\cite{DoyleCells}\\\hline
28&Astroglia&Adult (precise age not given)&\cite{DoyleCells}\\\hline
29&Astroglia&Adult (precise age not given)&\cite{DoyleCells}\\\hline
30&Astrocytes&P7-P8&\cite{ChungCells}\\\hline
31&Astrocytes&P17&\cite{ChungCells}\\\hline
32&Astrocytes&P17&\cite{ChungCells}\\\hline
33&Mixed Neurons&P7&\cite{ChungCells}\\\hline
34&Mixed Neurons&P16&\cite{ChungCells}\\\hline
35&Mature Oligodendrocytes&P17&\cite{ChungCells}\\\hline
36&Oligodendrocytes&P7&\cite{ChungCells}\\\hline
37&Oligodendrocyte Precursors&P7&\cite{ChungCells}\\\hline
\end{tabular}
\caption{Description of cell-type-specific samples (I).}
\label{metadataTable1}
\end{table}
\begin{table}
\centering
\begin{tabular}{|m{0.07\textwidth}|m{0.44\textwidth}|m{0.3\textwidth}|m{0.15\textwidth}|}
\hline
\textbf{Index}&\textbf{Description}&\textbf{Age of animal}&\textbf{Reference}\\\hline
38&Pyramidal Neurons, Callosally projecting, P3&P3&\cite{ArlottaCells}\\\hline
39&Pyramidal Neurons, Callosally projecting, P6&P6&\cite{ArlottaCells}\\\hline
40&Pyramidal Neurons, Callosally projecting, P14&P14&\cite{ArlottaCells}\\\hline
41&Pyramidal Neurons, Corticospinal, P3&P3&\cite{ArlottaCells}\\\hline
42&Pyramidal Neurons, Corticospinal, P6&P6&\cite{ArlottaCells}\\\hline
43&Pyramidal Neurons, Corticospinal, P14&P14&\cite{ArlottaCells}\\\hline
44&Pyramidal Neurons, Corticotectal, P14&P14&\cite{ArlottaCells}\\\hline
45&Pyramidal Neurons&$\sim$P60&\cite{foreBrainTaxonomy}\\\hline
46&Pyramidal Neurons&$\sim$P60&\cite{foreBrainTaxonomy}\\\hline
47&Pyramidal Neurons&$\sim$P60&\cite{foreBrainTaxonomy}\\\hline
48&Pyramidal Neurons&$\sim$P60&\cite{foreBrainTaxonomy}\\\hline
49&Pyramidal Neurons&$\sim$P60&\cite{foreBrainTaxonomy}\\\hline
50&Pyramidal Neurons&$\sim$P60&Unpublished\\\hline
51&Tyrosine Hydroxylase Expressing&$\sim$P45&Unpublished\\\hline
52&Purkinje Cells&$\sim$P45&Unpublished\\\hline
53&Glutamatergic Neuron (not well defined)&$\sim$P60&\cite{foreBrainTaxonomy}\\\hline
54&GABAergic Interneurons, VIP+&$\sim$P60&\cite{foreBrainTaxonomy}\\\hline
55&GABAergic Interneurons, VIP+&$\sim$P60&\cite{foreBrainTaxonomy}\\\hline
56&GABAergic Interneurons, SST+&$\sim$P60&\cite{foreBrainTaxonomy}\\\hline
57&GABAergic Interneurons, SST+&$\sim$P60&\cite{foreBrainTaxonomy}\\\hline
58&GABAergic Interneurons, PV+&$\sim$P60&\cite{foreBrainTaxonomy}\\\hline
59&GABAergic Interneurons, PV+&$\sim$P60&\cite{foreBrainTaxonomy}\\\hline
60&GABAergic Interneurons, PV+, P7&P7&\cite{OkatyCells}\\\hline
61&GABAergic Interneurons, PV+, P10&P10&\cite{OkatyCells}\\\hline
62&GABAergic Interneurons, PV+, P13-P15&P15&\cite{OkatyCells}\\\hline
63&GABAergic Interneurons, PV+, P25&P25&\cite{OkatyCells}\\\hline
64&GABAergic Interneurons, PV+&$\sim$P45&\cite{OkatyCells}\\\hline
\end{tabular}
\caption{Description of cell-type-specific samples (II).}
\label{metadataTable2}
\end{table}
\clearpage

\section{Tables of anatomical origin of the cell-type-specific samples}
For each of the cell-type-specific samples analyzed in this study, the
following two tables give a brief description of the cell types, the
region from which those cell types were extracted according to the
coarsest (or \bigTwelve) version of the Allen Reference Atlas, and the
finest region to which it can be assigned according to the data
provided in the studies
\cite{OkatyCells,RossnerCells,CahoyCells,DoyleCells,ChungCells,ArlottaCells,HeimanCells,foreBrainTaxonomy}.\\
\begin{table}
\centering
\begin{tabular}{|m{0.05\textwidth}|m{0.39\textwidth}|m{0.18\textwidth}|m{0.38\textwidth}|}
\hline
\textbf{Index}&\textbf{Description}&\textbf{{\footnotesize{Region in the ARA (\bigTwelve)}}}&\textbf{Finest label in the ARA}\\\hline
1&Purkinje Cells&Cerebellum&       Cerebellar cortex \\\hline
2&Pyramidal Neurons&Cerebral cortex& Primary motor area; Layer 5 \\\hline
3&Pyramidal Neurons&Cerebral cortex& {\small{Primary somatosensory area; Layer 5}} \\\hline
4&A9 Dopaminergic Neurons&Midbrain& Substantia nigra\_ compact part \\\hline
5&A10 Dopaminergic Neurons&Midbrain& Ventral tegmental area \\\hline
6&Pyramidal Neurons&Cerebral cortex& Cerebral cortex; Layer 5 \\\hline
7&Pyramidal Neurons&Cerebral cortex&Cerebral cortex; Layer 5 \\\hline
8&Pyramidal Neurons&Cerebral cortex&Cerebral cortex; Layer 6\\\hline
9&Mixed Neurons&Cerebral cortex& Cerebral cortex \\\hline
10&{\footnotesize{Motor Neurons, Midbrain Cholinergic Neurons}}&Midbrain& Peduncolopontine nucleus\\\hline
11&Cholinergic Projection Neurons&Pallidum& Pallidum\_ ventral region\\\hline
12&{\footnotesize{Motor Neurons, Cholinergic Interneurons}}&Medulla& Spinal cord\\\hline
13&Cholinergic Neurons&Striatum&Striatum \\\hline
14&Interneurons&Cerebral cortex& Cerebral cortex\\\hline
15&Drd1+ Medium Spiny Neurons&Striatum& Striatum\\\hline
16&Drd2+ Medium Spiny Neurons&Striatum& Striatum\\\hline
17&Golgi Cells&Cerebellum& Cerebellar cortex\\\hline
18&Unipolar Brush cells (some Bergman Glia)&Cerebellum&Cerebellar cortex \\\hline
19&Stellate Basket Cells&Cerebellum& Cerebellar cortex\\\hline
20&Granule Cells&Cerebellum& Cerebellar cortex\\\hline
21&Mature Oligodendrocytes&Cerebellum& Cerebellar cortex\\\hline
22&Mature Oligodendrocytes&Cerebral cortex& Cerebral cortex\\\hline
23&Mixed Oligodendrocytes&Cerebellum& Cerebellar cortex\\\hline
24&Mixed Oligodendrocytes&Cerebral cortex& Cerebral cortex\\\hline
25&Purkinje Cells&Cerebellum& Cerebellar cortex\\\hline
26&Neurons&Cerebral cortex& Cerebral cortex\\\hline
27&Bergman Glia&Cerebellum& Cerebellar cortex\\\hline
28&Astroglia&Cerebellum& Cerebellar cortex\\\hline
29&Astroglia&Cerebral cortex&  Cerebral cortex\\\hline
30&Astrocytes&Cerebral cortex&  Cerebral cortex\\\hline
31&Astrocytes&Cerebral cortex& Cerebral cortex \\\hline
32&Astrocytes&Cerebral cortex& Cerebral cortex \\\hline
33&Mixed Neurons&Cerebral cortex& Cerebral cortex \\\hline
34&Mixed Neurons&Cerebral cortex& Cerebral cortex \\\hline
35&Mature Oligodendrocytes&Cerebral cortex& Cerebral cortex \\\hline
36&Oligodendrocytes&Cerebral cortex&  Cerebral cortex\\\hline
37&Oligodendrocyte Precursors&Cerebral cortex& Cerebral cortex \\\hline
\end{tabular}
\caption{Anatomical origin of the cell-type-specific samples (I).}
\label{metadataAnatomyTable1}
\end{table}

\begin{table}
\centering
\begin{tabular}{|m{0.05\textwidth}|m{0.39\textwidth}|m{0.18\textwidth}|m{0.38\textwidth}|}
\hline
\textbf{{\footnotesize{Index}}}&\textbf{Description}&\textbf{{\footnotesize{Region in the ARA (\bigTwelve)}}}&\textbf{Finest label in the ARA}\\\hline
38&\footnotesize{Pyramidal Neurons, Callosally projecting, P3}&Cerebral cortex& Cerebral cortex \\\hline
39&\footnotesize{Pyramidal Neurons, Callosally projecting, P6}&Cerebral cortex& Cerebral cortex  \\\hline
40&\footnotesize{Pyramidal Neurons, Callosally projecting, P14}&Cerebral cortex& Cerebral cortex\\\hline
41&\footnotesize{Pyramidal Neurons, Corticospinal, P3}&Cerebral cortex& Cerebral cortex \\\hline
42&\footnotesize{Pyramidal Neurons, Corticospinal, P6}&Cerebral cortex& Cerebral cortex \\\hline
43&\footnotesize{Pyramidal Neurons, Corticospinal, P14}&Cerebral cortex& Cerebral cortex \\\hline
44&\footnotesize{Pyramidal Neurons, Corticotectal, P14}&Cerebral cortex& Cerebral cortex \\\hline
45& Pyramidal Neurons&Cerebral cortex& Cerebral cortex, Layer 5 \\\hline
46& Pyramidal Neurons&Cerebral cortex& Cerebral cortex, Layer 5 \\\hline
47& Pyramidal Neurons&Cerebral cortex& {\small{Primary somatosensory area; Layer 5}} \\\hline
48&Pyramidal Neurons&Cerebral cortex& {\small{Prelimbic area and Infralimbic area; Layer 5 (Amygdala)}} \\\hline
49&Pyramidal Neurons&Hippocampal region& Ammon's Horn; Layer 6B\\\hline
50&Pyramidal Neurons&Cerebral cortex& Primary motor area\\\hline
51&{\footnotesize{Tyrosine Hydroxylase Expressing}}&Pons& Pontine central gray\\\hline
52&Purkinje Cells&Cerebellum& Cerebellar cortex \\\hline
53&\footnotesize{Glutamatergic Neuron (not well defined)}&Cerebral cortex& Cerebral cortex; Layer 6B (Amygdala)\\\hline
54&GABAergic Interneurons, VIP+&Cerebral cortex& Prelimbic area and Infralimbic area\\\hline
55&GABAergic Interneurons, VIP+&Cerebral cortex& Primary somatosensory area\\\hline
56&GABAergic Interneurons, SST+&Cerebral cortex&Prelimbic area and Infralimbic area \\\hline
57&GABAergic Interneurons, SST+&Hippocampal region& Ammon's Horn\\\hline
58&GABAergic Interneurons, PV+&Cerebral cortex& Prelimbic area and Infralimbic area\\\hline
59&GABAergic Interneurons, PV+&Thalamus& {\small{Dorsal part of the lateral geniculate complex}}\\\hline
60&GABAergic Interneurons, PV+, P7&Cerebral cortex& Primary somatosensory area \\\hline
61&GABAergic Interneurons, PV+, P10&Cerebral cortex& Primary somatosensory area\\\hline
62&\footnotesize{GABAergic Interneurons, PV+, P13-P15}&Cerebral cortex&Primary somatosensory area \\\hline
63&GABAergic Interneurons, PV+, P25&Cerebral cortex& Primary somatosensory area\\\hline
64&GABAergic Interneurons, PV+&Cerebral cortex& Primary motor area\\\hline
\end{tabular}
\caption{Anatomical origin of the cell-type-specific samples (II).}
\label{metadataAnatomyTable2}
\end{table}

\end{document}